\documentclass[aps,prd,twocolumn,nofootinbib]{revtex4-1}

\usepackage[utf8]{inputenc}
\usepackage{soul,xcolor}

\usepackage{graphicx}
\usepackage{epsfig}
\usepackage{amssymb}
\usepackage{color}
\usepackage{epstopdf}
\usepackage{dcolumn}
\usepackage{relsize}
\usepackage{amsmath}
\usepackage{bm}
\usepackage{dcolumn}
\usepackage{txfonts}
\usepackage{physics}
\usepackage{slashed}
\usepackage{float}
\usepackage[mathscr]{euscript}
\usepackage{lipsum}
\usepackage{microtype}

\def\V{\overline{V}}
\def\H{\overline{H}}

\usepackage{graphicx,graphics,epsfig}
\usepackage{amsmath,amssymb,slashed,dcolumn,amsfonts,slashed,
}
\usepackage{comment}
\usepackage{multirow}
\usepackage{mathptmx}      

\def\Xint#1{\mathchoice
   {\XXint\displaystyle\textstyle{#1}}%
   {\XXint\textstyle\scriptstyle{#1}}%
   {\XXint\scriptstyle\scriptscriptstyle{#1}}%
   {\XXint\scriptscriptstyle\scriptscriptstyle{#1}}%
   \!\int}
   
\def\XXint#1#2#3{{\setbox0=\hbox{$#1{#2#3}{\int}$}
     \vcenter{\hbox{$#2#3$}}\kern-.5\wd0}}

\def\dashint{\Xint-}

\begin{document}

\title{Isospectral scattering for relativistic equivalent Hamiltonians on a
  coarse momentum grid} \author{Mar{\'\i}a G\'omez-Rocha
}\email{mgomezrocha@ugr.es}\affiliation{Departamento de F\'{\i}sica
  At\'omica, Molecular y Nuclear and Instituto Carlos I de F{\'\i}sica
  Te\'orica y Computacional \\ Universidad de Granada, E-18071
  Granada, Spain.}

\author{Enrique Ruiz Arriola}\email{earriola@ugr.es}
\affiliation{Departamento de F\'{\i}sica At\'omica, Molecular y
  Nuclear and Instituto Carlos I de F{\'\i}sica Te\'orica y
  Computacional \\ Universidad de Granada, E-18071 Granada, Spain.}

\date{\today}

\begin{abstract} 
  \rule{0ex}{3ex} \vskip1cm The scattering phase-shifts are invariant
  under unitary transformations of the Hamiltonian. However, the
  numerical solution of the scattering problem that requires to
  discretize the continuum violates this phase-shift invariance among
  unitarily equivalent Hamiltonians.  We extend a newly found
  prescription for the calculation of phase shifts which relies only
  on the eigenvalues of a relativistic Hamiltonian and its
  corresponding Chebyshev angle shift.  We illustrate this procedure
  numerically considering $\pi\pi$, $\pi N$ and $NN$ elastic
  interactions which turns out to be competitive even for small number
  of grid points.
\end{abstract}


\maketitle


\section{Introduction}

Hadronic reactions at intermediate energies provide a working and
phenomenological scheme to access to the corresponding dynamical
interactions from scattering experiments and their
corresponding partial wave analysis in terms of phase-shifts. Even in
the simplest cases the conditions of relativity and unitarity are
mandatory requirements, while the description of bound states and
resonances requires a non-perturbative approach. Within a Lagrangian
and covariant field theoretical setup all these demands are best
encapsulated within the Bethe-Salpeter equation
(BSE)~\cite{Salpeter:1951sz} (see Ref.~\cite{Nakanishi:1969ph} for an
early review), where the interaction is defined by a two-particle
irreducible four-point function.  In practice, this object needs to be
truncated, depends on the renormalization scale and is itself
off-shell ambiguous as there is a reparameterization freedom in the
definition of the fields~\cite{Chisholm:1961tha,Kamefuchi:1961sb} (see
e.g. ~\cite{Nieves:1999bx} for an explicit discussion of low energy
interactions). The BSE is an integral 4D equation and hence presents
not only many practical but also challenging conceptual mathematical
challenges because scattering is naturally formulated in Minkowsky
space and truncated exchange interactions display an intricate
singularity structure~\cite{Levine:1967zza} so that a full solution
has only been found
recently~\cite{Carbonell:2013kwa,Carbonell:2014dwa}.

Due to all these complications the conventional approach to the two body
relativistic problem has been the study of judicious 3D reductions of
the BSE closer in spirit to the Lippmann-Schwinger
equation~\cite{Lippmann:1950zz} in the non relativistic case (see e.g.
\cite{Landau:1990qp,Gross:1993zj} for elementary discussions), but
preserving the unitarity character of the scattering amplitude. This
viewpoint leads to Quasioptical or Quasipotential models proposed long
ago~\cite{Logunov:1963yc}.  Among the many different proposals and
variants based on this idea it is worth mentioning the
Blankenbecler-Sugar equation~\cite{Blankenbecler:1965gx},
the Kadyshevsky equation~\cite{Kadyshevsky:1967rs} and the Gross
spectator equation~\cite{Gross:1969rv,Gross:1982nz}. While any of these
schemes has their advantages and disadvantages, our main results and
formulas, however, can be easily extended to these other schemes with minor
modifications.

In this paper we will choose for definiteness the Kadyshevsky
equation~\cite{Kadyshevsky:1967rs} which befits a Hamiltonian
formulation in quantum field theory. The usefulness of the Hamiltonian
approach, besides providing a compelling physical picture, relies on
the explicit use of a Hilbert space and becomes more evident when
dealing with the few-body problem, where one expects to determine
binding energies of multihadron systems in terms of their mutual
interactions. Unfortunately, only in few cases, such as e.g. separable
potentials, can one provide an analytical or semi-analytical solution
of the relativistic two-body scattering problem and in this case one
employs a numerical inversion method which implies a discretization
procedure on a given momentum grid~\cite{Haftel:1970zz}. From a
physical point of view, the introduction of a momentum grid
corresponds to add an external interaction or to introduce a
restriction on the Hilbert which constraints the energy levels of the
system. A well known example corresponds to impose boundary conditions
at a spherical box with finite volume and radius $R$ which provides an
equidistant momentum grid for large box sizes, $p_n \sim \pi
n/R$~\cite{Fukuda:1956zz} or equidistant
energies~\cite{DeWitt:1956be}. Another example which will be relevant
in this paper corresponds to diagonalizing in a Laguerre
basis~\cite{Heller:1974zz} which yields a Gauss-Chebyshev momentum
grid (See \cite{Deloff:2006hd} for a comprehensive and self-contained
exposition on Chebyshev methods.). This so-called
$L^2$-methods~\cite{Reinhardt:1972zz} have clear computational
advantages, but quite generally, basic properties of scattering such
as the the intertwining property of the Moller wave operators does not
hold~\cite{muga1989stationary} and is only recovered in the continuum
limit.

One important aspect within the Hamiltonian approach and relevant to
the present study is the notion of equivalent
potentials~\cite{Ekstein:1960xkd,Monahan:1971zc}, i.e. the fact that
unitarily equivalent Hamiltonians produce identical phase-shifts,
hence they are referred to as phase-equivalent potentials. Because the
eigenvalues of the Hamiltonian $H$ are invariant under $H \to U H
U^\dagger$ with $U U^\dagger = U^\dagger U=1$ we will also talk about
isospectral phase-shifts, namely those that fulfill
\begin{eqnarray}
  \delta_{l,H} (p) = \delta_{l,UHU^\dagger} (p)
\label{eq:unit-eq}  
\end{eqnarray}
where $l$ is the angular momentum, $p$ the CM momentum, $H$ is the
Hamiltonian and $U$ an arbitrary unitary transformation. On a broader
context, this is the counterpart of the Lagrangian field
reparameterization of the
BSE~\cite{Chisholm:1961tha,Kamefuchi:1961sb,Nieves:1999bx}. A
characterization for equivalent relativistic Hamiltonians has been
proposed in Ref.~\cite{Polyzou:2010eq}. It is perhaps not so
well-known that the numerical methods employed to invert the
scattering matrix equation generally violate this unitary equivalence,
namely a unitary transformation of the Hamiltonian on the grid does
not yield the {\it same} phase-shift, see Eq.~(\ref{eq:unit-eq}). The
effect disappears when the grid is sufficiently fine or equivalently
when the number of grid points becomes large.  This violation has been
illustrated explicitly in the non-relativistic
case~\cite{Arriola:2014nia,Arriola:2016fkr} and will also be shown to occur in the present work.

The question is that while one expects that with a fine grid the
continuum limit will eventually and effectively be recovered and hence
the isospectral invariance of the phase-shifts, spectral methods based
on the eigenvalues provide themselves a natural and invariant
definition of the phase-shift. These methods based on the Fredholm
determinant originally proposed by DeWitt~\cite{DeWitt:1956be} ( see
also \cite{Fukuda:1956zz}) and improved by
others~\cite{Reinhardt:1972zz} (see e.g. \cite{alhaidari2008j} for a
review and references therein).  However, while these methods are by
construction isospectral for any number of grid points they are not
necessarily accurate. In a recent letter~\cite{Gomez-Rocha:2019xum} we
have provided a method which is both isospectral and accurate for a
coarse grids in the non-relativistic case. In this paper we analyze
the consequences of such a method for the relativistic situation and
illustrate it with several low energy, $S,P$ and $D$ phases for
$\pi\pi$, $\pi N$ and $NN$.


The present paper is organized as follows. We will review this issue
and will use for definiteness the Kadyshevsky equation in
Section~\ref{sec:kad} and we review some of its properties including a
proof of isospectrality.  The solution of the scattering equations
requires a momentum grid which may be implemented with the
Gauss-Chebyshev quadrature in three different ways none of them
complying with the isospectrality requirement~\ref{sec:grid}.  In
section~\ref{sec:shifts} we analyze three isospectral definitions of
the scattering phase shifts based on the energy-shift, the momentum
shift and the Chebyshev angle shift which specifically depend on the
mass of the particles.  In Section~\ref{sec:num} we present our
numerical results for some separable $\pi\pi$, $\pi N$ and $NN$ model
interactions. Finally, in Section~\ref{sec:concl} we come to the
conclusions and provide some outlook for future work.

\section{Relativistic scattering: The Kadyshevsky approach}
\label{sec:kad}

\subsection{Generalities}

In this section we review some relevant quantities for completeness
and in order to fix our notation and conventions. Elementary
discussions may be found in
textbooks~\cite{Landau:1990qp,Gross:1993zj}. The Kadyshevsky equation
in the CM frame with $\sqrt{s}$ CM energy and in the equal mass case
reads~\cite{Kadyshevsky:1967rs} ~\footnote{The case of two different
  masses corresponds to replace $E_q^2 \to E_q \omega_q$ and $\sqrt{s}=2 E_q
  \to E_q + \omega_q$ with $E_q= \sqrt{M^2+q^2}$ and $\omega_q =
  \sqrt{m^2 + q^2 }$. We will keep the equal mass case because the
  formulas are much simpler for presentation purposes and will return
  to this situation when analyzing the $\pi N$ case.}
\begin{eqnarray}
T(\vec p', \vec p, \sqrt{s}) =  V(\vec p', \vec p)  + \int \frac{d^3 q}{(2\pi)^3}
\frac{V(\vec p',\vec q)}{4 E_q^2} 
\frac{T(\vec q,\vec p, \sqrt{s})}{\sqrt{s}-2 E_q + i \epsilon} \nonumber \\
\end{eqnarray}
where the potential is symmetric $V(\vec p', \vec p)=V(\vec p, \vec
p')$ and energy {\it independent}. These two conditions are necessary
in order to check unitarity, since 
\begin{eqnarray}
T(\vec p', \vec p, \sqrt{s})-T(\vec p, \vec p', \sqrt{s})^* &=& 
\int \frac{d^3 q}{(2\pi)^3} 2 \pi i \delta (\sqrt{s}-2 E_q ) \nonumber \\
&\times& \frac{T(\vec p',\vec q,\sqrt{s})T(\vec q,\vec p,\sqrt{s})^*
}{4 E_q^2} \ .
\end{eqnarray}

A residual ambiguity of the Kadyshevsky equation has been discussed in
Ref.~\cite{Yaes:1971vw} and the 3D-reduction of the BS equation with a
separable kernel has been addressed~\cite{Frohlich:1982yn}. The
3D-reduction of the relativistic three-body Faddeev equation
associated to the this quasipotential was proposed
afterwards~\cite{Vinogradov:1971ne}.  As compared to other
approaches~\cite{polivanov1964spectral}, this particular 3-D reduction
satisfies a Mandelstam representation, i.e. a double dispersion
relation both in the invariant mass $s$ and momentum $t$ Mandelstam
variables~\cite{Skachkov:1970ia}. The appearance of spurious
singularities has been addressed in the different approaches in
Ref.~\cite{Yaes:1973kj}. In addition, the Kadyshevsky equation also
lacks spurious singularities in the related three-body
problem~\cite{Garcilazo:1984rx}.  Actually, there has been already
some work with this equation for the case of $\pi\pi$, $N\pi$ and $NN$
scattering~\cite{Mathelitsch:1986ez} for separable potentials where
the lowest partial waves corresponding to $S$, $P$ and $D$ angular
momenta have been fitted which will be discussed below in more detail.

\subsection{Partial waves}

This 3D scheme has the advantage that besides enabling a relativistic
Hamiltonian interpretation for the scattering problem they also become
amenable to numerical analysis since at the partial waves level they
reduce to 1D linear integral equations. Using rotational
invariance~\footnote{We restrict ourselves to central isotropic
  interactions. The important case of tensor anisotropic potentials
  leading to coupled channels presents some differences and
  complications and will be discussed in a separate publication.}
\begin{eqnarray}
  T(\vec p', \vec p, \sqrt{s}) = 4 \pi \sum_{lm} Y_{lm} (\hat p') Y_{lm}
  (\hat p)^* T_l (p',p, \sqrt{s}) \ .
\end{eqnarray}
At the partial waves level and for spin zero equal mass particles we get
\begin{eqnarray}
  T_l (p',p,\sqrt{s}) &=&  V_l(p',p)  \nonumber \\
  &+& \int_0^\infty dq \, \frac{q^2}{4 E_q^2} 
  \frac{V_l(p',q)T_l (q,p, \sqrt{s})}{\sqrt{s}-2 E_q + i \epsilon}  \ , 
  \label{Eq:Kad} 
\end{eqnarray}
where $+i \epsilon$ implements the original Feynman boundary condition
of the BSE and corresponds to outgoing spherical waves, $E_q =
\sqrt{q^2+m_\pi^2}$ and on the mass shell one has $\sqrt{s} = 2
\sqrt{p^2+m_\pi^2}$ with $p$ the CM momentum. For a real potential
this equation satisfies the two-body unitarity condition, so that the
phase-shift is given by
\begin{eqnarray}
  T_l^{-1}(p,p,\sqrt{s})= - \frac{\pi p}{8 E_p} \left[\cot
      \delta_l (p) -i \right] \ .
\label{eq:phLSt}    
\end{eqnarray}
Alternatively we may define the reaction matrix $R_l$
\begin{eqnarray}
  T_l^{-1}(p,p,\sqrt{s}) = R_l^{-1}(p,p,\sqrt{s}) + i \frac{\pi p}{8 E_p} \ ,
\end{eqnarray}
so that 
\begin{eqnarray}
  -\tan \delta_l(p) = \frac{\pi}{8} \frac{p}{E_p} R_l (p,p,\sqrt{s} ) 	\ ,
  \label{eq:phLS}
\end{eqnarray}
where the corresponding reaction matrix satisfies the equation 
\begin{eqnarray}
R_l (p',p,\sqrt{s}) &=& V_l(p',p) \nonumber \\ &+& \dashint_0^\infty dq \, \frac{q^2}{4 E_q^2} 
\frac{V_l(p',q) R_l (q,p, \sqrt{s})}{\sqrt{s}-2 E_q }  \ ,
\label{eq:Kad-PV}
\end{eqnarray}
where the principal value has been introduced in the integral. As it is
well known we can implement the principal value by means of a subtraction
using the trivial identity
\begin{eqnarray}
\dashint_{0}^\infty \frac{
  2k_0 dp}{p^2-k_0^2} = \dashint_{-\infty}^\infty \frac{dp}{p-k_0}= 0  \ ,
\end{eqnarray}
whence follows the integration rule
\begin{eqnarray}
\dashint_{0}^\infty dp \frac{ f(p) 
}{2 E_0- 2 E_k} = \int_{0}^\infty dp \left[ \frac{f(p) 
  }{2 E_0- 2 E_p} -  \frac{f(k_0) E_0}{k_0^2-p^2} \right] \ ,
\end{eqnarray}
where $\sqrt{s}= 2 E_0 = \sqrt{k_0^2+m^2}$. Using this we get 
\begin{eqnarray}
&& R_l (p',p,\sqrt{s}) = V_l(p',p) + \int_0^\infty dq  \nonumber \\ && \, \left[ \frac{q^2}{4 E_q^2} 
  \frac{V_l(p',q) R_l (q,p, 2 E_0)}{2 E_0-2 E_q }-
\frac{k_0^2}{4 E_0} 
  \frac{V_l(p',k_0) R_l (k_0,p, 2 E_0)}{k_0^2-q^2} \right]
  \ . \nonumber \\ 
\label{eq:Kad-Sub}  
\end{eqnarray}
In the continuum the
Eqs.~(\ref{Eq:Kad}) , (\ref{eq:Kad-PV}) and (\ref{eq:Kad-Sub}) are
fully equivalent, but discretized versions provide different results,
all of them violating the isospectrality of the phase-shifts, as will
be shown in Section~\ref{sec:equiv}.

Note that for our normalization convention in the spherical basis we
have the closure relation
\begin{eqnarray}
  1 = \int_0^\infty dq \frac{q^2}{4 E_q^2} | q \rangle \langle q | \ .
  \label{eq:closure}
\end{eqnarray}
As it is well known,  bound states appear as poles of the scattering matrix.
This allows to define a Hamiltonian in the CM system,
\begin{eqnarray}
  H \Psi_l(p)  \equiv 2 E_p \Psi(p) + \int_0^\infty dq \frac{q^2}{4 E_q^2} V_l(p,q)
  \Psi_l (q)  \ ,
\end{eqnarray}
so that the homogeneous Kadyshevsky equation reads
\begin{eqnarray}
  H \Psi_l(p)    = \sqrt{s} \Psi_l(p) \ .
\end{eqnarray}
While this equation is usually meant to solve for the bound state problem,
we will actually show below how it can also be used to solve the scattering 
problem on a finite momentum grid.

\subsection{Scattering equivalence}
\label{sec:equiv}

One of the most remarkable features of quantum scattering is the lack
of uniqueness of the interaction; under unitary transformations of the
Hamiltonian the S-matrix, or equivalently the phase-shifts remain
invariant. In this section we remind of this fact by considering the
continuum limit first. We will then see that its discretized
counterpart through a finite momentum grid {\it does not} preserve
this symmetry if the corresponding phase-shifts are defined as in
Eq.~(\ref{eq:phLSgrid}).

In operator form $V(\vec p' , \vec p) \equiv \langle \vec p' | V |
\vec p \rangle$ and $T(\vec p' , \vec p, \sqrt{s}) \equiv \langle \vec
p' | T(\sqrt{s}) | \vec p \rangle$ and the Kadyshevsky equation
written as a Lippmann-Schwinger reads
\begin{eqnarray}
T &=& V + V G_0 T \\ &=&  V + V G V \\ &=& V \left(1-G_0 V \right)^{-1} =
\left(1-V G_0 \right)^{-1} V
\label{wq:KEop}
\end{eqnarray}
which we write alternatively in equivalent forms and have defined
$G^{-1}= \sqrt{s}+ i \epsilon -H = G_0^{-1}-V$. Within this
Hamiltonian framework, in the continuum, we consider a unitary
transformation $U$ of the Hamiltonian $H$, given by $H \to \tilde H =
U H U^\dagger \equiv H_0 + \tilde V $ where $\tilde V = U H U^\dagger
-H_0$. Taking the exponential representation of a unitary operator
$U=e^{i \xi}$ with $\xi=\xi^\dagger$ a self-adjoint operator, for an
infinitesimal transformation we have to lowest order $U= 1 + i \xi +
{\cal O} (\xi^2) $ and hence $\Delta V = i [\xi , H]$. If we take the
form $T^{-1}= V^{-1}-G_0$ we have, $\Delta V^{-1}= - V^{-1} \Delta V
V^{-1}$ and similarly for $T$ so that
\begin{eqnarray}
\Delta T &=& T V^{-1} \Delta V V^{-1} T \nonumber \\ &=& \left( 1 -
G_0 V \right)^{-1} \Delta V \left( 1 - V G_0 \right)^{-1} \nonumber
\\ &=& G_0^{-1} G [\xi, H] G G_0^{-1} \nonumber \\ &=& ( 1 + T G_0)
\xi G_0^{-1}- G_0^{-1} \xi (1+G_0 T)
\label{eq:deltaT}
\end{eqnarray}
where we have used the Eqs.~(\ref{wq:KEop}). Thus, taking matrix
elements and because of the external factors $G_0^{-1}$ we get in the
limit $\epsilon \to 0$ at the on shell point $2 E_p=2 E_p'= \sqrt{s}$
the result
\begin{eqnarray}
\Delta T (\vec p', \vec p)\Big|_{2E_p=2E_p'=\sqrt{s}} = 0
\end{eqnarray}
Thus, for a given generator $\xi=\xi^+$ we have that
\begin{eqnarray}
\Delta V= i [\xi,H] \implies \Delta
\delta_l (p)=0 \, .
\end{eqnarray}
or equivalently, for finite transformations $\delta_{l,H} (p)=
\delta_{l,UH U^\dagger} (p)$.

\section{Discretization schemes and scattering inequivalence}
\label{sec:grid}

\subsection{Momentum grid}

There are only few cases where the scattering equations can be solved
analytically.  The momentum grid discretization introduces both an
infrared $\Delta p$ as well as an ultraviolet numerical cut-off,
$\Lambda_{\rm num}$. In our previous work we used a Gauss-Chebyshev
grid~\cite{Gomez-Rocha:2019xum} for interactions which have a fast
fall-off.  However, the kind of hadronic interactions we will be
dealing with here to illustrate our method have long tails in
momentum.  Thus, we consider a Gauss-Chebyshev quadrature which is
re-scaled in such a way that we distinguish two subdivisions within
the $[0,\infty)$ integration range. Namely, half of the grid points
  are arranged within interval $[0, \Lambda_{1/ 2}]$, and the other
  half are distributed along the $[\Lambda_{1/ 2},\infty )$. The
    parameter $\Lambda_{1/ 2}$ is chosen in order to select the region
    of interest. In this way, the long-tails effects are broadly taken
    into account and at the same time the physical region of interest
    is covered with an enough density of points. This allows us to
    study the region of interest in detail, without neglecting
    long-tails effects.  The grid differs then from the
    Gauss-Chebyshev parametrization used in our non-relativistic
    $NN$-scattering study~\cite{Gomez-Rocha:2019xum}, and is given
    by\footnote{One could alternatively use $p_n={2 \over \pi}
      \tan^{-1}z_n$ as it is done by Haftel and
      Tabakin~\cite{Haftel:1970zz}.}
\begin{eqnarray} 
p_n &=& { 1 + z_n \over 1 - z_n } \ , 
\label{eq:pn}
\\
w_n & = & { 2 \Lambda_{ {N \over 2}} \over (1 - z_n)^2} d z_n \ ,
\label{eq:dpn}
\end{eqnarray}
with
\begin{eqnarray}
z_n & = & - \cos \left[{\pi\over N} \left( n - {1\over 2}\right)\right] \ , 
\label{eq:zn} 
\\
dz_n & = & {\pi \over N} \sin \left[{\pi\over N} \left( n - {1\over 2}\right)\right] \ .
\label{eq:dzn}
\end{eqnarray}
where $n=1, \dots, N$. The parameter $\Lambda_{1/2}$ selects the interval $[0, \Lambda_{ 1/2}]$ that contains the first ${N \over 2}$ points. The lowest and highest momenta in the grid are 
\begin{eqnarray}
p_{\rm min} &=& p_1 = {1 -  \cos \left({\pi\over 2 N} \right) \over 1 +  \cos \left({\pi\over 2 N} \right)} \ ,  \\ 
p_{\rm max} &=& p_N =  { 1 - \cos \left[\pi \left( 1 - {1\over 2N}\right)\right] \over 1 + \cos \left[\pi \left( 1 - {1\over 2N}\right)\right] } \ .
\end{eqnarray}
For a large grid and for $n \ll N$ we have $p_n= \Lambda (\pi n/2N)^2/2$
which differs from the spherical box quantization. 
The integration rule becomes  
\begin{eqnarray}
\int_0^\infty dp f(p) \to \sum_{n=1}^N w_n f(p_n) \ .
\end{eqnarray}
On the momentum grid, the Hamiltonian is defined as
\begin{eqnarray}
H \Psi_n \equiv 2 E_n \Psi_n 
  + \sum_k w_k \frac{p_k^2}{4 E_k^2} V_{n,k} \Psi_k 
\end{eqnarray}
where $\Psi_n \equiv \langle p_n | \Psi \rangle = \Psi(p_n)$ and
$V_{nk}=V(p_n,p_k)$ . The closure relation on the grid is given by
\begin{eqnarray}
\sum_n |p_n \rangle \frac{w_n p_n^2}{4 E_n^2} \langle p_n | = {\bf 1} \ . 
\end{eqnarray}
While these factors are ubiquitous, they are a bit annoying because
the hermiticity  does not correspond to invariance under 
interchange of files and rows. Therefore we define the barred basis
\begin{eqnarray}
|p_n \rangle \equiv \frac{\sqrt{w_n}p_n}{2 E_n} | \overline{p_n} \rangle 
\end{eqnarray}
so that the barred Hamiltonian reads 
\begin{eqnarray}
  \overline{H}_{nk} &=&  (\frac{\sqrt{w_n}p_n}{2 E_n})^{-1}  H_{nk}  \frac{\sqrt{w_k}p_k}{2 E_k} \\ &=&  2 E_n \delta_{nk} + \overline{V}_{nk}  
\end{eqnarray}
where the barred potential reads 
\begin{eqnarray}
  \overline{V}_{nk} &=&  \frac{\sqrt{w_n}p_n}{2 E_n}  V_{nk}  \frac{\sqrt{w_k}p_k}{2 E_k} 
\end{eqnarray}
which are obviously Hermitean, $\overline{H}_{nk} = \overline{H}_{kn}$
and $\overline{V}_{nk} = \overline{V}_{kn}$. Within this so that an
infinitesimal unitary transformation generates a change $\Delta V= i
[\xi,H]$ on the grid, which in the partial waves barred basis reads
\begin{eqnarray}
  \Delta \overline{V}_{nk} = - \Delta \overline{V}_{kn}= i \sum_{l=1}^N \left[ \xi_{kl} \overline{H}_{ln} -\overline{H}_{kl} \xi_{ln} \right]
\end{eqnarray}
where we have dropped the angular momentum $l$ for simplicity.  We can
then proceed to discuss the discretization of Eqs.~(\ref{Eq:Kad}) ,
(\ref{eq:Kad-PV}) and (\ref{eq:Kad-Sub}) which basically fall into two
categories: schemes where just the grid points are needed and schemes
where additional observation points are added.

It is worth noticing that unlike standard solution methods, where the
energy, $\sqrt{s}$, and momentum, $p$, grids are independent from each
other (see e.g.~\cite{Landau:1990qp}), here we will address versions
of the scattering equation which invoke {\it only} momentum grid
points. However, as it was shown in
\cite{Arriola:2014aia,Arriola:2016fkr} for the non-relativistic case,
this definition of the phase-shift is not invariant under unitary
transformations on the finite momentum grid. The phase inequivalence
goes away in the continuum limit $\Delta p \to 0$ corresponding to $N
\to \infty$.  It must also be said that the numerical problem can be
also formulated following the Haftel-Tabakin
procedure~\cite{Haftel:1970zz}, which provides a value of the reaction
matrix at any point outside the momentum grid (the so-called
observation point). However, in order to consider a family of
scattering-equivalent Hamiltonians, which are known in a given
momentum grid, the calculation of matrix elements at points outside
the grid, would require some extrapolation.

\subsection{Scattering amplitude on the grid}

In order to illustrate the lack of isospectrality in the finite
momentum grid, let us consider the discretized version of the equation
Eq.~(\ref{Eq:Kad}) with a finite $\epsilon$ and an arbitrary energy
$e=\sqrt{s}= 2 \sqrt{p^2+m^2}$. This corresponds to take matrix
elements of the operator form, so that
\begin{eqnarray}
  T_{nm}(\sqrt{s}) = V_{nm} + \sum_{k=1}^N \frac{w_k p_k^2}{4 E_k^2}\frac{V_{nk} T_{km}(\sqrt{s})}{\sqrt{s}-2E_k+ i \epsilon}
\label{eq:kad-2}
\end{eqnarray}
which in the barred basis becomes
\begin{eqnarray}
\overline{T}_{nm}(\sqrt{s}) = \overline{V}_{nm} + \sum_{k=1}^N \frac{\overline{V}_{nk} \overline{T}_{km}(\sqrt{s})}{\sqrt{s}-2E_k+ i \epsilon}
\label{eq:kad-2}
\end{eqnarray}
Let us remind that the meaning of this equation is to take the
continuum limit before the limit $\epsilon \to 0$. In practice, this
corresponds to assume $w_n / \epsilon \ll 1$ and a practical
consequence is the strict loss of unitarity since the delta function
on the grid becomes smeared as a Lorentz function. Nonetheless, we may
take the prescription ($K1$)
\begin{eqnarray}
  {\rm Re }[T_l^{-1}(2E_n)]_{nn} = - \frac{\pi p_n}{8 E_n} \cot
  \delta_l^{\rm K1} (p_n)
\end{eqnarray}
which corresponds to the real part of Eq.~(\ref{eq:phLSt})
on the grid. 
In any case, under a unitary finite dimensional transformation the
chain of relations leading to Eq.~(\ref{eq:deltaT}) follow, and thus
in the momentum grid we have (for finite $\epsilon$ and unrestricted
summation)
\begin{eqnarray}
\Delta \overline{T}_{nn} (2 E_n) &=& \sum_{m \neq n} \frac{4 (E_n-E_m)
  \epsilon}{4(E_n-E_m)^2+\epsilon^2} \xi_{nm} \overline{T}_{nm} (2 E_n)
\end{eqnarray}
which is non-vanishing, unless the continuum limit is taken. Although
the solution based on this method is not terribly accurate it serves
the purpose of illustrate our point.  We have also numerically checked
that for particular unitary transformations $U$ inducing the change $V
\to \tilde V \equiv U H U^\dagger - H_0$ the phases from the
Eq.~(\ref{eq:kad-2}) are indeed {\it not} invariant, unless a large
number of grid points is considered.

\begin{figure}[t]
\includegraphics[scale=0.5]{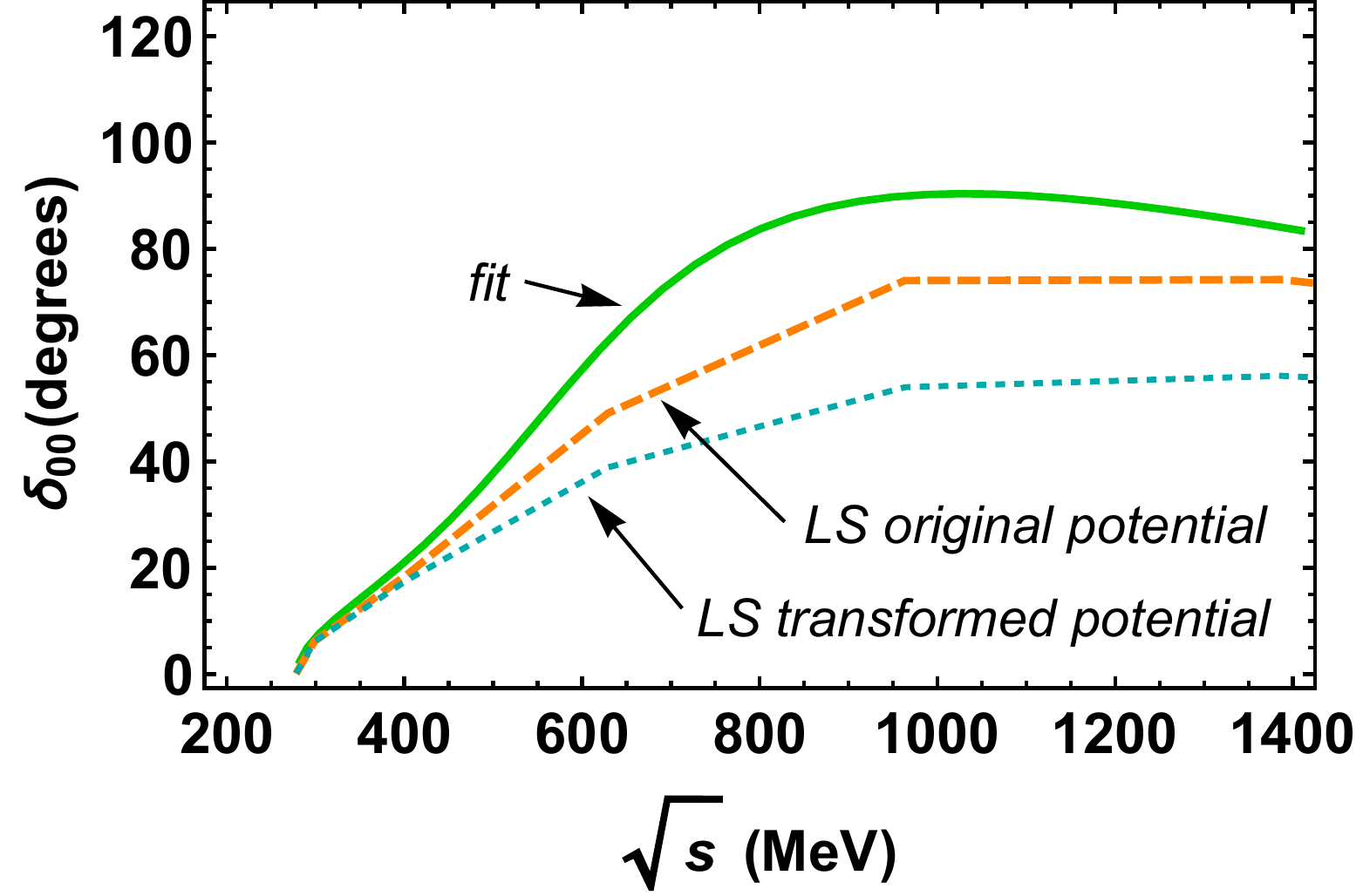}
\caption{Comparison of results obtained using the discretized
  scattering equation for the reaction matrix with $N=25$ points using
  the prescription $K2$ for the 00 $\pi\pi$ phase-shift (see main text) and its
  evolved result after a uniparametric family of unitary operators
  according to $\xi = - i [H_0,V]$.}
\label{Fig:LSevolved}
\end{figure}

\subsection{Reaction matrix on the grid}

The scattering problem for the reaction matrix associated with the
Kadyshevsky equation for the half-off shell reaction matrix on the
grid reads (the limit $\epsilon \to 0$ is already taken) 
\begin{eqnarray}
  R_{nm}  =  V_{nm} + \sum_{k \neq m}^N  V_{nk} w_k \frac{p_k^2}{4 E_k^2} \frac{1}{2 E_m - 2E_k} R_{km} \ ,
\label{eq:kad-hos}
\end{eqnarray}
where $R_{nm} \equiv r(p_n,p_m, 2E_m)$ and the restricted sum,
$\sum_{k \neq m}$, implements in the momentum grid the principal value
prescription. This problem can directly be solved by $N$ matrix
inversions for every single energy $E_n$ in the grid, whence the
phase-shift can be extracted using Eq.~(\ref{eq:phLS}) evaluated on
the grid points (prescription K2),
\begin{eqnarray}
  -\tan \delta_l^{\text{K2}}(p_n) = \frac{\pi p_n}{8 E_n} R_{nn} \ .
  \label{eq:phLSgrid}
\end{eqnarray}
Our arguments, apply equally well to the discretized form of
Eq.(\ref{eq:Kad-PV}) as given by Eq.~\ref{eq:kad-hos} and in
Figure~\ref{Fig:LSevolved} we show for definiteness a particular case
obtained by generating a uniparametric family of unitary operators
according to $\xi = - i [H_0,V]$ so that the infinitesimal change
$\Delta V = [[H_0,V],H] \Delta s$ and we integrate from $s=0$ to $s=10$  fm$^2$~(see
e.g. Ref.~\cite{Gomez-Rocha:2019zkz} and references therein).

\subsection{Scattering on the grid with observation points}

Finally, let us consider the original approach of Haftel and Tabakin
for Eq.~(\ref{eq:Kad-Sub}), where in addition to the grid points,
$p_1, \dots, p_N$, the notion of observation point, say $k_0 \neq
p_n$, is introduced. The algorithm to find the phases is given by
the equation 
\begin{eqnarray}
  R(p,k_0,2E_0) &=& V(p,k_0) + \sum_{k=1}^N  \frac{w_k p_k^2}{4 E_k^2} \frac{V(p,p_k) R(p_k,k_0,2E_0)}{2 E_0 - 2 E_k} \nonumber \\
  &-& \sum_{k=1}^N  \frac{w_k k_0^2}{4 E_0} \frac{V (p,k_0) R(k_0,k_0,2E_0)}{k_0^2-p_k^2} \nonumber \\ 
  &=& \sum_{k=0}^N  V(p,p_k) R (p_k,k_0,2E_0) D_k 
\end{eqnarray}
where $E_k=\sqrt{p_k^2+m^2}$ and $E_0=\sqrt{k_0^2+m^2}$. Taking
$p=p_n$ {\it and } $p=k_0$ one generates $N+1$ equations. To ease the
notation we define $\rho_{n0} = R(p_n,k_0,2 E_0)$ and $\rho_{00} =
R(k_0,k_0,2 E_0)$, so that the equations read
\begin{eqnarray}
  \rho_{n0} &=& V_{n0} + \sum_{k=1}^N D_k V_{nk} \rho_{k0} + D_0 V_{n0} \rho_{n0} \\
  \rho_{00} &=& V_{00} + \sum_{k=1}^N D_k V_{0k} \rho_{k0} + D_0 V_{00} \rho_{00} 
\end{eqnarray}
where 
\begin{eqnarray}
D_k = \left\{ \begin{array}{lr}
        \frac{w_k p_k^2}{4 E_k^2} \frac{1}{2 E_0 - 2 E_k}, & \text{for } 1 \leq k\leq N\\
\frac{w_k k_0^2}{4 E_0} \frac{1}{k_0^2 - p_k^2}  , & \text{for }  k=0\
        \end{array} \right. 
\end{eqnarray}
In the continuum $D_0$ vanishes, but on the finite grid it actually
improves the accuracy. The solution is given by $R(k_0,k_0,2E_0)$, similarly to
Eq.~(\ref{eq:phLS}) (prescription K3), namely
\begin{eqnarray}
  -\tan \delta_l^{\text{K3}}(k_0) = \frac{\pi k_0}{8 E_0} R_l
  (k_0,k_0,2E_0) \ .
  \label{eq:phHTgrid}
\end{eqnarray}
The question if we can check whether the calculated phase-shift, or
$\rho_{00}=R (k_0,k_0, 2E_0)$, at the observation point $k_0$ is
isospectral or not, i.e. under the changes $\Delta V= i [\xi,H]$ on
the grid requires to distinguish two relevant cases, depending on whether
the observation point is included or not in the unitary transformation.

We sketch here a perturbative proof that isospectrality does not hold.
In perturbation theory, and going to the barred basis we get to second
order
\begin{eqnarray}
\overline{\rho}_{00} &=& \overline{V}_{00} + \sum_{k=1}^N
\overline{D}_k \overline{V}_{0k}^2 + \overline{D}_0
\overline{V}_{00}^2 + {\cal O} (V^3)
\end{eqnarray}
so that because in any case $\Delta \overline{V}_{00}=0$ and
\begin{eqnarray}
  \Delta \V_{0k} &=& - \Delta \V_{k0} \nonumber \\ 
  &=& i \sum_{l=1}^N \left( \xi_{0l} \H_{lk}
  -\H_{0l} \xi_{lk} \right) + i (\xi_{00} \H_{0k}-\H_{00} \xi_{0k}) 
\end{eqnarray}
where $\H_{0l}= \V_{0l}$ and $\H_{lk}= 2 E_l \delta_{lk}+\V_{lk}$ and 
we have 
\begin{eqnarray}
  \Delta \overline{\rho}_{00} &=& 2 \sum_{k=1}^N \frac{1}{2 E_0 - 2 E_k}
  \overline{V}_{0k} \Delta \overline{V}_{k0} + {\cal
    O} (V^3)
\end{eqnarray}
which is non-vanishing. Non-perturbatively we may take specific
unitary transformations. While the observation points can be chosen
arbitrarily, generally, we observe that close to the momentum grid
points the phase-shifts are particularly unstable against unitary
transformations either in the space ${\cal H}_N$ or ${\cal H}_{N+1}$.
We have also analyzed the case of an unitary uniparametric family
where infinitesimally $\Delta V=[ [H_0, V],H] \Delta
s$~\cite{Gomez-Rocha:2019zkz} using a grid of $N+1$ observation points
$k_n$ nested into the grid of $N$ momentum points $p_n$, i.e.  $k_0 <
p_1 < k_2 < p_2 < \dots < p_n < k_n$ which generates a $2N+1$
dimensional space which leads to similar results.

\section{Isospectral phase-shifts}
\label{sec:shifts}

The requirement of isospectrality naturally suggests to determine the
phase shifts from the spectrum of the Hamiltonian, a fact noted by
DeWitt~\cite{DeWitt:1956be} and Fukuda and Newton~\cite{Fukuda:1956zz}
long ago based on equidistant energy or momentum grids
respectively. Here we will present three different alternatives based
on the Gauss-Chebyshev grid whose performance will be analyzed in the
next section. On the momentum grid, the eigenvalues equation can be
written as~\footnote{The barred equations lead to identical
  eigenvalues.}
\begin{eqnarray}
H \Psi_n \equiv 2 E_n \Psi_n 
  + \sum_k w_k \frac{p_k^2}{4 E_k^2} V_{n,k} \Psi_k = \sqrt{s} \, \Psi_n  \ ,
\end{eqnarray}
where $\Psi_n \equiv \langle p_n | \Psi \rangle$. Denoting the $N$ eigenvalues and eigenfunctions as
\begin{eqnarray}
\Psi_{n,\alpha} \qquad \sqrt{s_{\alpha}} \equiv 2 E_\alpha \ ,
\end{eqnarray}
we write the energy in the form 
\begin{eqnarray}
E_{\alpha} = \sqrt{P_\alpha^2+m_\pi^2} \ ,
\end{eqnarray}
where $P_\alpha$ is the ``distorted'' momentum by the interaction.
As it was proposed in 
\cite{Fukuda:1956zz}  
and exemplified in \cite{Arriola:2014aia,Arriola:2016fkr} the phase-shift
can be identified as the momentum shift in units of the momentum resolution, namely~\footnote{We are assuming here that there are no bound states. For the bound state case, the formulas have to be modified in order to comply with Levinson's theorem and in \cite{Arriola:2014aia,Arriola:2016fkr}.} 
\begin{eqnarray}
\delta_n (P_n) = - \pi \frac{P_n-p_n}{\Delta p_n} = - \pi \frac{\Delta P_n}{\Delta p_n} \ .
\end{eqnarray} 
This prescription is a consequence of describing the scattering problem in a box and imposing the physical condition of a vanishing wave function in the wall (see \cite{Gomez-Rocha:2019xum} for a reexamination).  It is equivalent to a trapezoidal rule quadrature, and for a Chebyshev grid can be written as
\begin{eqnarray}
\delta_n^{\text{MS}} (P_n) = - \pi \frac{P_n-p_n}{w_n} \ ,
\label{eq:pshift}
\end{eqnarray} 
where the label MS stands for \textit{momentum-shift} formula.

Another prescription is given by DeWitt~\cite{DeWitt:1956be} which
relates the phase shifts with the energy-levels shift produced in the
stationary states of a system bound in a large spherical box, when a
finite-range perturbation is introduced. This is formulated in the following way
\begin{eqnarray}
\delta_n = - \pi \frac{\Delta E_n}{\Delta e} \ ,
\end{eqnarray}
where $\Delta E_n$ is the shift from the unperturbed to the perturbed energy levels and $\Delta e$ is the separation between levels in the unperturbed system. 
In terms of momentum-grid points the \textit{energy-shift} (ES) formula reads
\begin{eqnarray}
\delta_n^{\text{ES}} = - \pi \frac{\sqrt{p_n^2+m^2}}{p_n w_n}
\left( \sqrt{P_n^2+m^2} - \sqrt{p_n^2+m^2} \right) \ .
\label{eq:dewitt}
\end{eqnarray}

Note that in the ultrarelativistic case, i.e. for very small masses
Eqs.~(\ref{eq:phishift}) and (\ref{eq:dewitt}) are equivalent. This
situation holds in the $\pi\pi$ scattering case at intermediate
energies.

Based on DeWitt's argument, we have generalized the formula
Eq.~(\ref{eq:dewitt}) to any momentum grid in the non-relativistic
case~\cite{Gomez-Rocha:2019xum}, even in the case that the energy
levels are not equidistant. As an example, we consider the employed
momentum grid in this work, Eqs.~(\ref{eq:pn})-(\ref{eq:dzn}).  Using
the analogy between the energy levels of scattering states in a box
and the discretization given by a finite grid, and observing that the
equidistance happens in the argument of the cosine function, we have
prescribed~\cite{Gomez-Rocha:2019xum} the following
$\phi$-\textit{shift} formula based on the shift of such an angle, and
write:
\begin{eqnarray}
  \delta_n^{\Phi S} = - \pi \frac{\Phi_n- \phi_n}{d\phi_n} = - \pi \frac{\Delta \phi_n}{d\phi_n} \ .
\label{eq:phishift}  
\end{eqnarray}
where $\phi_n={\pi\over N}\left(n - {1\over 2}\right)$, $d\phi_n={\pi\over N}$, and the ``distorted'' angles $\Phi_n$ are calculated inverting Eqs.~(\ref{eq:pn})-(\ref{eq:dzn}) replacing $p_n$ by $P_n$. 

These three prescriptions, momentum-, energy- and $\phi$-shift, have
been considered in the analysis of $NN$ scattering using a
nonrelativistic toy model~\cite{Gomez-Rocha:2019xum}.  We will show
here again that the $\phi$-shift method prescription is the one that
best reproduces the solution in the continuum. As we will see in our
numerical study, the method gives reliable predictions even for a grid
with a small number of points. The generalization to any momentum grid
amounts to finding the variable that is distributed equidistantly
along the momentum grid.

Note that if we want the value of the phase-shift at $N$ single energy
values the inversion method requires $N$ matrix inversions, whereas in
the spectral shift methods the $N$ phases are obtained {\it at once}
in a single diagonalization.

\begin{figure*}[t]
\includegraphics[scale=0.45]{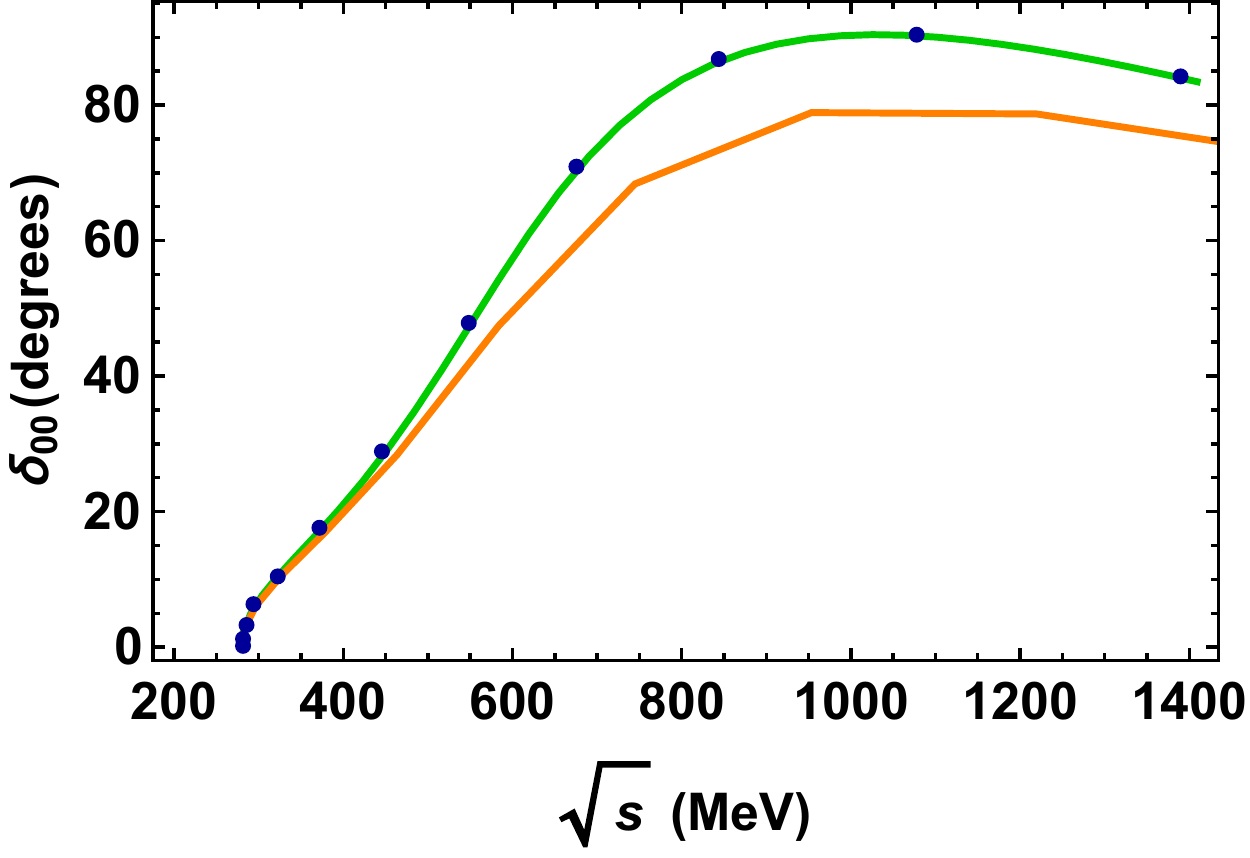}
\includegraphics[scale=0.45]{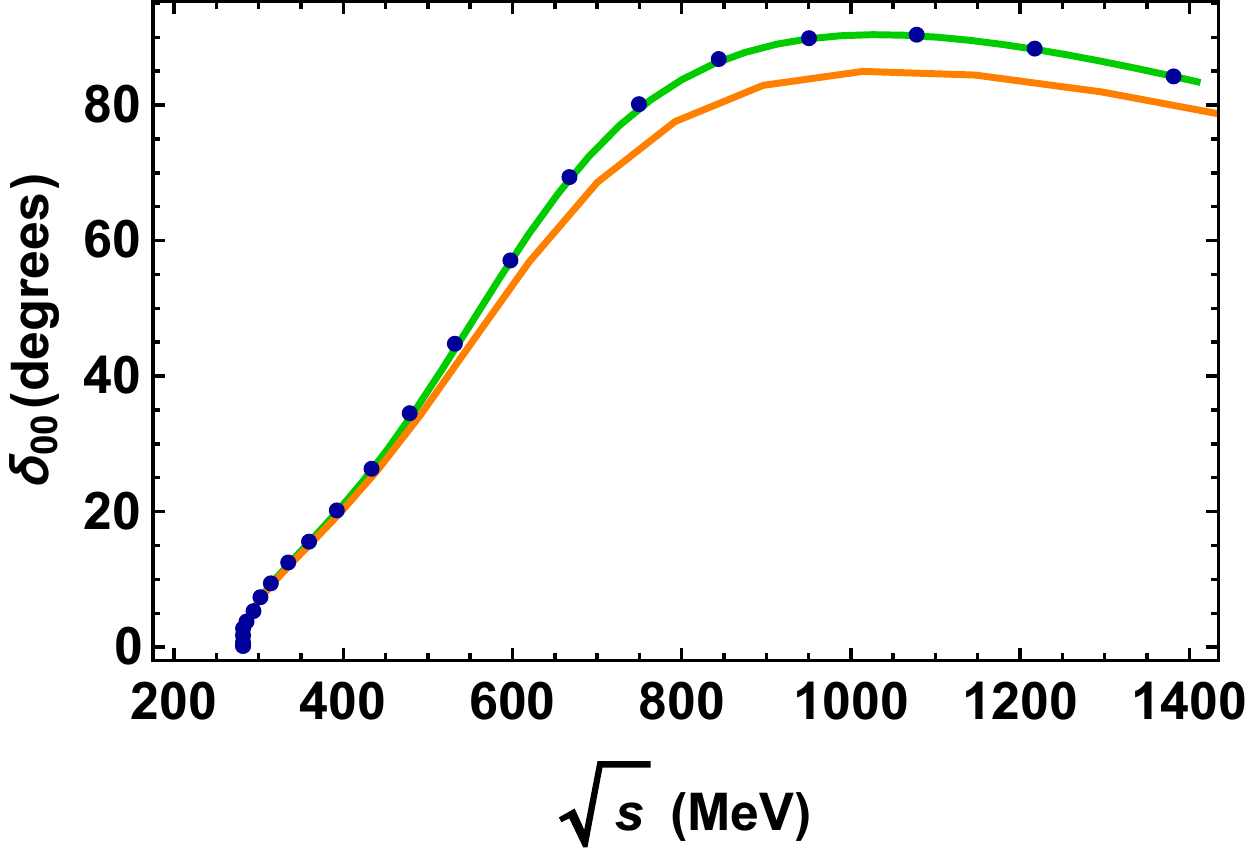}
\includegraphics[scale=0.45]{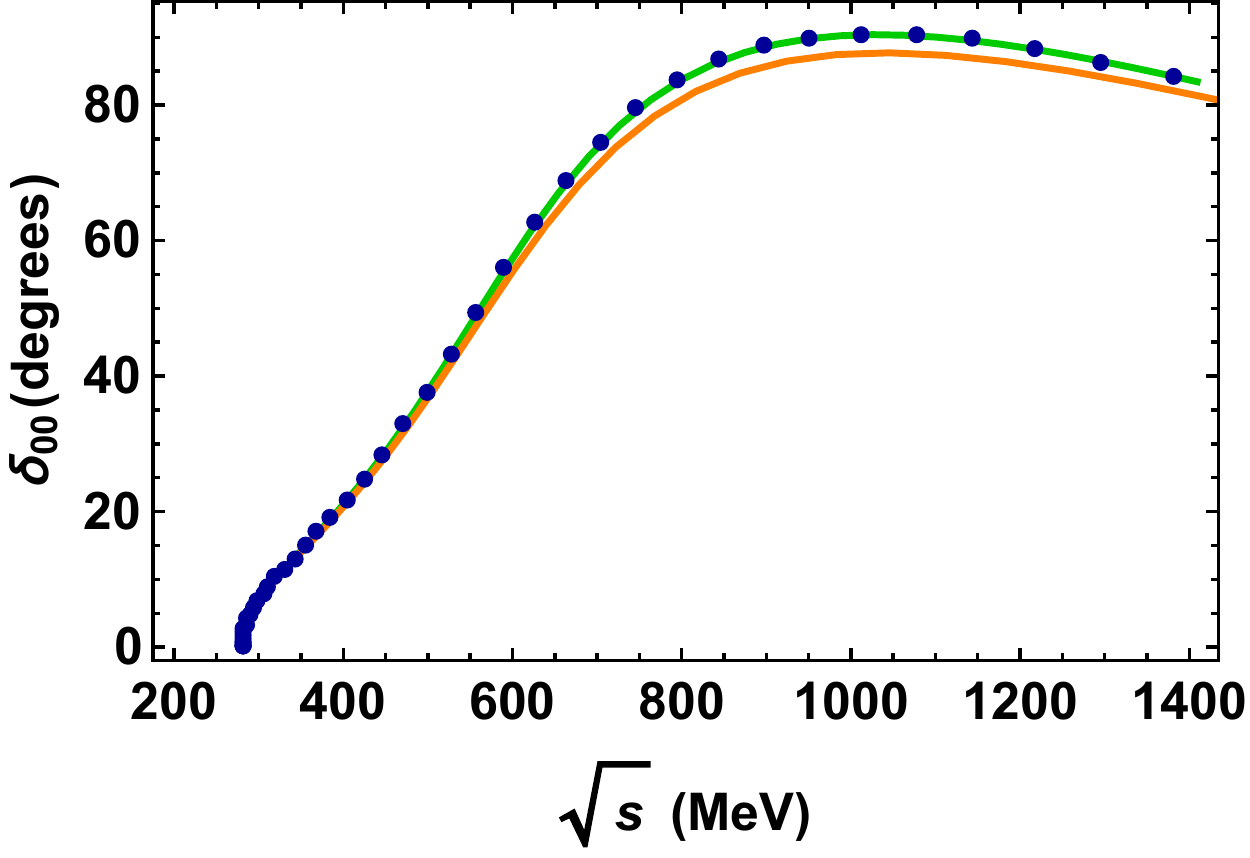}

\includegraphics[scale=0.45]{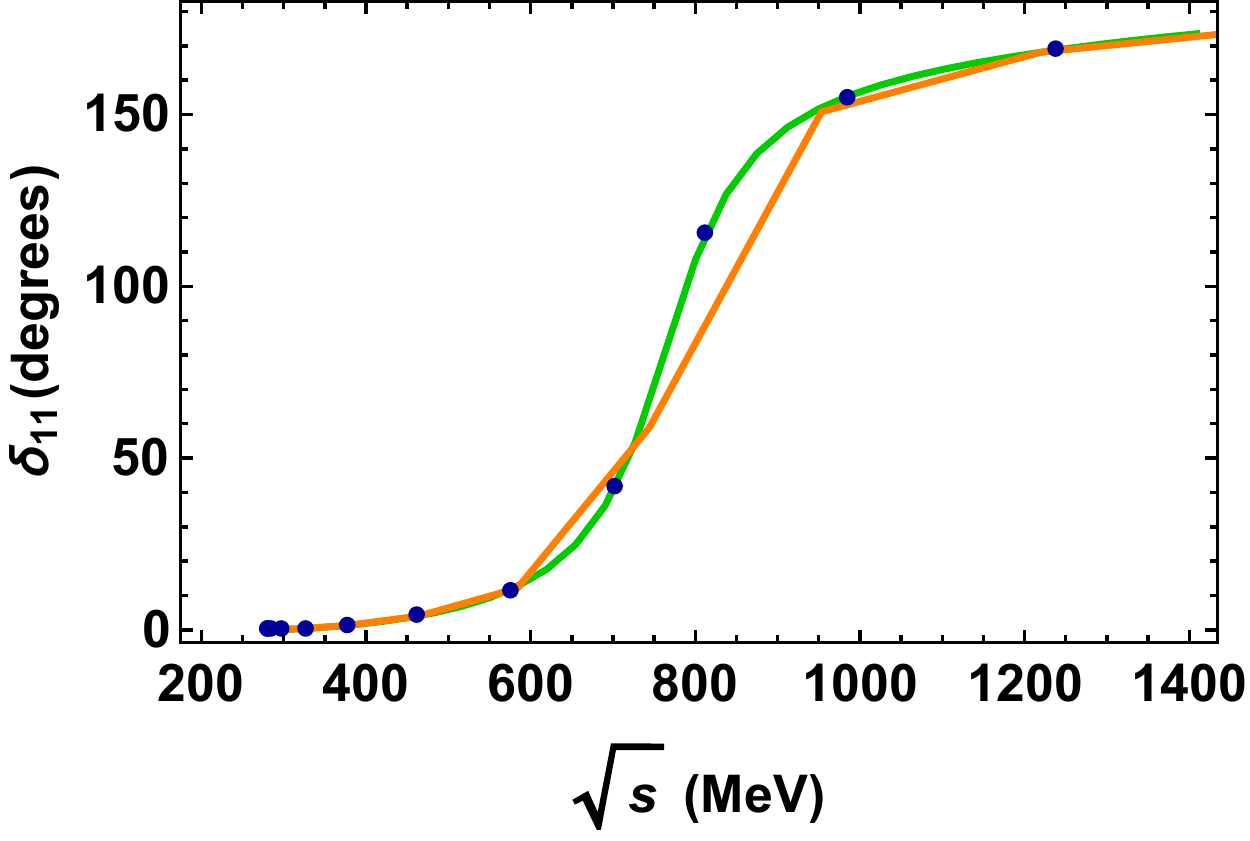}
\includegraphics[scale=0.45]{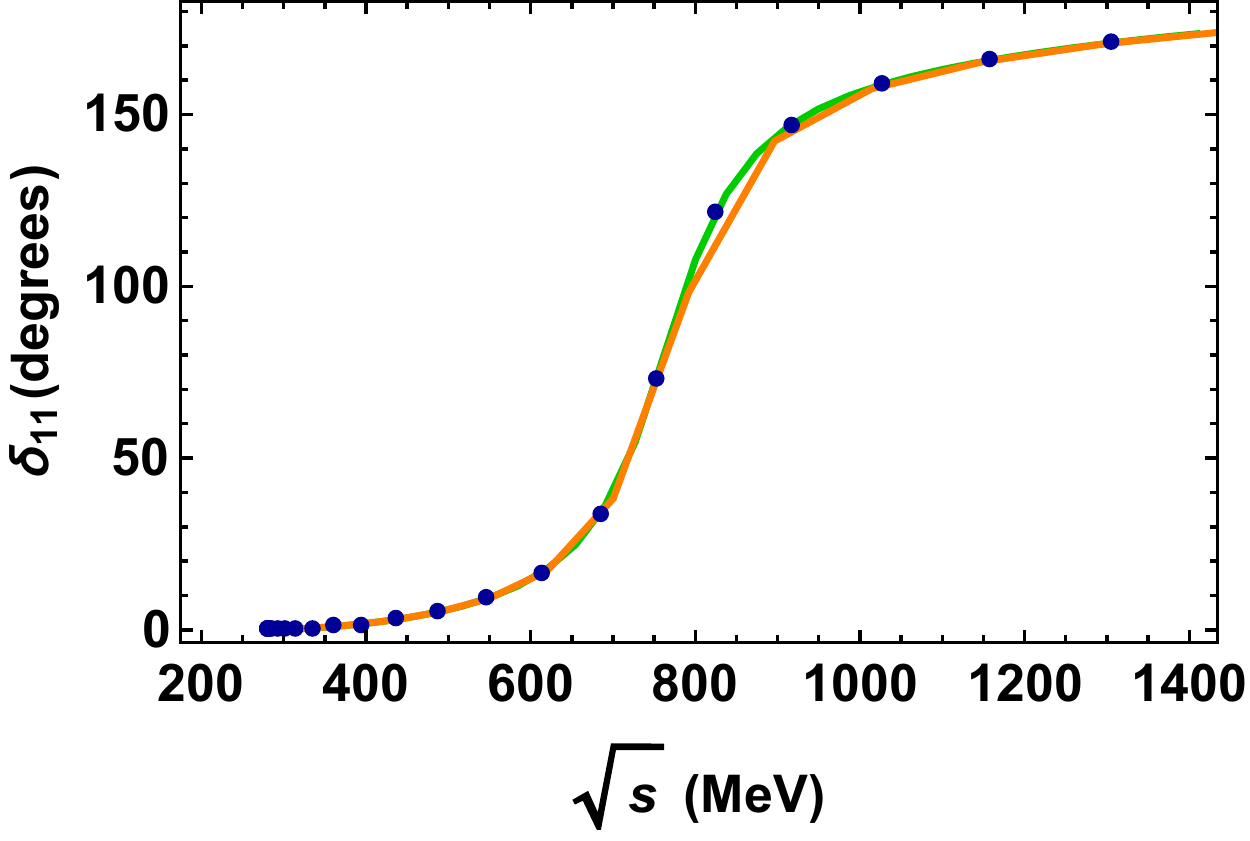}
\includegraphics[scale=0.45]{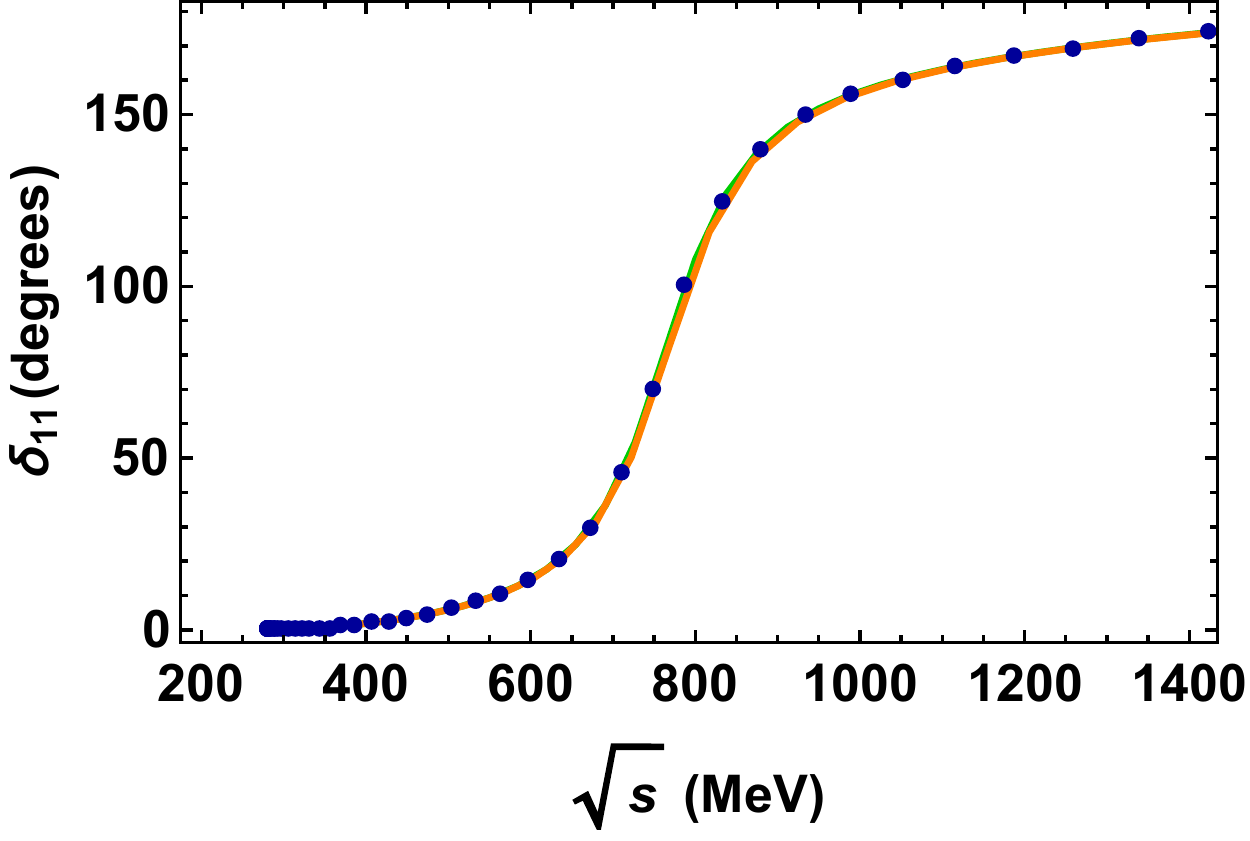}

\includegraphics[scale=0.45]{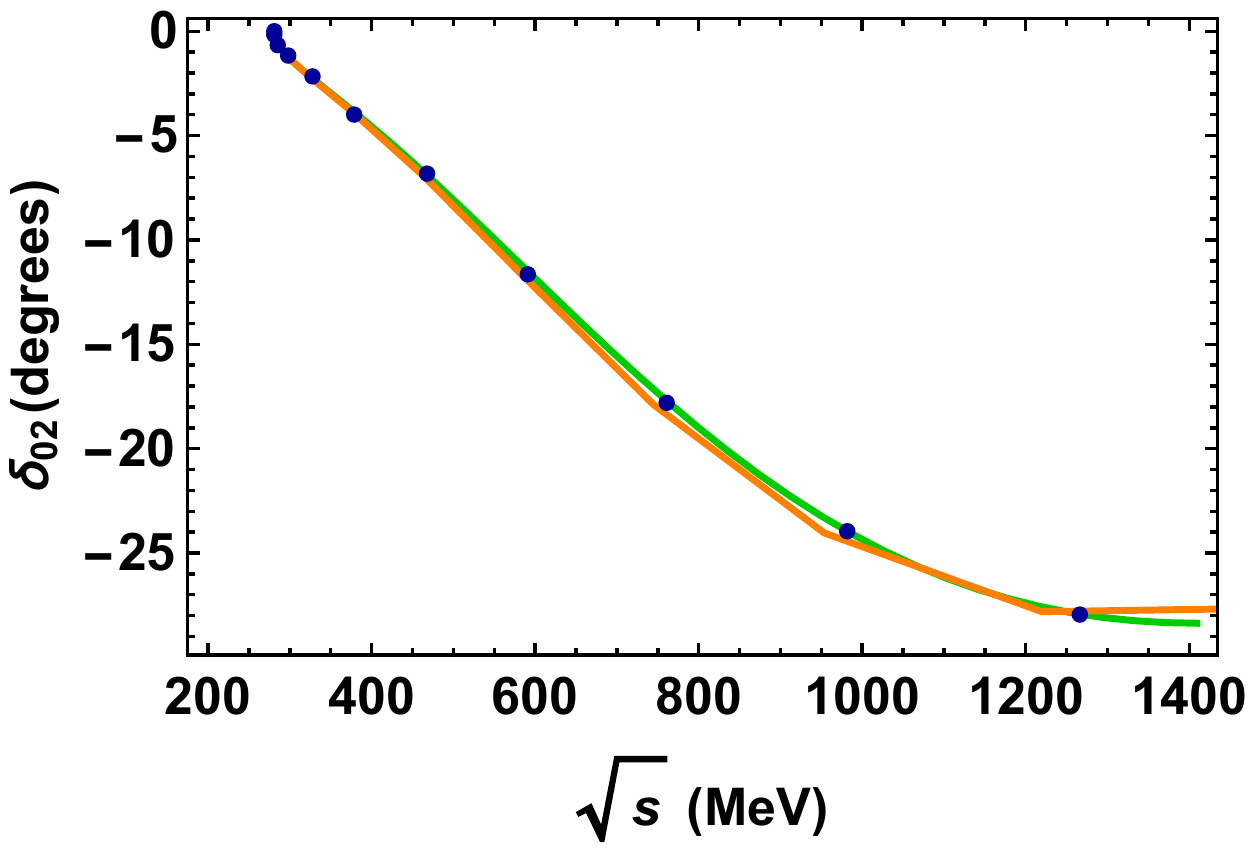}
\includegraphics[scale=0.45]{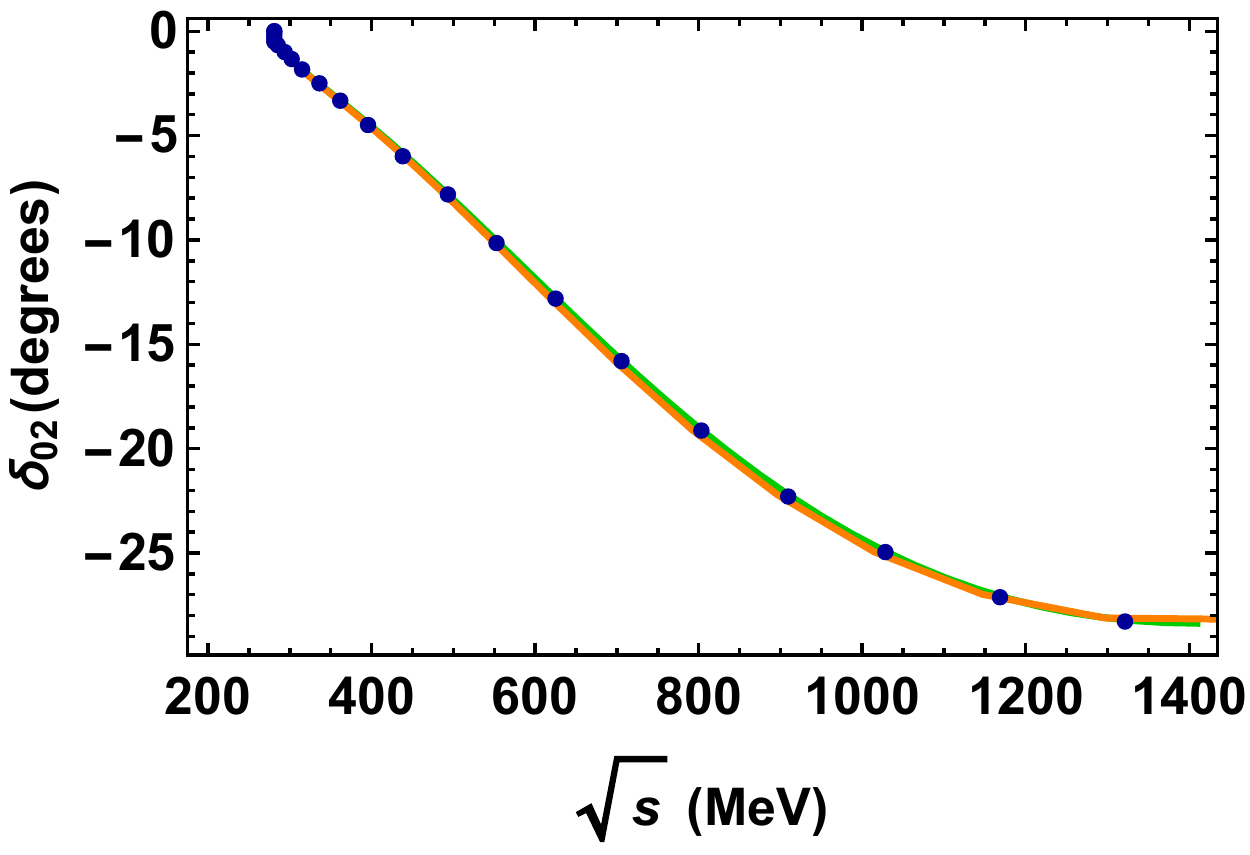}
\includegraphics[scale=0.45]{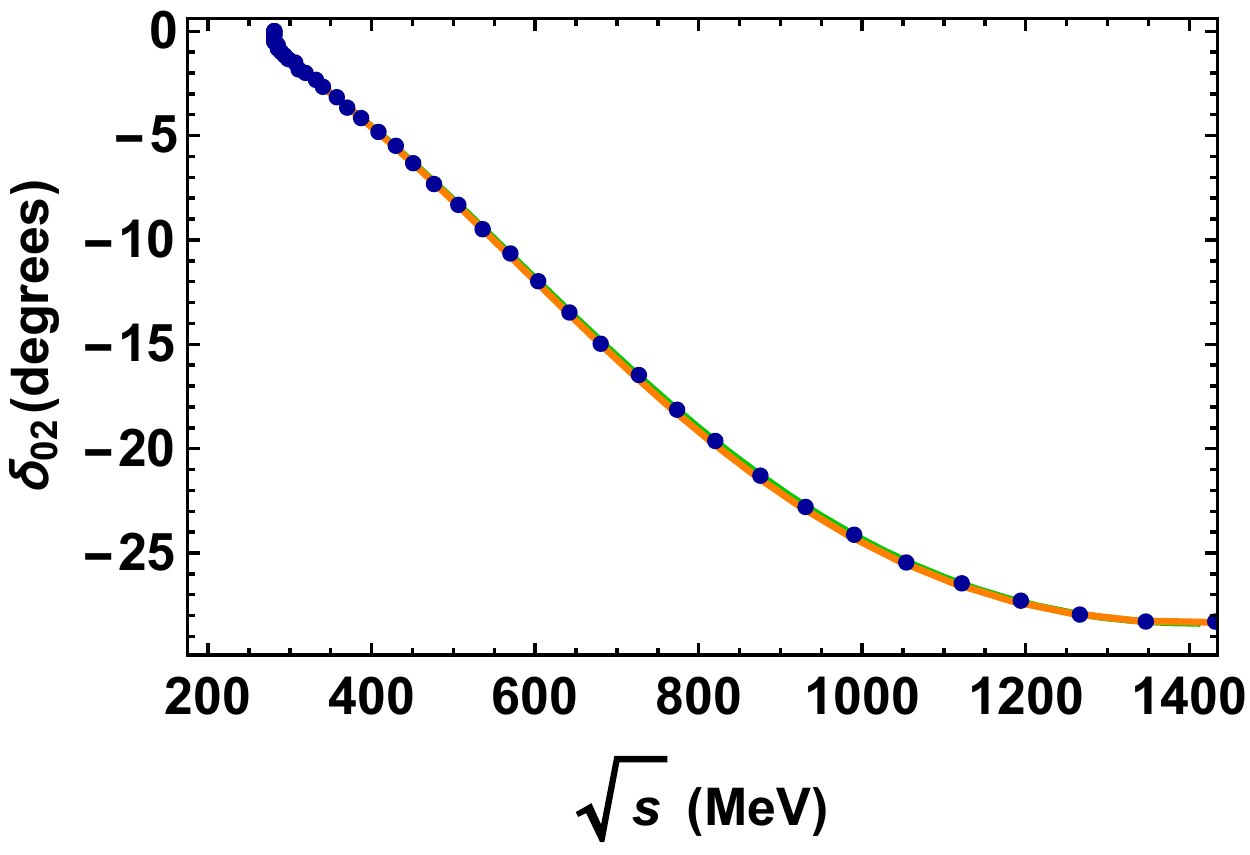}

\includegraphics[scale=0.45]{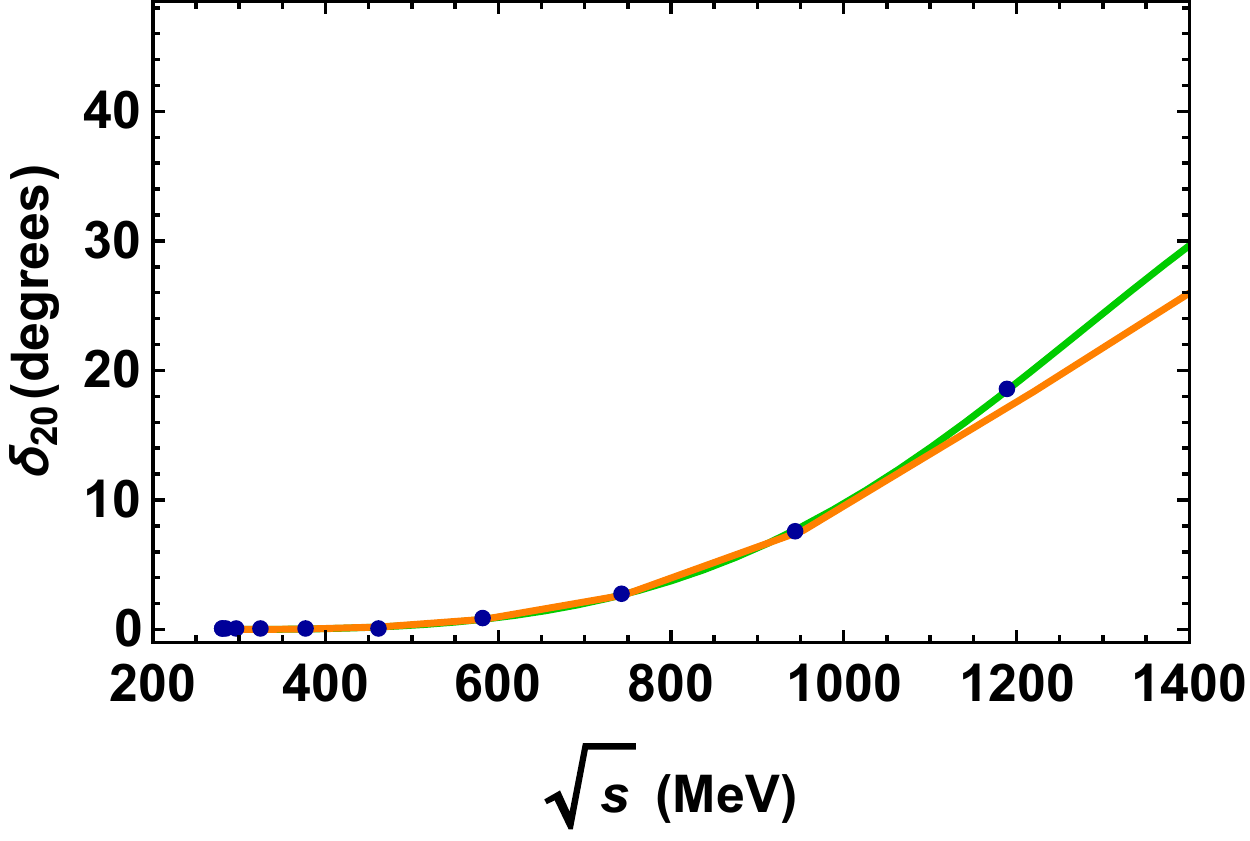}
\includegraphics[scale=0.45]{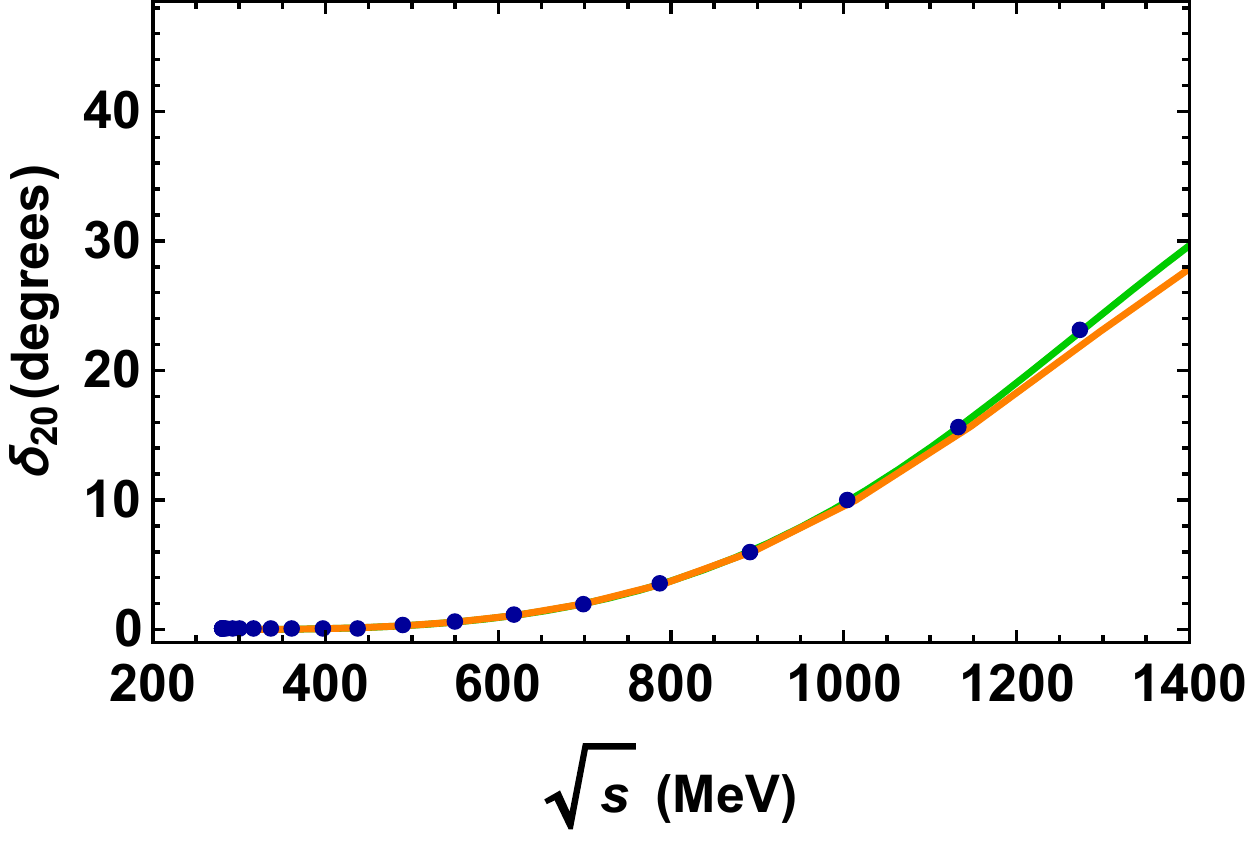}
\includegraphics[scale=0.45]{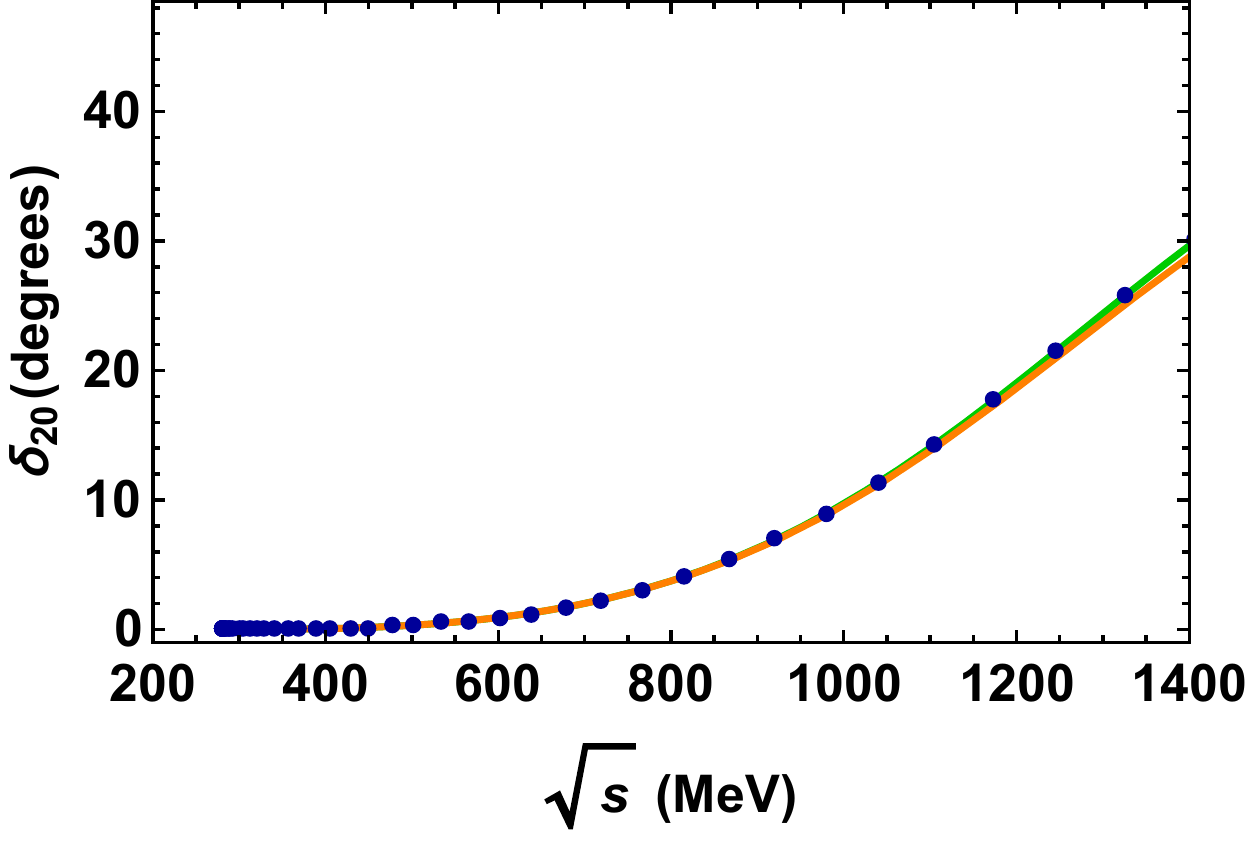}

\includegraphics[scale=0.45]{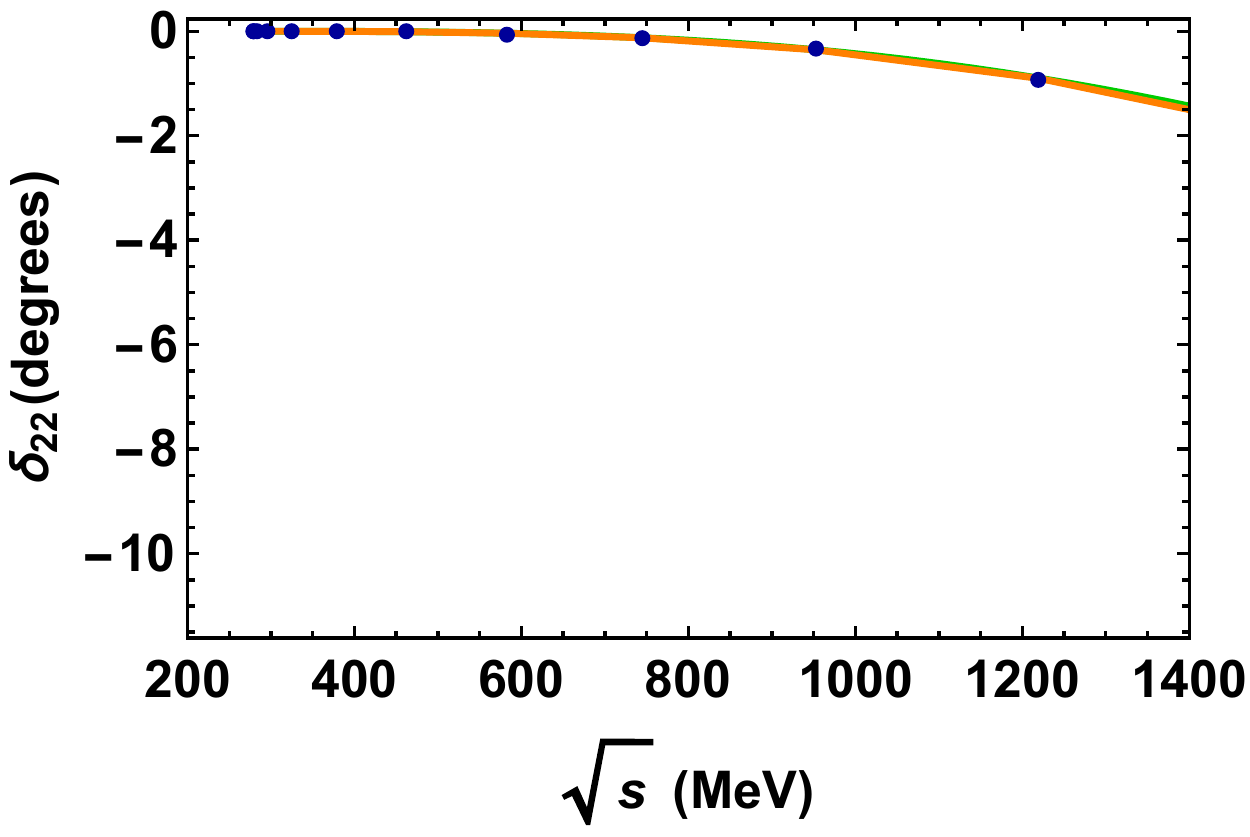}
\includegraphics[scale=0.45]{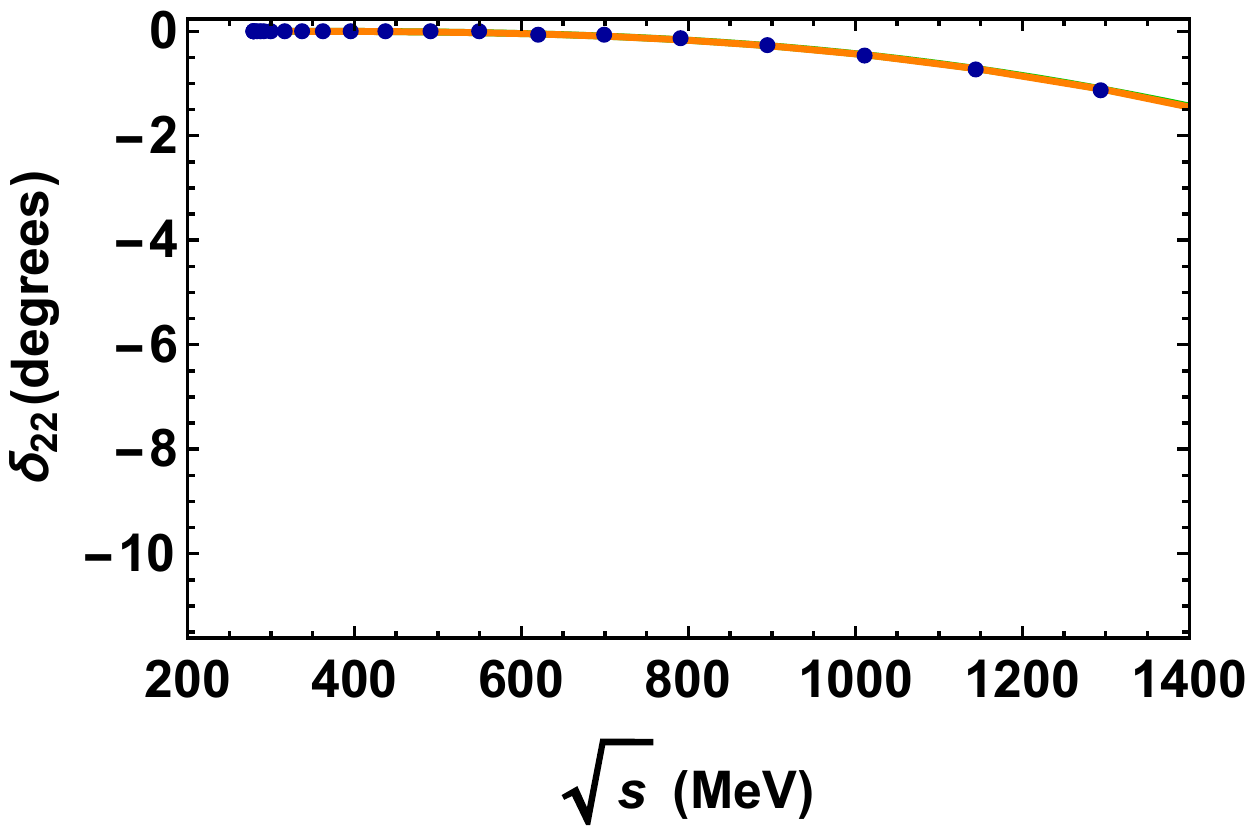}
\includegraphics[scale=0.45]{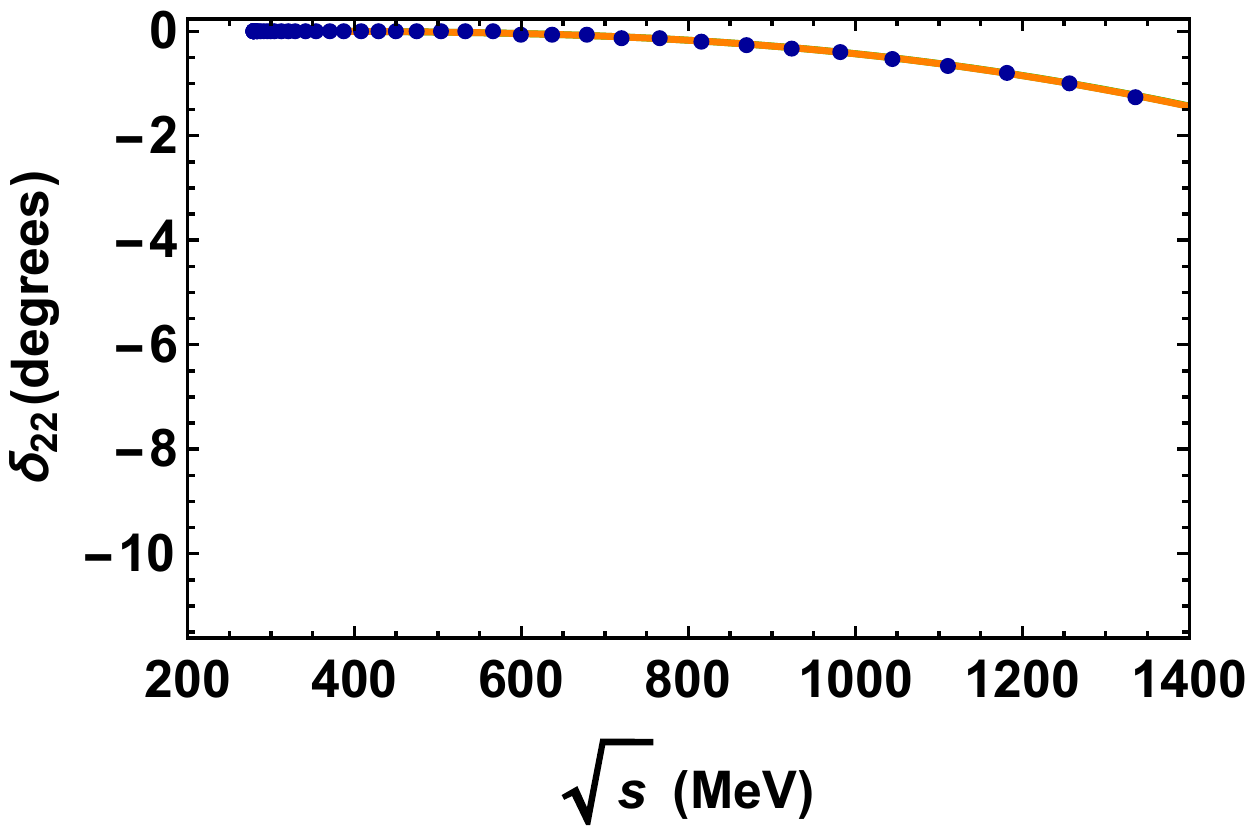}

\caption{(Color online) Phase shifts calculated using our $\phi$-shift prescription (blue dots) compared with the numerical fit (green, smooth line) and with the result obtained from the Lippmann-Schwinger equation (step-wise, orange line). Each column corresponds to the calculation made with a grid of $N=25$, 50, and 100 points, respectively. }
\label{Fig:comparisons}
\end{figure*}

\section{Numerical results}  
\label{sec:num}

The purpose of this numerical analysis is to study the predictive
power and the accuracy of the $\phi$-shift method in the relativistic
case of $\pi\pi$ scattering, in a similar way as it was done in the
case of $NN$-scattering using a nonrelativistic toy
model~\cite{Gomez-Rocha:2019xum}. For completeness and in order to
compare the mass effect in the different cases, we are going to
consider also several channels in $NN$ and $\pi N$ scattering to
illustrate the heavy and heavy-light systems respectively. We will use
here more realistic potentials than the ones used
in~\cite{Gomez-Rocha:2019xum}.

\subsection{Separable models}

For definiteness, we use the form of potentials
determined by the fit already carried out by H. Garcilazo and
L. Mathelitsch~\cite{Mathelitsch:1986ez} for the lowest partial waves
using separable potentials and upgrade the fitting parameters to the
newest phases reported by the most recent Madrid group 2011
analysis~\cite{GarciaMartin:2011cn}. The most remarkable feature of
these fits is the very long tail of the interaction, particularly for
the $P$-wave, which reaches up to 10 GeV.

Long tails in momentum space indicate large strengths
in configuration space. In fact the effect has been observed in the
Marchenko approach to the inverse scattering problem~\cite{Sander:1997br}.
The effect becomes milder when the interaction is coarse grained.

The long-tails feature is not just an artifact of the fit, as for instance the inverse scattering
problem in coordinate space provides very short-distance local
potentials~\cite{Sander:1997br}. Of course, the fact that the
potential is separable, of the form
\begin{eqnarray}
V_l(p',p)= \eta g_l(p') g_l(p) \ ,
\label{eq:vsep}
\end{eqnarray}
where $\eta= \pm 1$ facilitates the solution, reducing it to a simple
quadrature. Indeed, the Kadyshevsky equation, Eq.~(\ref{Eq:Kad}) is solved by the ansatz
\begin{eqnarray}
T_l(p',p, \sqrt{s})= g_l(p') g_l(p) T_l (\sqrt{s}) \ ,
\end{eqnarray}
and inserting this in Eq.~(\ref{Eq:Kad}) we get
\begin{eqnarray}
\left[T_l( \sqrt{s}) \right]^{-1} = 1 - \int_0^\infty dq \,
\frac{q^2}{4 E_q^2} \frac{ \eta  \left[g_l(q)\right]^2}{\sqrt{s}-2
  E_q } \ ,
\end{eqnarray}
yielding the final result 
\begin{eqnarray}
p \cot \delta_l(p) = - \frac{8 E_p}{\pi V_l(p,p)} \left[ 1 - \dashint_0^\infty dq \,
\frac{q^2}{4 E_q^2} \frac{V_l(q,q)}{\sqrt{s}-2 
  E_q } \right] \ ,
\nonumber \\ 
\end{eqnarray} 
whence the phase-shifts can directly be computed by any convenient
integration method for {\it any} value of the CM energy, $\sqrt{s}$.
Taking these values, we may then proceed to check the three different prescriptions, which only generate them on grid points. 
Form factors Eq.~(\ref{eq:vsep}) are
given in the Appendix~\ref{app:potentials}.

Following our preivous work~\cite{Gomez-Rocha:2019xum}, we consider the abbreviations $p$-shift, $E$-shift, and $\phi$-shift when refering to the \textit{momentum}-shift, \textit{energy}-shift and \textit{angle}-shift formulas, cf.  Eq.~(\ref{eq:pshift}), (\ref{eq:dewitt}) and (\ref{eq:phishift}), respectively.

%
%
%
%


%

\subsection{Dependence on the momentum grid and comparison with the standard method}

The first case we take into consideration in some detail is the $
\pi\pi$-scattering.  First of all, we study how our $\phi$-shift
results, calculated in a \textit{finite} momentum grid, differ from
the exact solution in the continuum and compare our results with the
procedure of solving the Lippmann-Schwinger (LS) like equation in the
same momentum grid (prescription K2).

In Figure~\ref{Fig:comparisons} we show our $\phi$-shift results (blue
dots), which turn out to lie exactly on the smooth, green line that
represents the exact solution. The LS calculation is represented by
the orange line. Each arrow in Fig.~\ref{Fig:comparisons} corresponds
to a different channel, and each column corresponds to a different
number of grid points used in the calculation, namely, $N=$25, 50,
100, respectively.

Similarly to what we observed in the non relativistic
case~\cite{Gomez-Rocha:2019xum}, the $\phi$-shift formula provides
excellent results in all cases, reproducing very accurately the exact
solution, even in the case of the grid with the smallest number of
points, $N=$25.  While the LS method converges to the continuum as the
number of points increases (the exception is the $P1$ wave, where the
LS turns out to predict values very accurately in the whole interval),
the $\phi$-shift results do not move away from the exact solution in
any visible way in the considered grids. Recall furthermore, that only
half of the points are inside the studied interval, while the other
half are distributed along the long tail of the potential.

Both methods turn out to be very similar and accurate in the case of
the $LI=S2$, $D0$ and $D2$ waves. This is foreseeable, since while in
the first two cases the phase shifts cover a wide range of values in a
short energy interval, in the last three channels, the phase shifts
remain rather small ($\delta_{02},\delta_{20}< \pm 30^o$ and
$\delta_{22}< 3^o$) in the same energy range.  Thus, perturbation
theory becomes applicable and the main difference is just a higher
order effect.

The $\phi$-shift method for calculating phase shifts
turns out to provide outstanding results in the $\pi\pi$-scattering
phase shifts. They are comparable or better than those provided by conventional
approaches.

\subsection{Comparison of the three different prescritpions}

In this section we calculate phase shifts using the three different prescriptions presented in Section~\ref{sec:shifts}.

When using the $\phi$-shift, Eq.~(\ref{eq:phishift}), or $E$-shift
prescription, Eq.~(\ref{eq:dewitt}), we may represent the results as a
function of the distorted momentum $P_n$, or as a function of the free
momentum $p_n$. The phase shifts $\delta (P_n)$ and $\delta (p_n)$
will acquire the same values but will be horizontally displaced from
each other by the momentum shift.  This ambiguity does not arise in
the $p$-shift case, since the phase shift is a function of the
interacting momentum by construction.
 
\begin{figure*}
\includegraphics[scale=0.45]{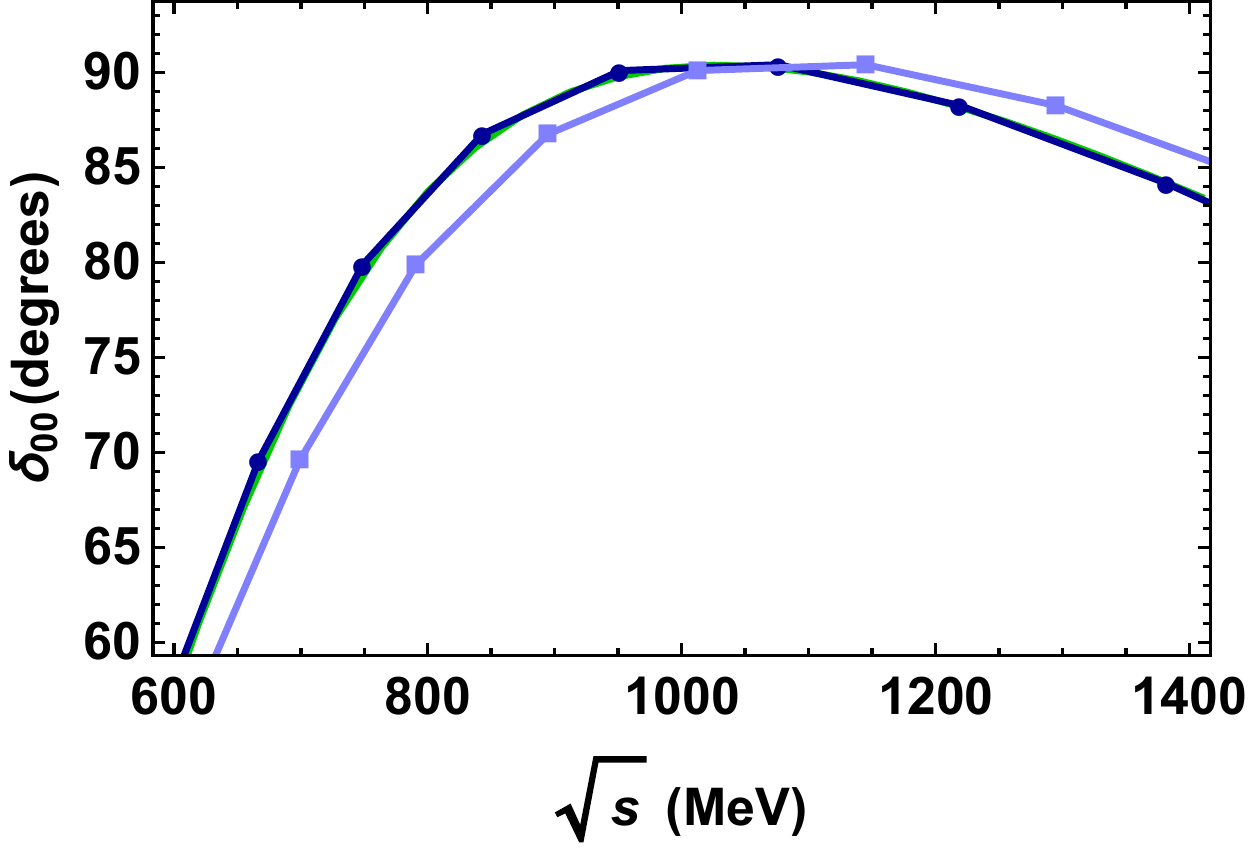}
\includegraphics[scale=0.46]{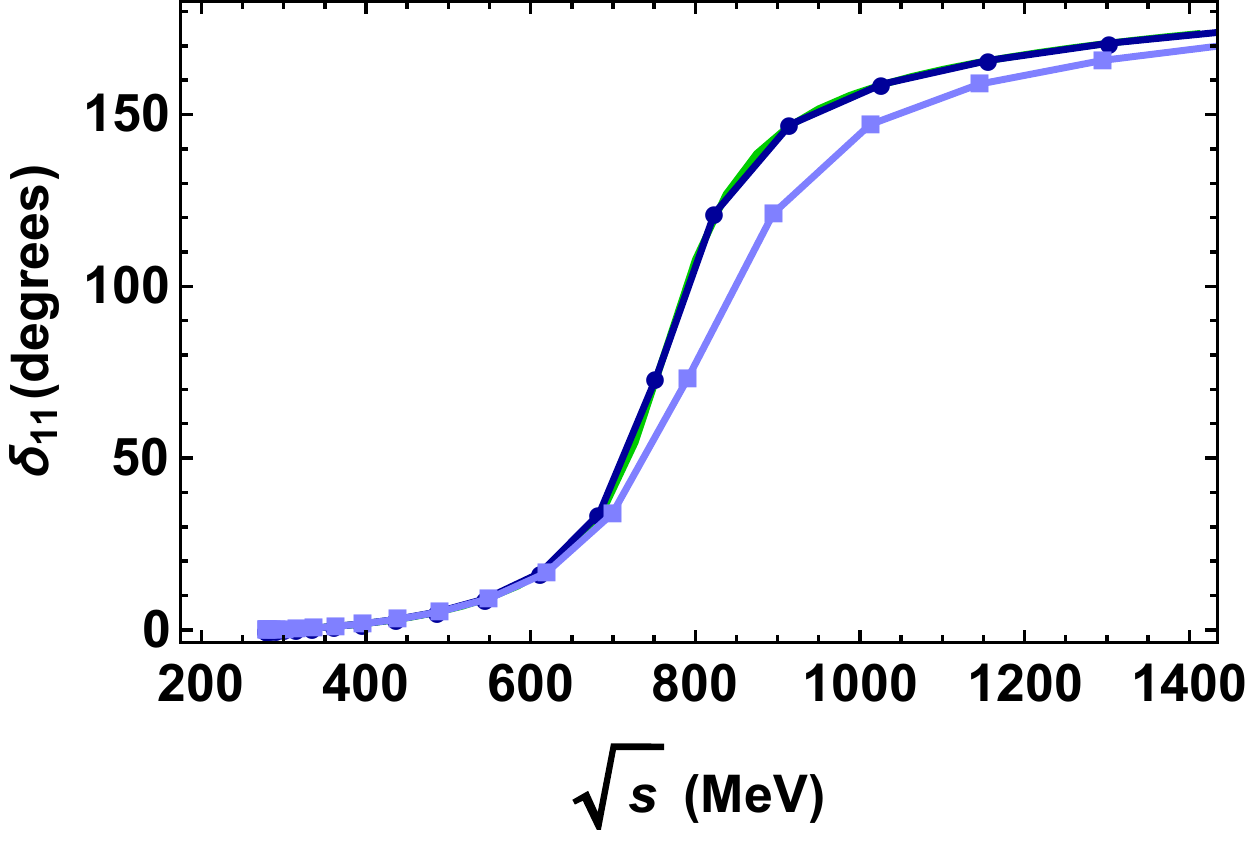}
\includegraphics[scale=0.46]{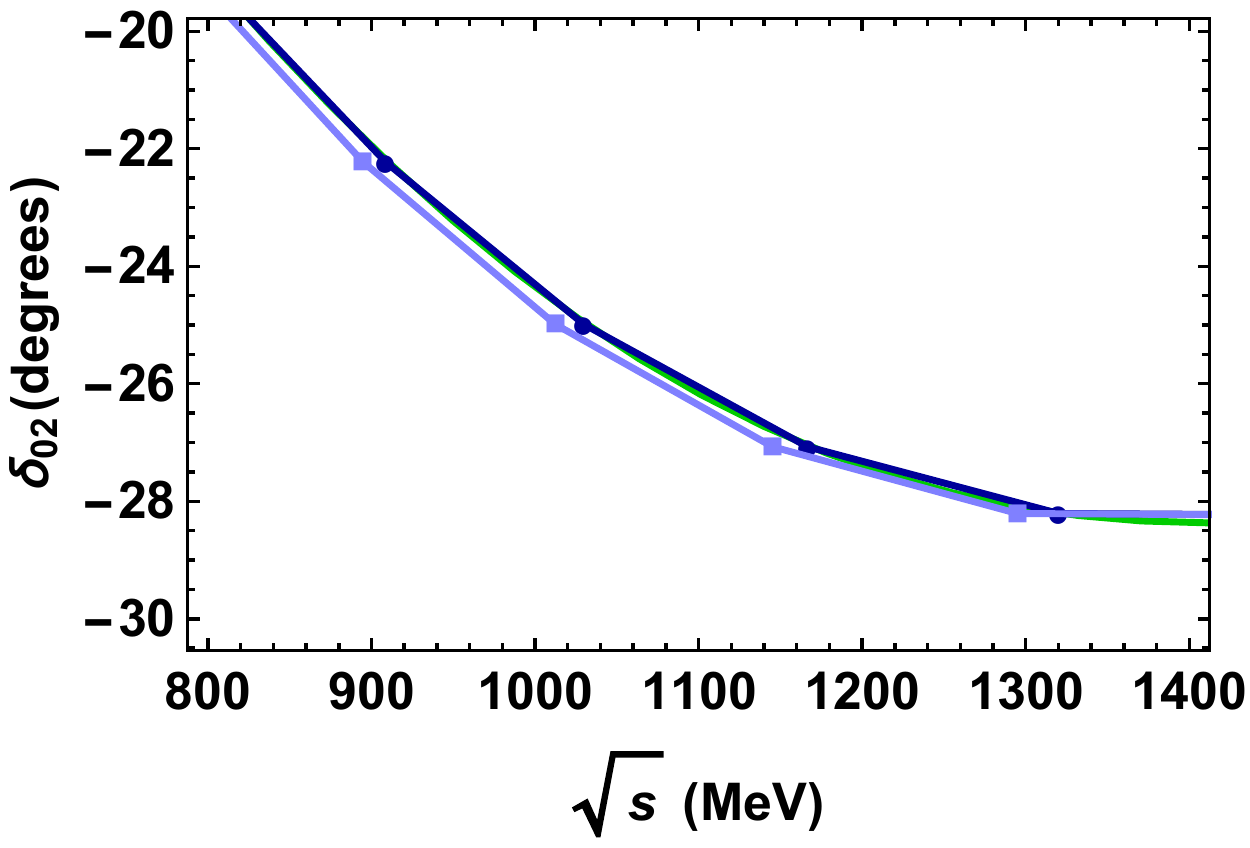}

\includegraphics[scale=0.45]{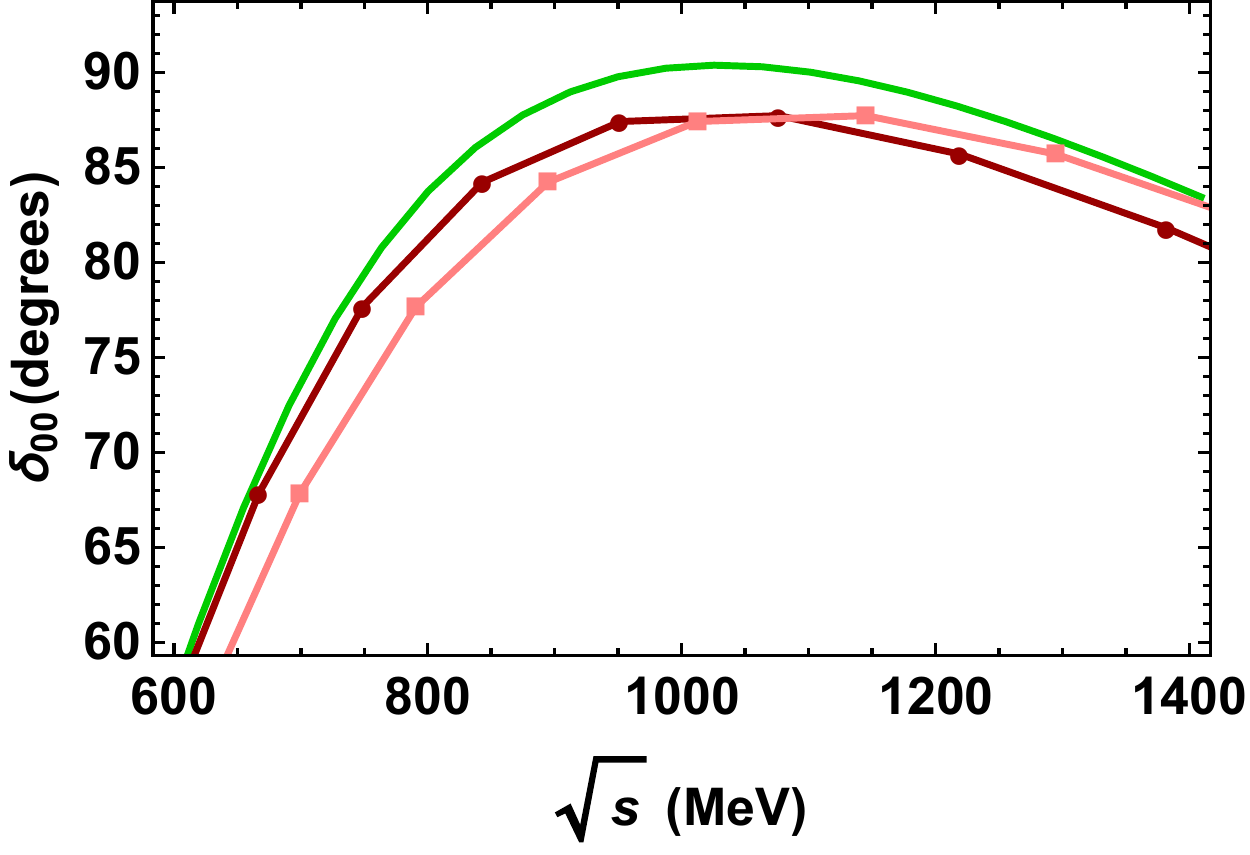}
\includegraphics[scale=0.46]{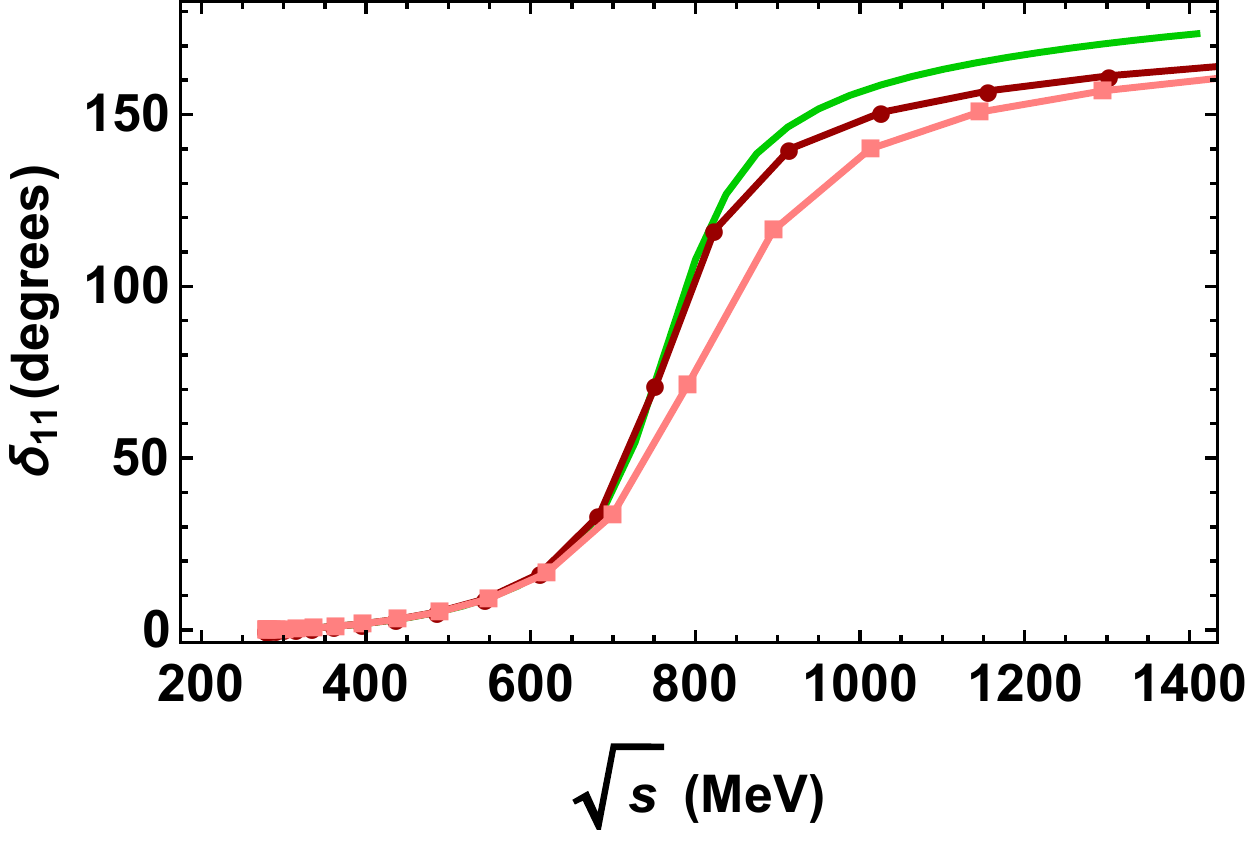}
\includegraphics[scale=0.46]{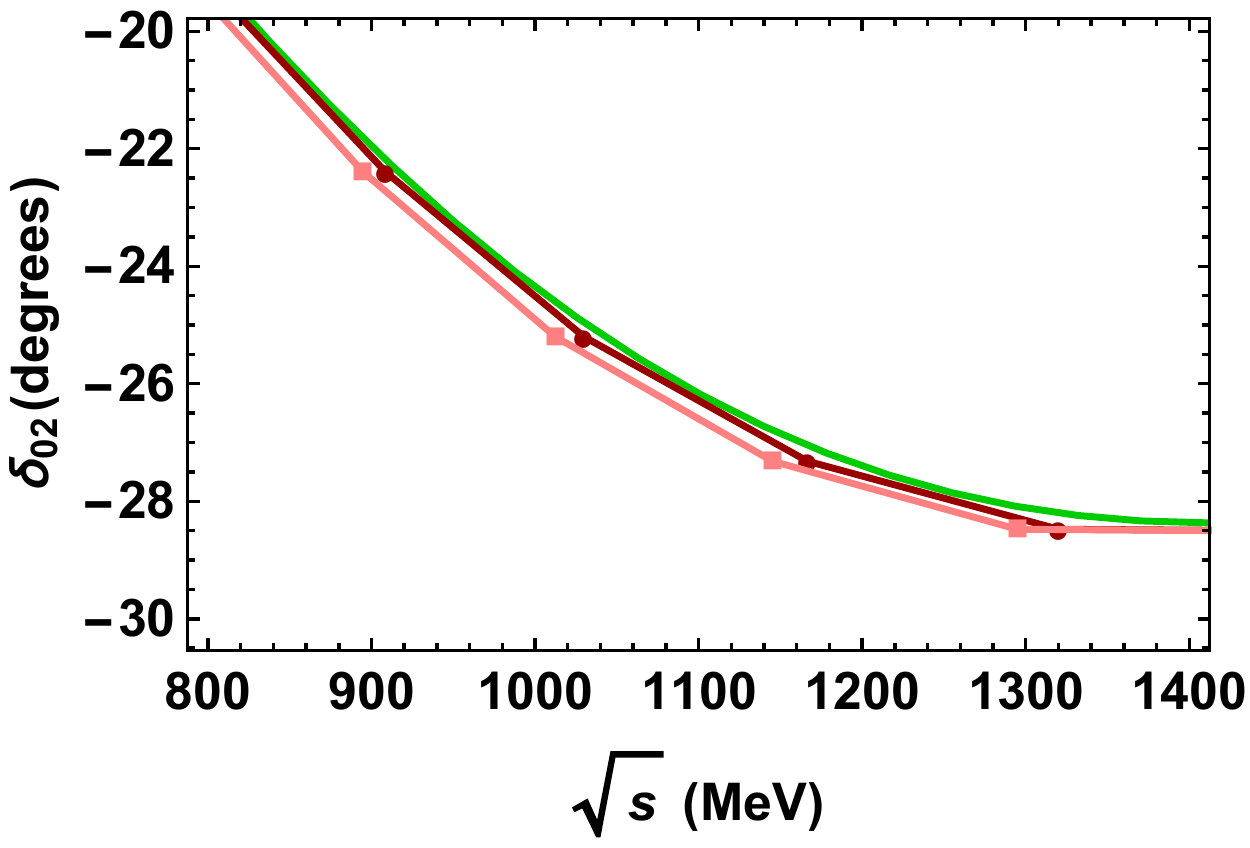}

\includegraphics[scale=0.46]{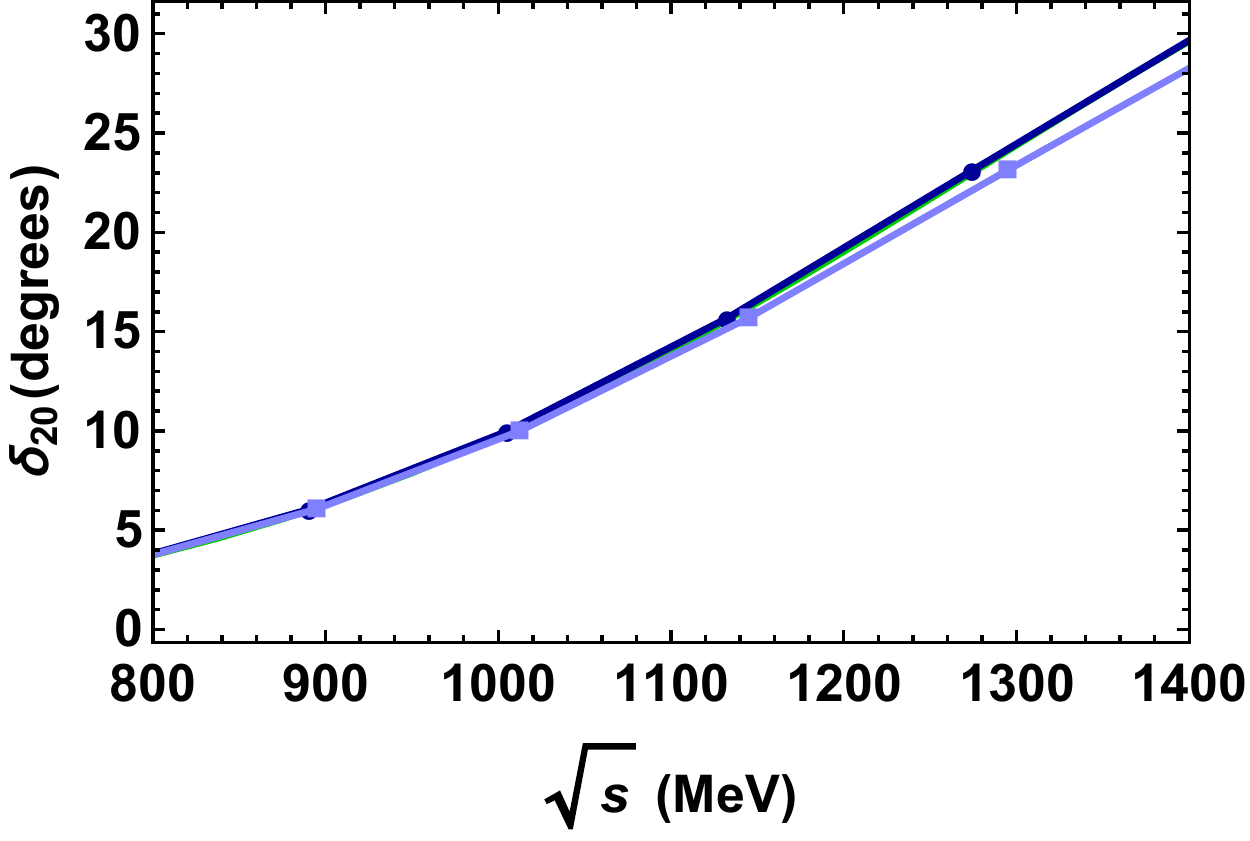}
\includegraphics[scale=0.46]{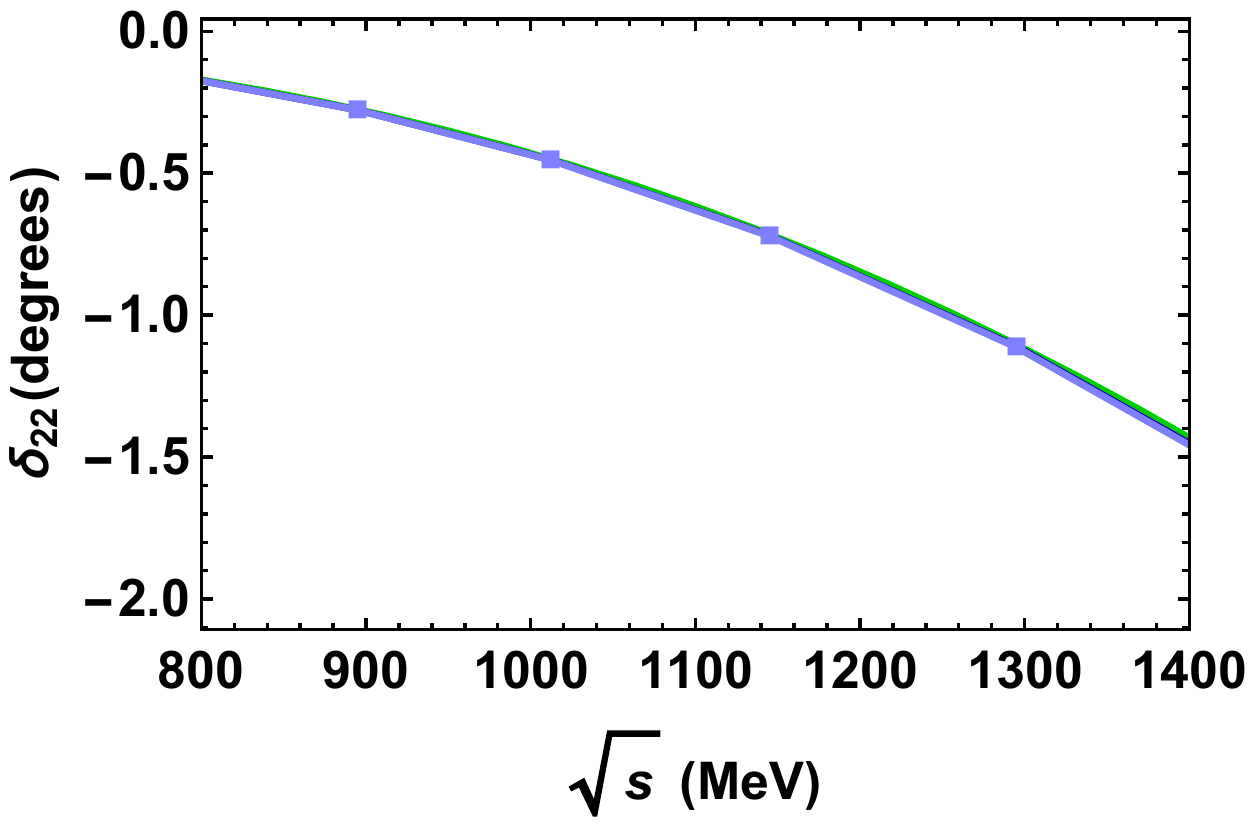}

\includegraphics[scale=0.46]{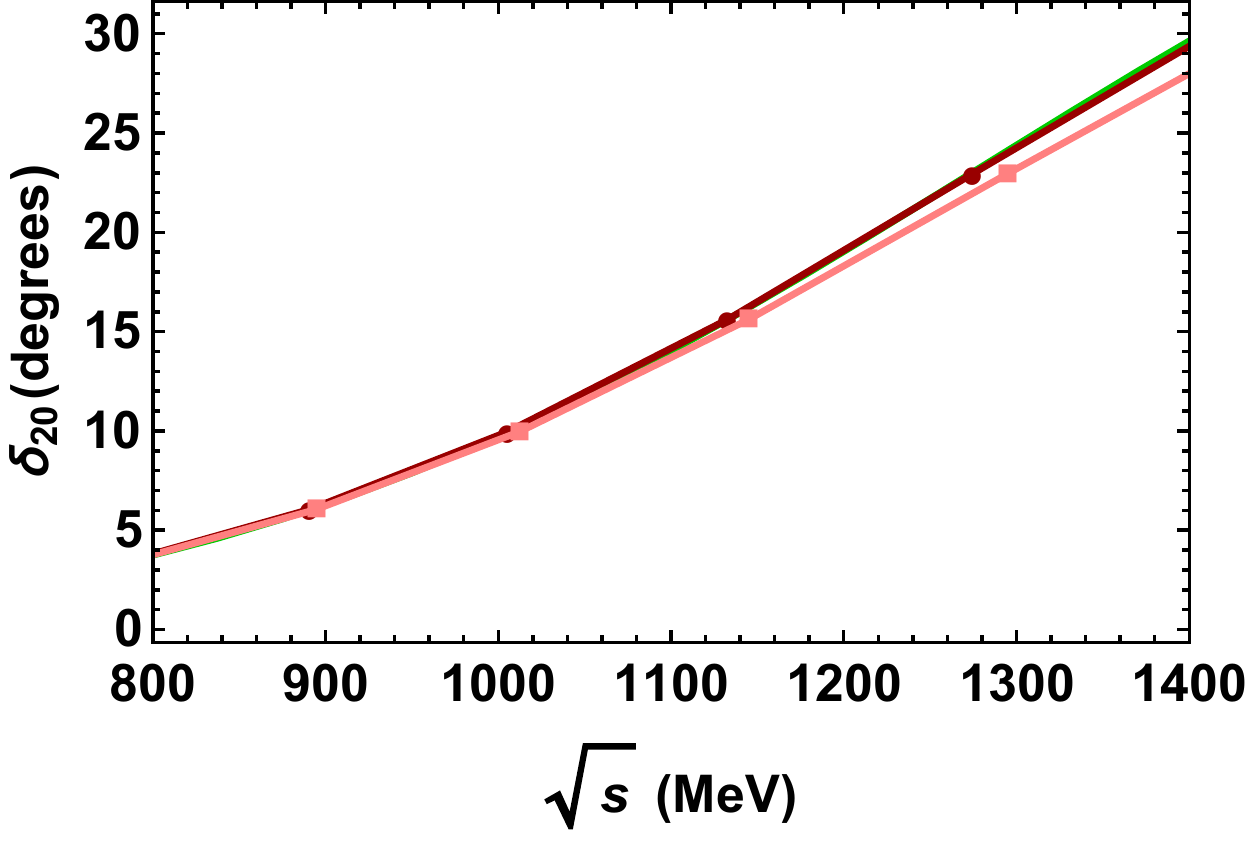}
\includegraphics[scale=0.46]{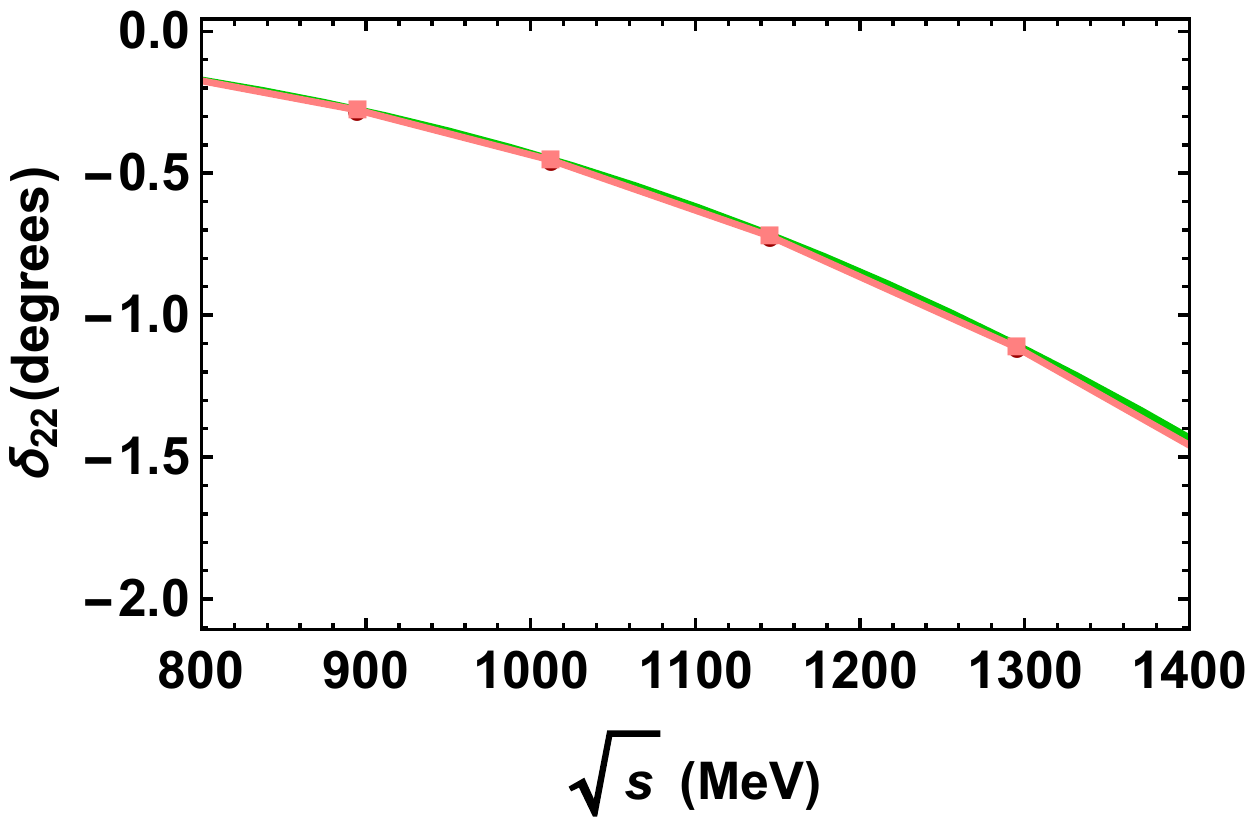}

\caption{(Color online) Phase shifts calculated using the $\phi$-shift, $E$-shift and $p$-shift methods for every channel in $\pi\pi$-scattering and compared with the exact solution (green, smooth line). The first and third lines (in blue) show the $\phi$-shift results and the second and forth lines (in red) show the $E$- or $p$-shift results, which are equal in this case due to the relativistic pion masses. In all cases the phase shifts are represented as a function of the distorted momentum (darker line with round markers), and as a function of the free momentum (lighter line with square markers). The calculation was made with a grid of $N=50$ points. } 
\label{Fig:prescriptions}
\end{figure*}

Fig.~\ref{Fig:prescriptions} shows two lines for every $\pi\pi$-scattering channel.  The
upper row (in blue) shows the phase shifts calculated using the
$\phi$-shift prescription, Eq.~\ref{eq:phishift}, while the lower row
(in red) shows the phase shifts according to the energy-shift
prescription, Eq.~(\ref{eq:dewitt}). The $p$-shift results are
numerically almost identical in this case to the $E$-shift ones, and
they are not depicted in an extra graphic. Phase shifts represented as
a function of the transformed momentum $P_n$ are plotted using a
darker line with round markers while phase shifts plotted as as
function of the free momentum $p_n$ are given by a lighter line with
square markers. All these lines are compared with the exact
calculation represented by the green line without markers.  In some
cases, we have chosen a reduced interval, in such a way that the
difference between lines is more visible.


We observe in Figure~\ref{Fig:prescriptions} that in all cases, the phase shifts represented as a function of the interacting momentum $P_n$ lie closer to the exact solution. This was already pointed out in the nonrelativistic case studied in~\cite{Gomez-Rocha:2019xum}. 
Observing the first and third rows (blue) in Figure~\ref{Fig:prescriptions}, we see that our $\phi$-shift results totally overlap the green line which is not even visible. 
The $E$-shift (as well as $p$-shift) prescription given in the second  and forth rows (red) yields values that lie always below those provided by the momentum-shift one. 
In all cases, $\phi$- and $E$-shift, the phase shifts represented as a function of the free momentum $p_n$ (light line with square markers) appear displaced according to the momentum shift: to the right for attractive interactions, and to the left for repulsive ones. Indeed, the $P_n-p_n$ is negative for attractive interactions and positive for repulsive ones. 
Since the $p$-shift formula prescribes that the phase shift is a function of the interacting momentum, and the $E$-shift formula reproduces it in this case of very light masses, we assume that taking the interacting momentum as the independent variable is the most adequate option. 

It was already explained in~\cite{Gomez-Rocha:2019xum}, that both the $E$- and $p$-shift prescriptions are actually an approximation of the $\phi$-shift formula. Indeed, the $E$-shift formula implies an equal-distance separation of energy levels, alike the $p$-shift formula implies an equal-distance separation in momentum space. The Gauus-Chebyshev grid employed here does not satisfy those conditions. Instead, the equidistant separation occurs in the Chebyshev angle. Therefore, the adequate formula for our grid is the $\phi$-shift.  Nevertheless, we have seen that still the $E$- and $p$-shift formulas turn out to be a very good approximation, since the obtained results analyzed for $N=50$ points are comparable or even better than those obtained through the standard LS equation.

\subsection{ Heavy masses and non-equal masses }

\begin{figure*}

\includegraphics[scale=0.45]{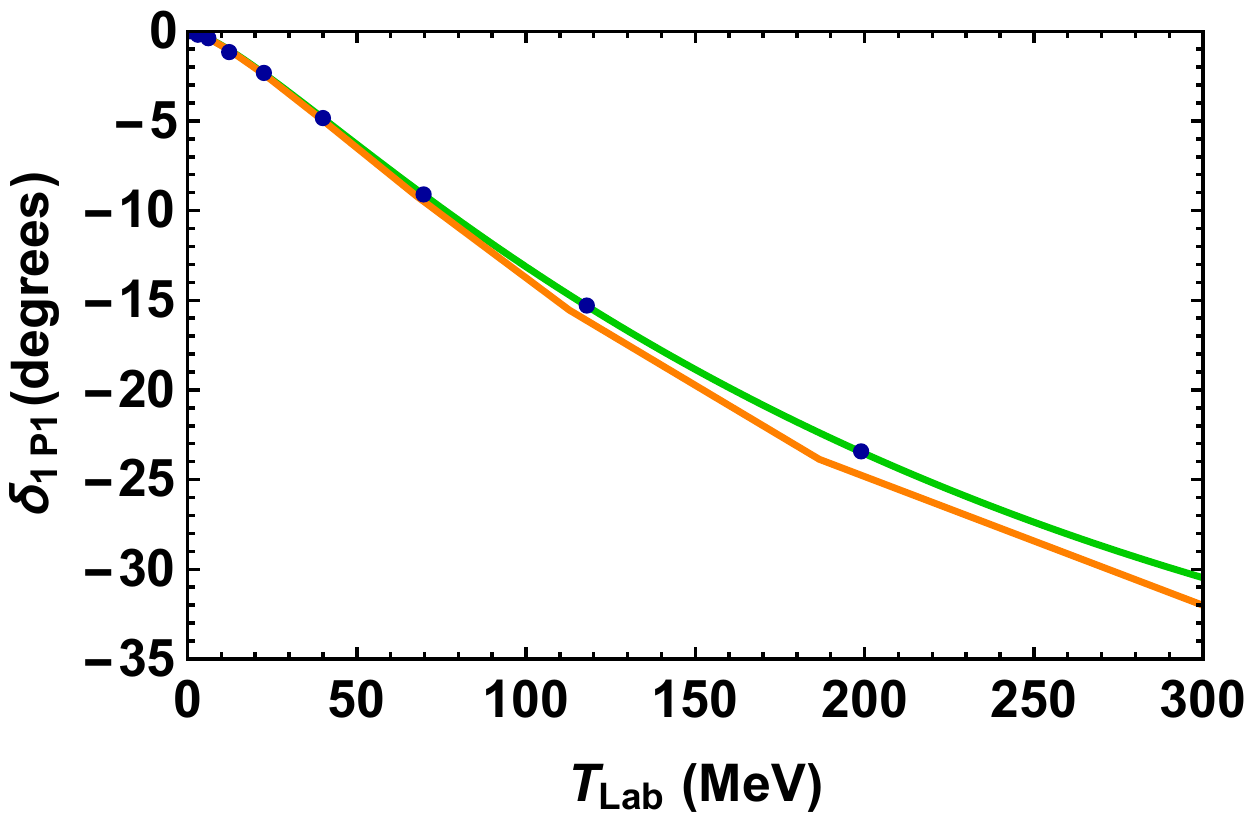}
\includegraphics[scale=0.45]{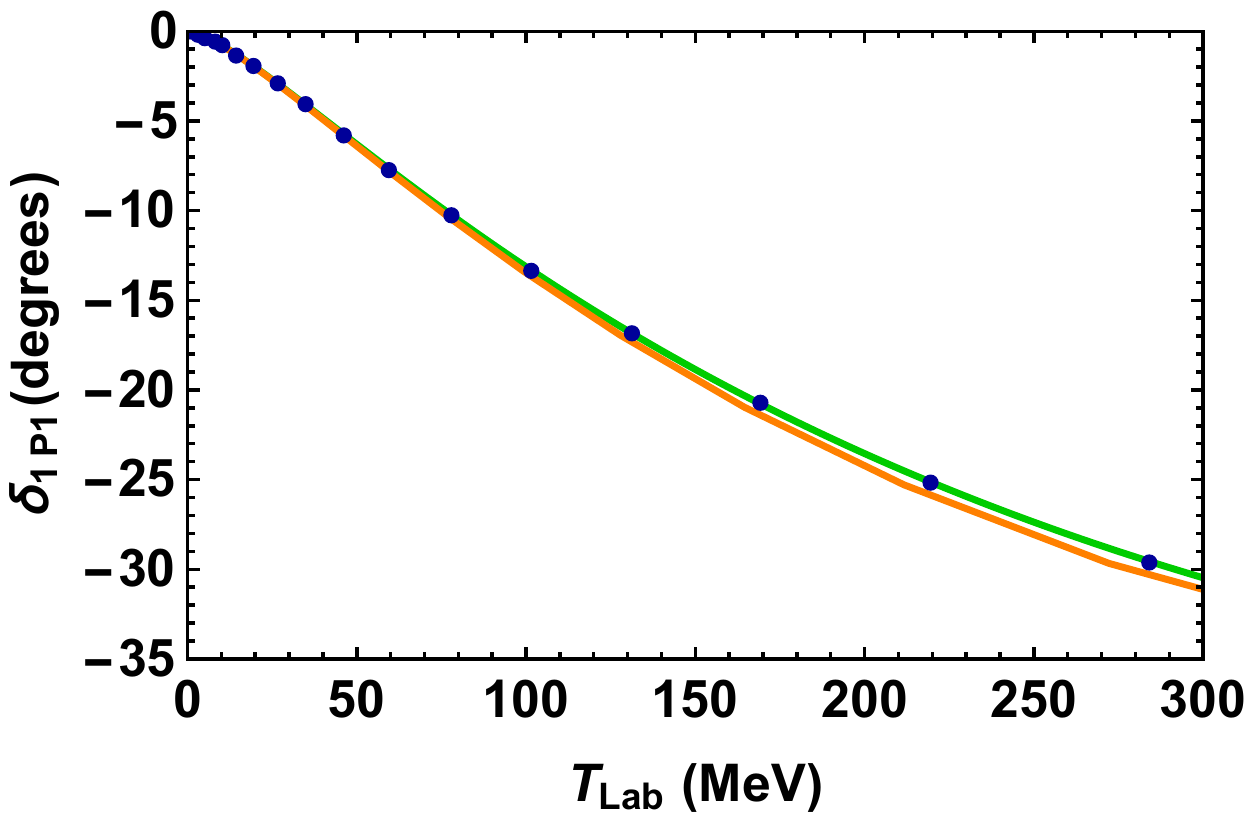}
\includegraphics[scale=0.45]{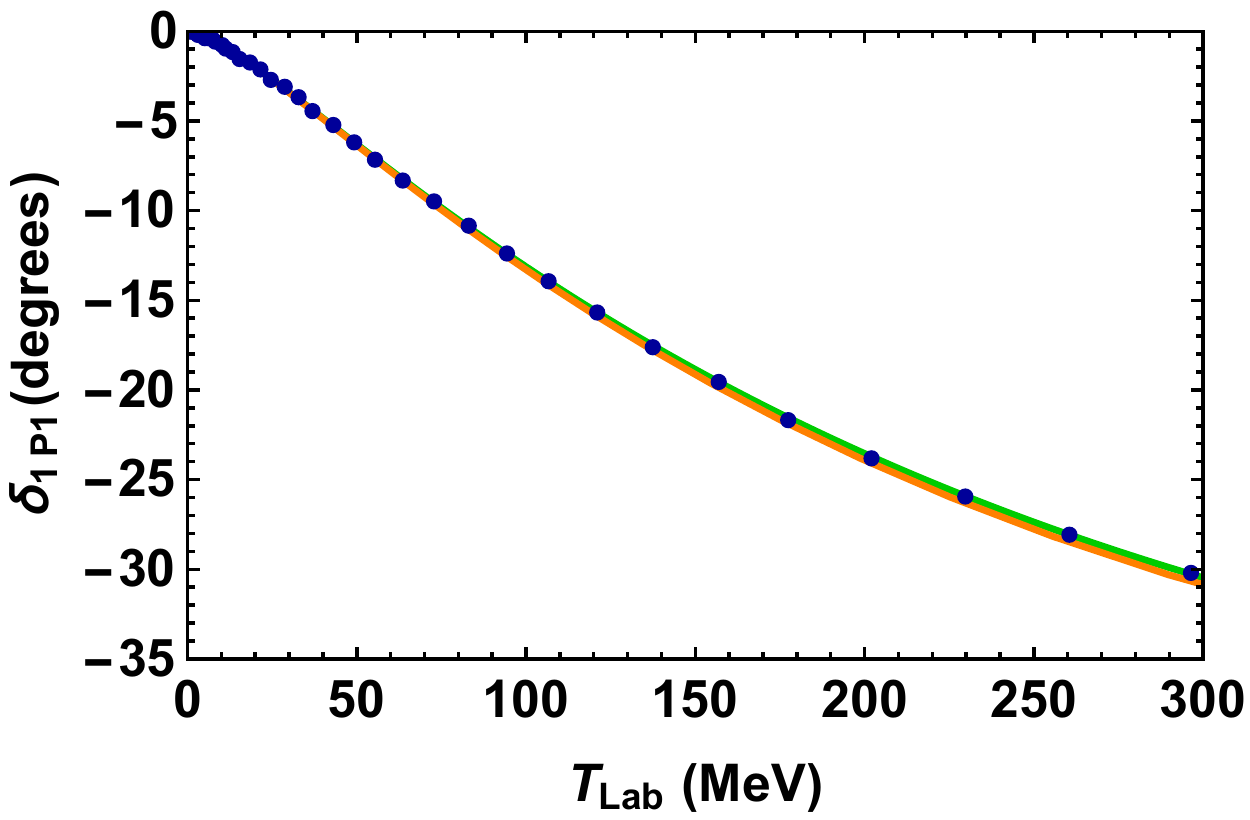}

\includegraphics[scale=0.45]{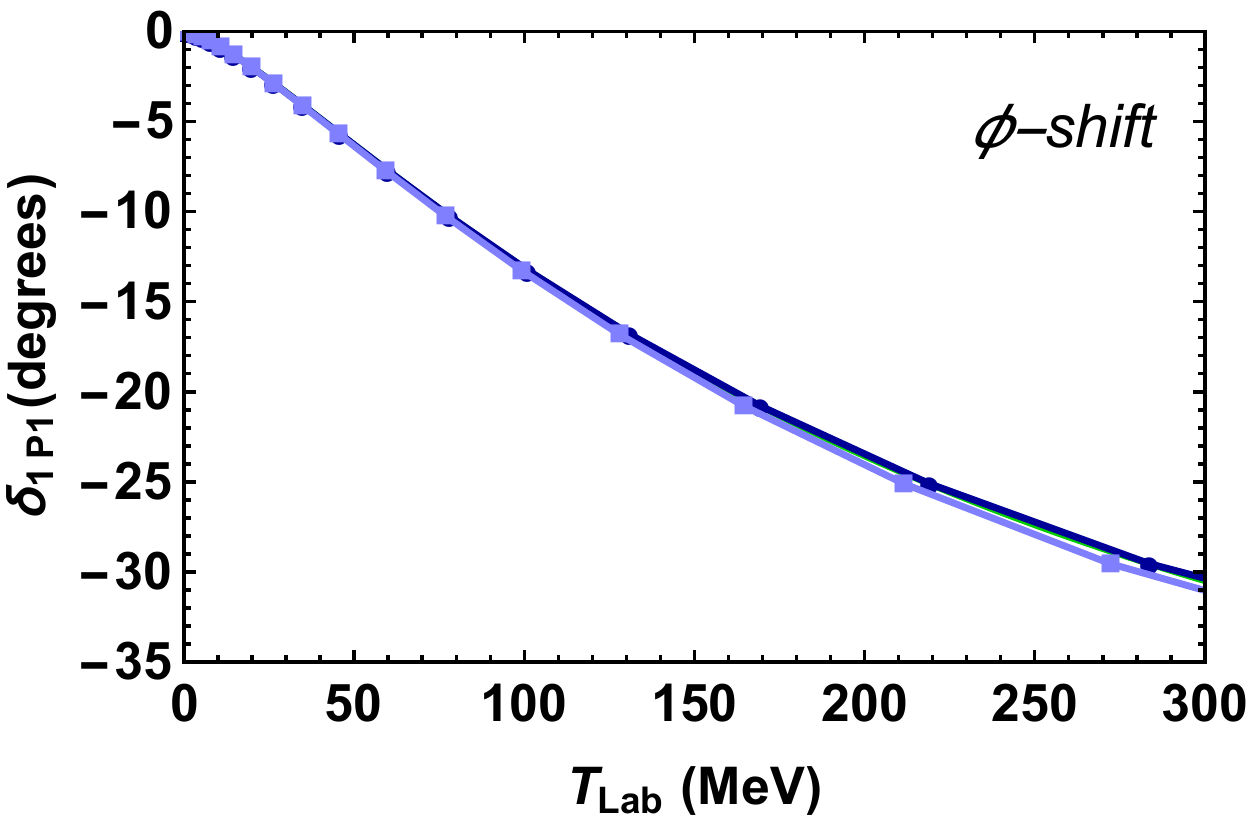}
\includegraphics[scale=0.45]{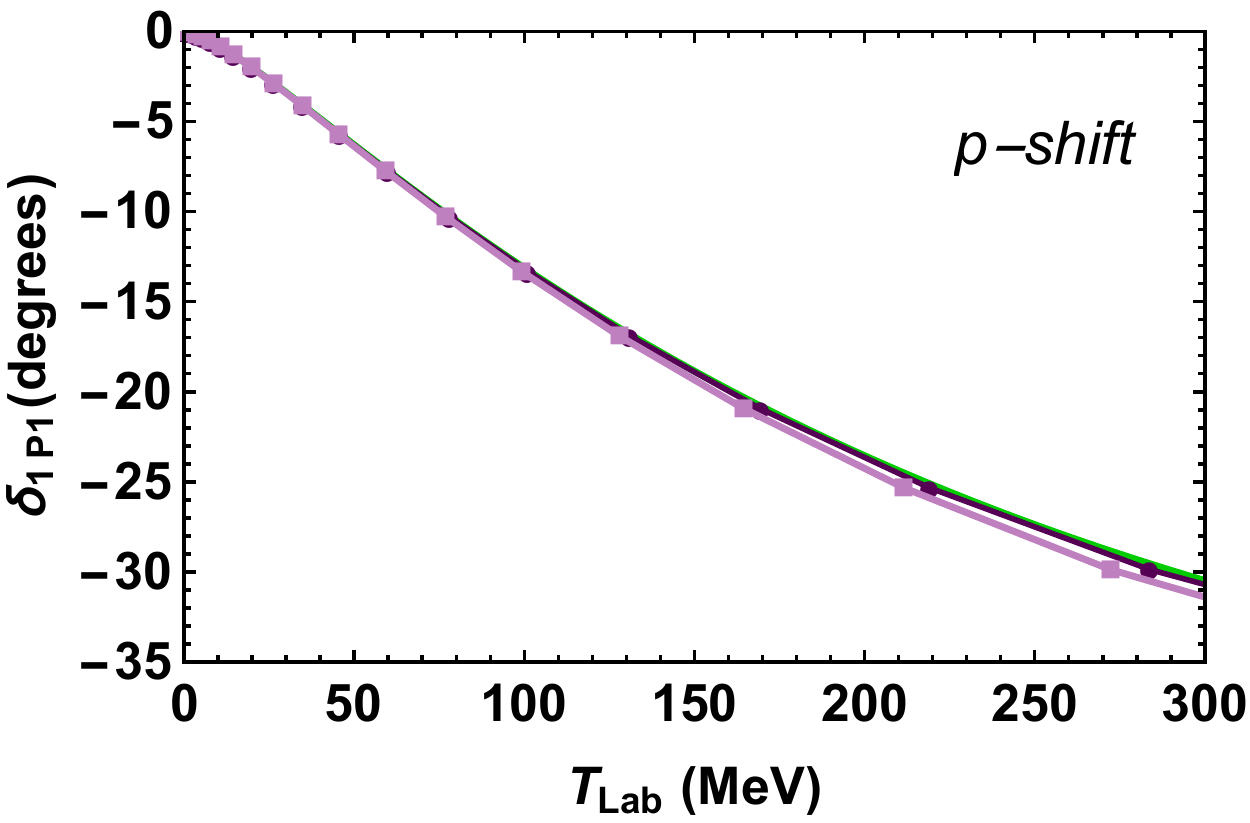}
\includegraphics[scale=0.45]{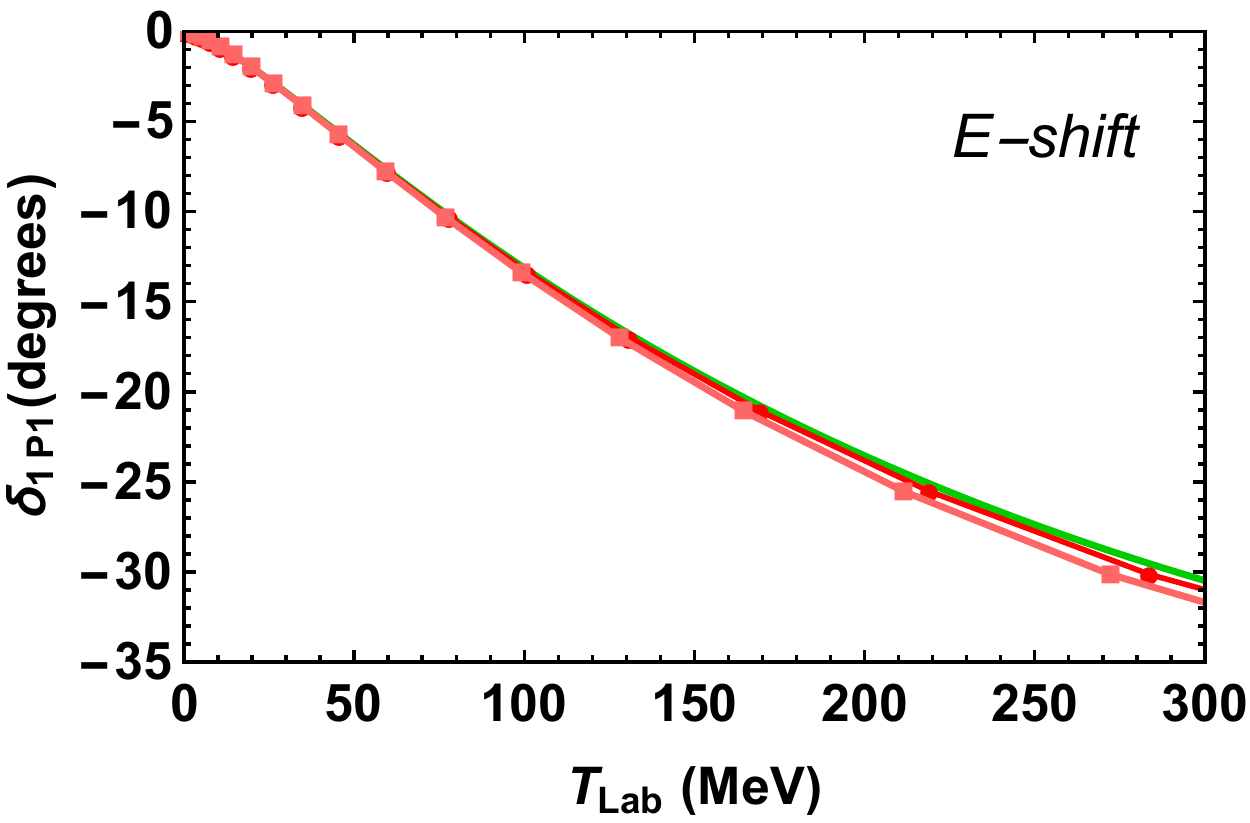}

\caption{ (Color online) $NN$-scattering phase shifts for the $ ^1 P_1$ channel. Upper row: comparison of our $\phi$-shift results (blue dots) with the exact solution (green, smooth line) and the LS result (orange line) calculated in a grid of $N=25$, 50 and 100, respectively. Lower row: phase shifts calculated in a grid of $N=$50 points using the three different prescriptions as labeled in the corner. In each figure of the lower row the phase shifts are represented as a function of the distorted momentum (darker line with round markers) and as a function of the free momentum (lighter line with square markers). }
\label{fig:1p1}
\end{figure*}

\begin{figure*}

\includegraphics[scale=0.45]{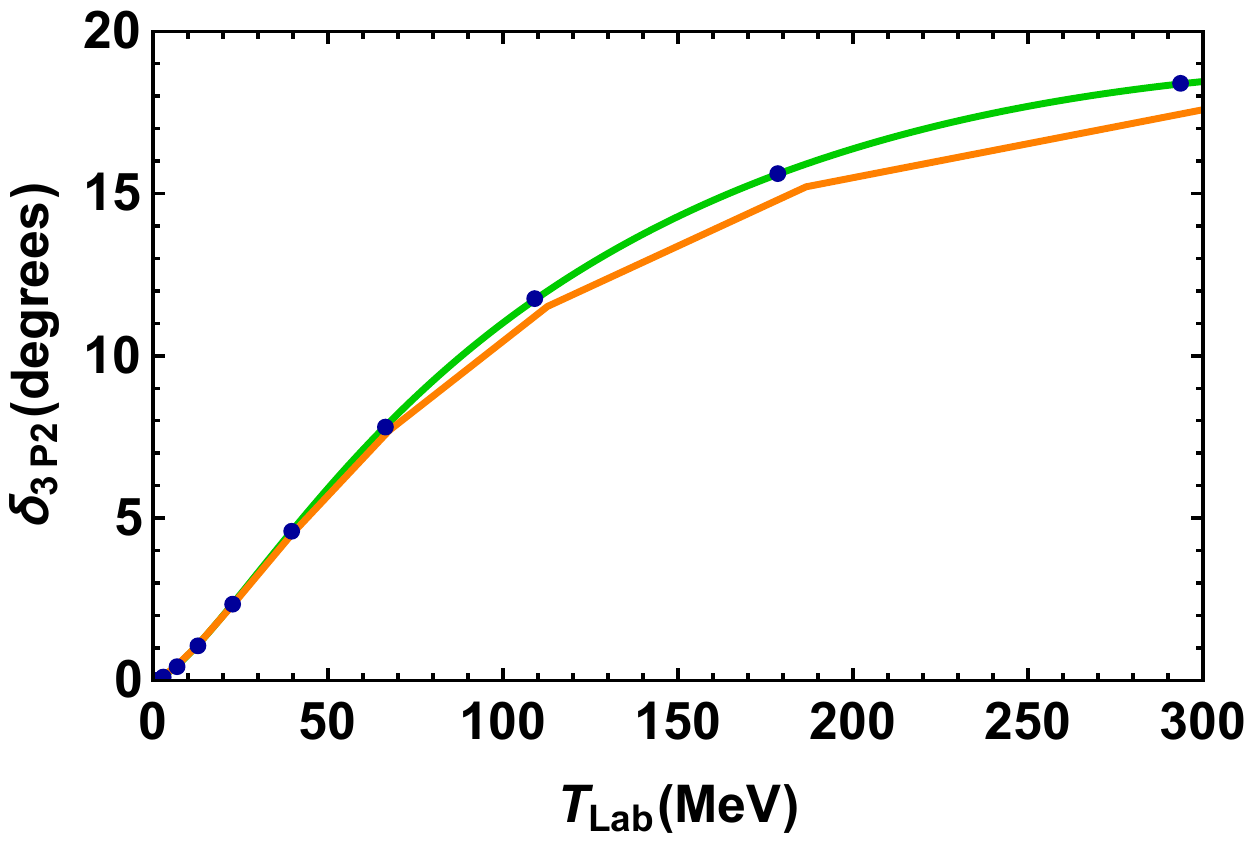}
\includegraphics[scale=0.45]{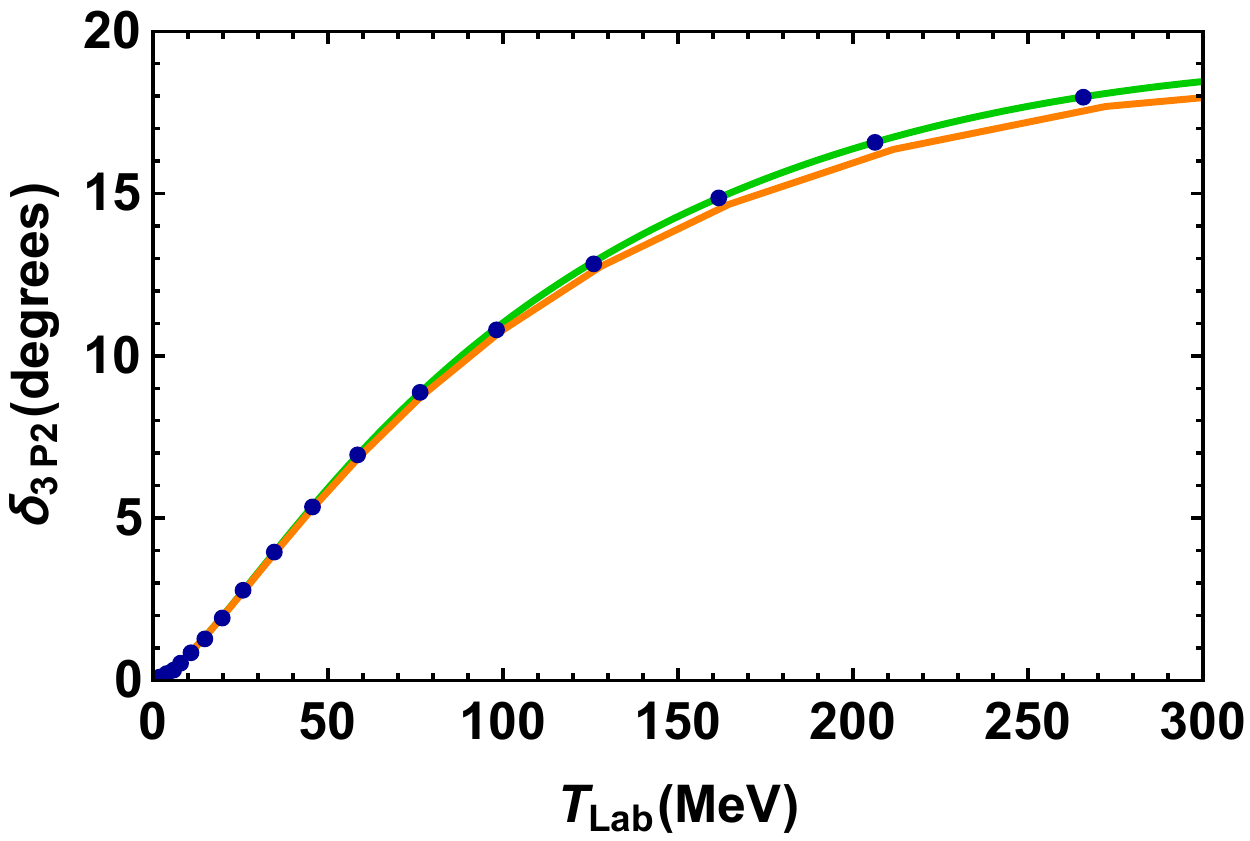}
\includegraphics[scale=0.45]{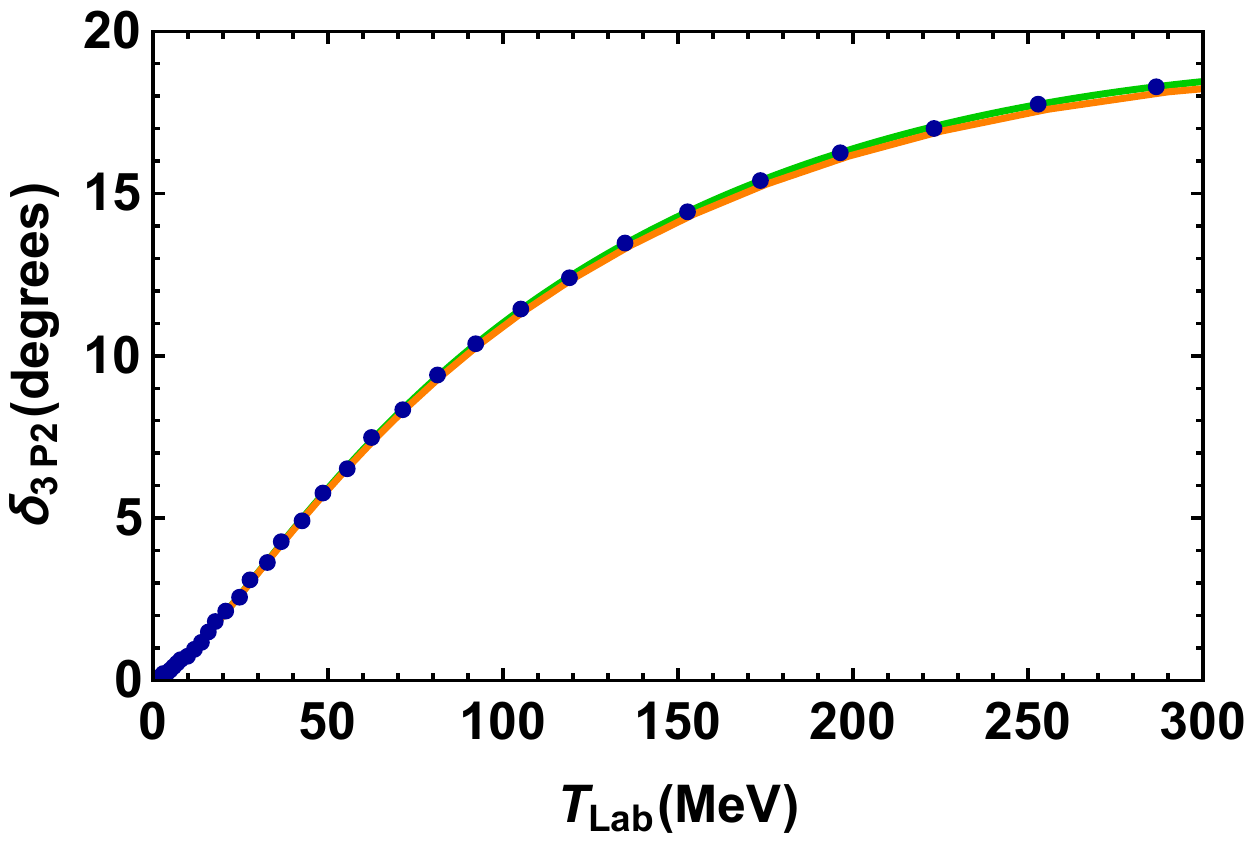}

\includegraphics[scale=0.42]{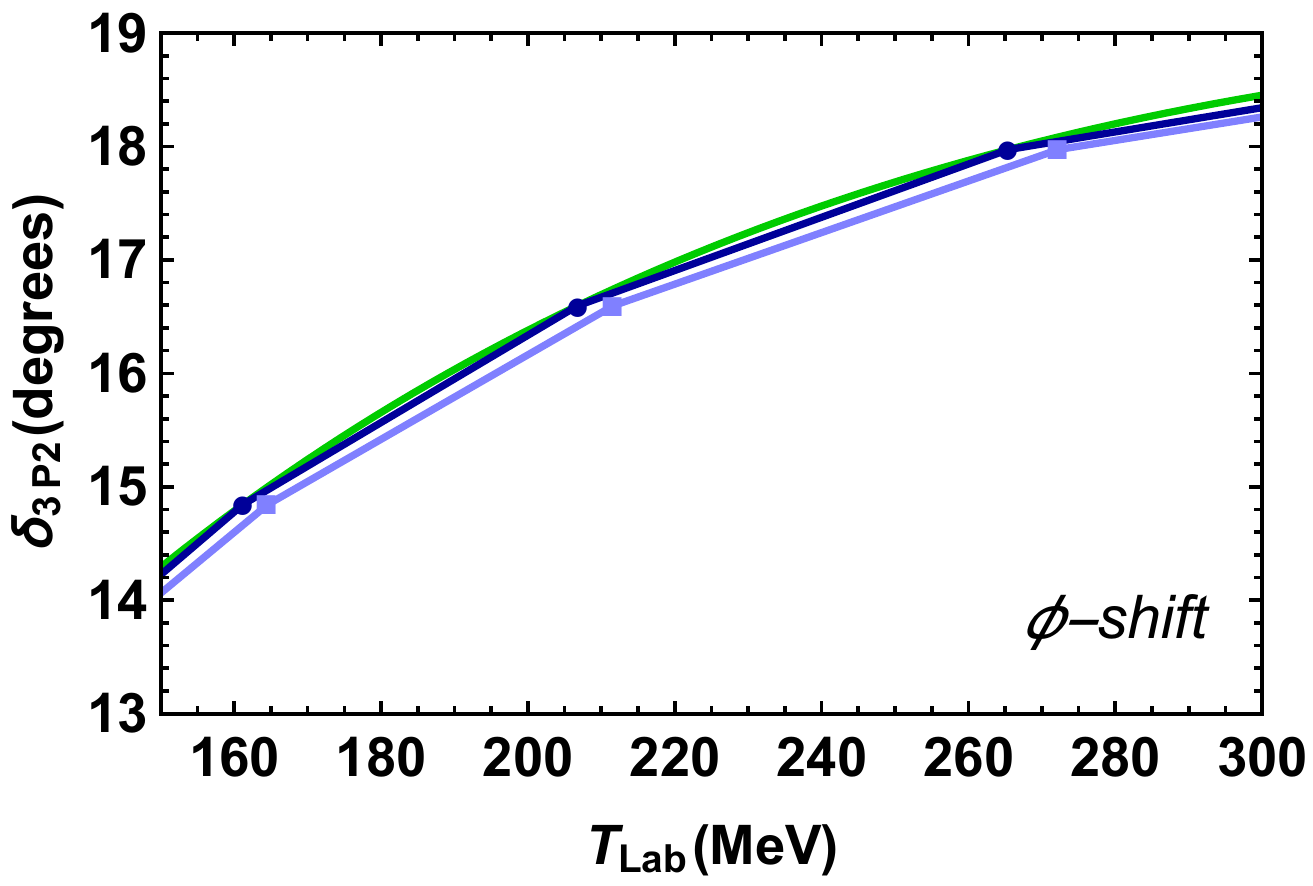}
\includegraphics[scale=0.42]{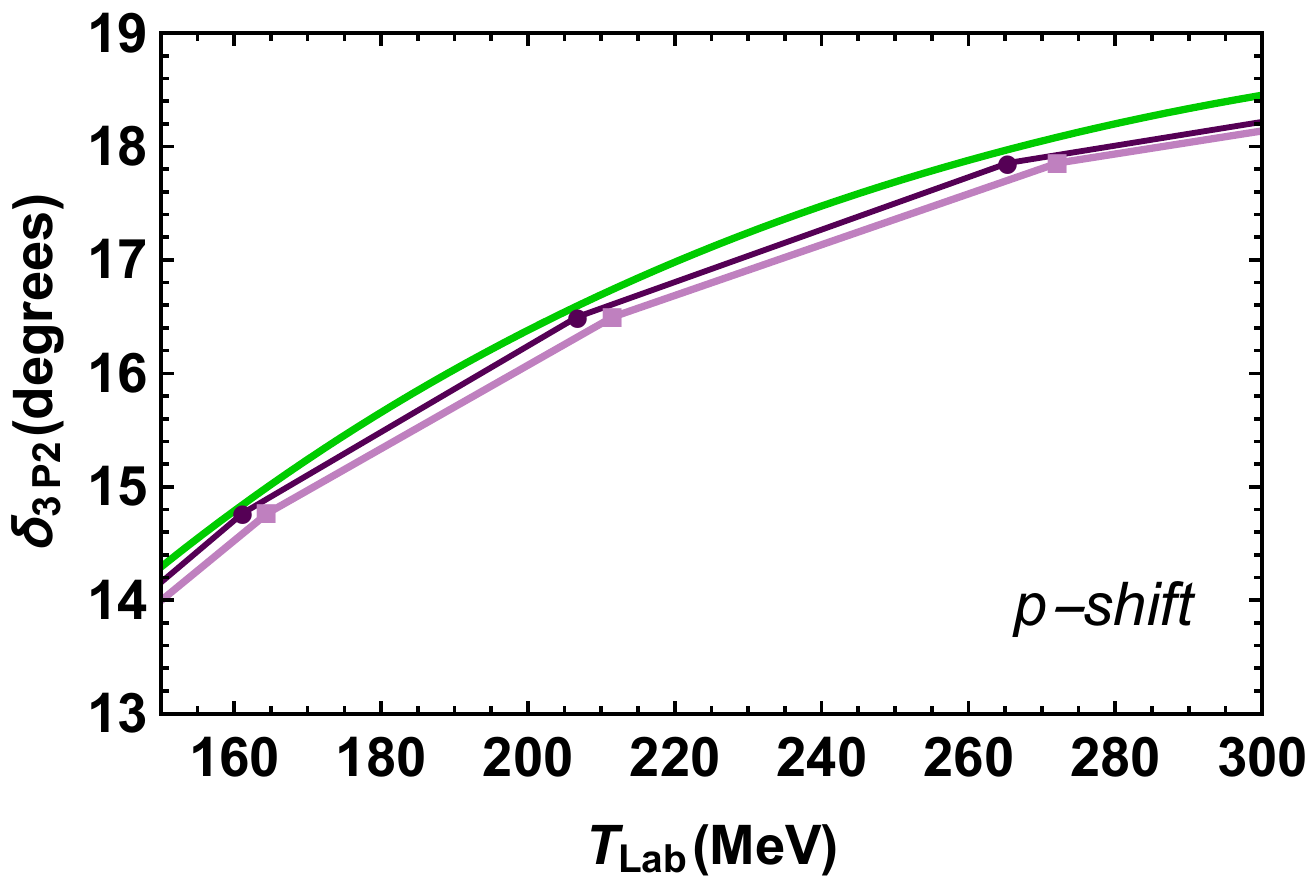}
\includegraphics[scale=0.42]{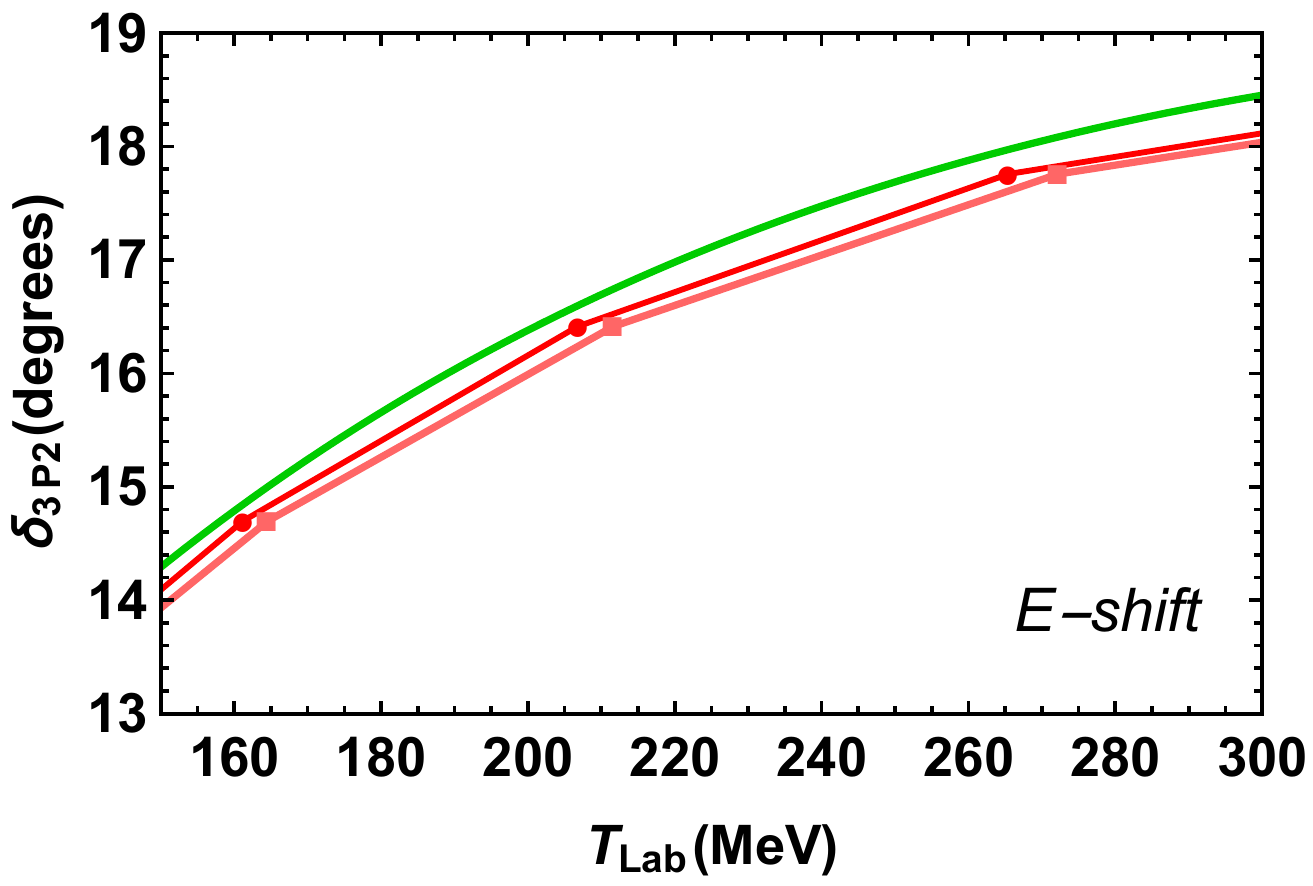}

\caption{ The same as in Figure~\ref{fig:1p1} but for the $ ^3 P_2$ channel.}
\label{fig:3p2}
\end{figure*}

\begin{figure*}

\includegraphics[scale=0.45]{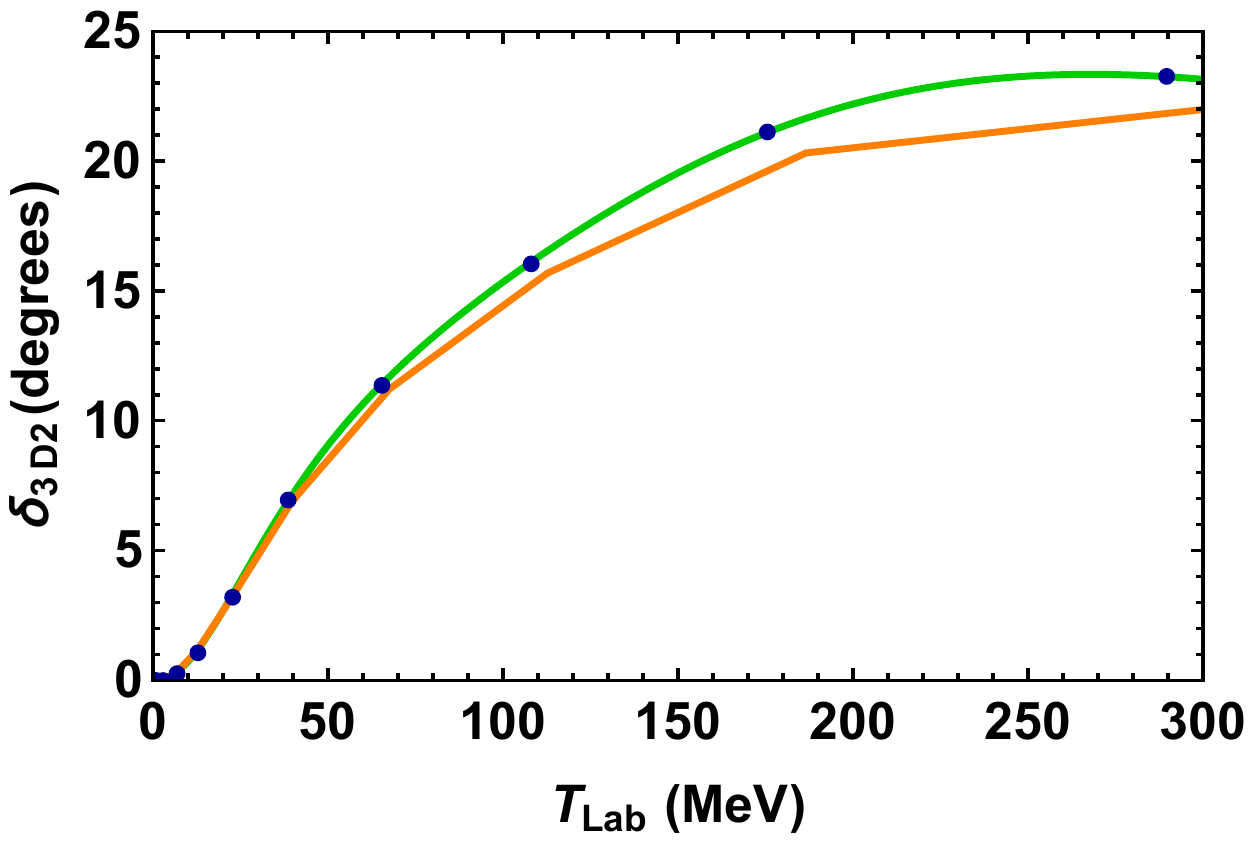}
\includegraphics[scale=0.45]{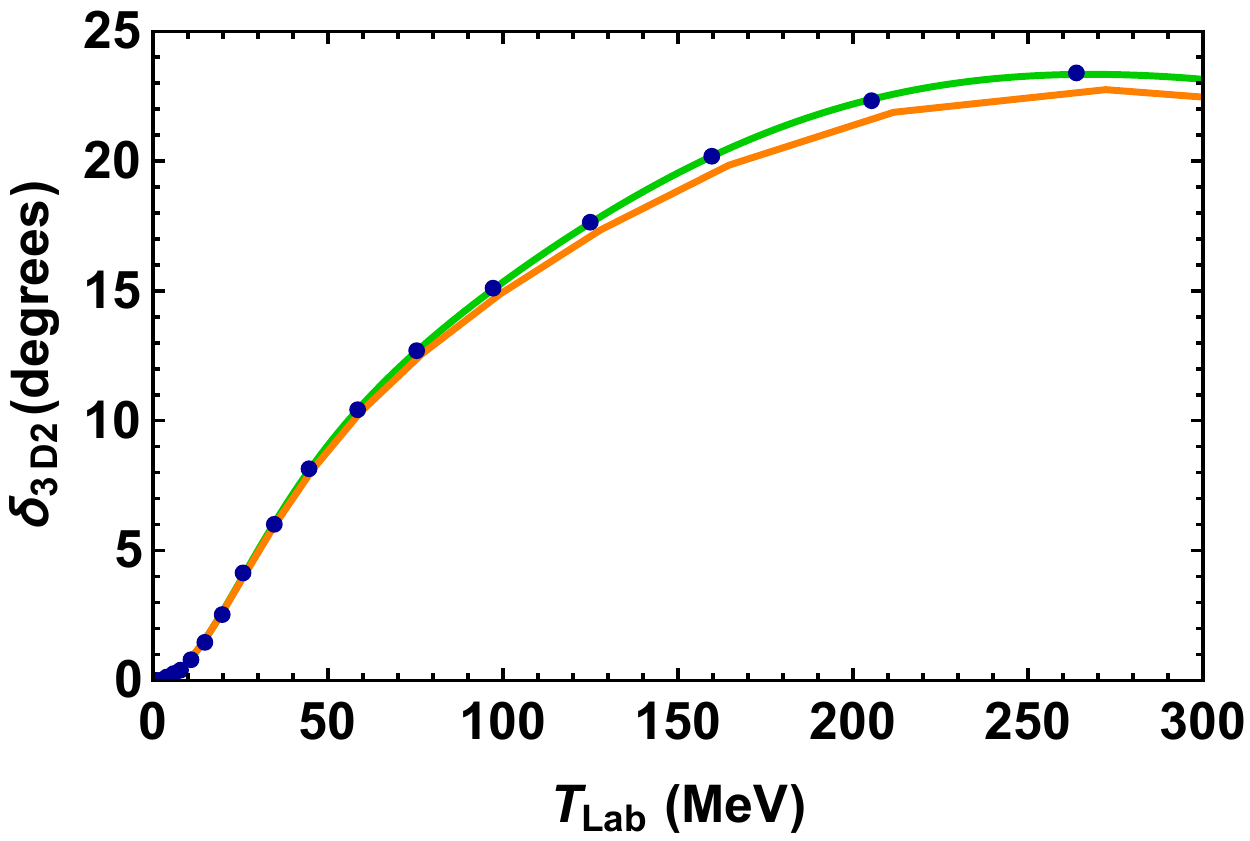}
\includegraphics[scale=0.45]{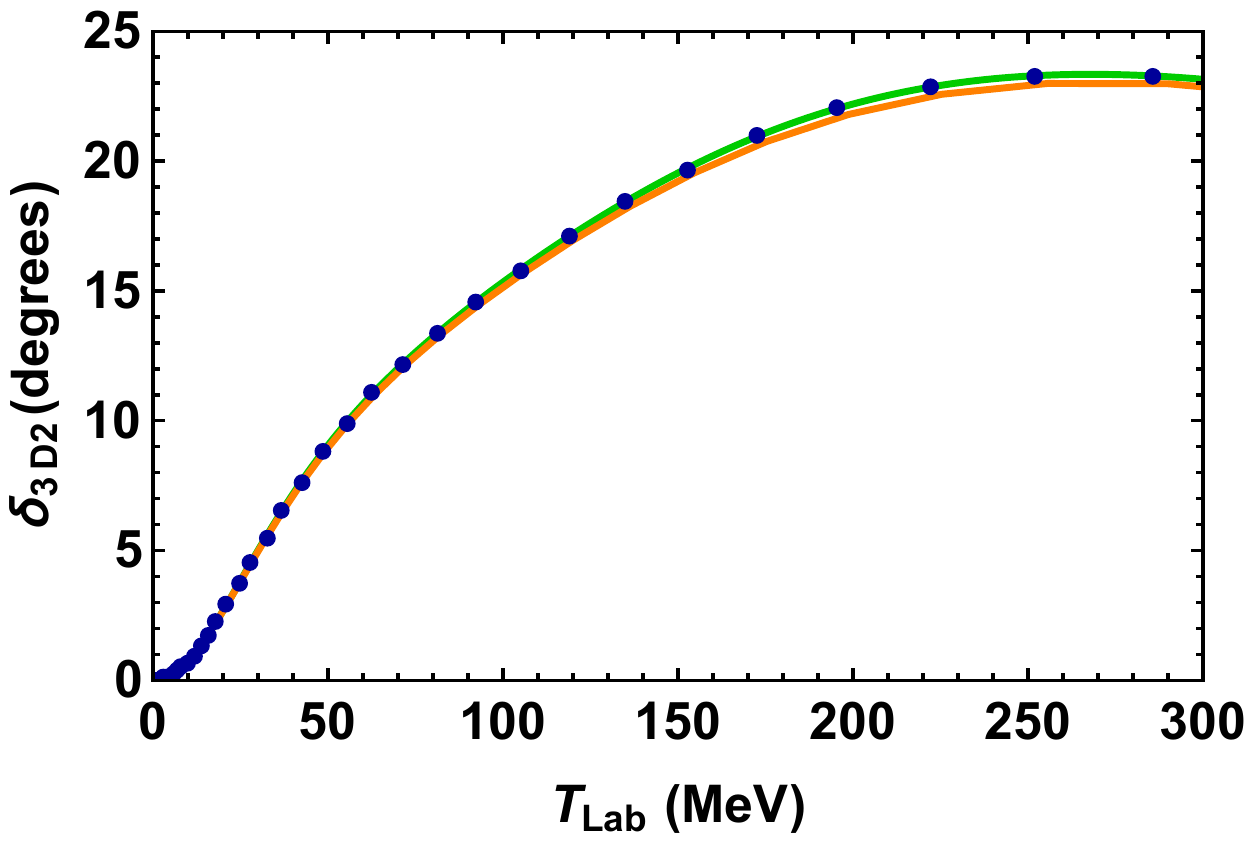}

\includegraphics[scale=0.45]{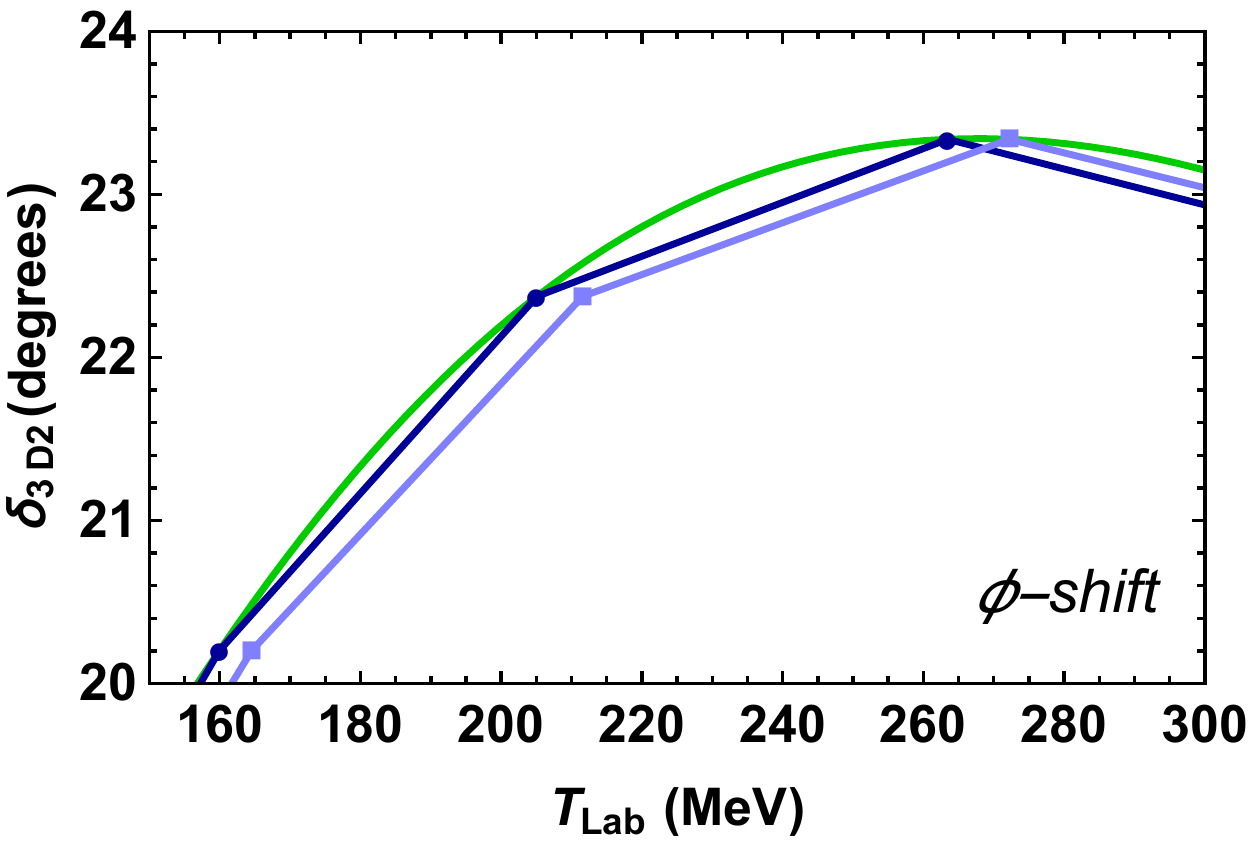}
\includegraphics[scale=0.45]{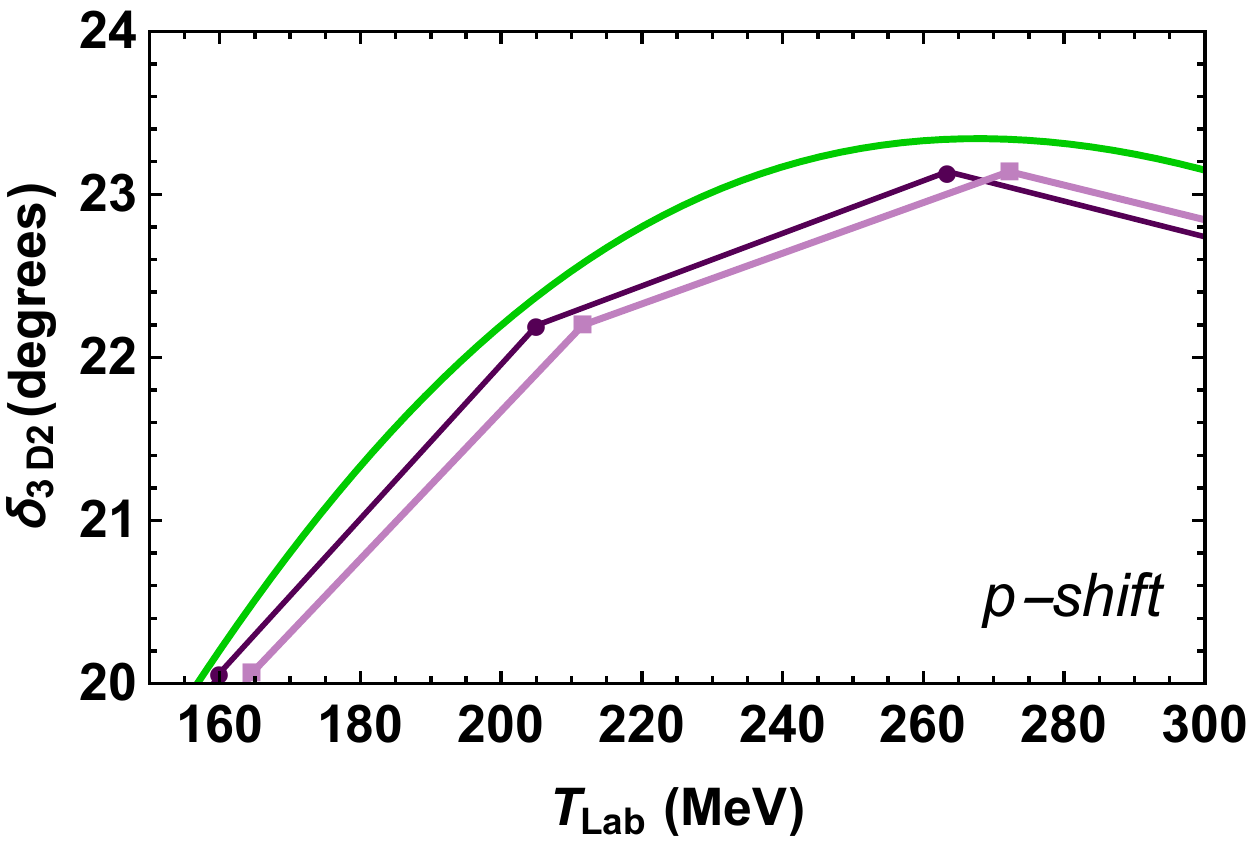}
\includegraphics[scale=0.45]{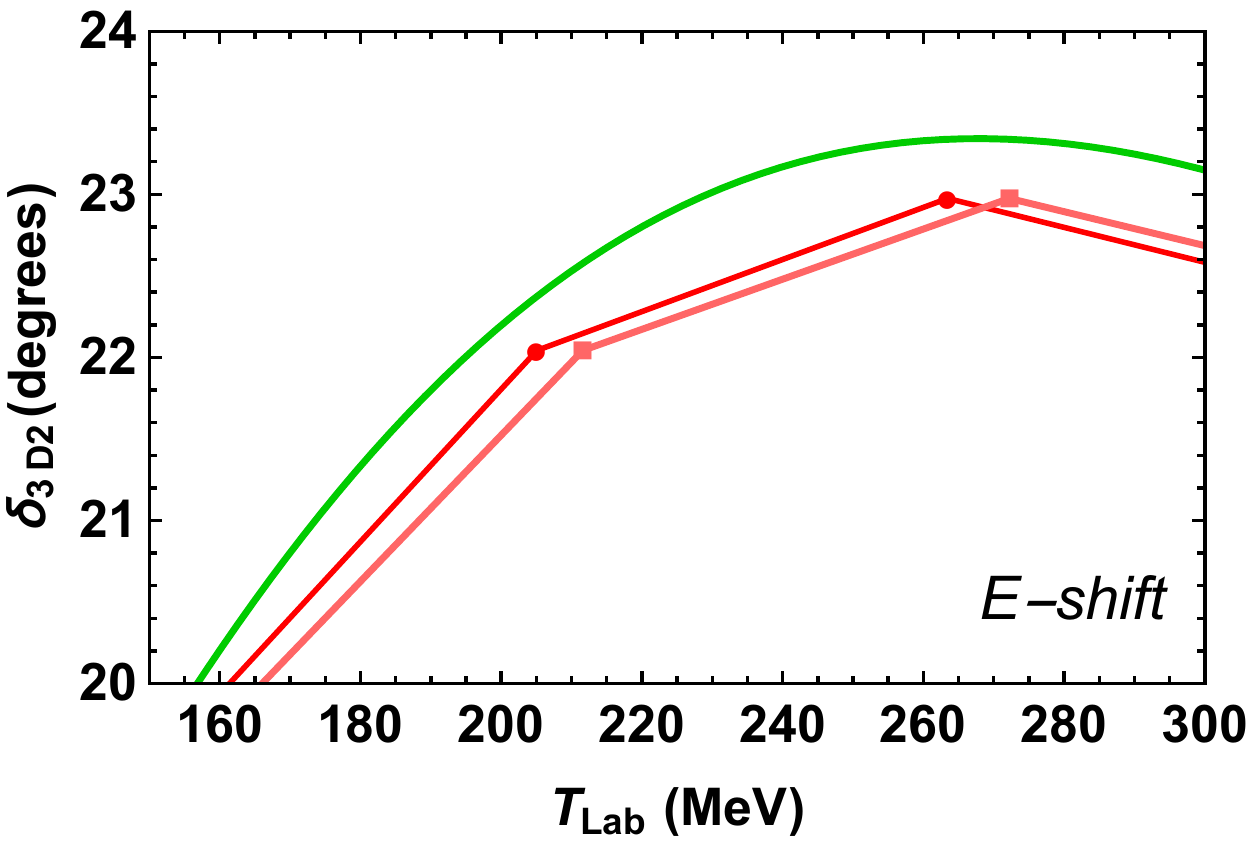}

\caption{ The same as in Figure~\ref{fig:1p1} but for the $ ^3 D_2$ channel.}
\label{fig:3d2}
\end{figure*}

\begin{figure*}

\includegraphics[scale=0.45]{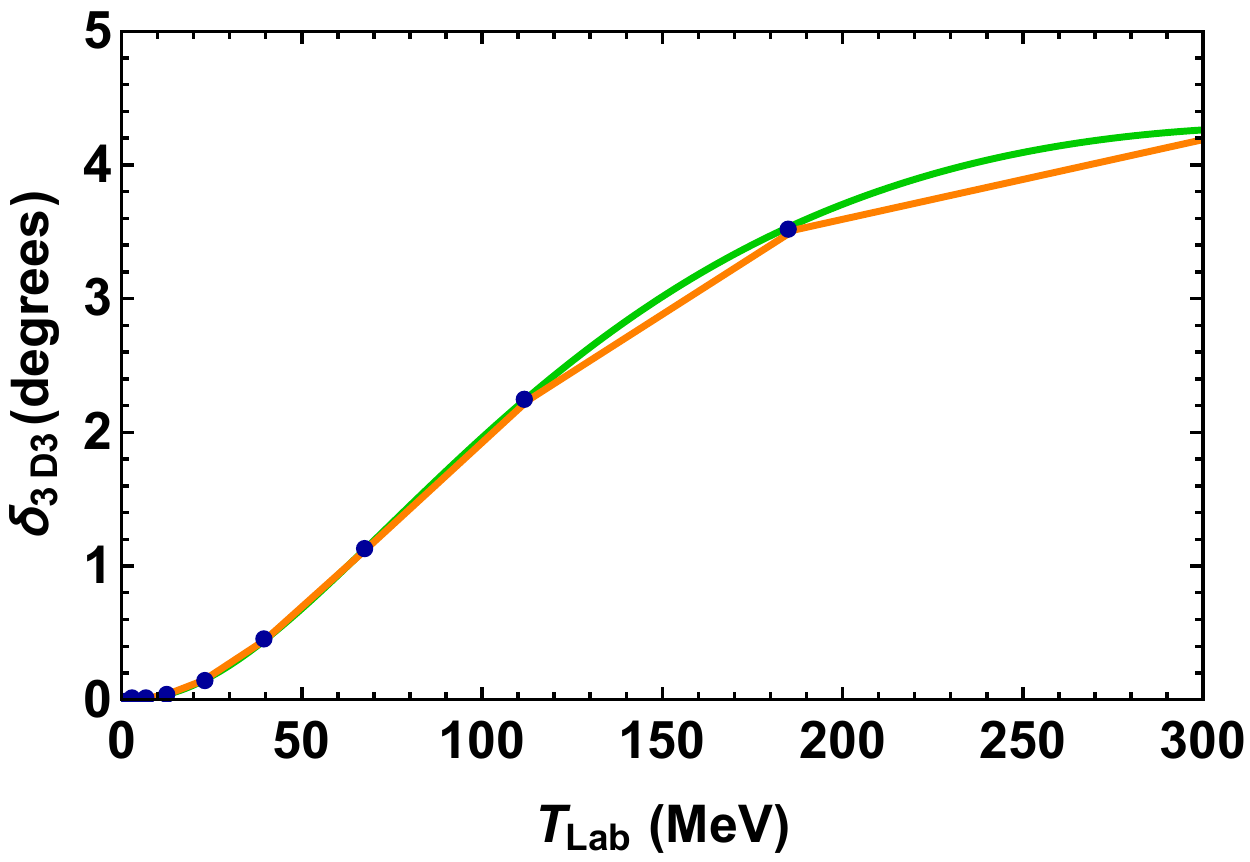}
\includegraphics[scale=0.45]{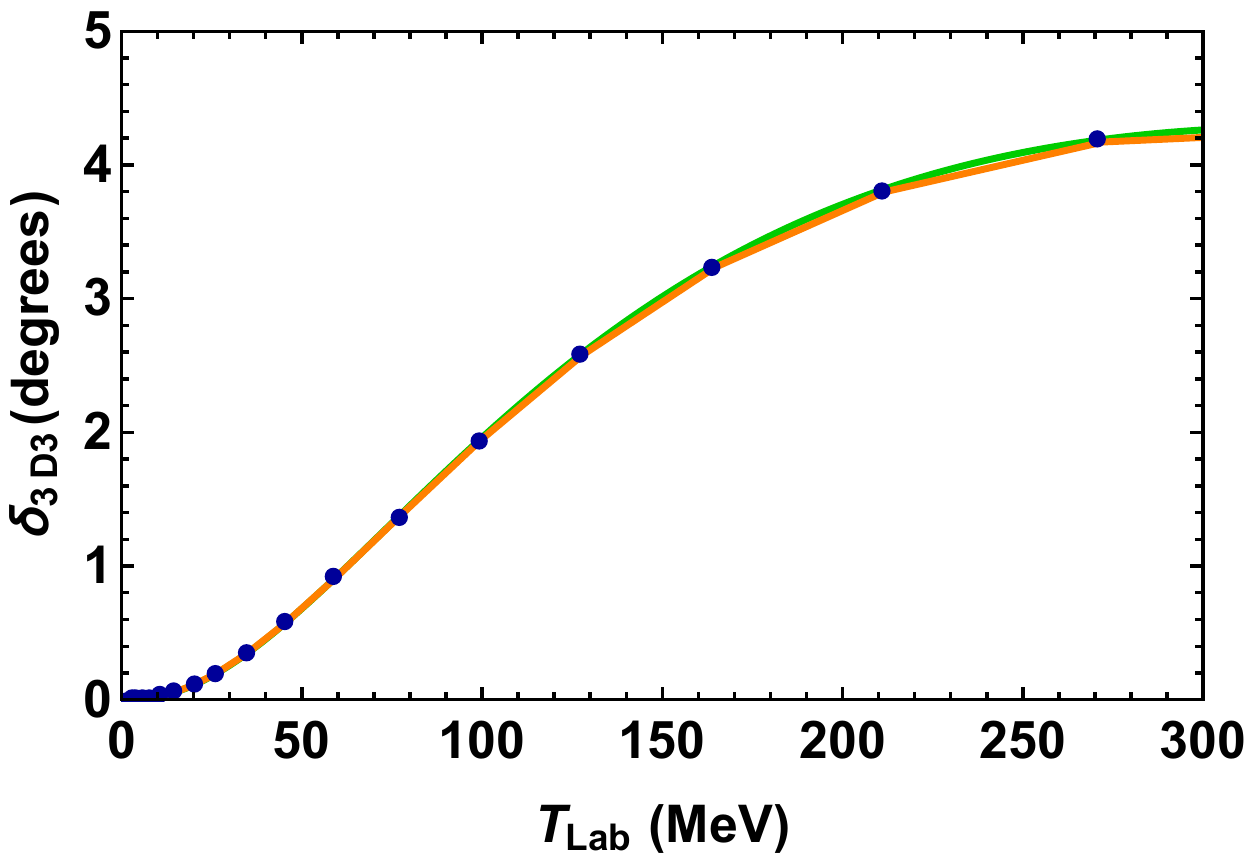}
\includegraphics[scale=0.45]{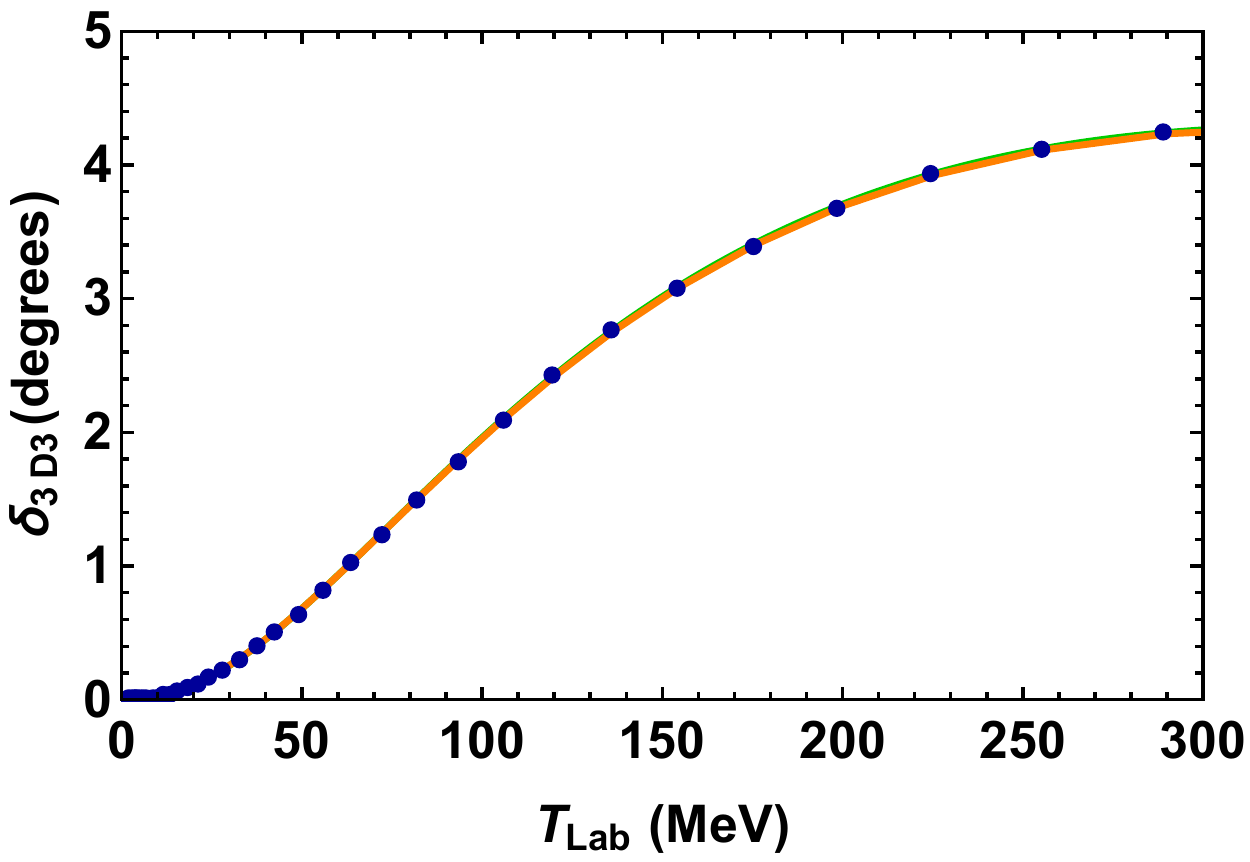}

\includegraphics[scale=0.45]{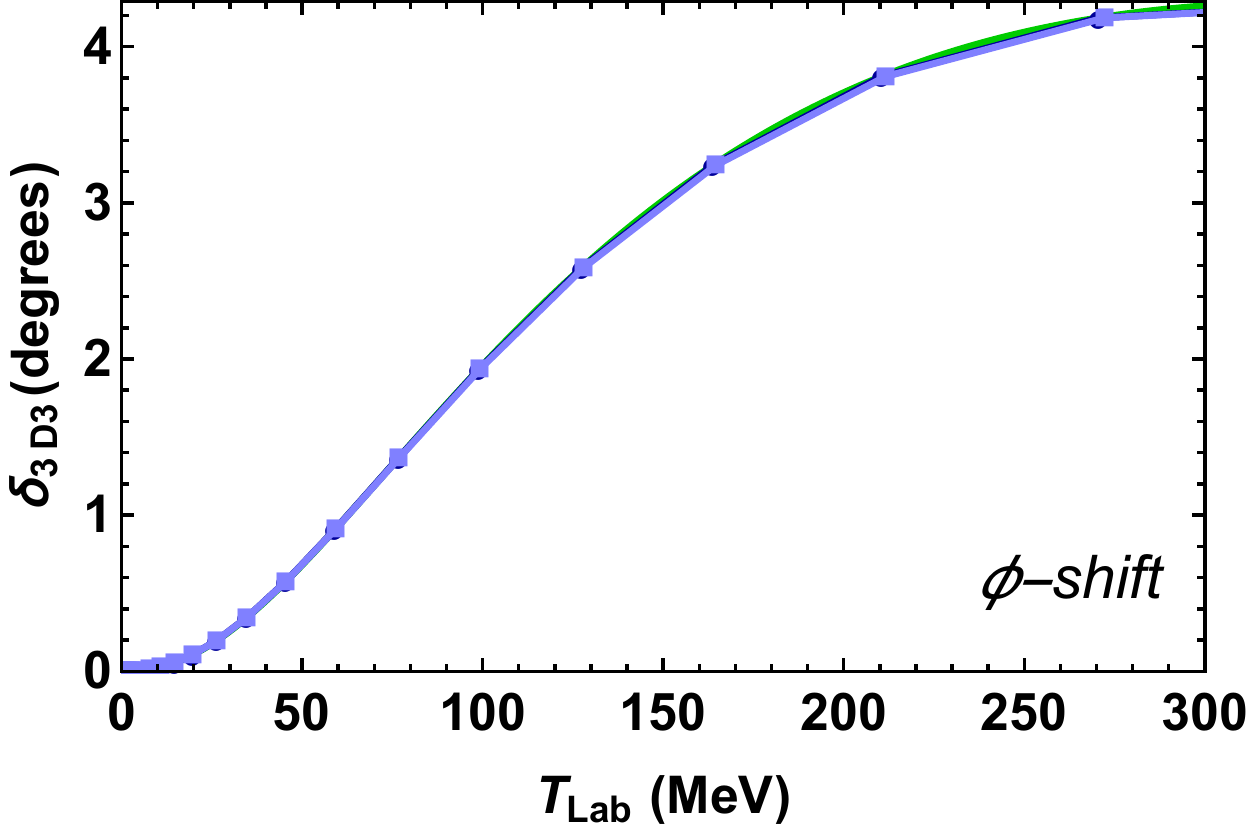}
\includegraphics[scale=0.45]{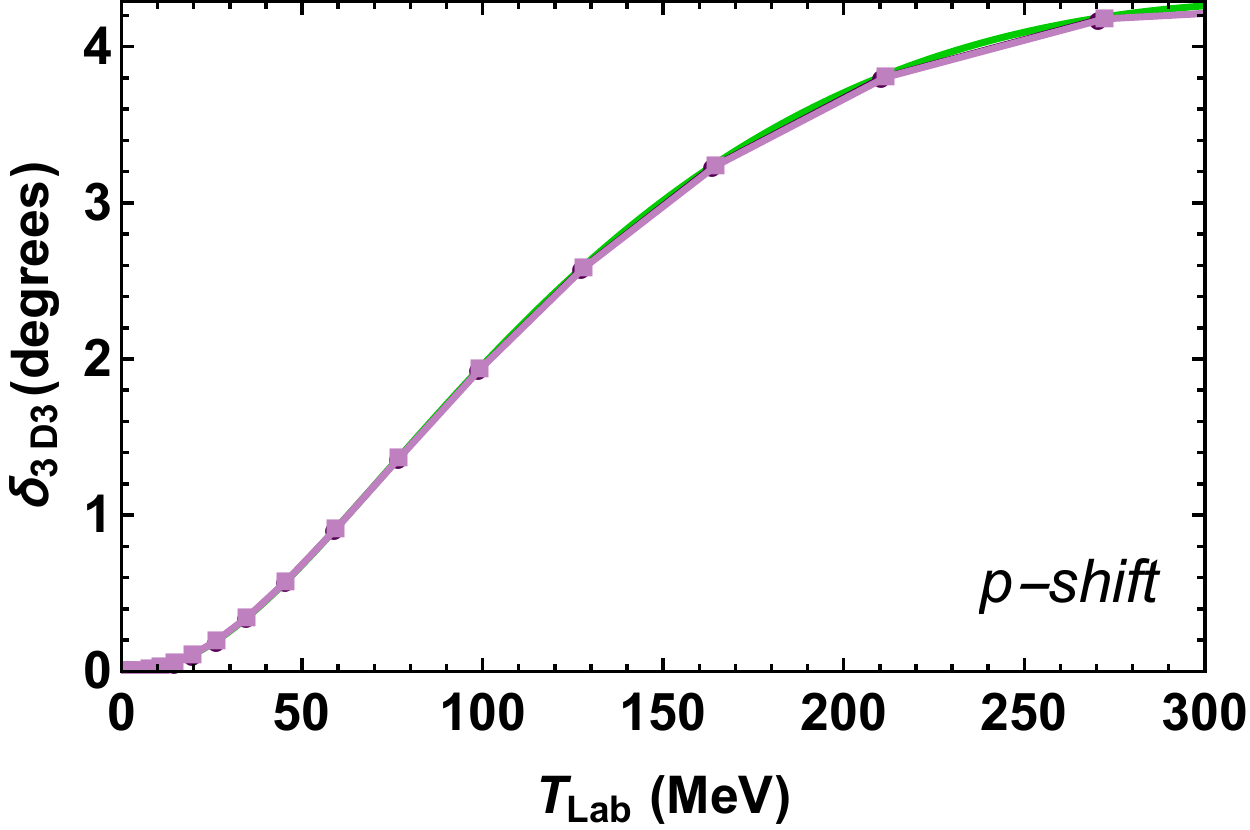}
\includegraphics[scale=0.45]{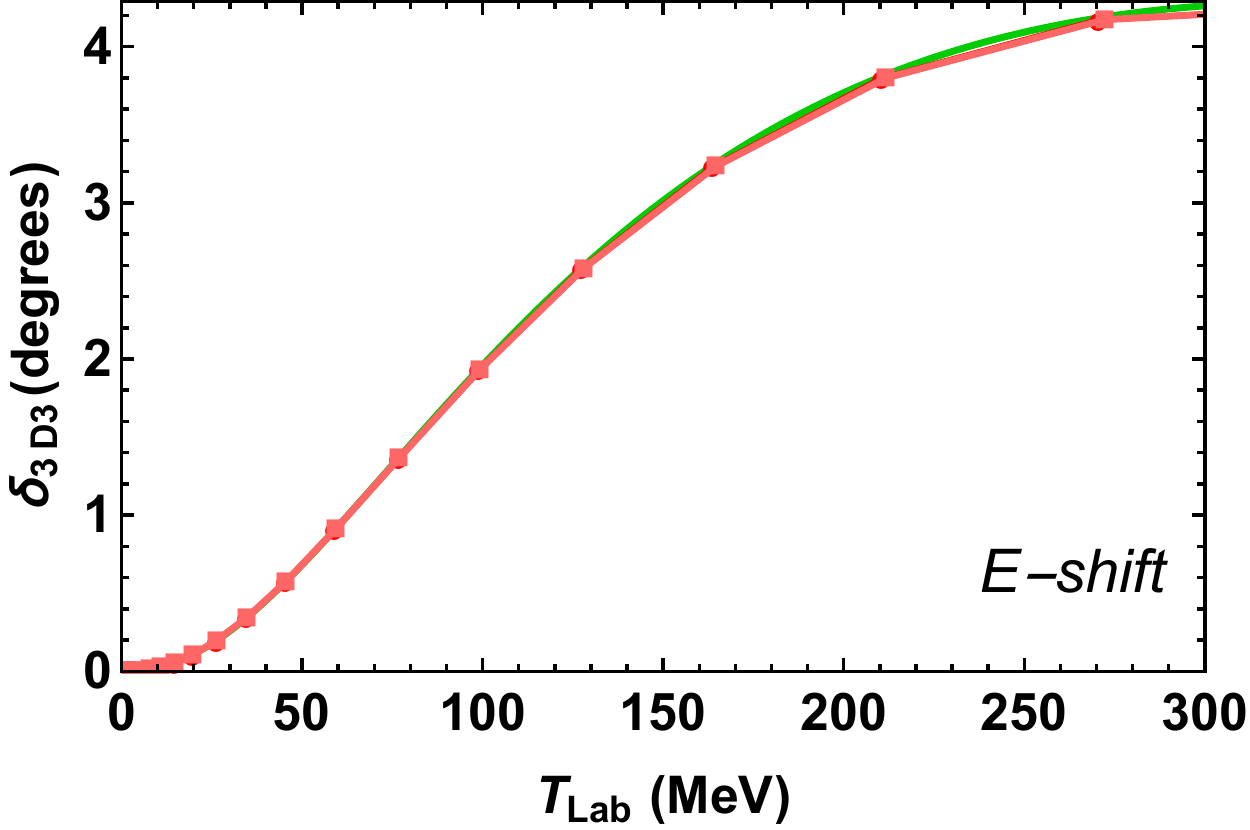}

\caption{ The same as in Figure~\ref{fig:1p1} but for the $ ^3 D_3$ channel.}
\label{fig:3d3}
\end{figure*}

Figures~\ref{fig:1p1}-\ref{fig:3d3} show the obtained resuls for $NN$-scattering, where the form factors for separable potentials are taken from~\cite{Mathelitsch:1986ez} and are given in Appendix~\ref{app:potentials}. The phase shifts are plotted as a function of $T_{\text{Lab}}$.

The first row of each of these figures shows the $\phi$-shift result,
compared with the LS results and with the exact solution for a grid of
$N=25$, 50, and 100 points, respectively.  The second row shows the
result obtained using the $\phi$-, $p$- and $E$-shift, as labeled in
the corner, for a momentum grid of $N=50$ points. In this case, the
proton mass is not negligible, and hence Eqs.~(\ref{eq:pshift}) and
(\ref{eq:dewitt}) are no longer equivalent, and the numerical
difference can be appreciated (see e.g. Figure~\ref{fig:3d2}). In
analogy to what has been done in the $\pi\pi$ analysis, we use a
darker like with round markers to represent the results when using the
interacting momentum as the implicit independent variable, and a
lighter line with square markers when we use the free momentum as the
independent variable in $T_{\text{Lab}}$.  We have selected in some
cases an interval where the difference between lines is more visible.

In the studied interval, $ 0 \leq T_{\text{Lab}}\leq 300$, the phase shifts do not reach values higher than around 30 degrees, so that there are no abrupt changes in the curves and, as a consequence, the deviation from results obtained in one or other method is not significant. 

\begin{figure*}

\includegraphics[scale=0.45]{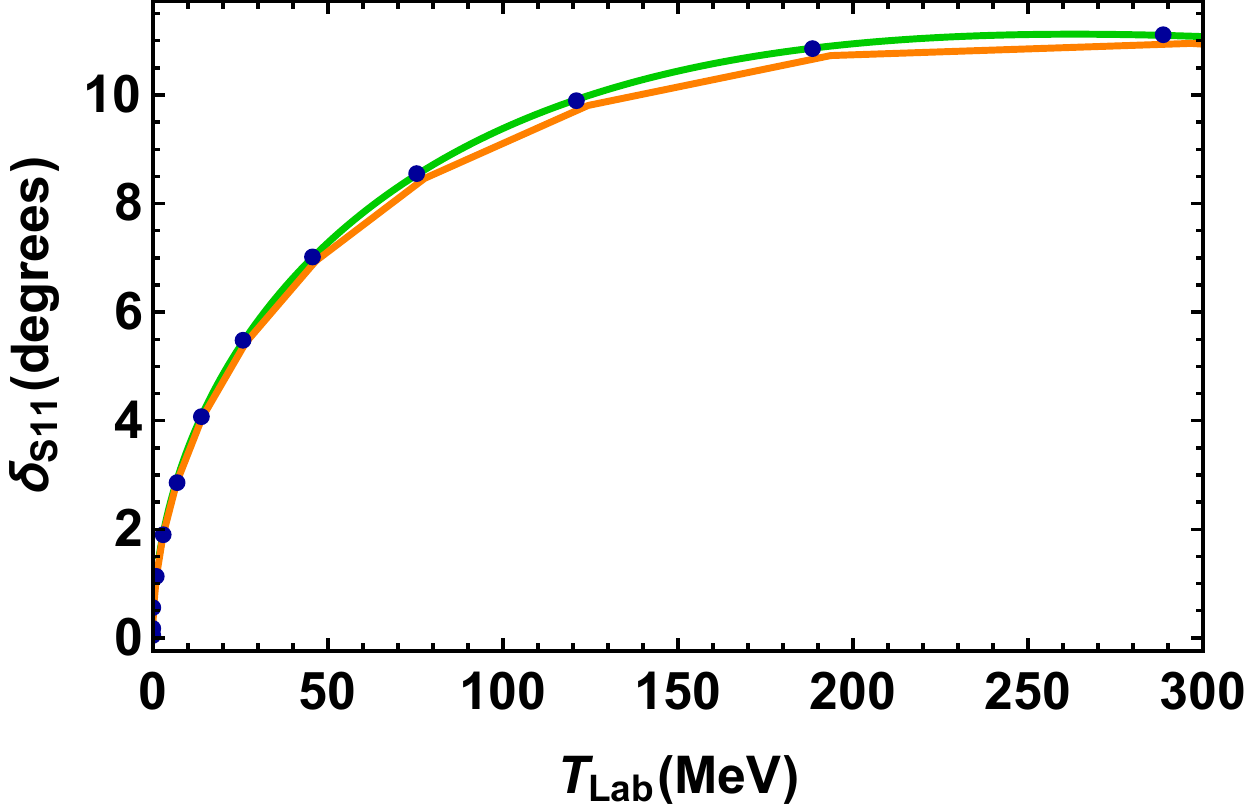}
\includegraphics[scale=0.45]{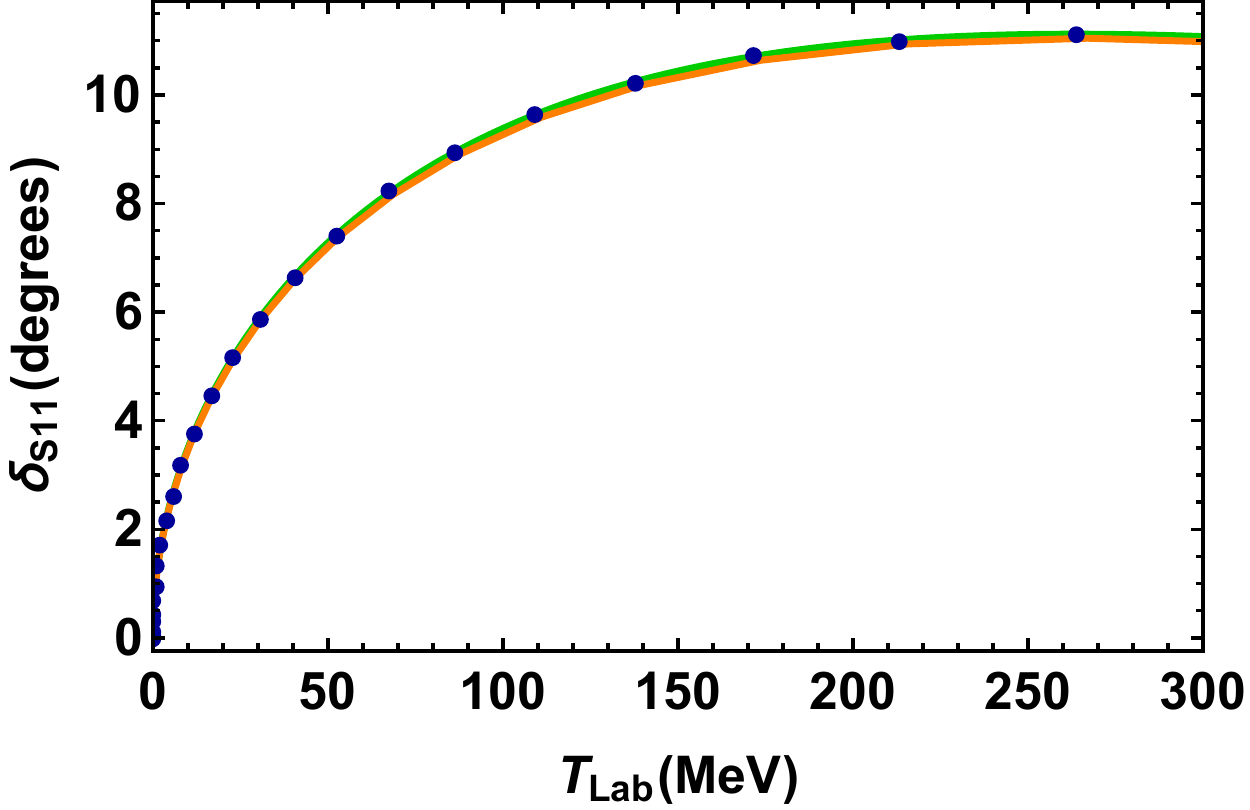}
\includegraphics[scale=0.45]{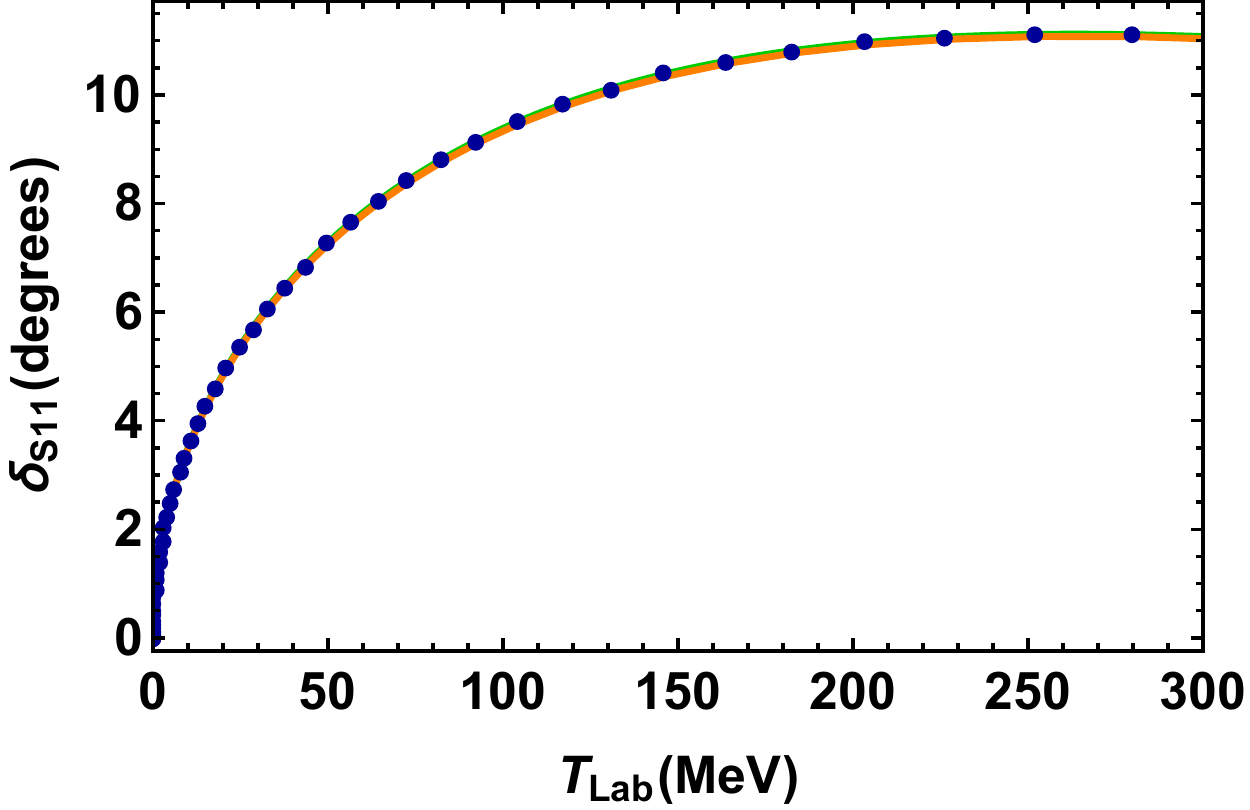}

\includegraphics[scale=0.45]{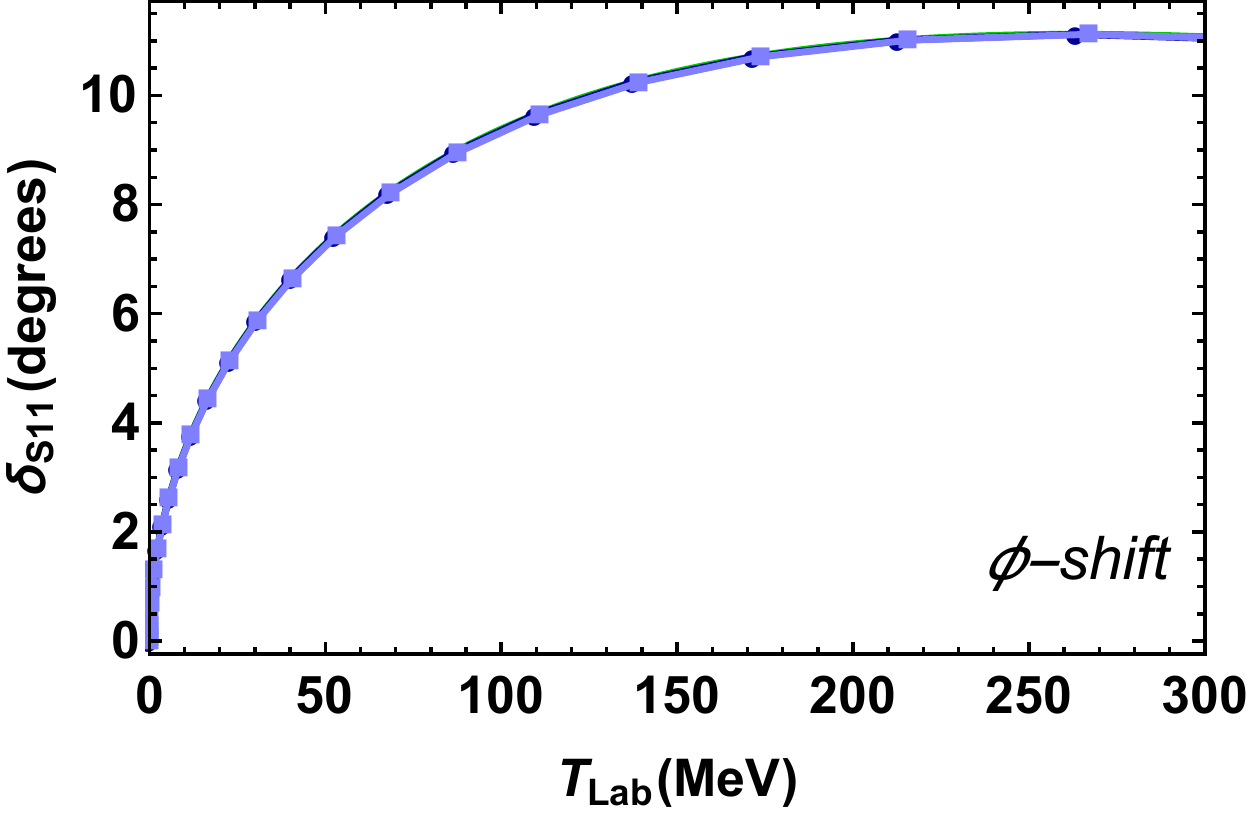}
\includegraphics[scale=0.45]{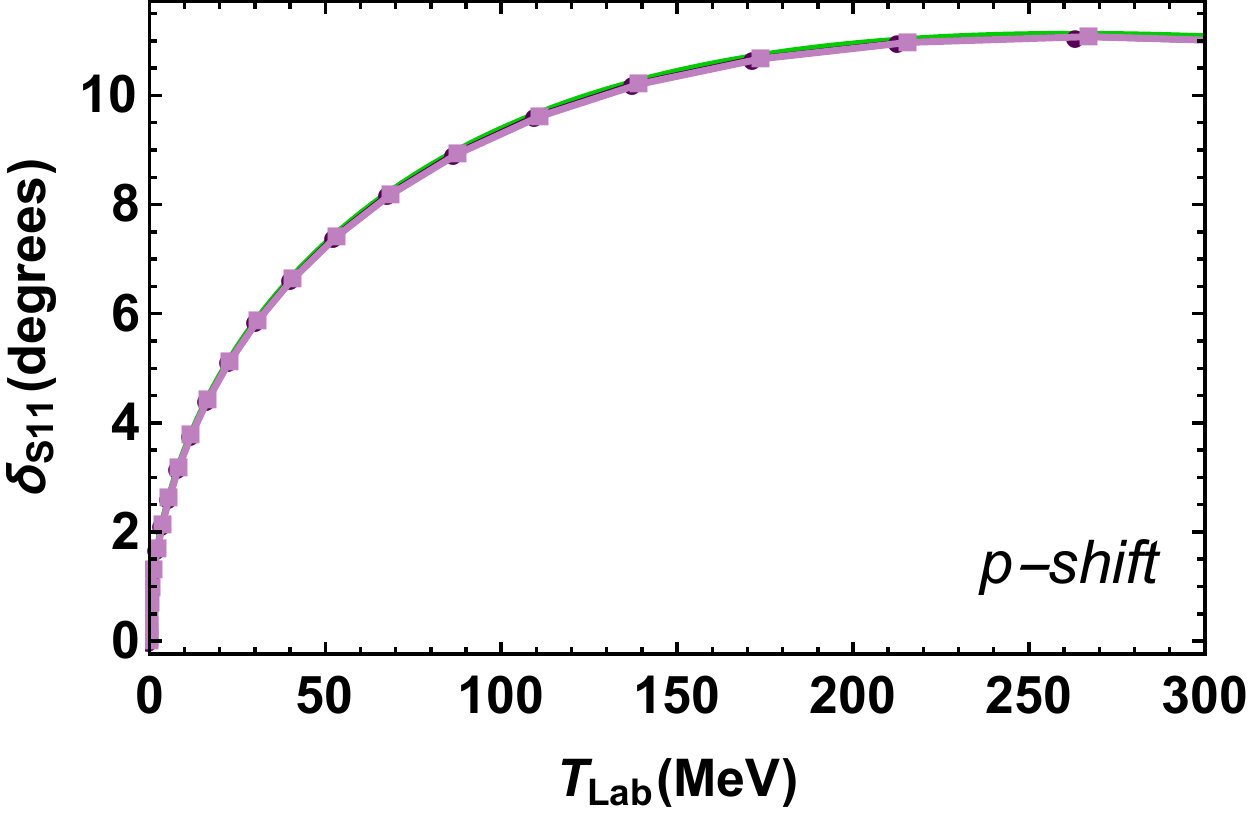}
\includegraphics[scale=0.45]{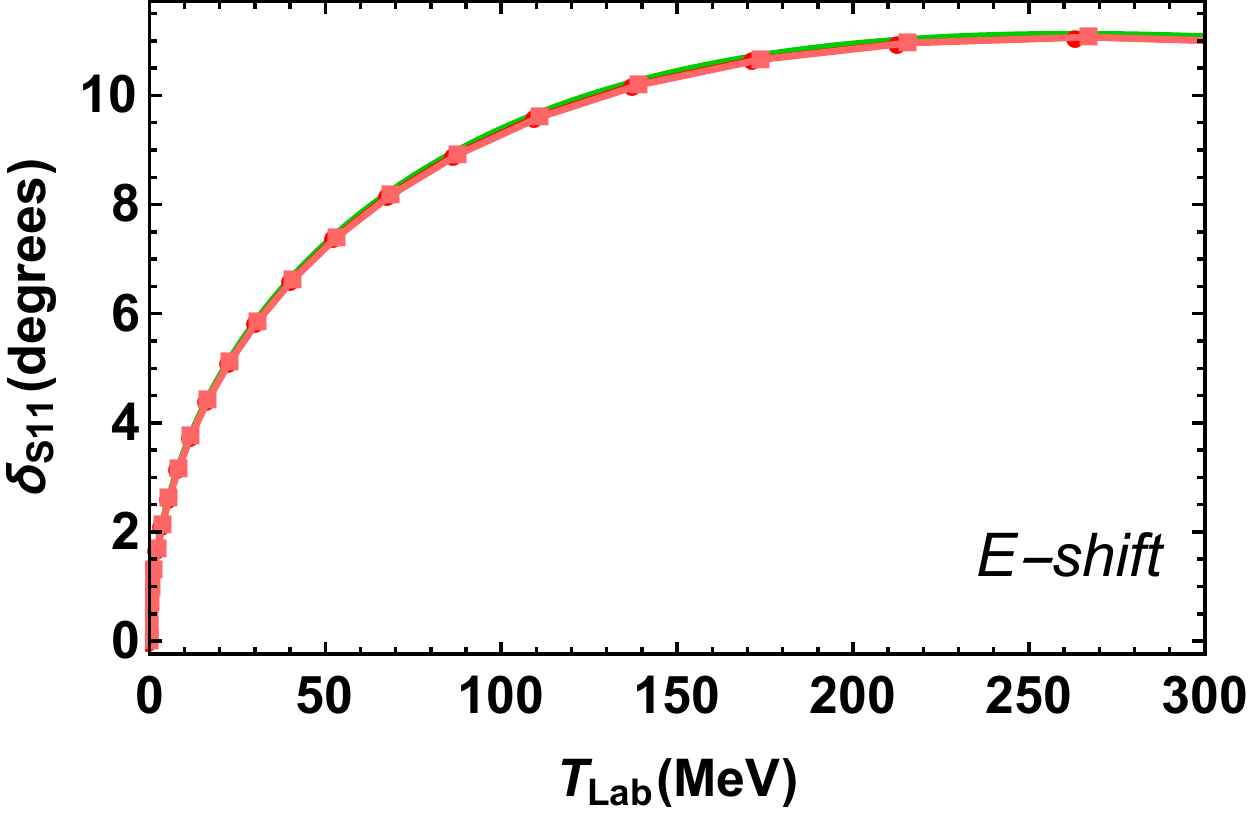}

\caption{ The same as in Figure~\ref{fig:1p1} but for $\pi N$ scattering in the $S_{11}$ channel.}
\label{fig:pins11}
\end{figure*}

\begin{figure*}

\includegraphics[scale=0.45]{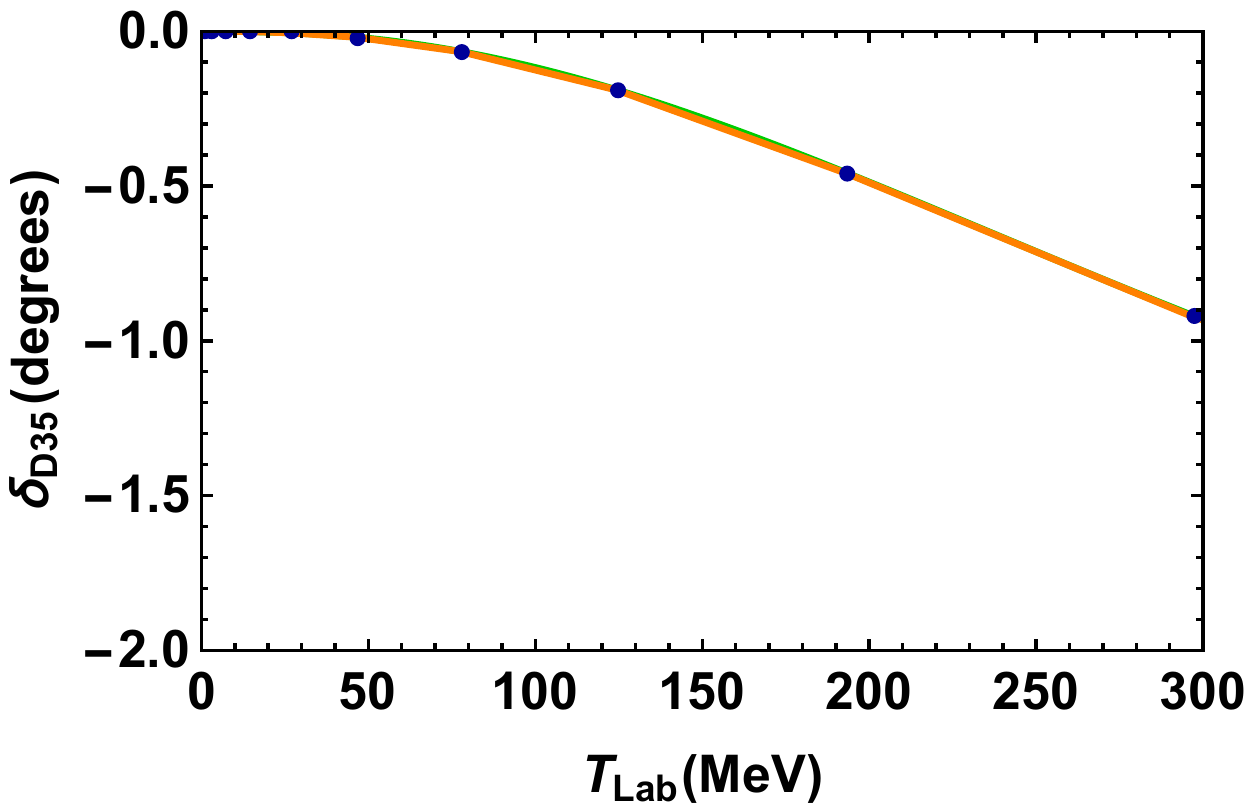}
\includegraphics[scale=0.45]{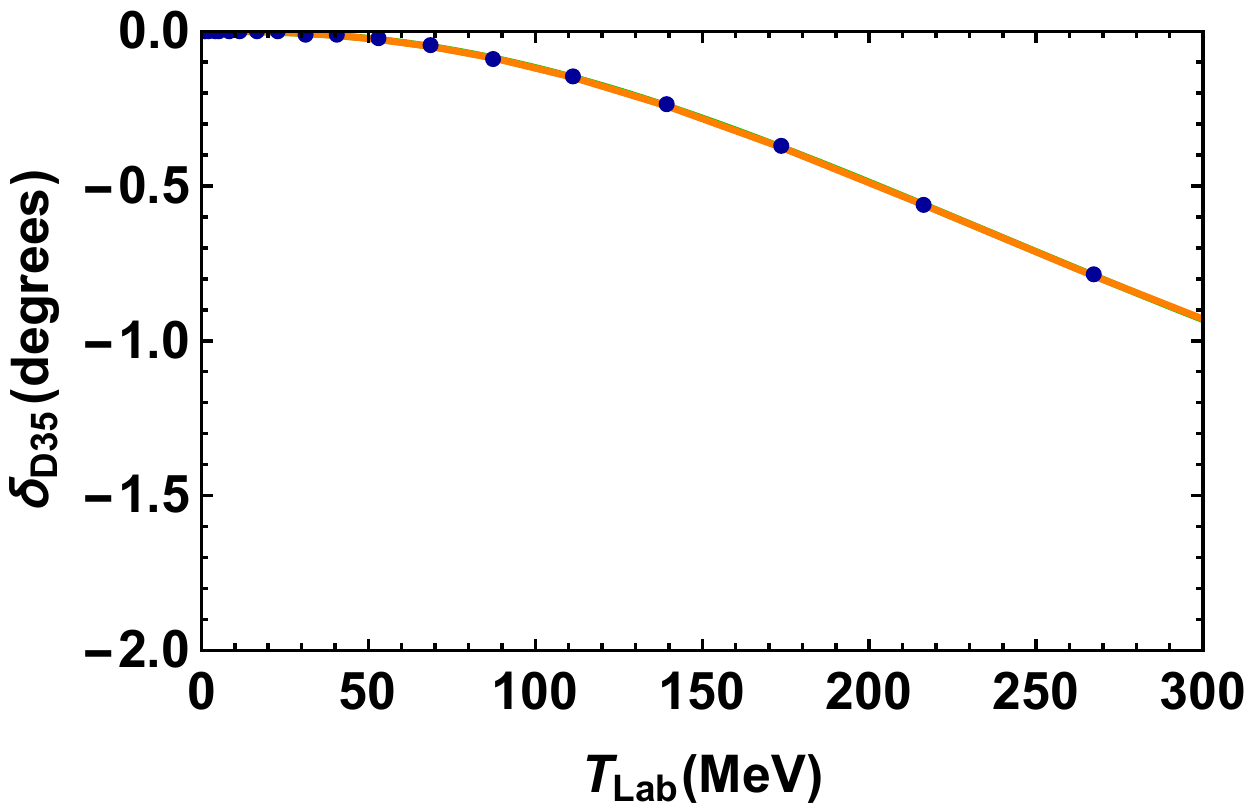}
\includegraphics[scale=0.45]{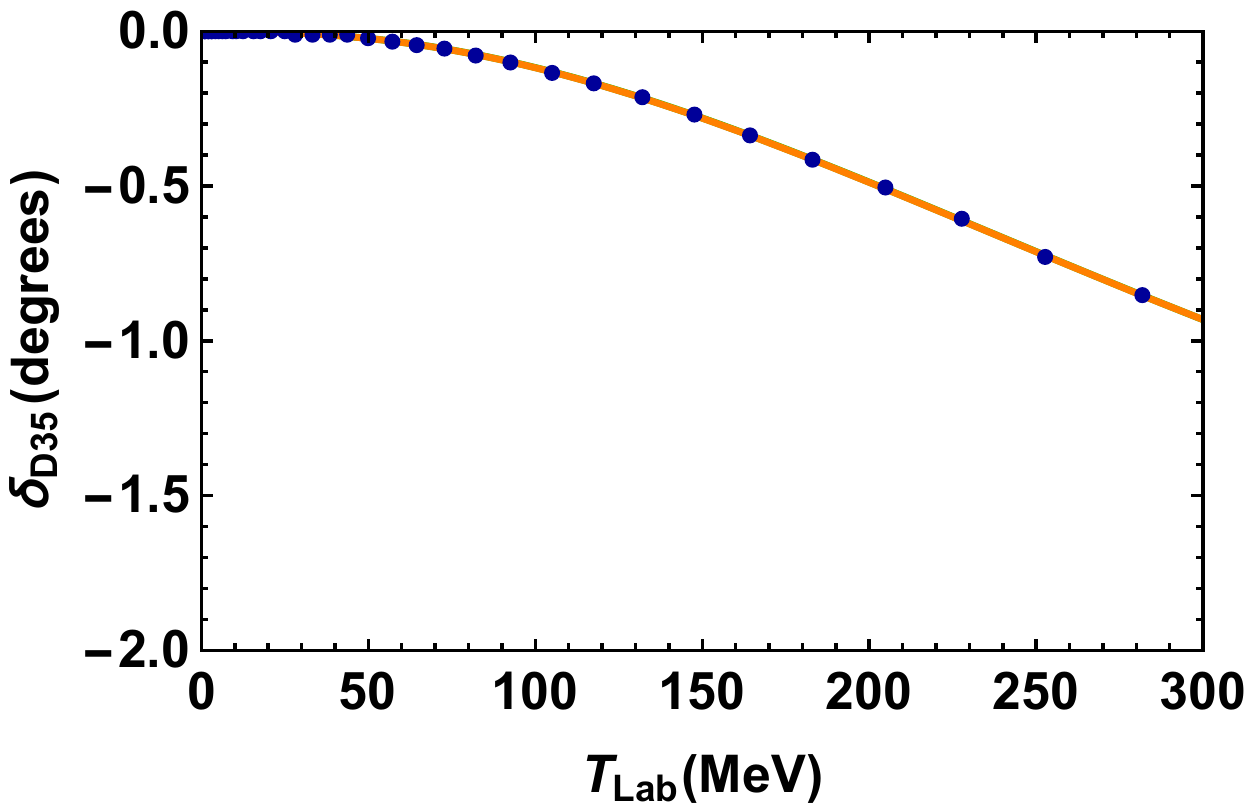}

\includegraphics[scale=0.45]{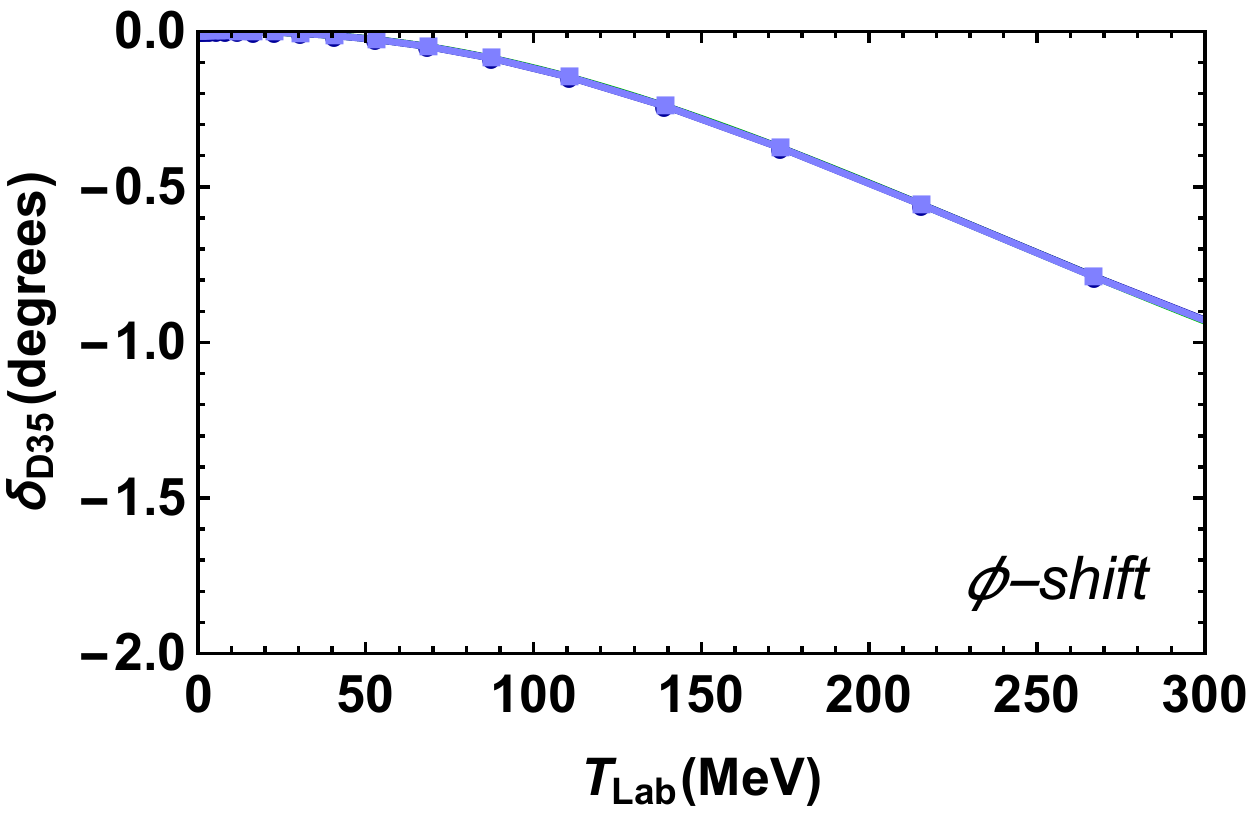}
\includegraphics[scale=0.45]{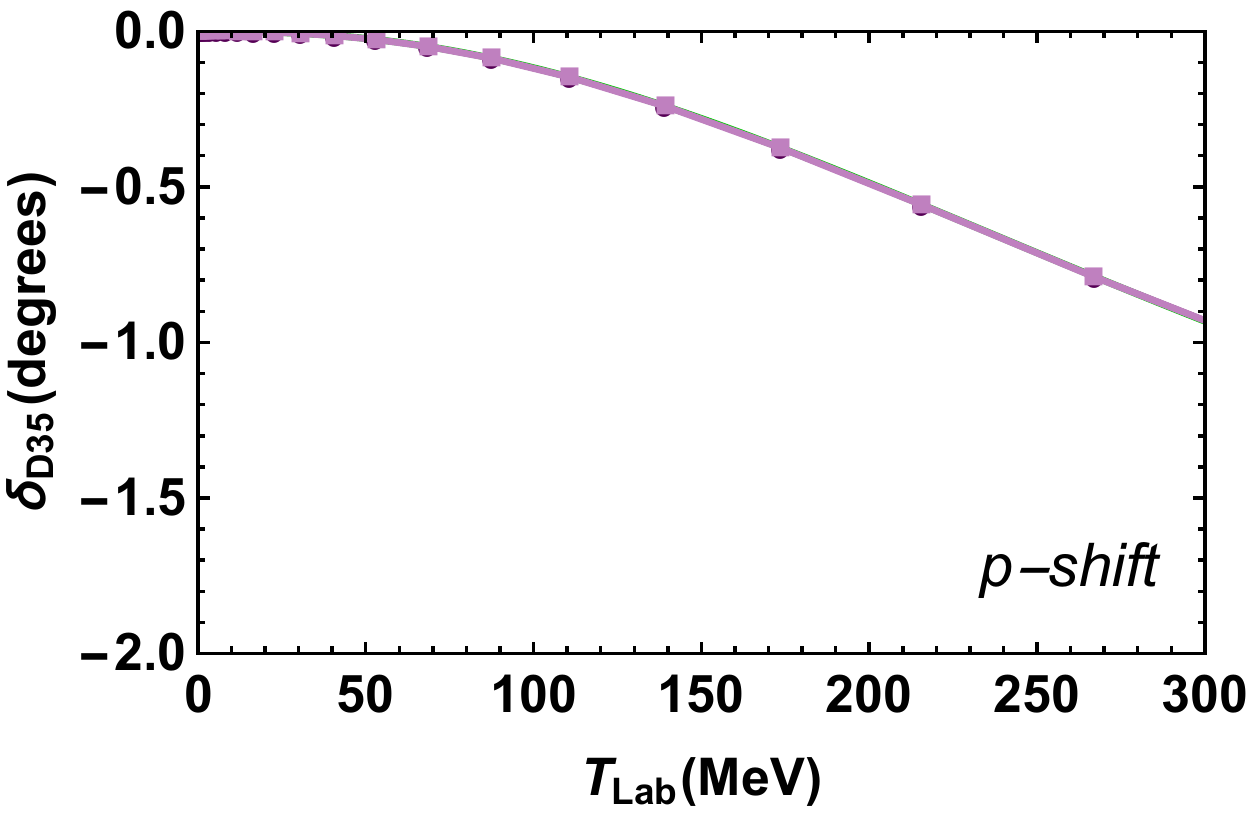}
\includegraphics[scale=0.45]{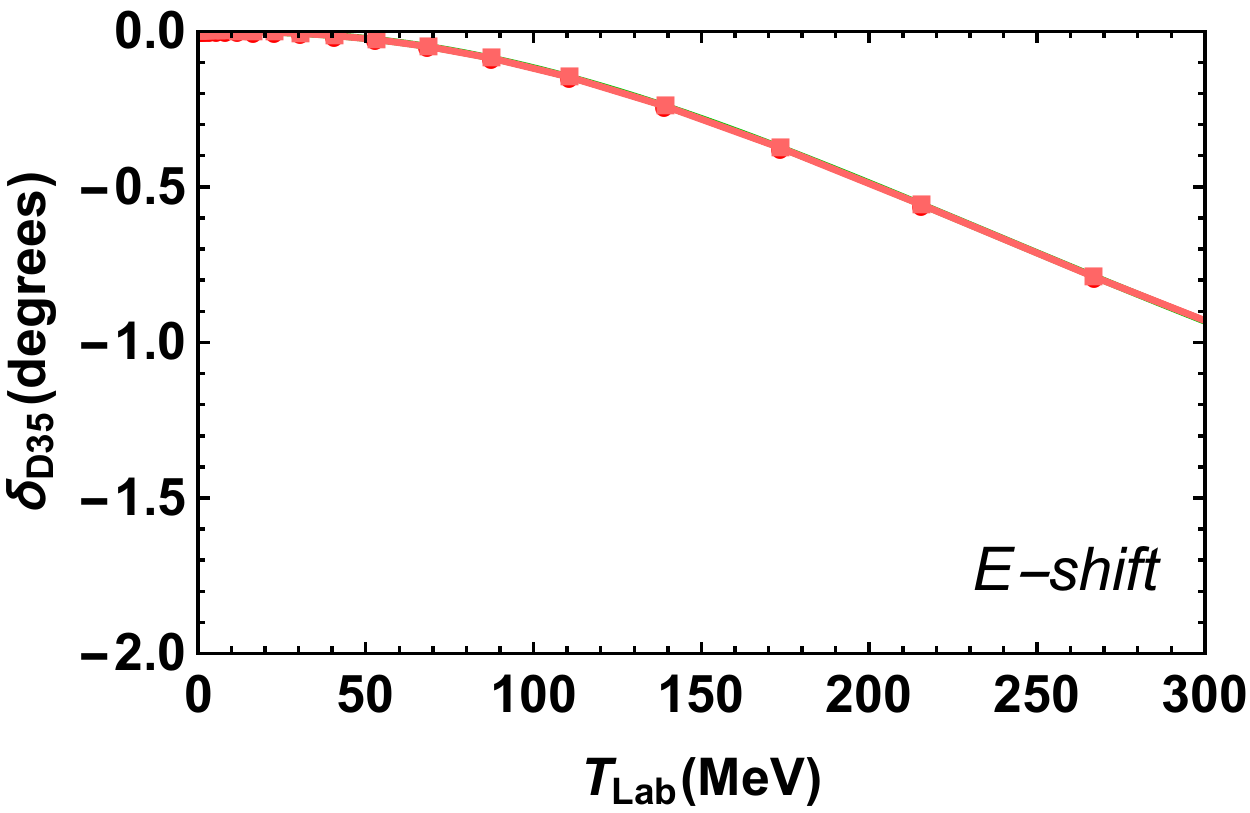}

\caption{ The same as in Figure~\ref{fig:1p1} but for $\pi N$ scattering in the $D_{35}$ channel.}
\label{fig:pind35}
\end{figure*}

Figures~\ref{fig:pins11}-\ref{fig:pind35} show the phase shifts calculated for $\pi N$ scattering. 
Alike in the $NN$ case, Eqs.~(\ref{eq:pshift}) and (\ref{eq:dewitt}) are not equivalent due to the large mass of the proton involved. But one can hardly appreciate the difference from the numerical results due to the very small range of values that the phase shifts take in most of the channels, with the exception of the $P_{33}$ wave, which reaches from 0 to around 120 degrees. 

\begin{figure*}

\includegraphics[scale=0.45]{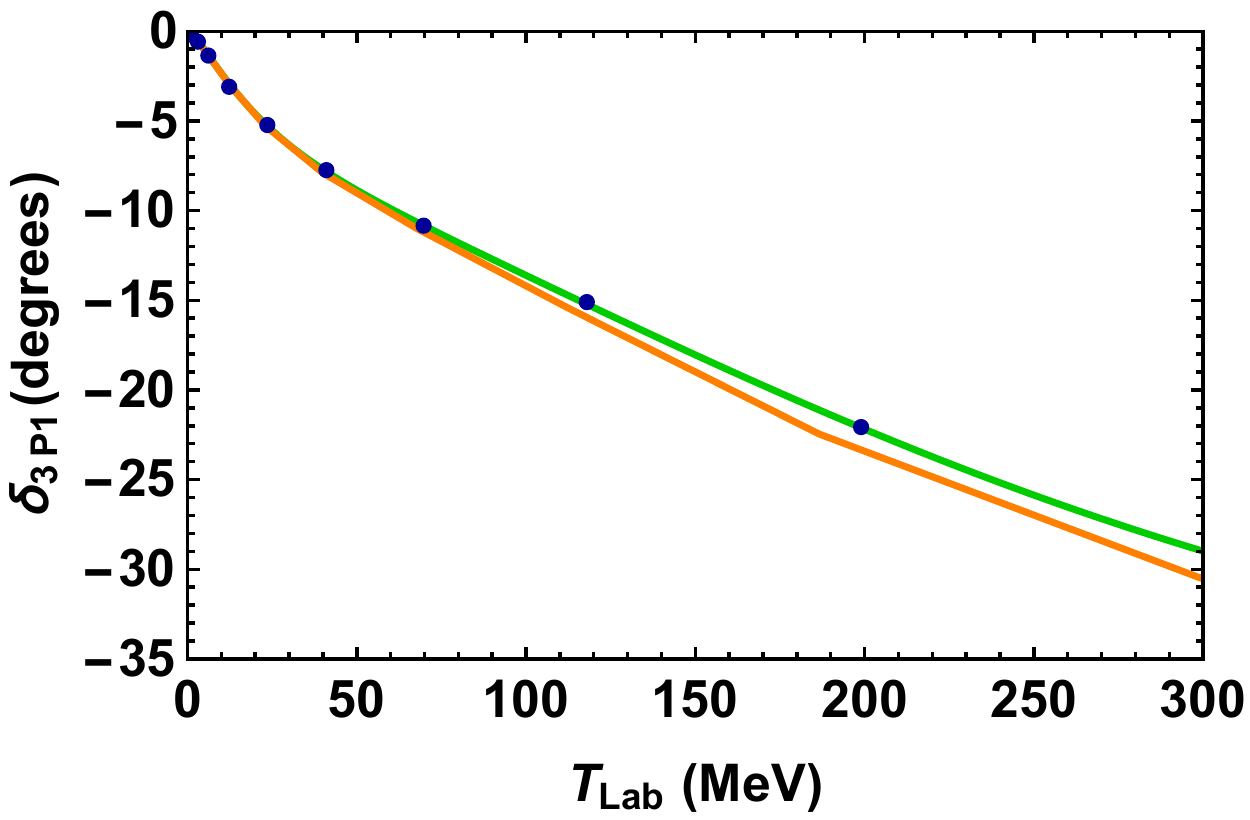}
\includegraphics[scale=0.45]{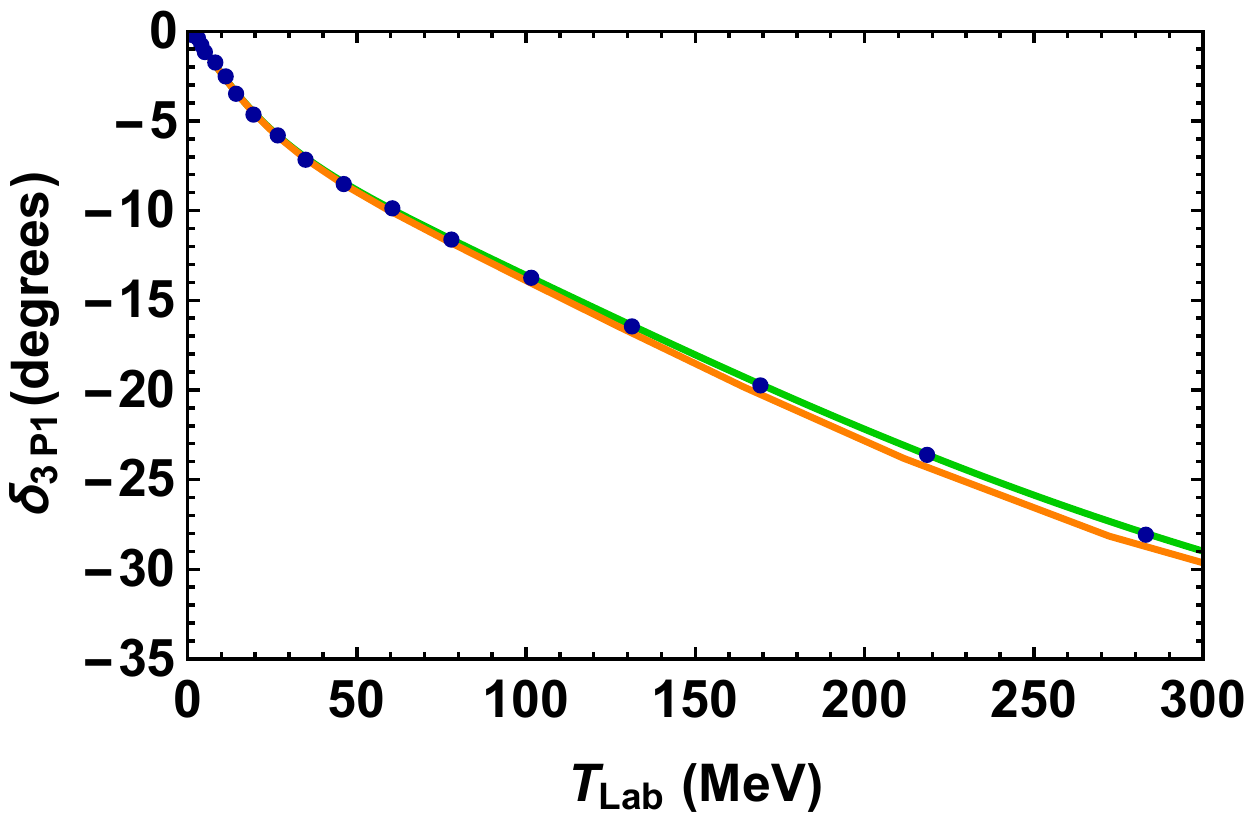}
\includegraphics[scale=0.45]{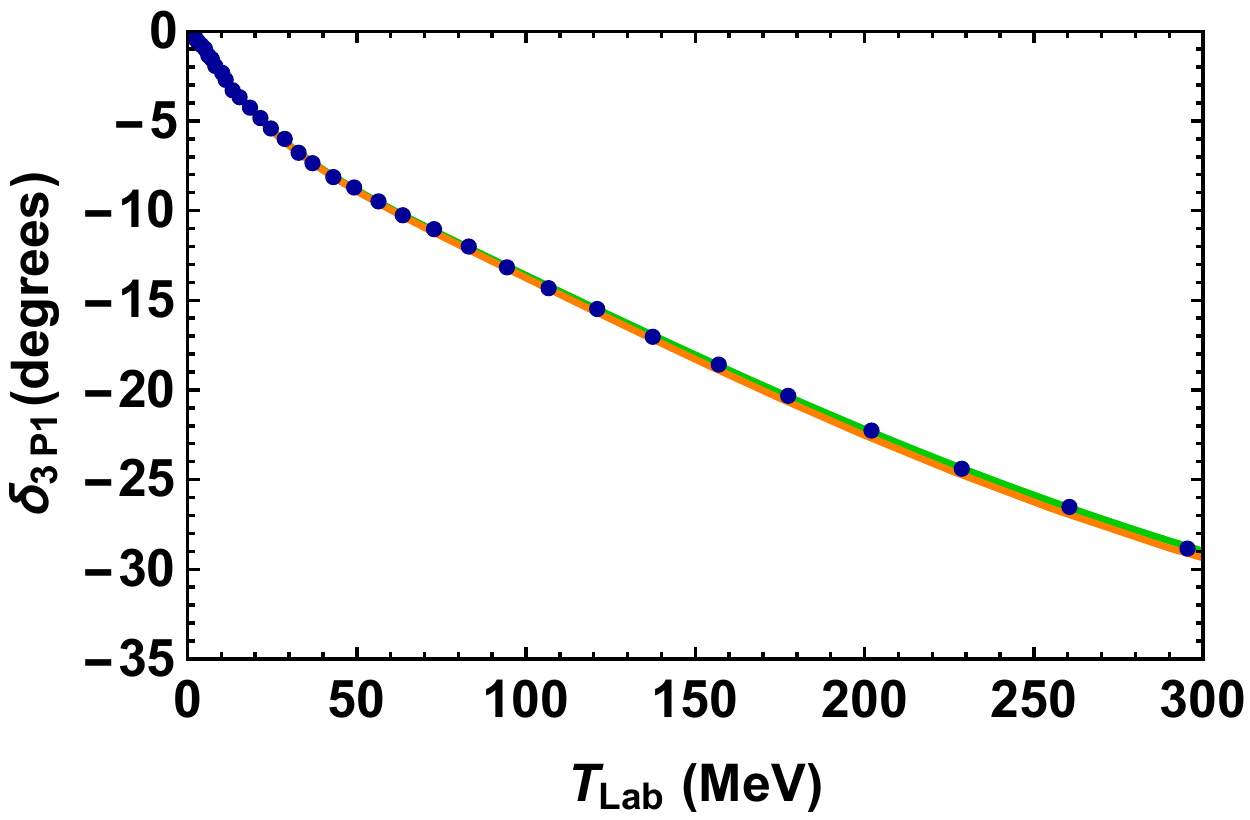}

\includegraphics[scale=0.45]{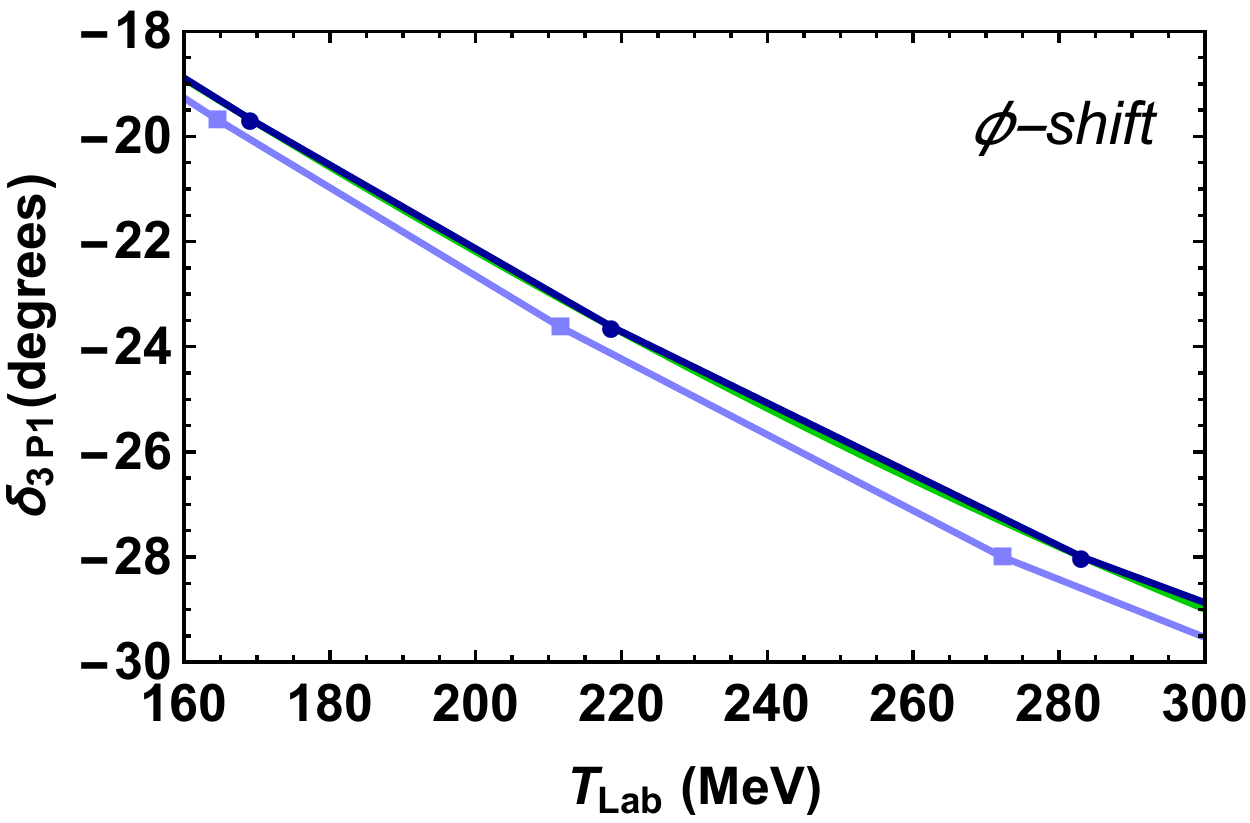}
\includegraphics[scale=0.45]{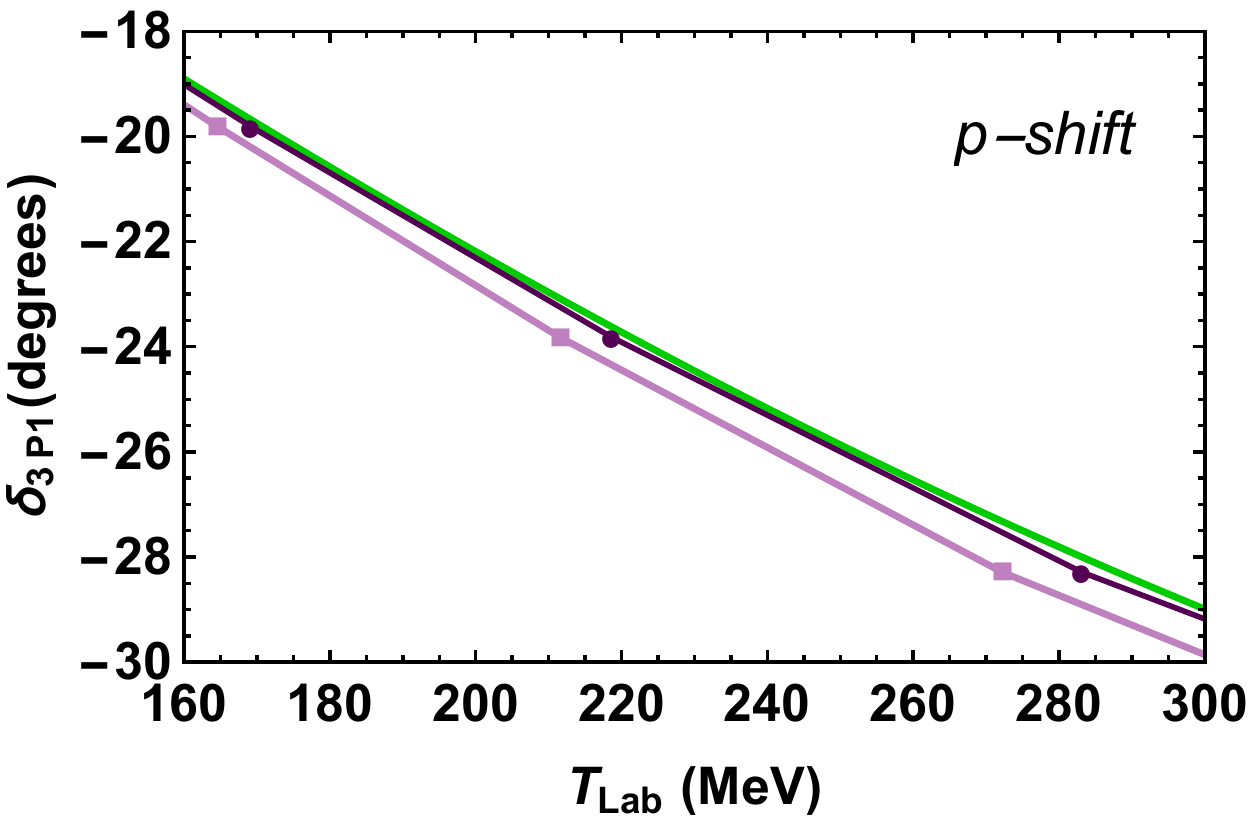}
\includegraphics[scale=0.45]{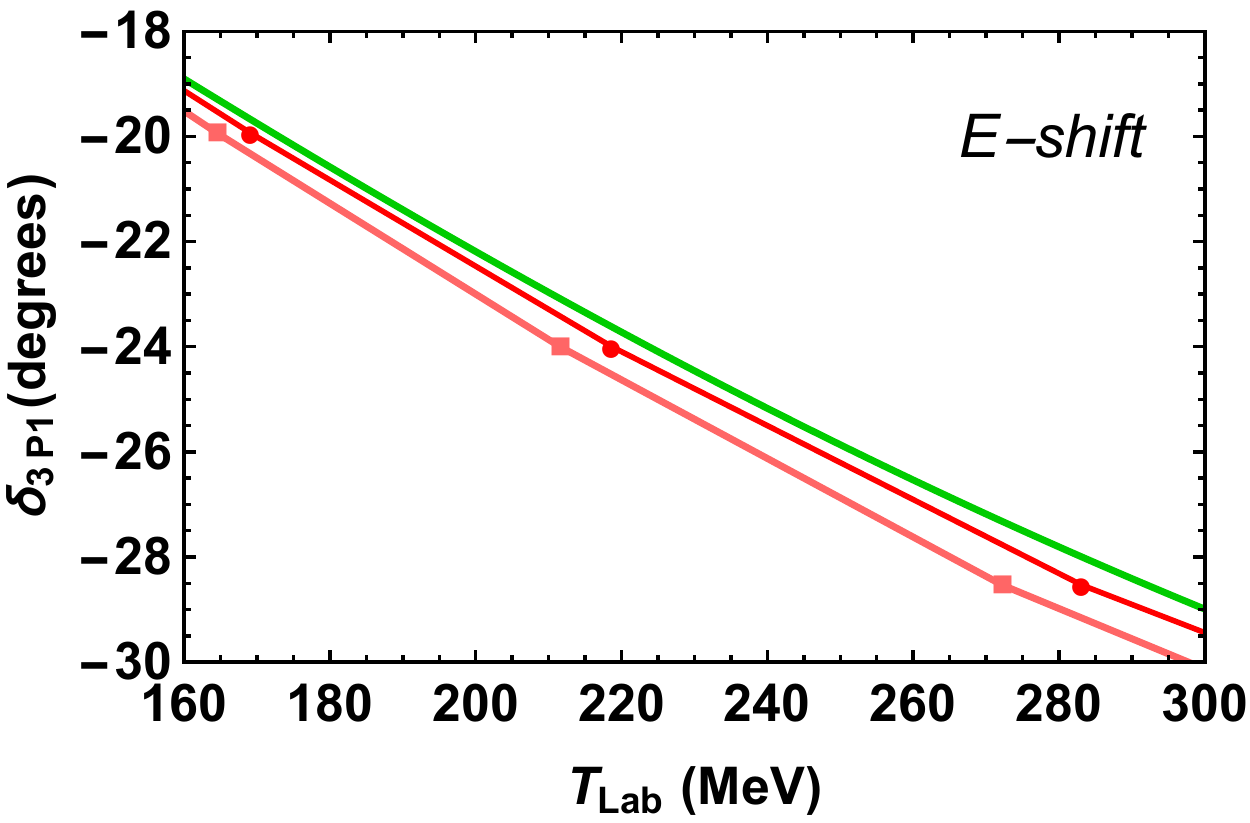}

\caption{ The same as in Figure~\ref{fig:1p1} but for the $ ^3 P_1$ channel.  }
\end{figure*}

\begin{figure*}

\includegraphics[scale=0.45]{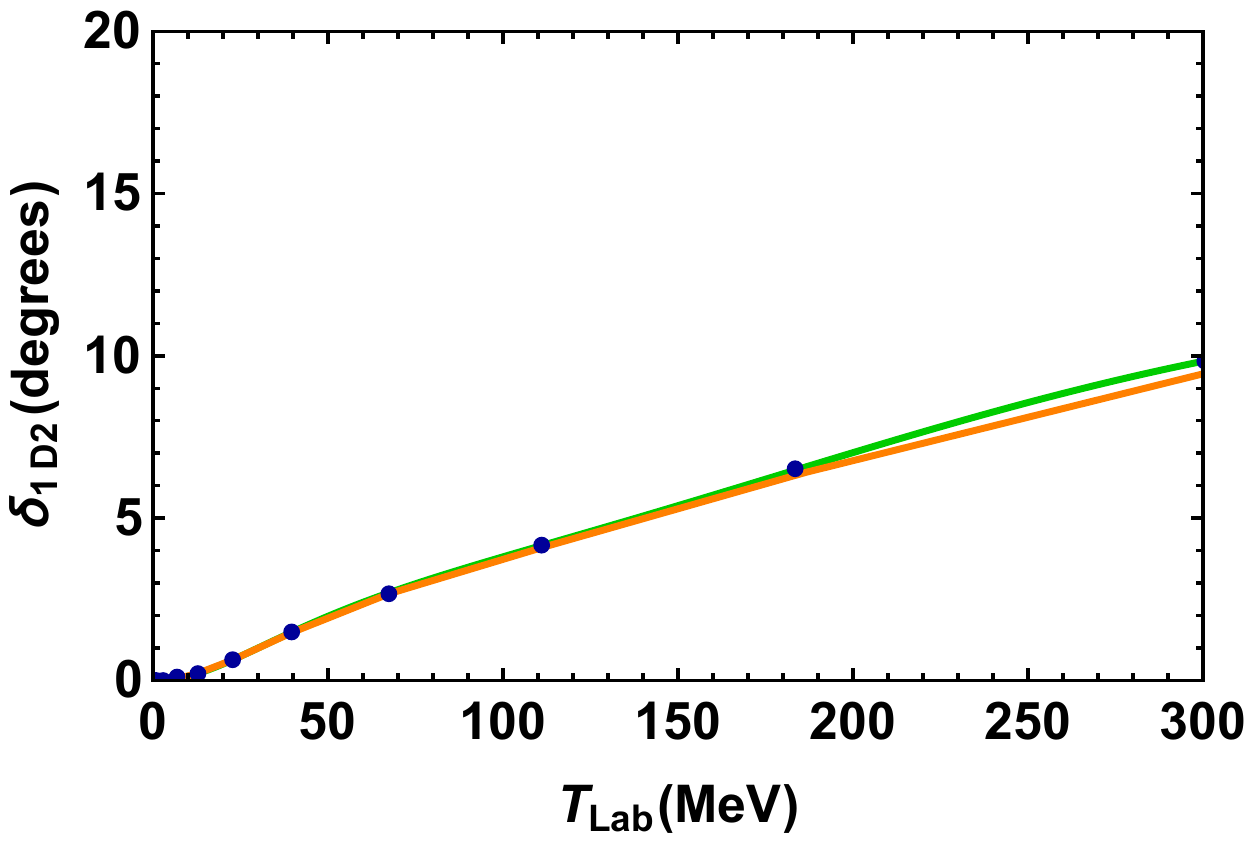}
\includegraphics[scale=0.45]{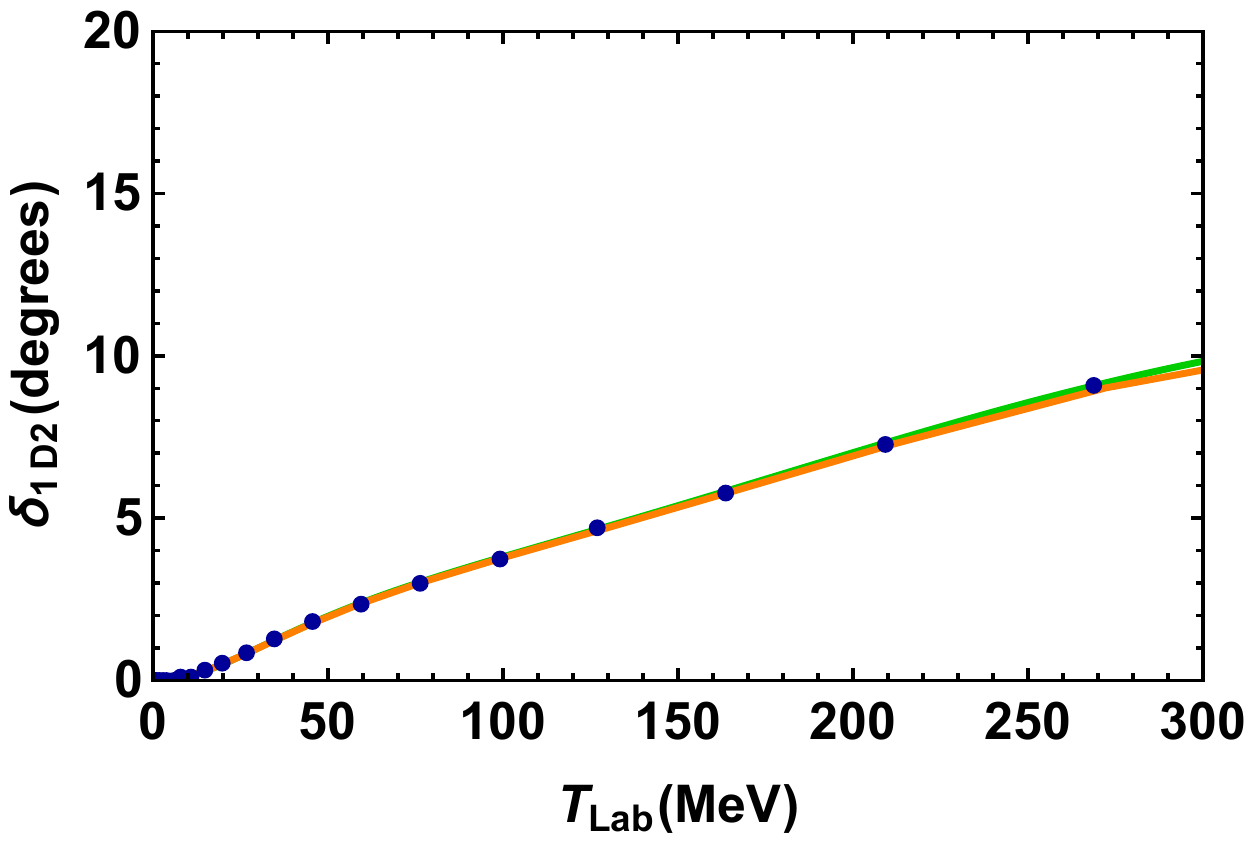}
\includegraphics[scale=0.45]{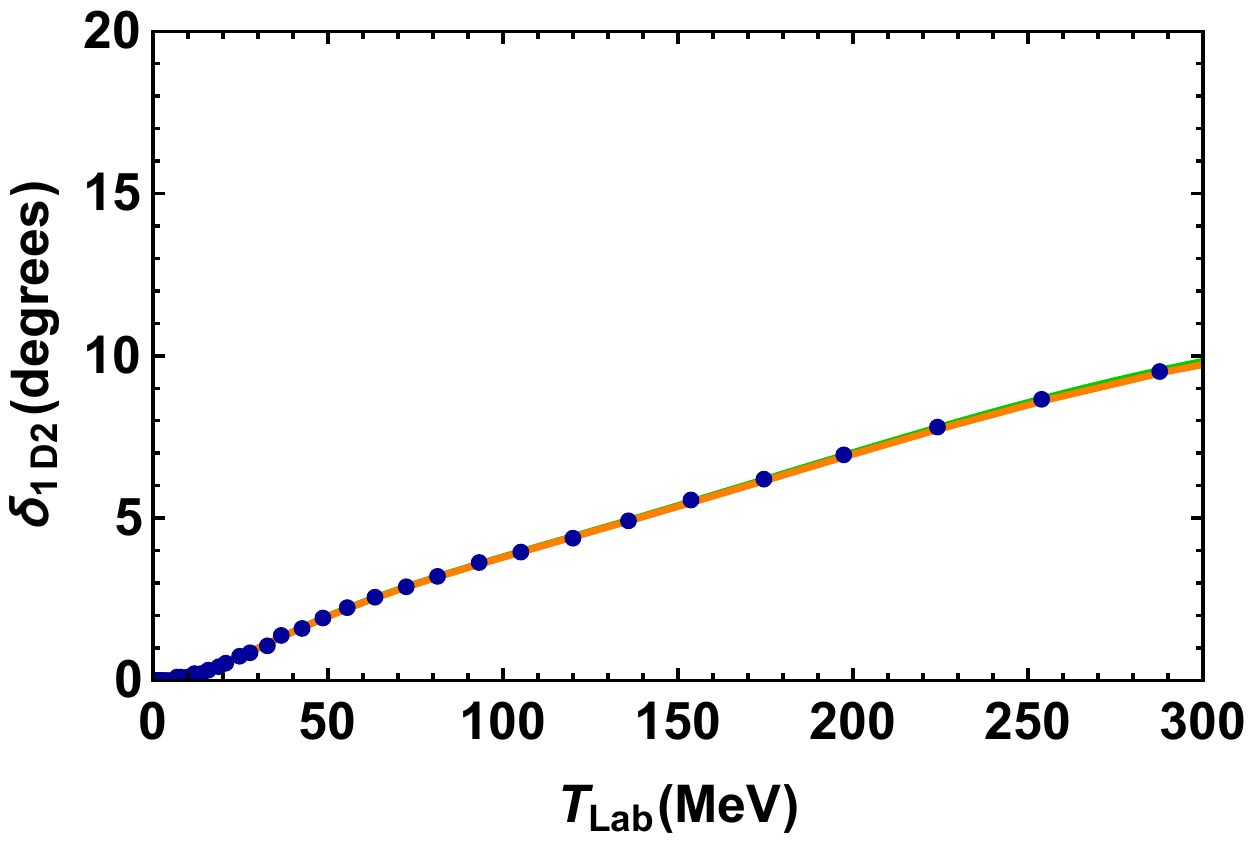}

\includegraphics[scale=0.45]{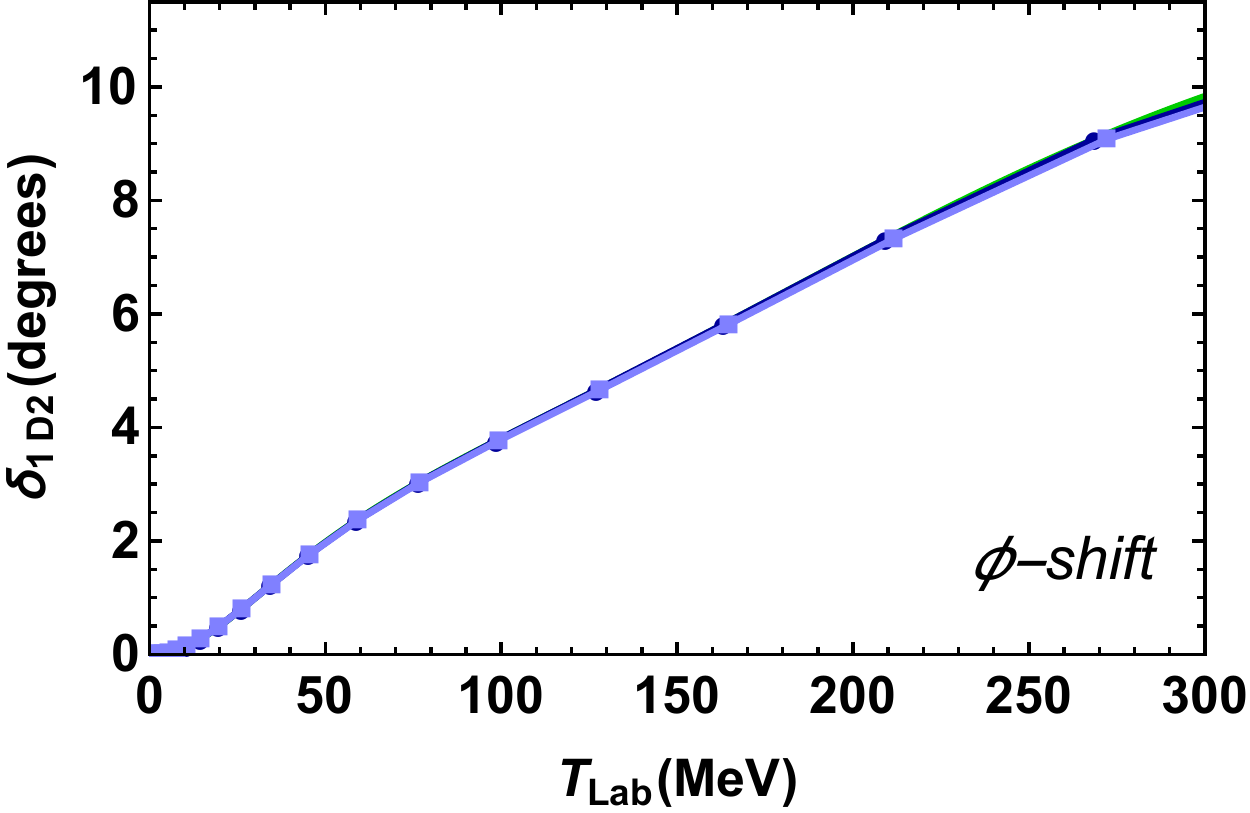}
\includegraphics[scale=0.45]{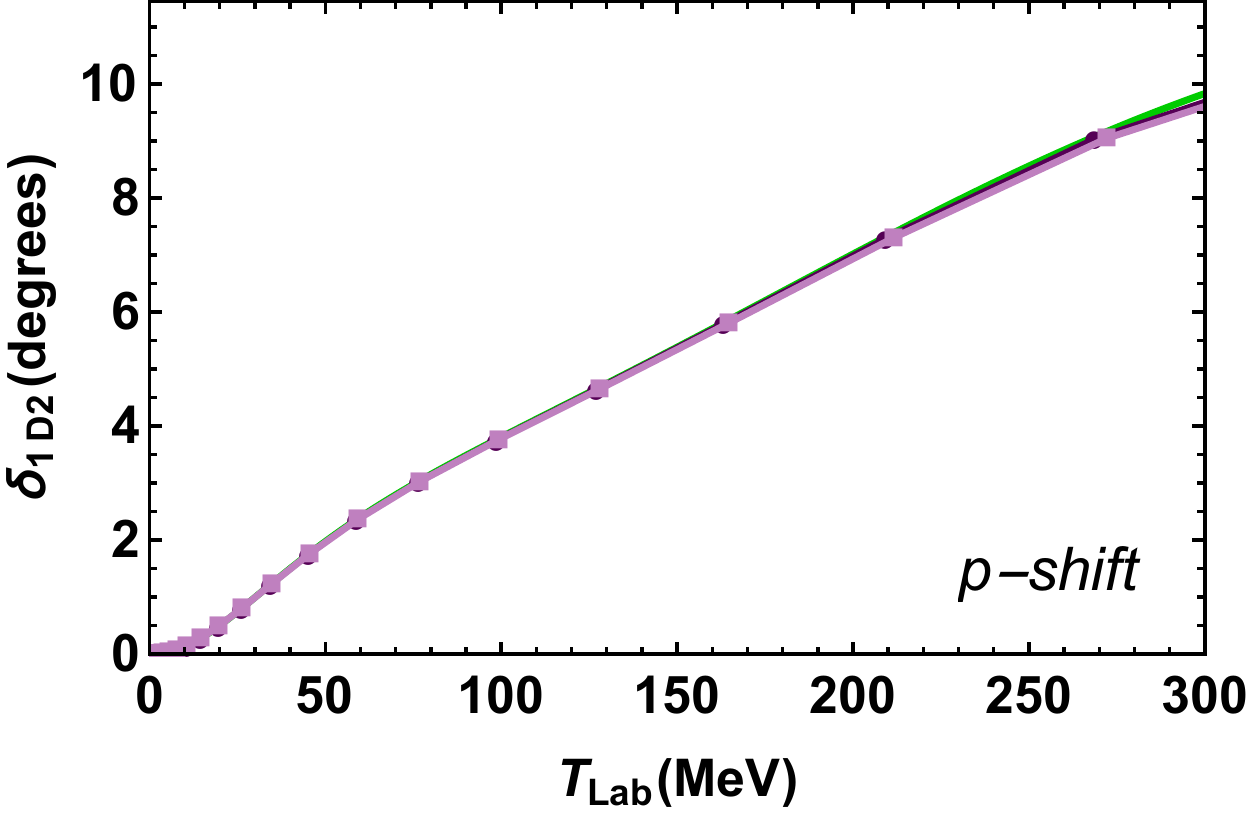}
\includegraphics[scale=0.45]{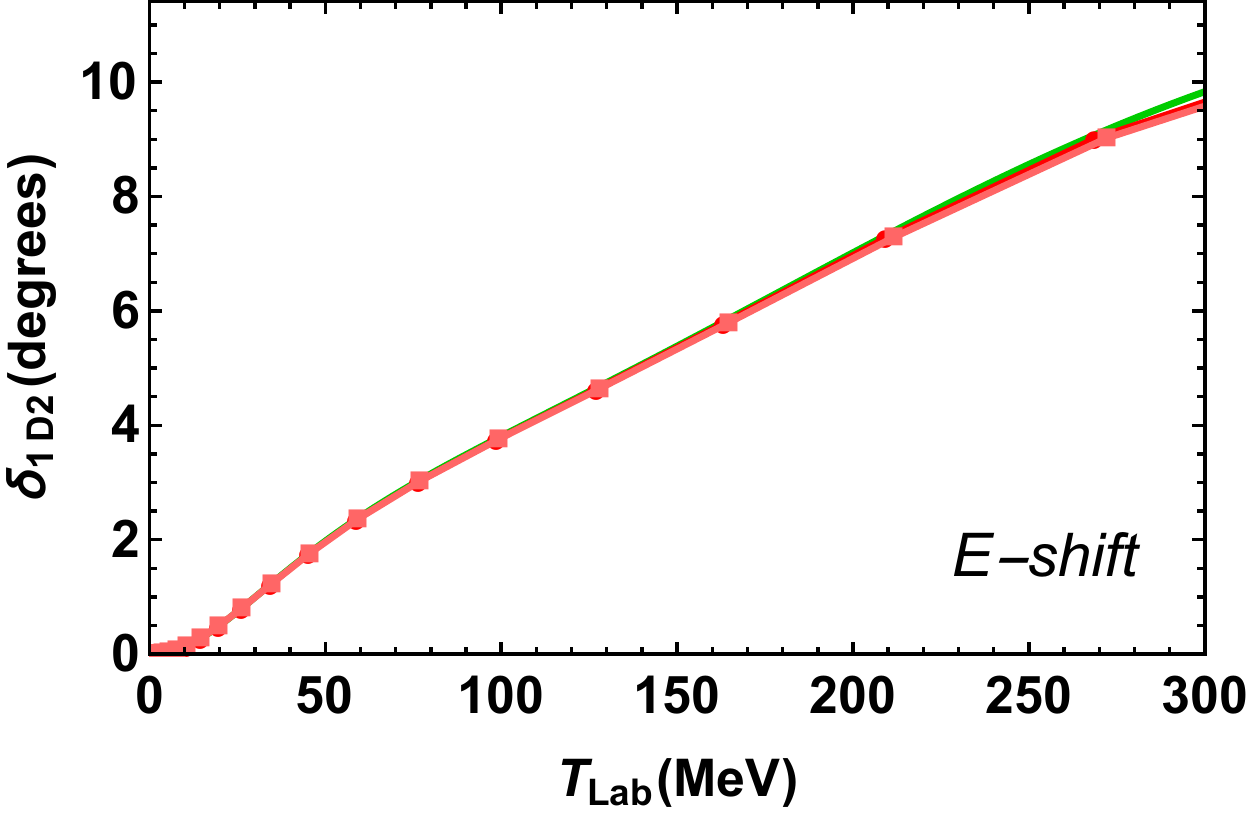}

\caption{ The same as in Figure~\ref{fig:1p1} but for the $ ^1 D_2$ channel.}
\end{figure*}

\begin{figure*}

\includegraphics[scale=0.45]{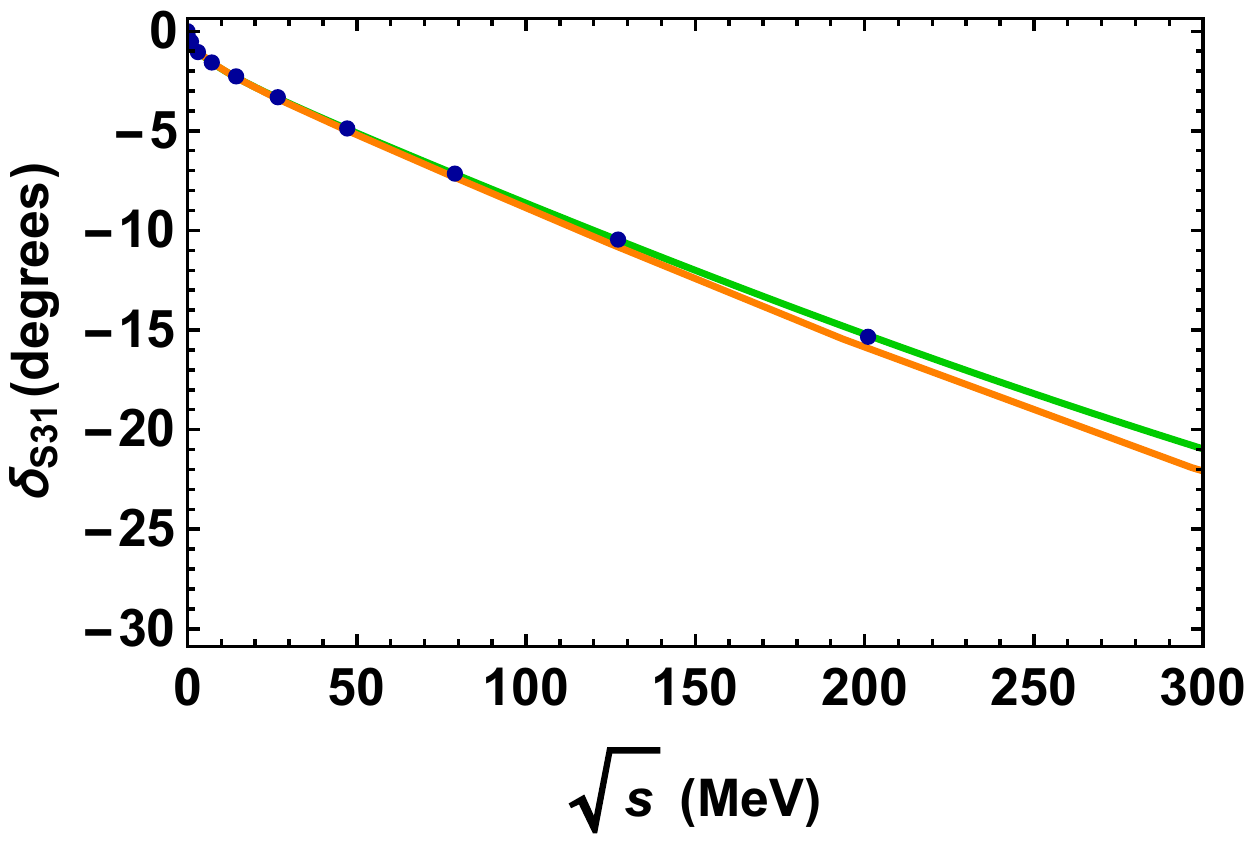}
\includegraphics[scale=0.45]{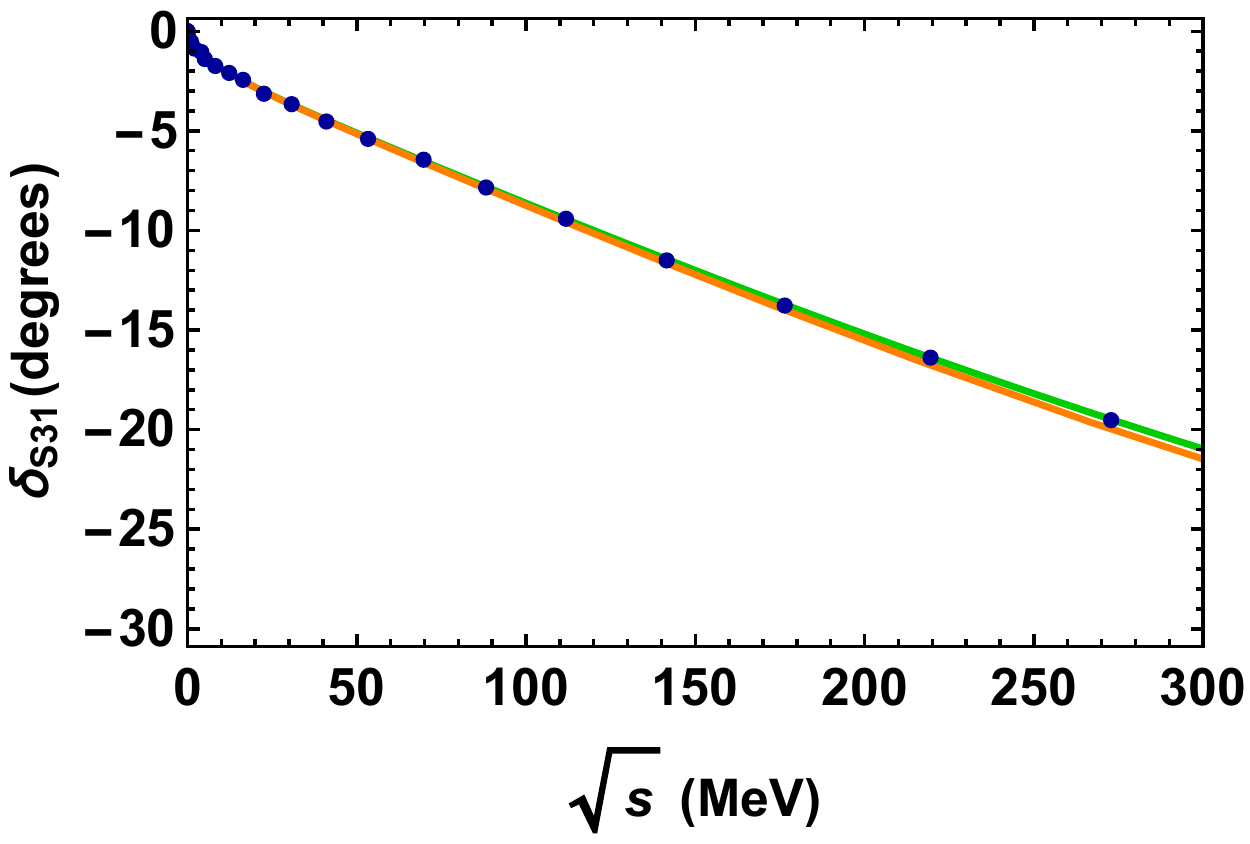}
\includegraphics[scale=0.45]{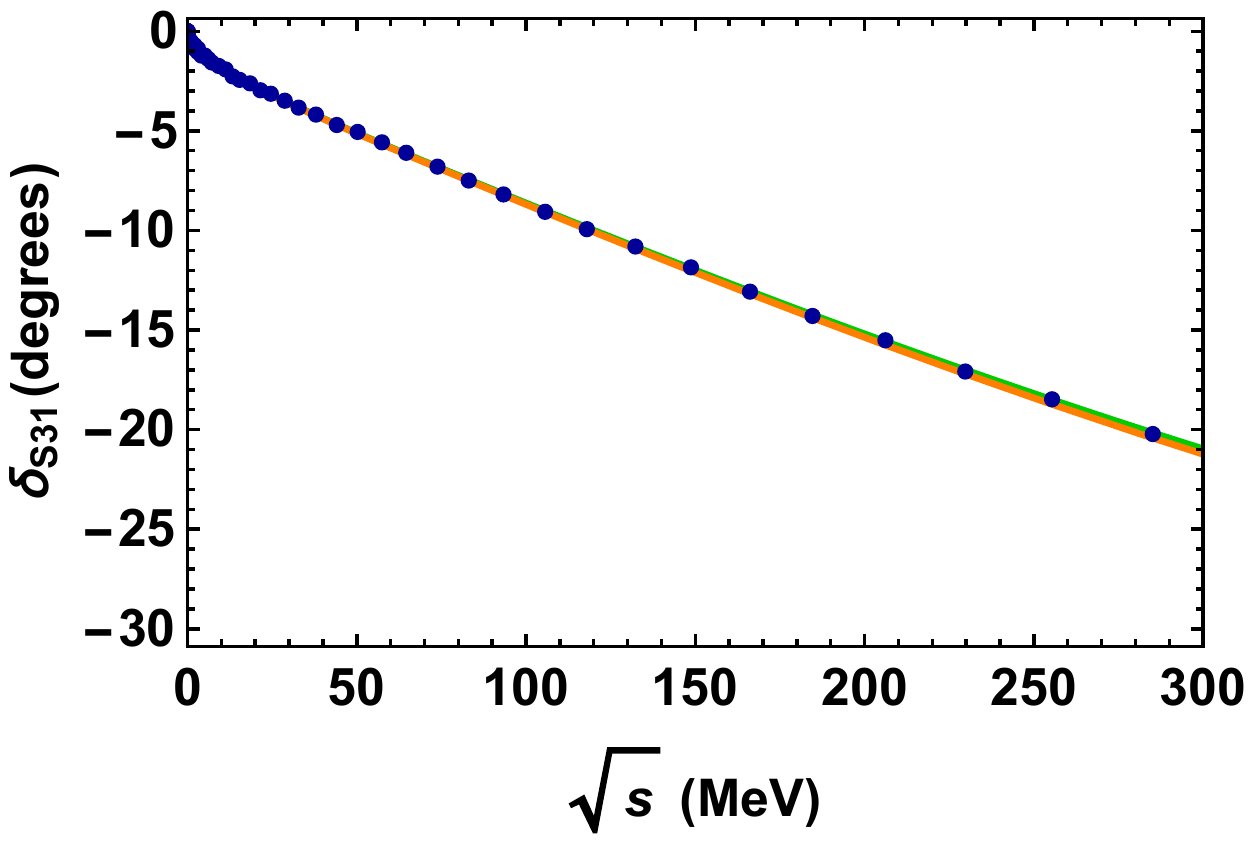}

\includegraphics[scale=0.45]{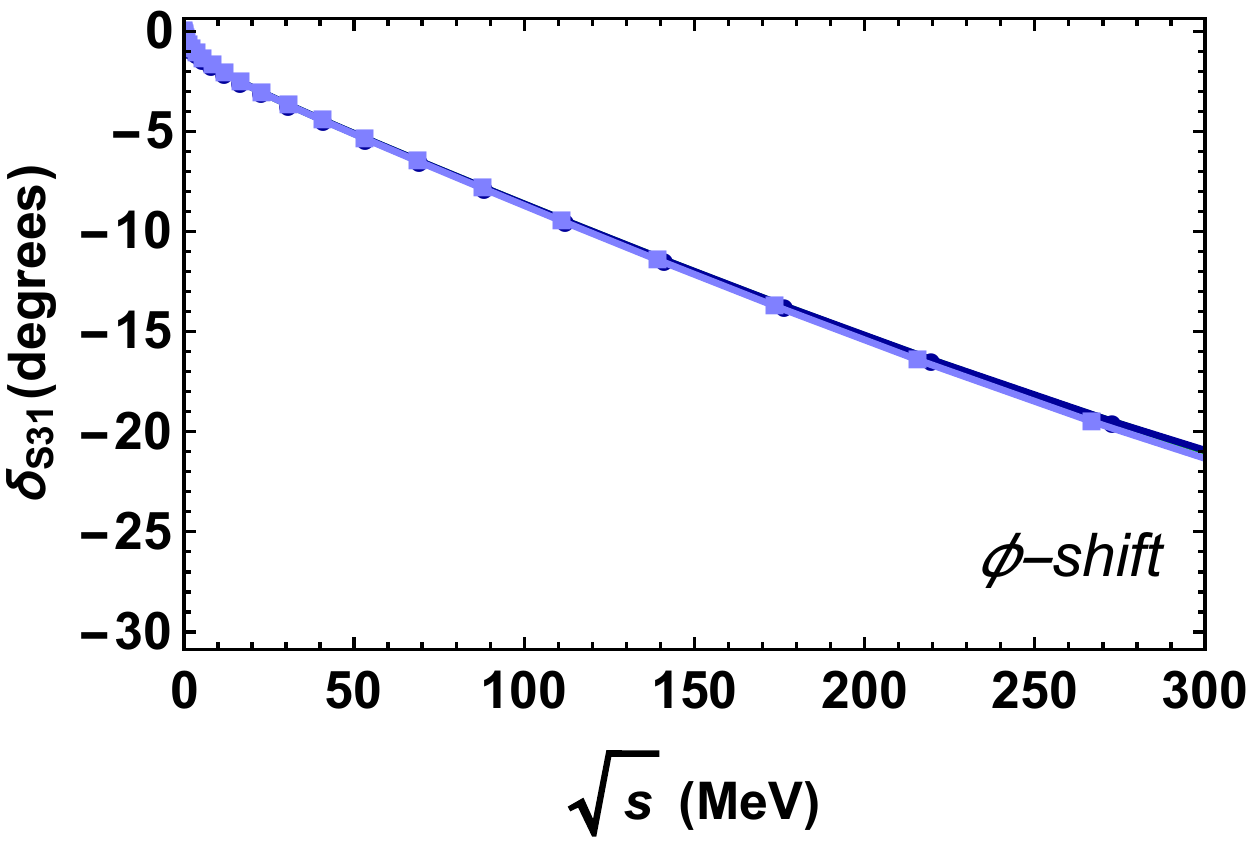}
\includegraphics[scale=0.45]{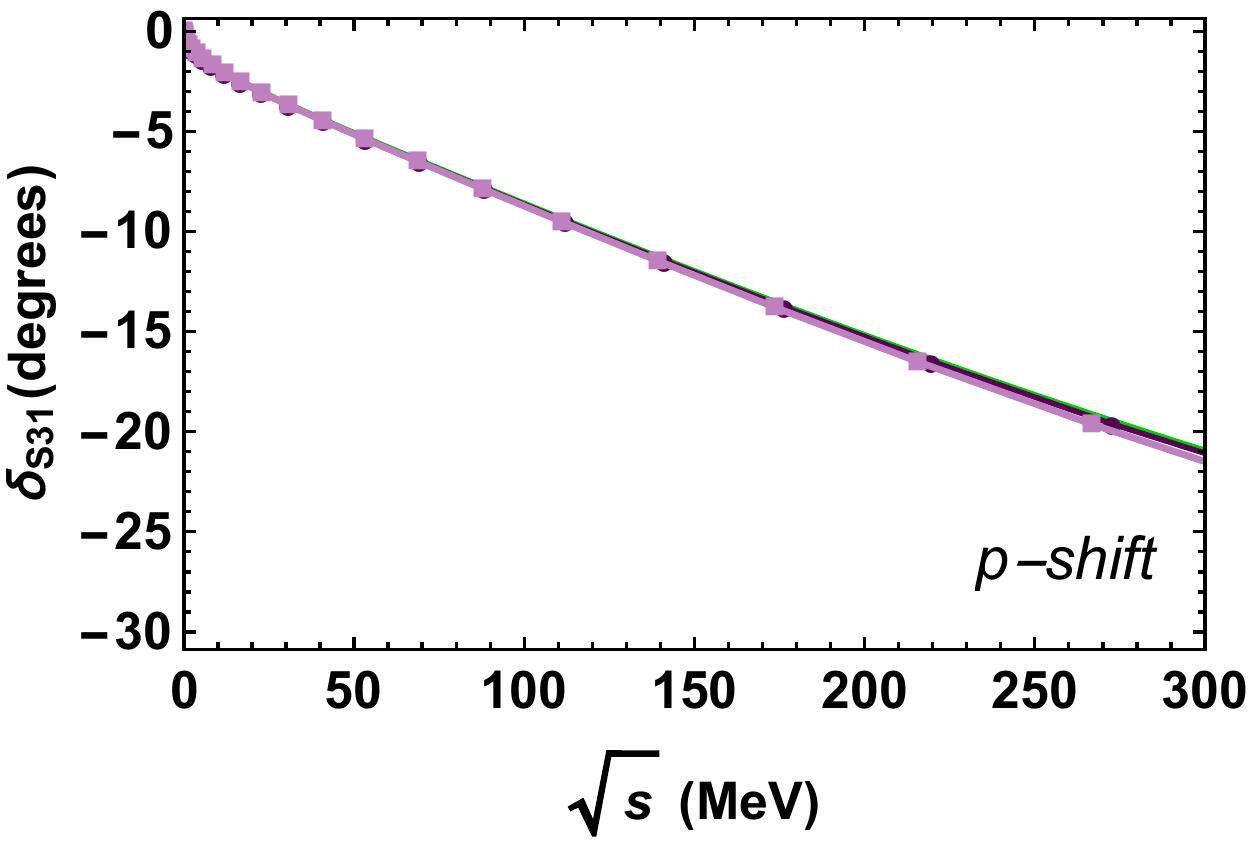}
\includegraphics[scale=0.45]{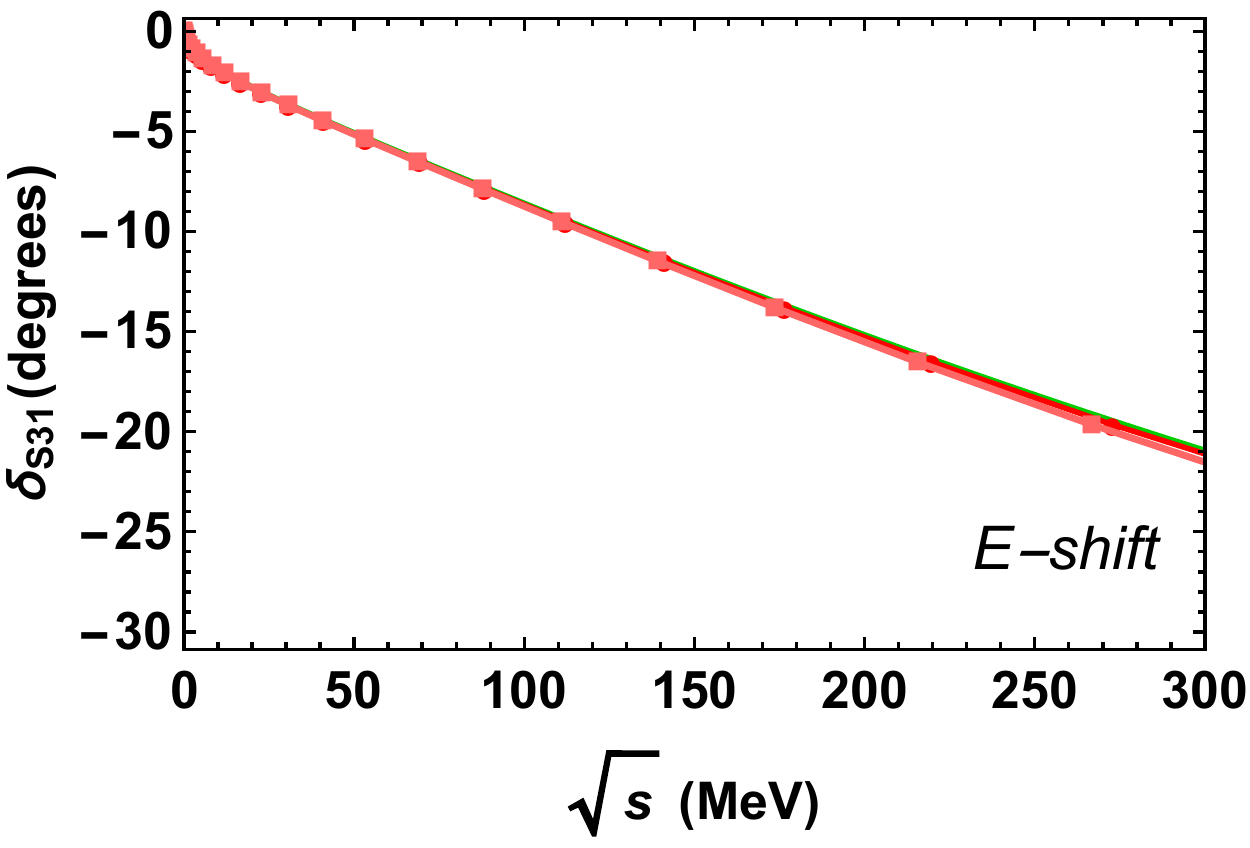}

\caption{ The same as in Figure~\ref{fig:1p1} but for $\pi N$ scattering in the $S_{31}$ channel.}
\end{figure*}

\begin{figure*}

\includegraphics[scale=0.45]{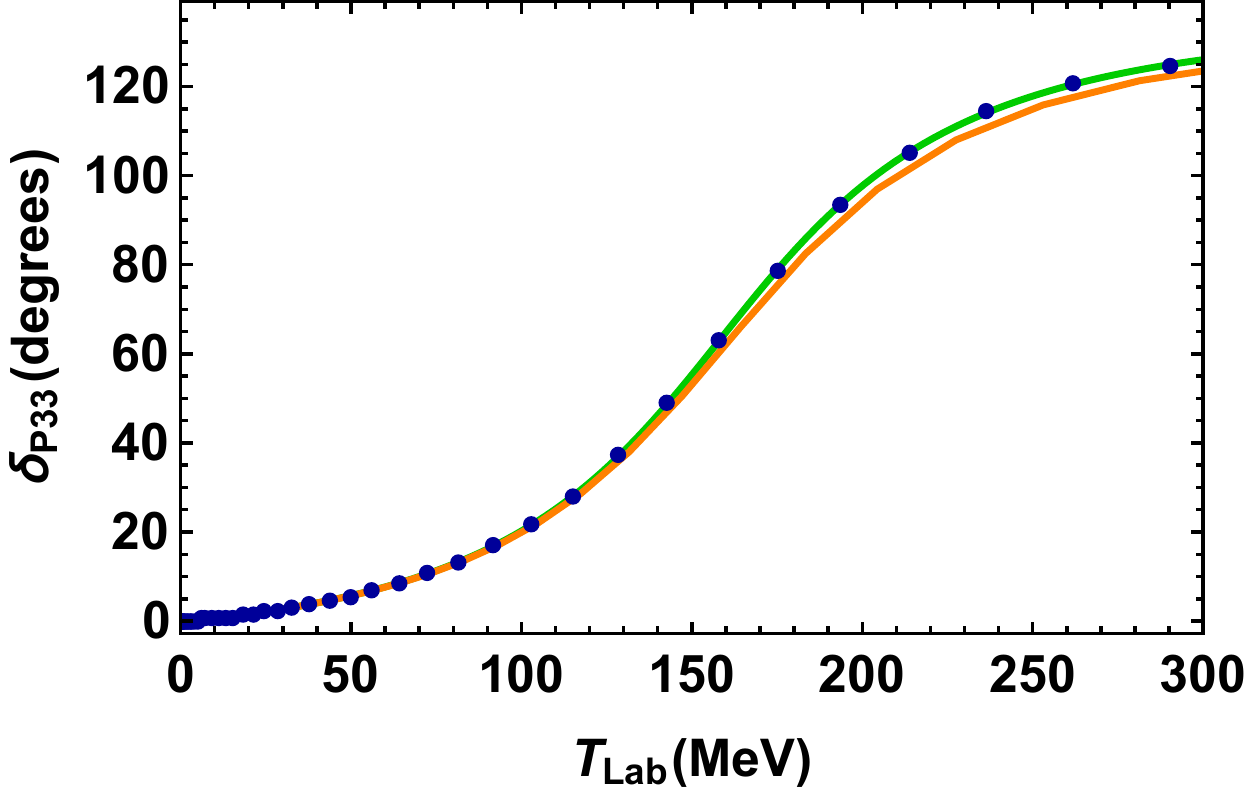}
\includegraphics[scale=0.45]{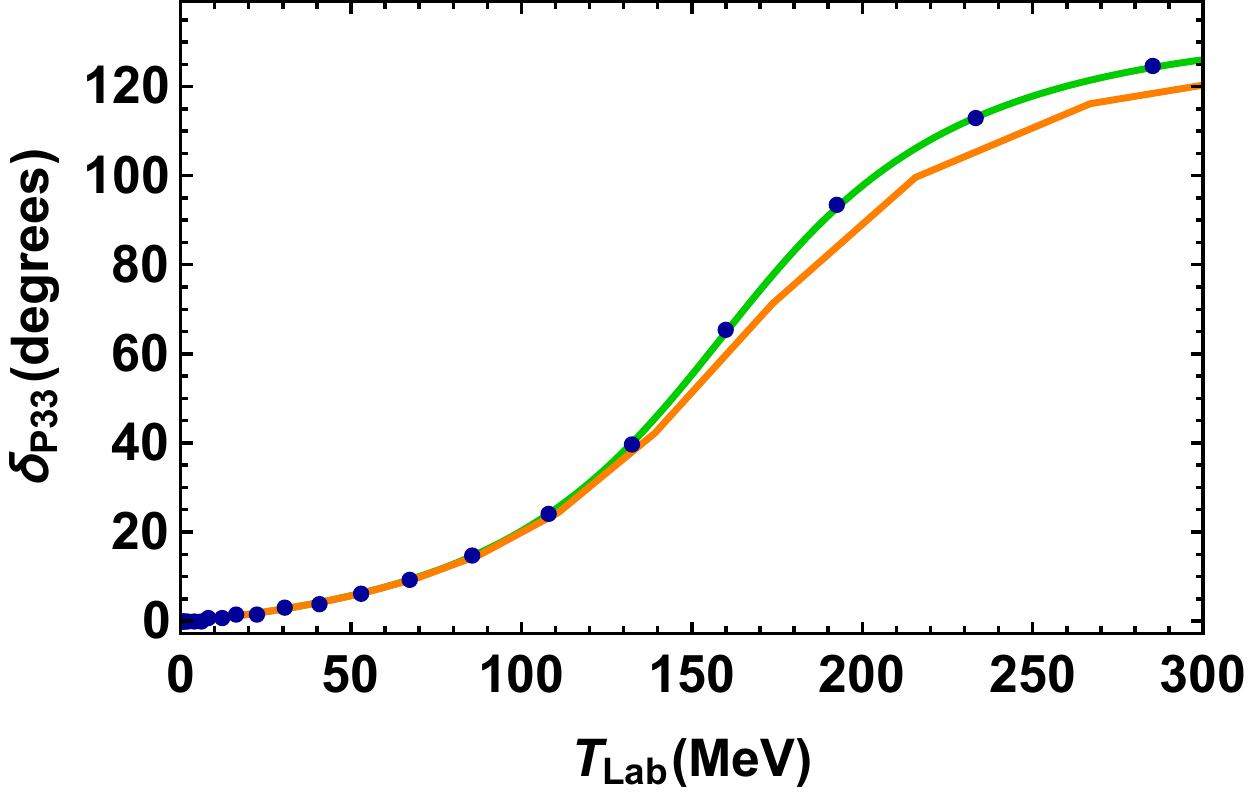}
\includegraphics[scale=0.45]{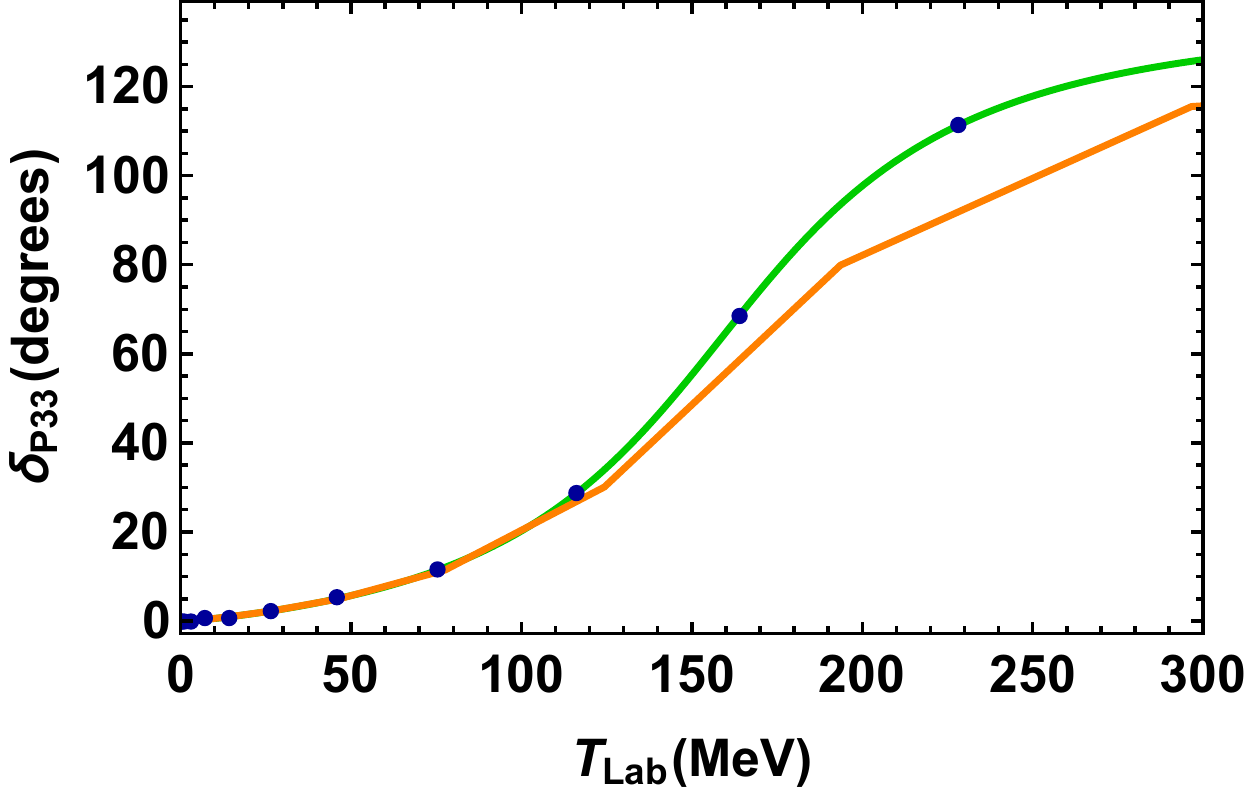}

\includegraphics[scale=0.45]{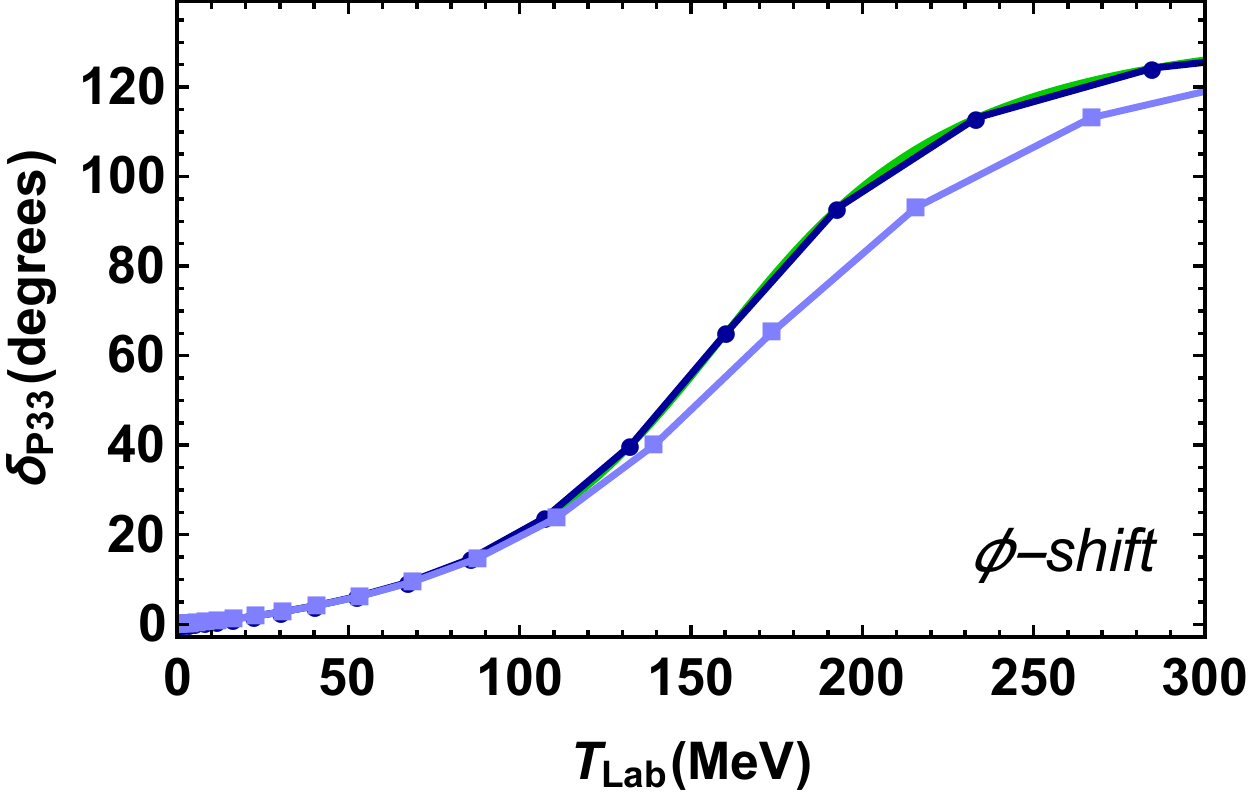}
\includegraphics[scale=0.45]{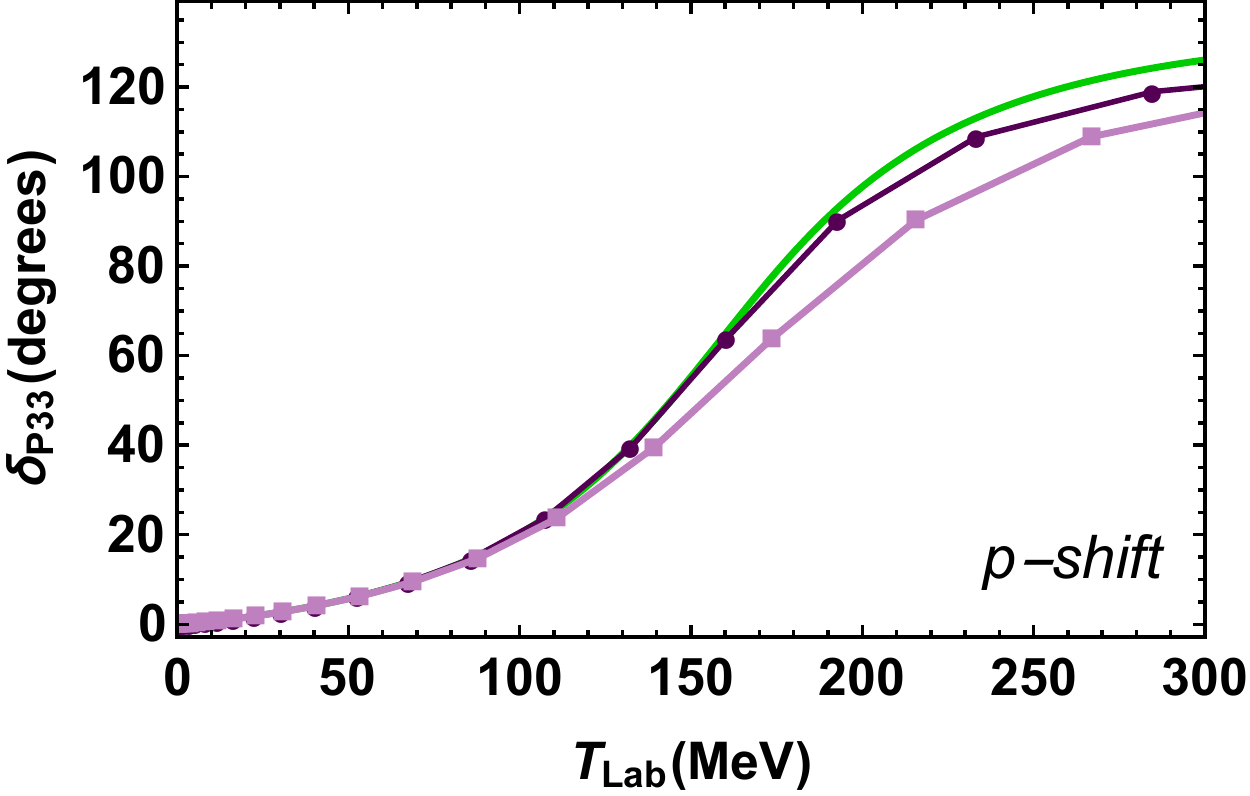}
\includegraphics[scale=0.45]{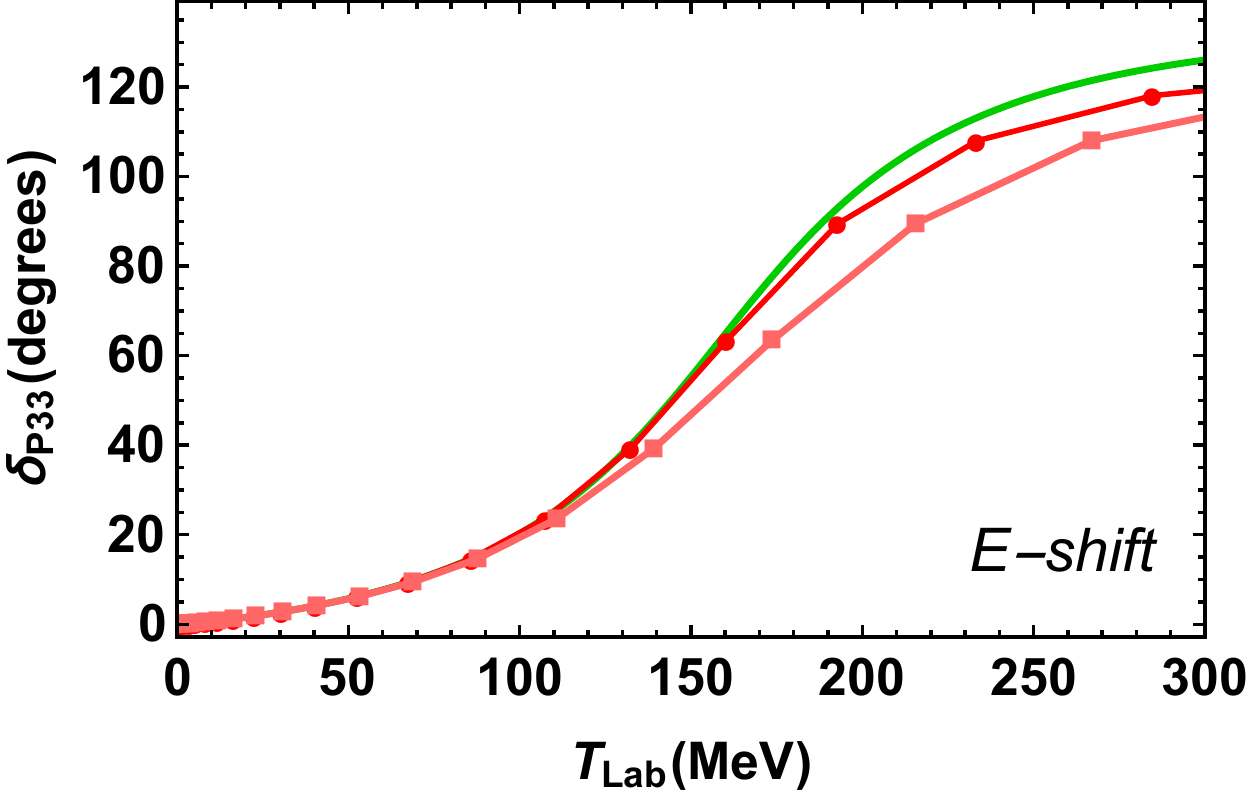}

\caption{ The same as in Figure~\ref{fig:1p1} but for $\pi N$ scattering in the $P_{33}$ channel.}
\end{figure*}

\begin{figure*}

\includegraphics[scale=0.45]{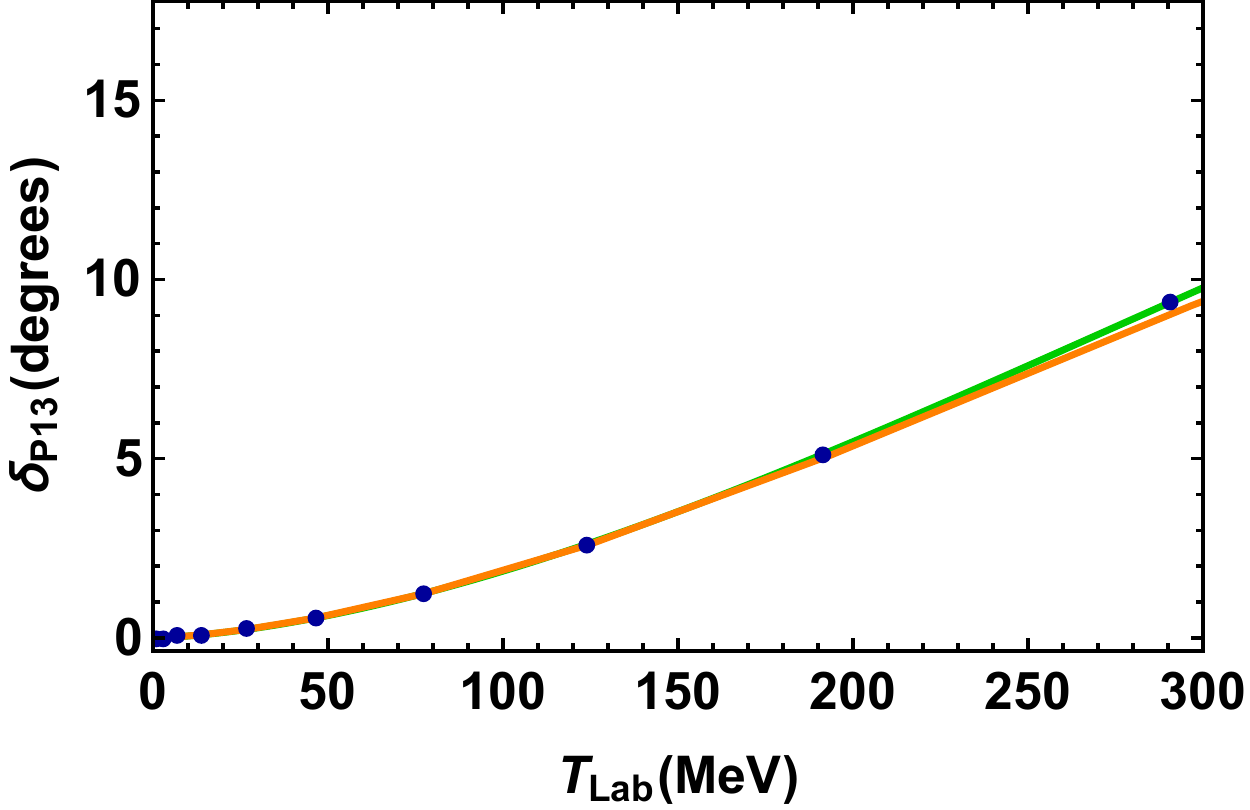}
\includegraphics[scale=0.45]{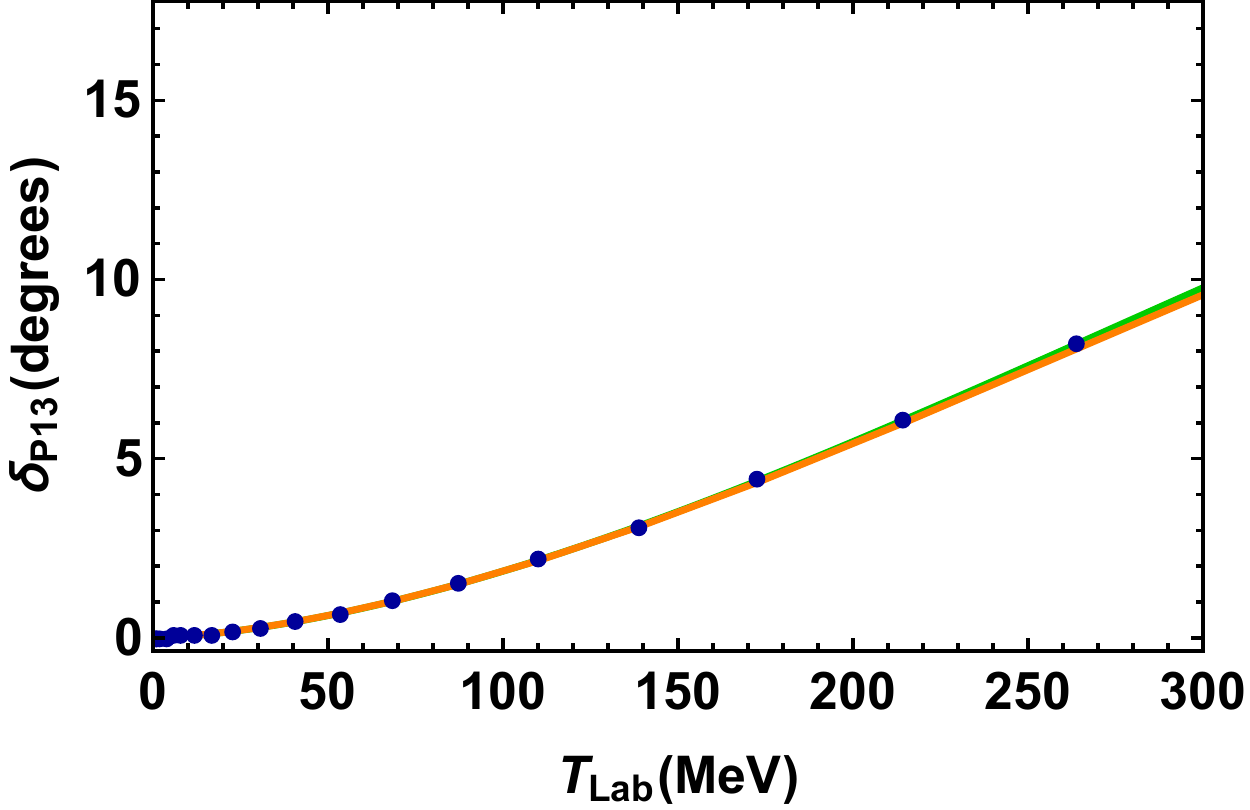}
\includegraphics[scale=0.45]{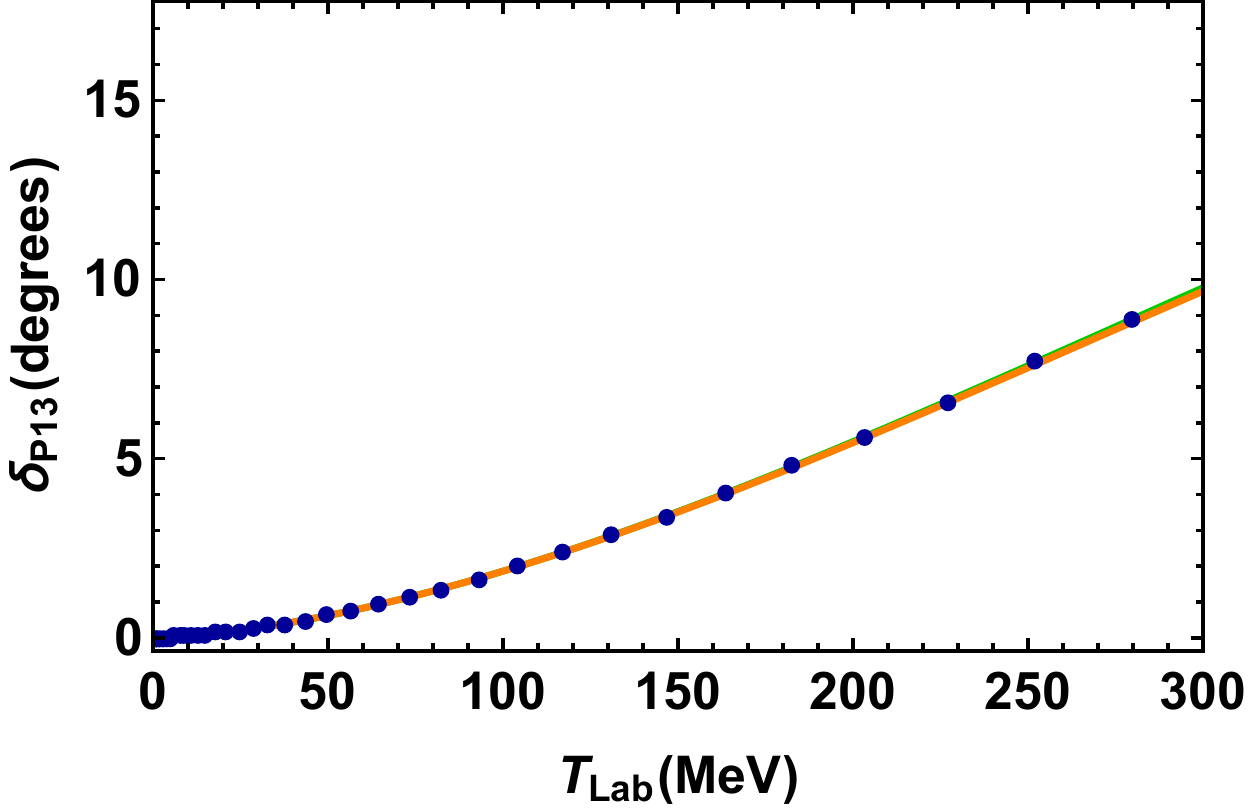}

\includegraphics[scale=0.45]{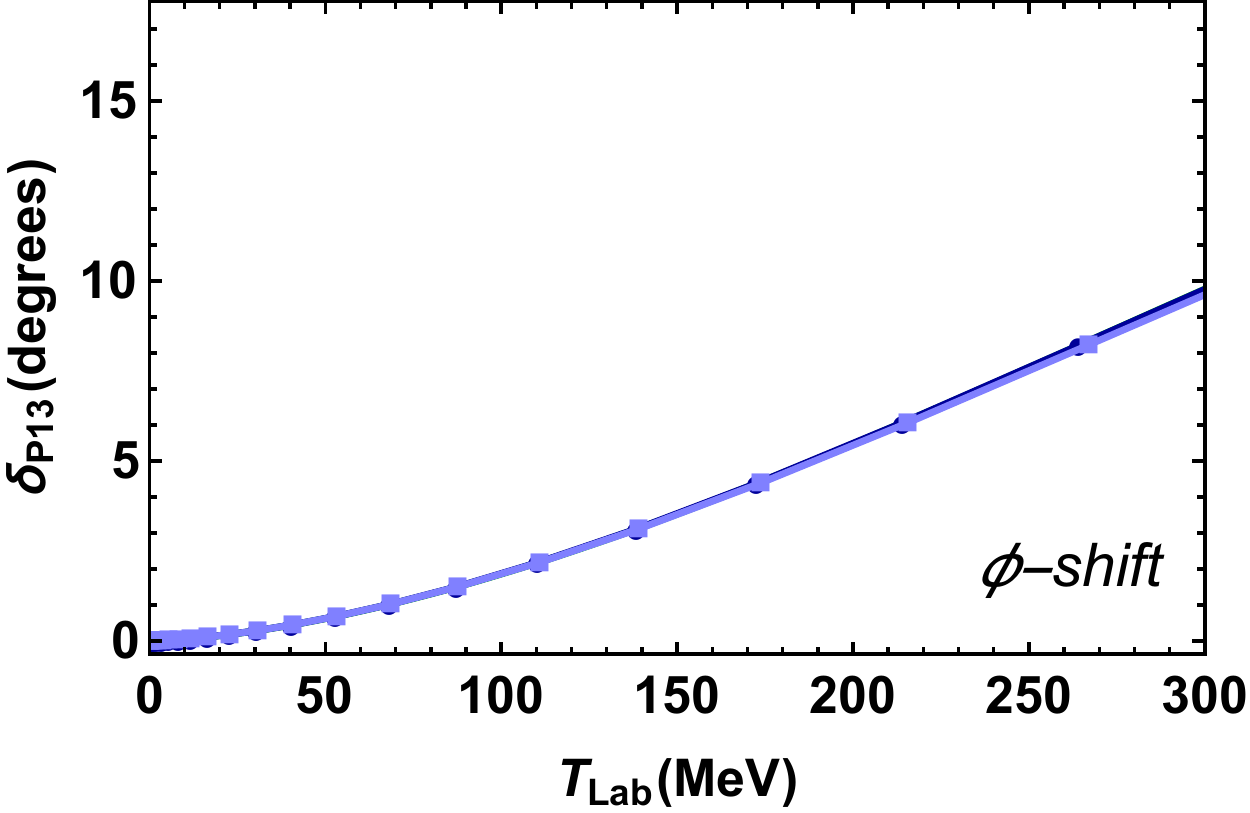}
\includegraphics[scale=0.45]{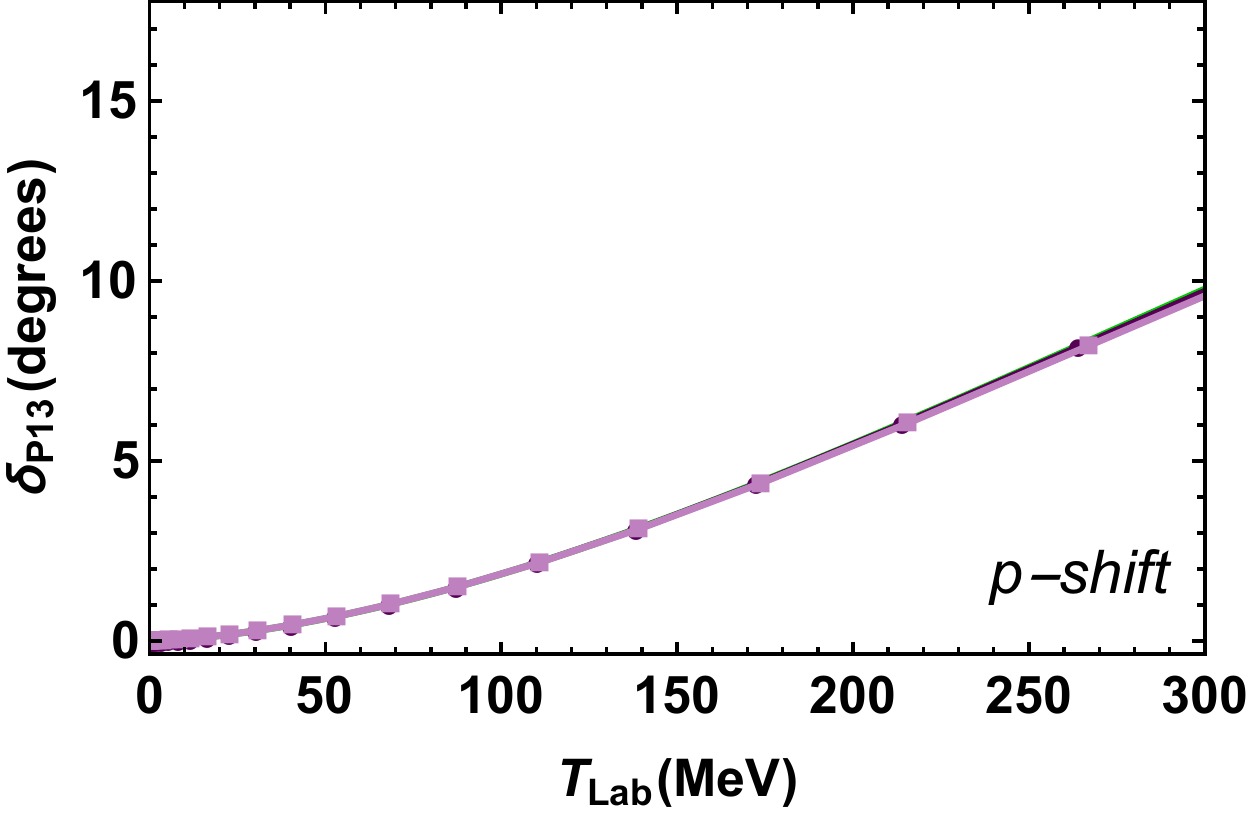}
\includegraphics[scale=0.45]{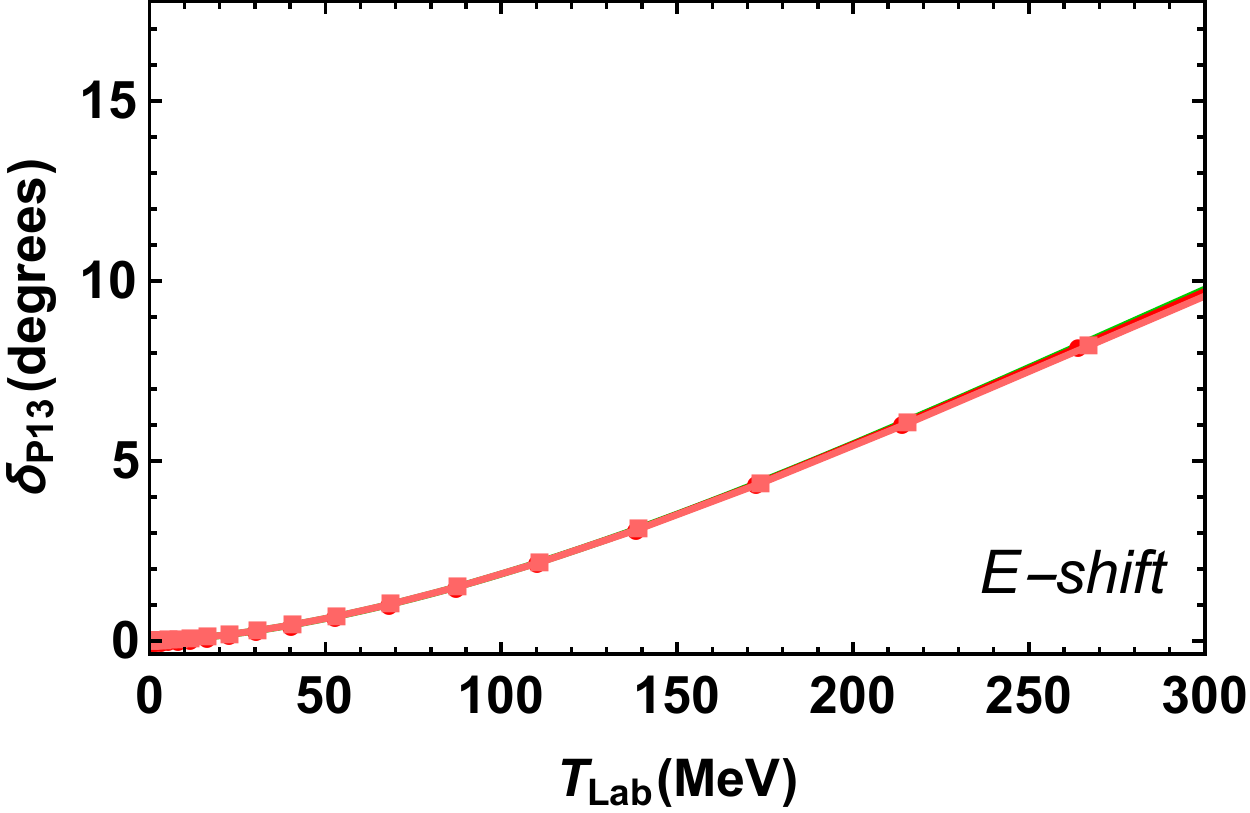}

\caption{ The same as in Figure~\ref{fig:1p1} but for $\pi N$ scattering in the $P_{13}$ channel.}
\end{figure*}

\begin{figure*}

\includegraphics[scale=0.45]{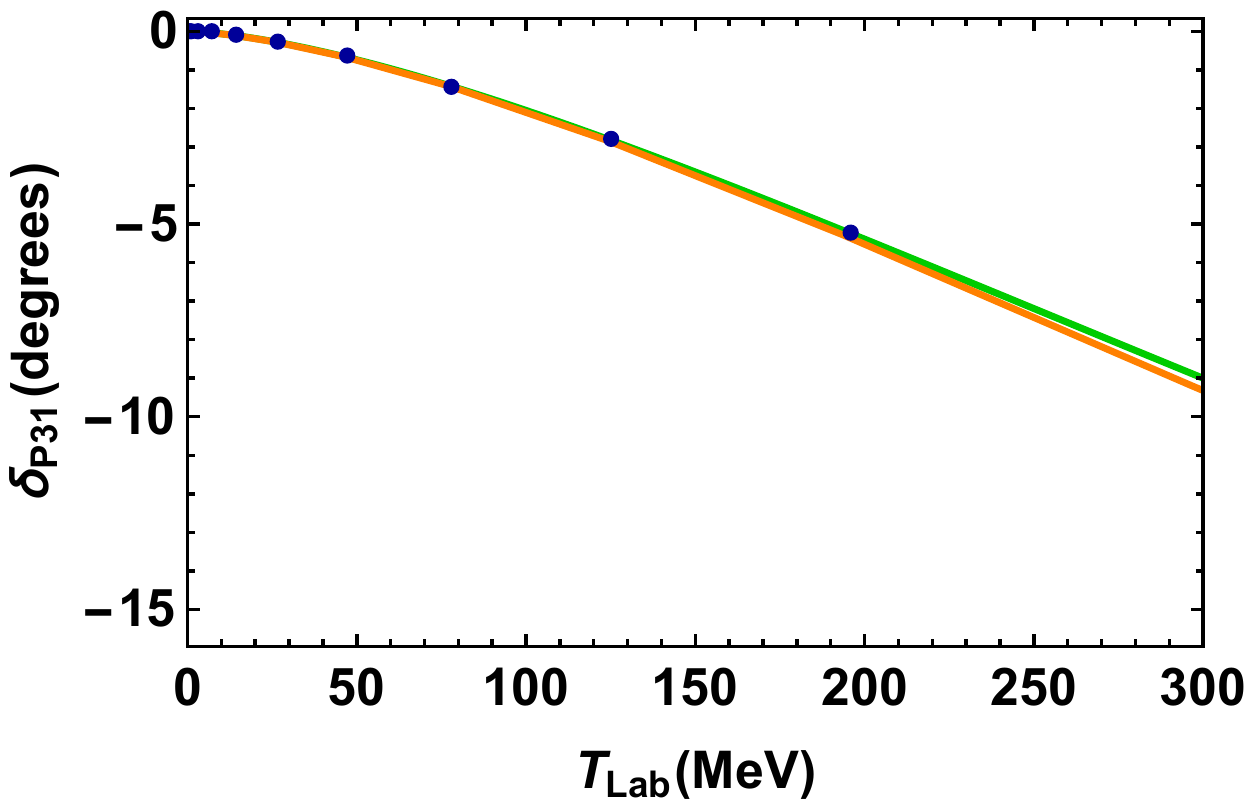}
\includegraphics[scale=0.45]{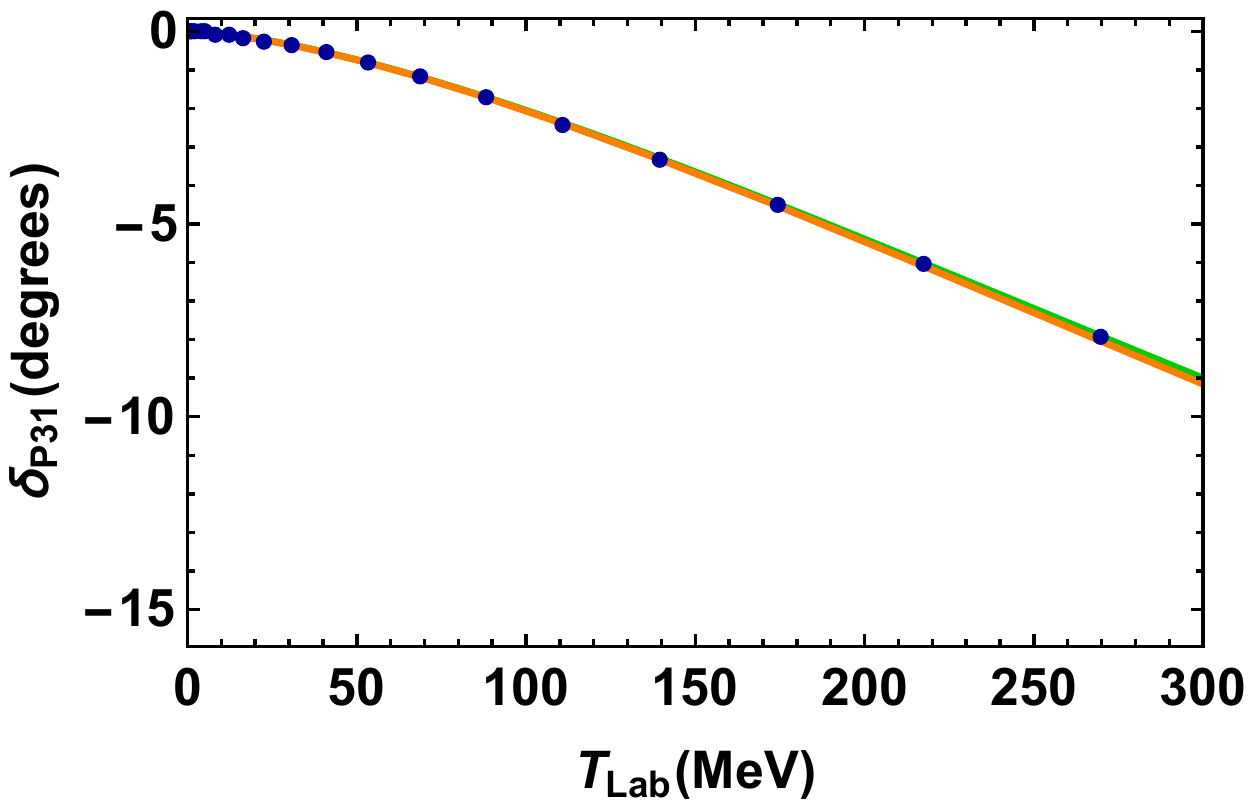}
\includegraphics[scale=0.45]{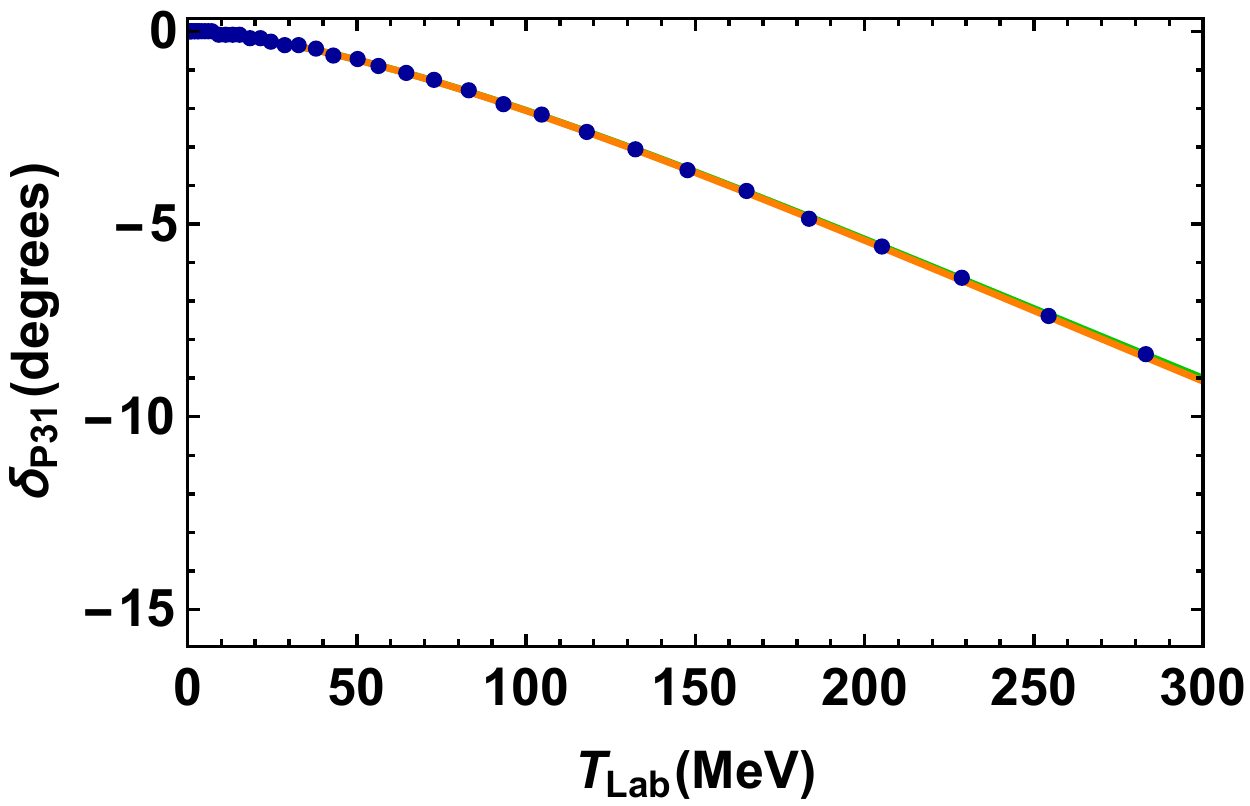}

\includegraphics[scale=0.45]{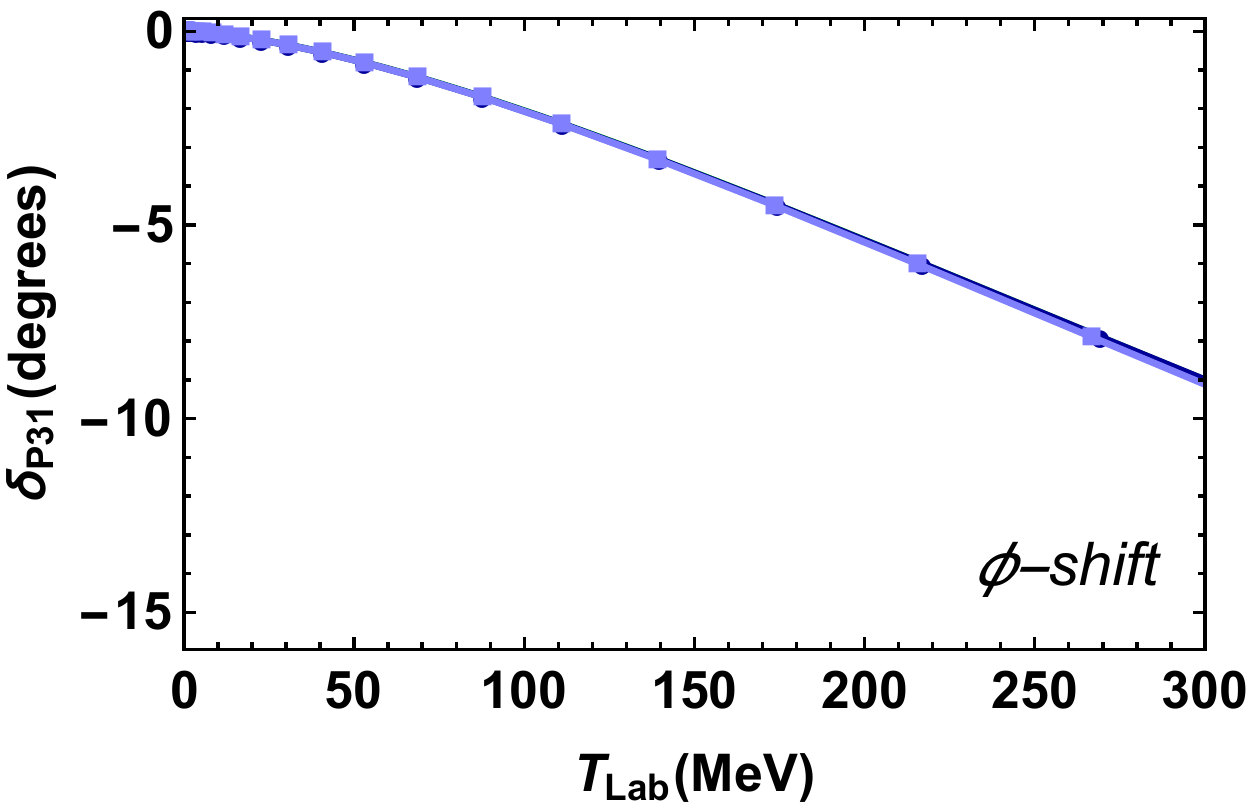}
\includegraphics[scale=0.45]{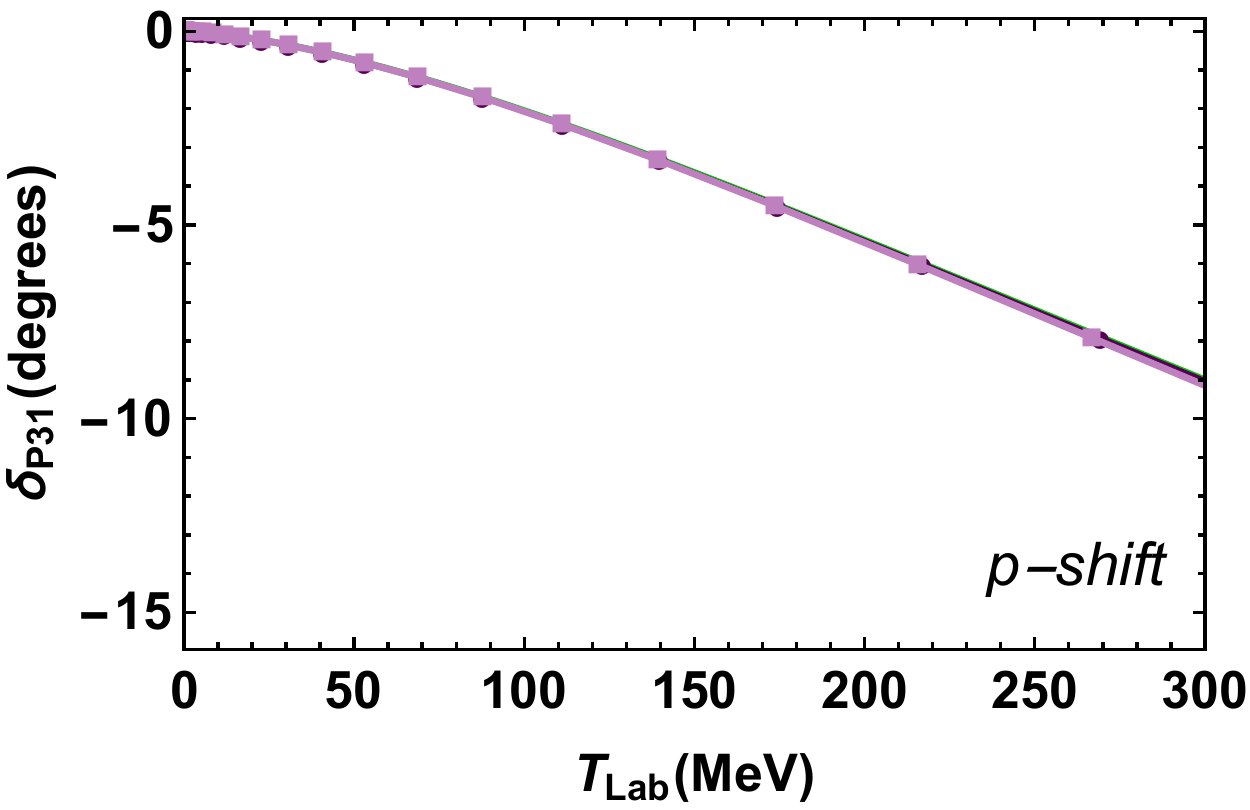}
\includegraphics[scale=0.45]{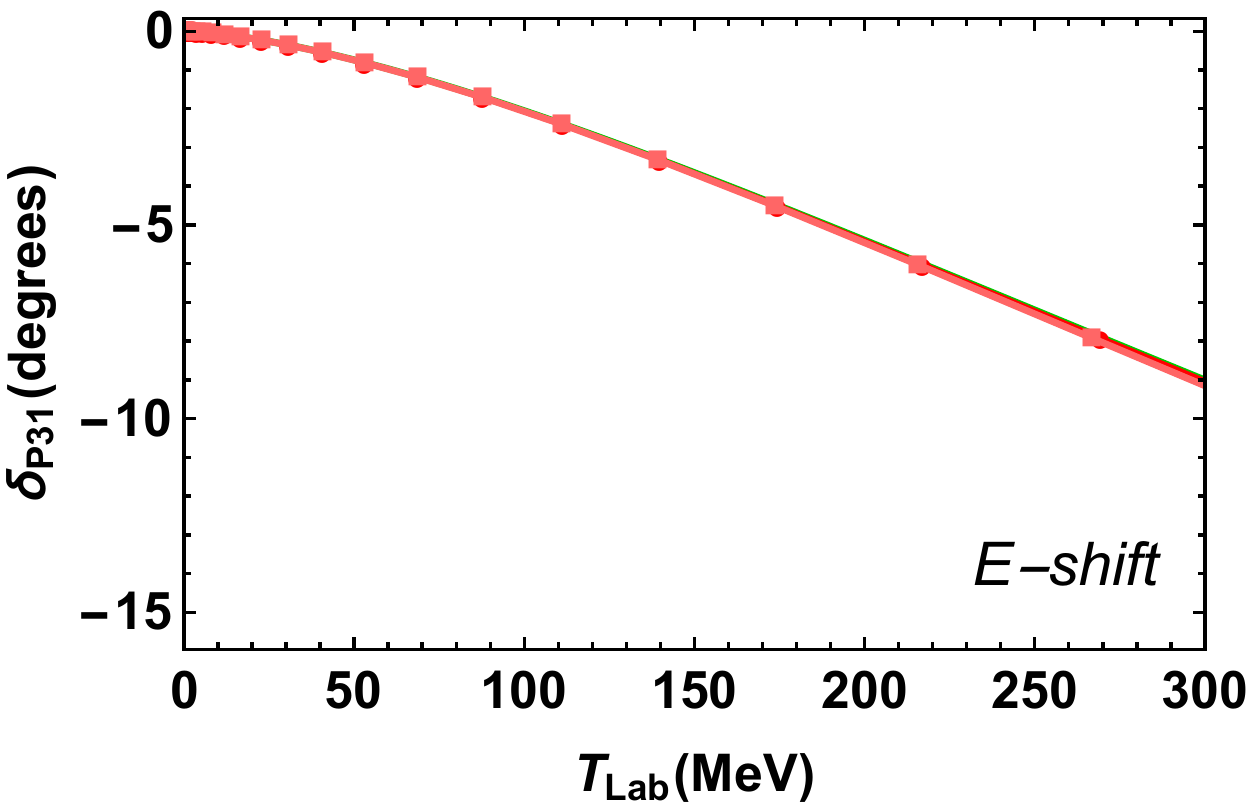}

\caption{ The same as in Figure~\ref{fig:1p1} but for $\pi N$ scattering in the $P_{31}$ channel.}
\end{figure*}

\begin{figure*}
\includegraphics[scale=0.45]{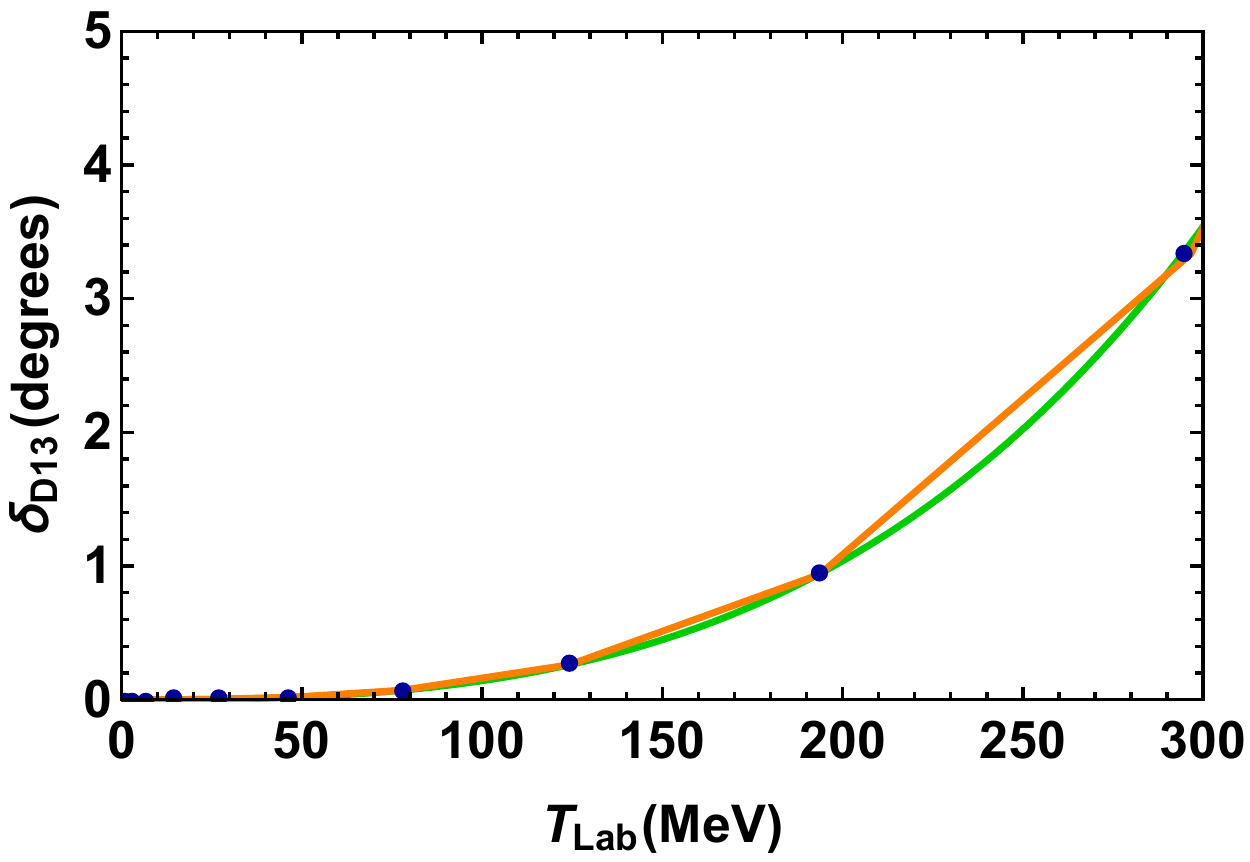}
\includegraphics[scale=0.45]{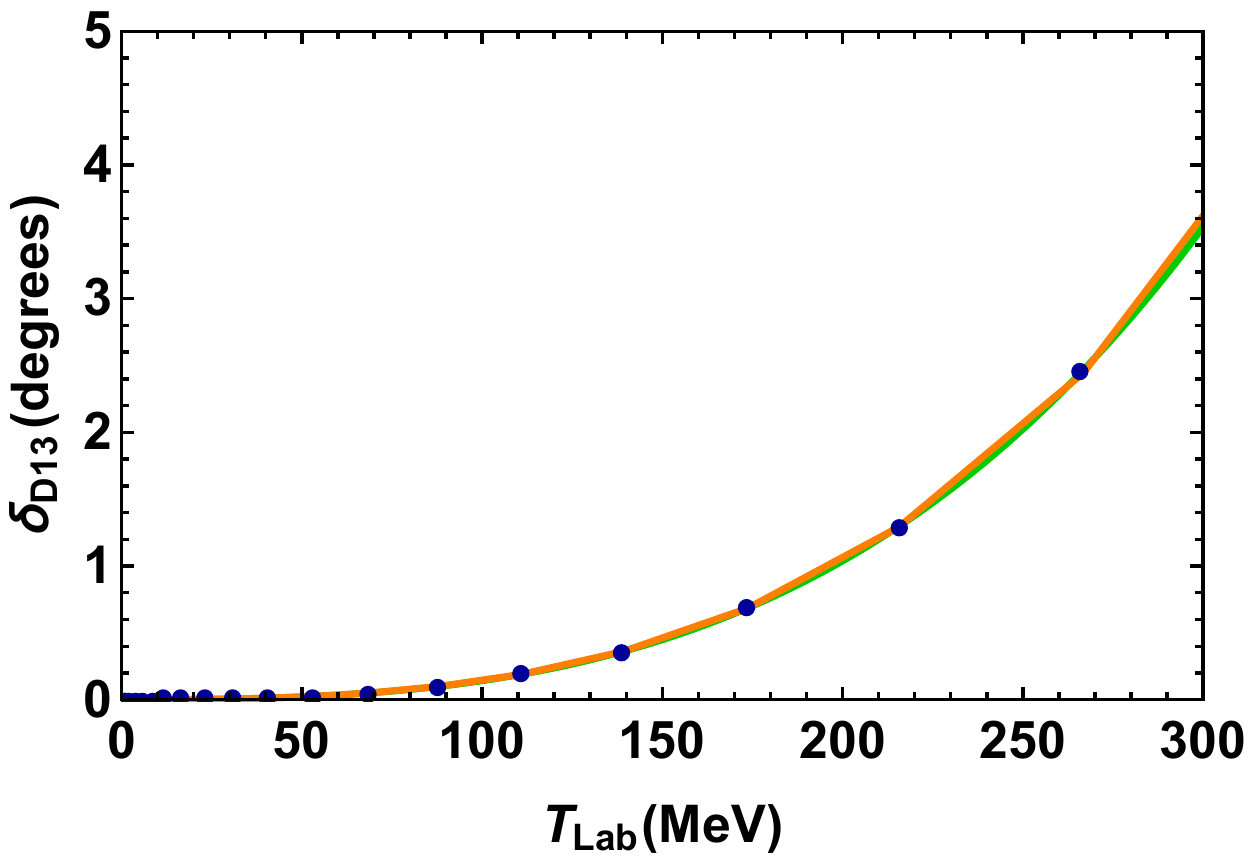}
\includegraphics[scale=0.45]{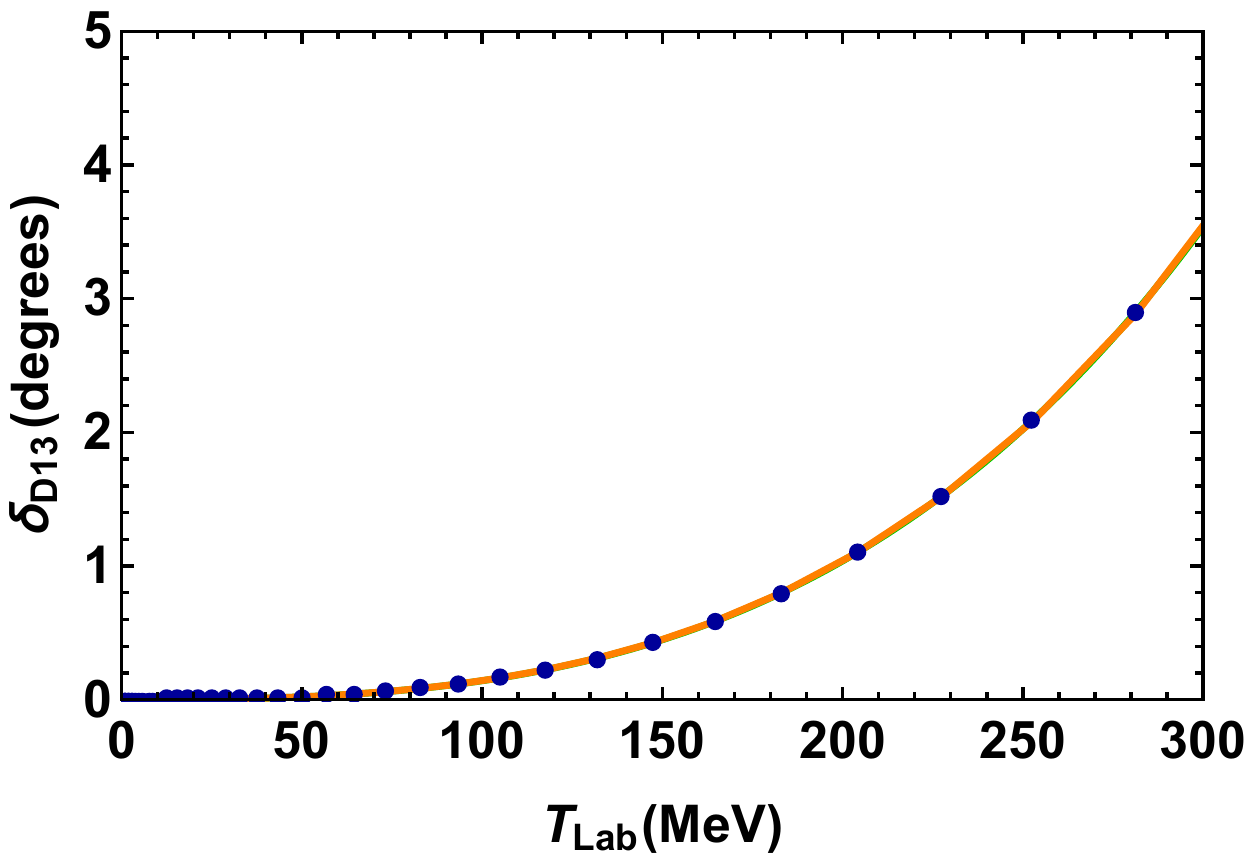}

\includegraphics[scale=0.45]{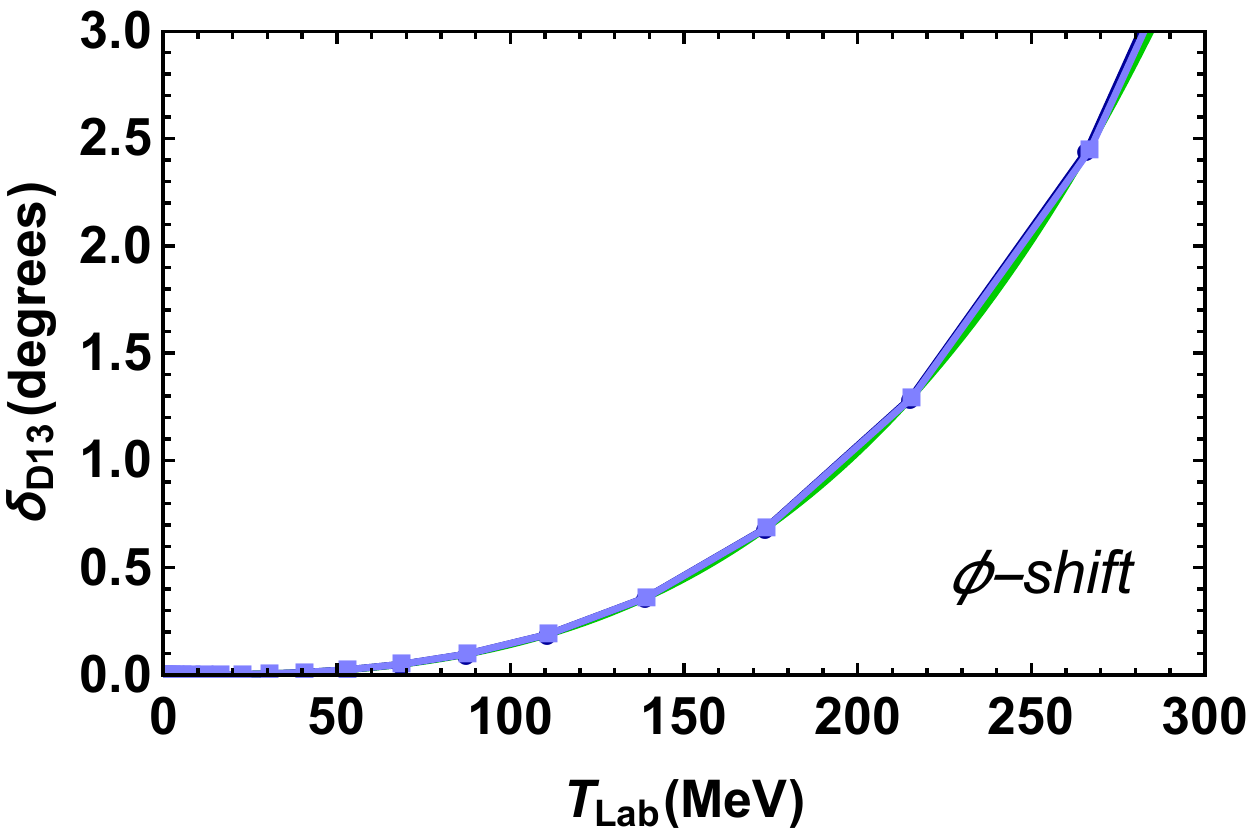}
\includegraphics[scale=0.45]{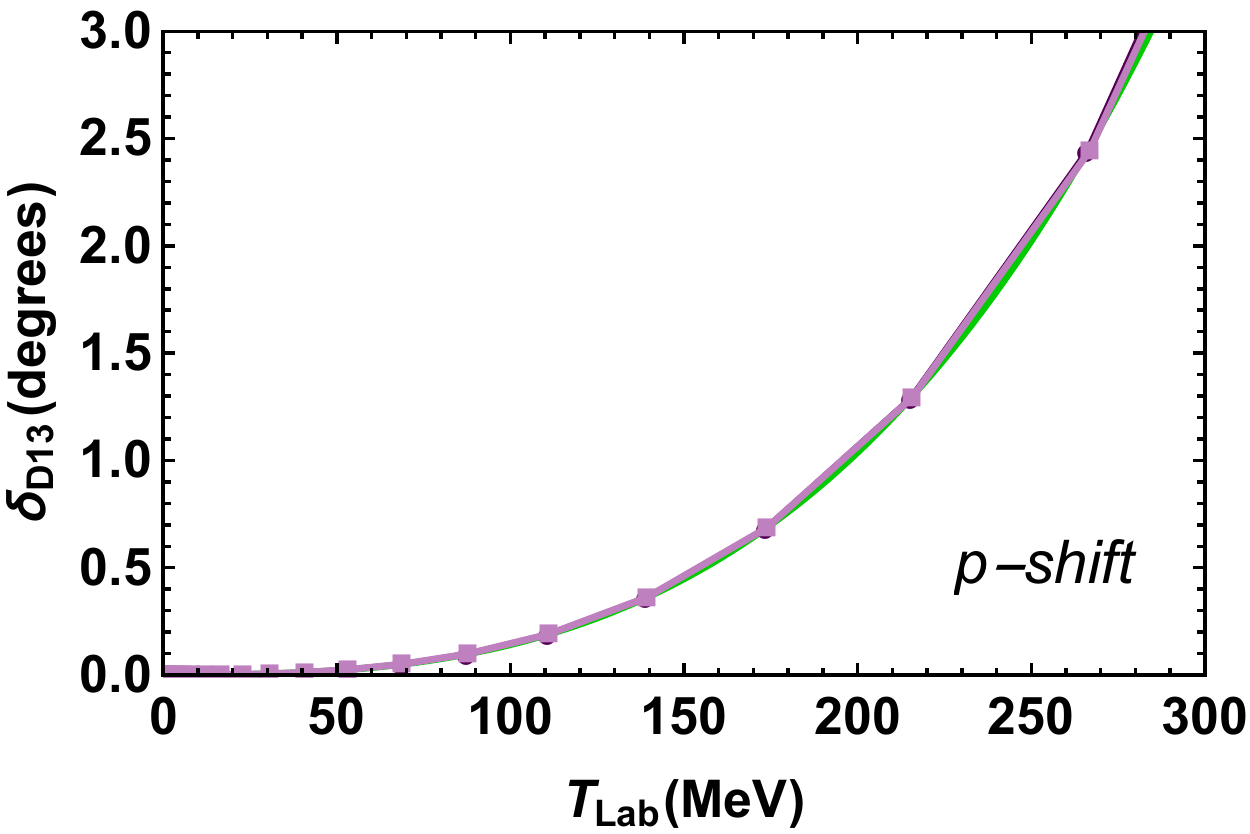}
\includegraphics[scale=0.45]{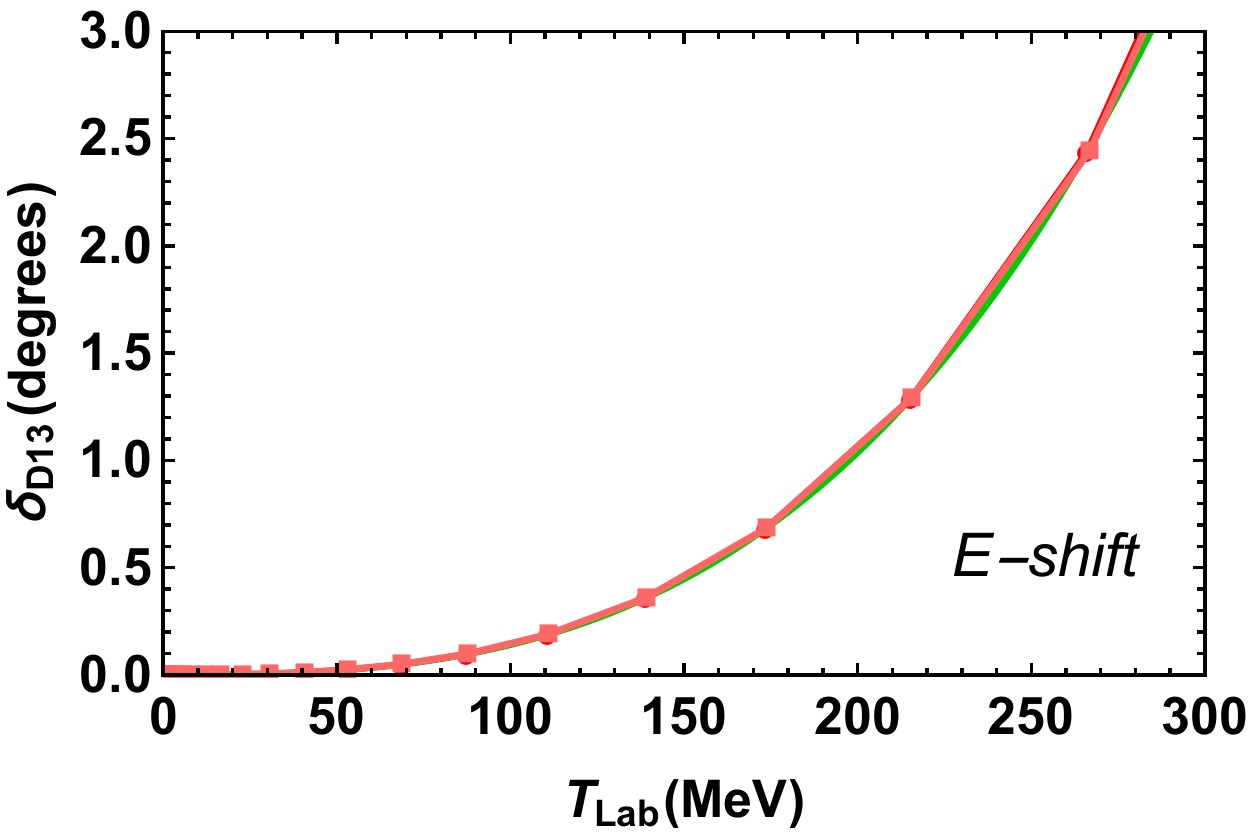}

\caption{ The same as in Figure~\ref{fig:1p1} but for $\pi N$ scattering in the $D_{13}$ channel.  }
\end{figure*}

\begin{figure*}

\includegraphics[scale=0.45]{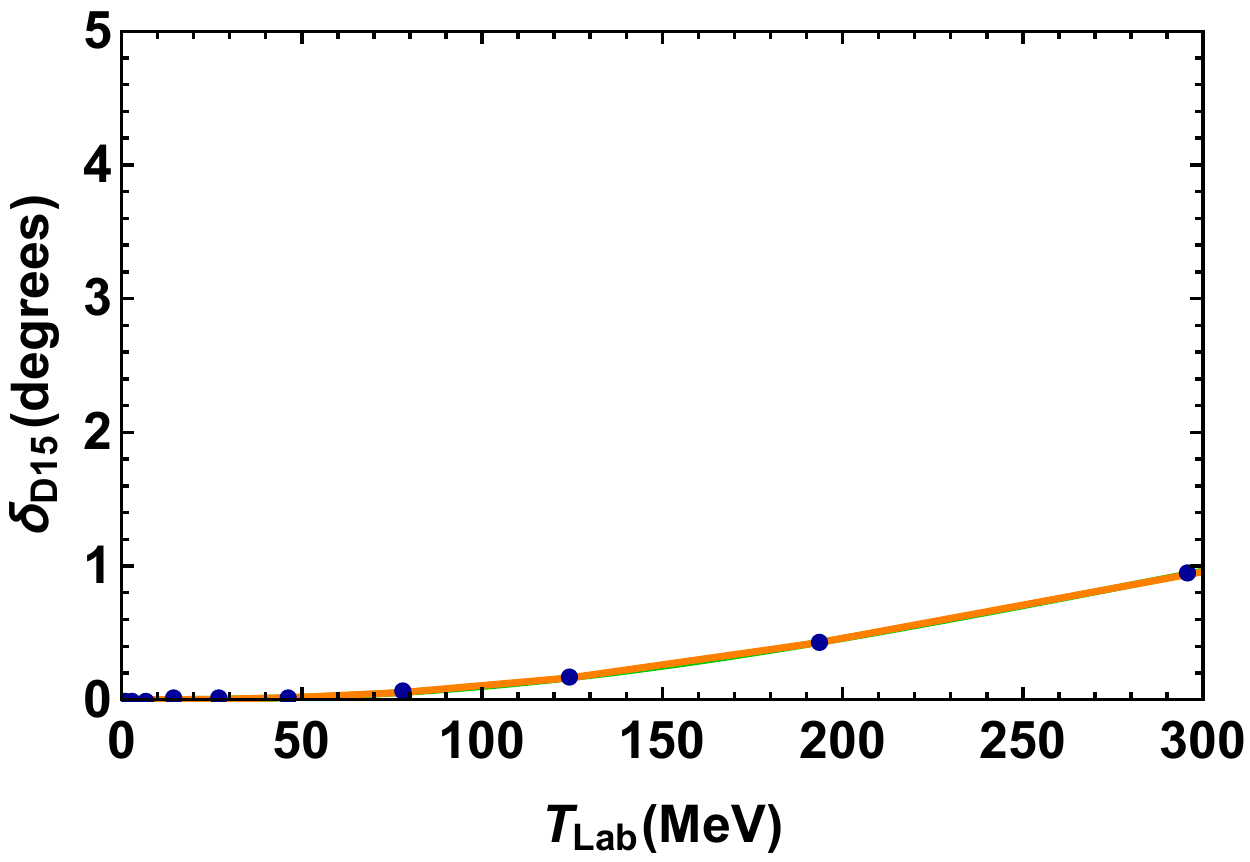}
\includegraphics[scale=0.45]{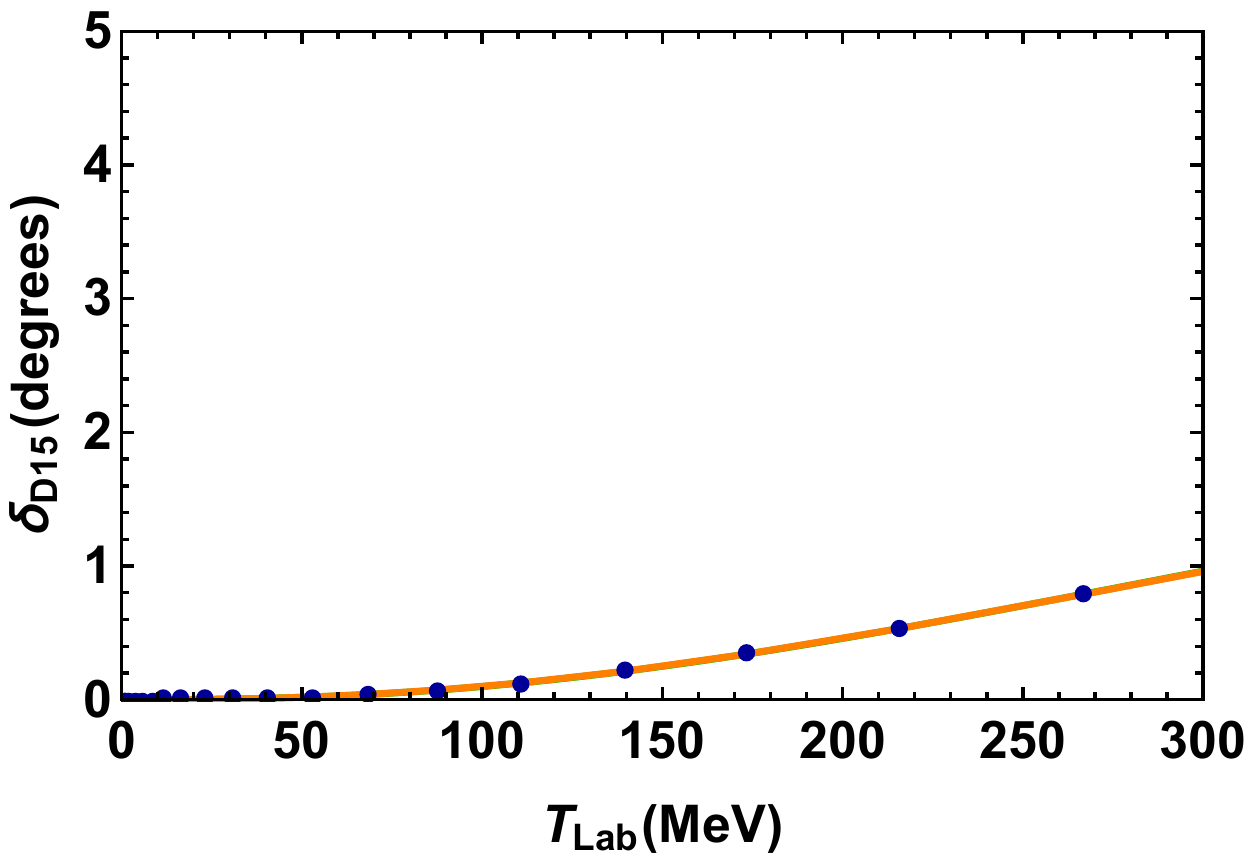}
\includegraphics[scale=0.45]{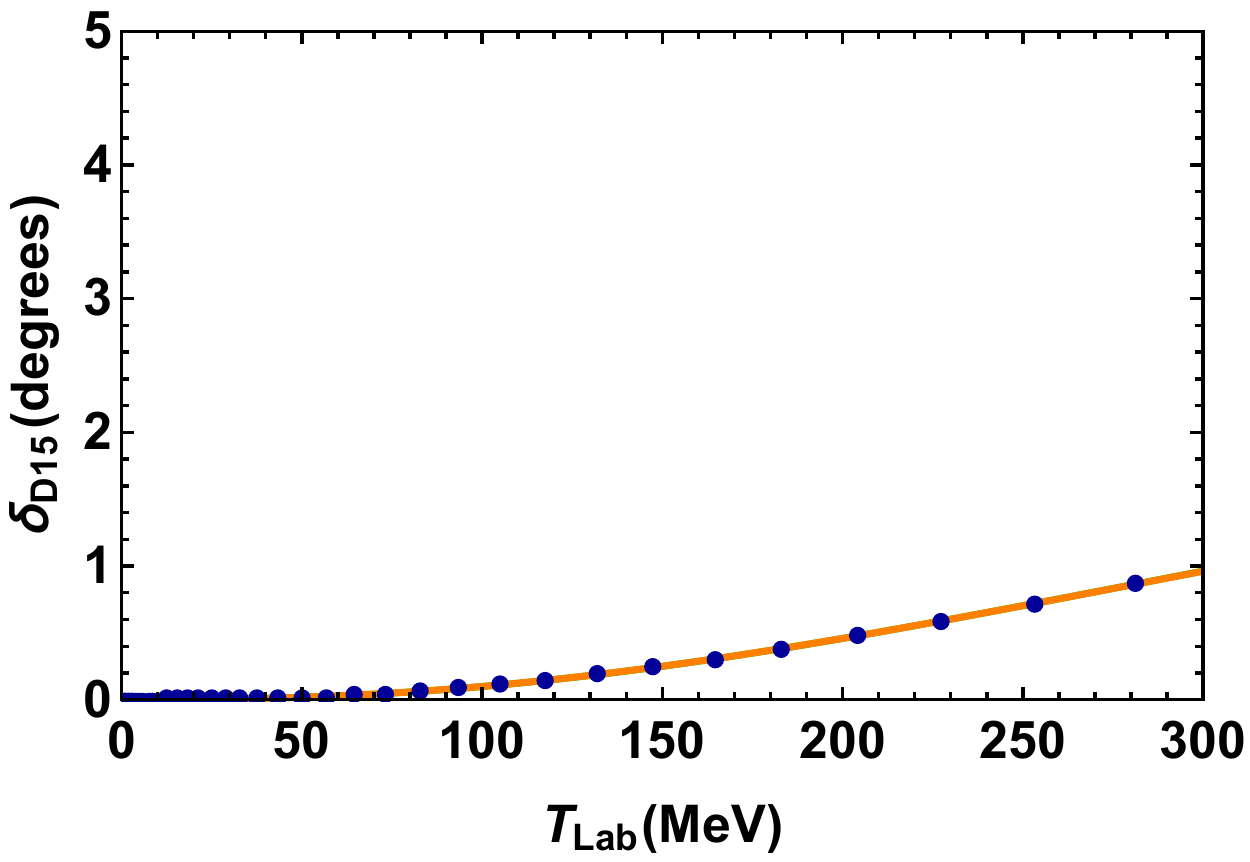}

\includegraphics[scale=0.45]{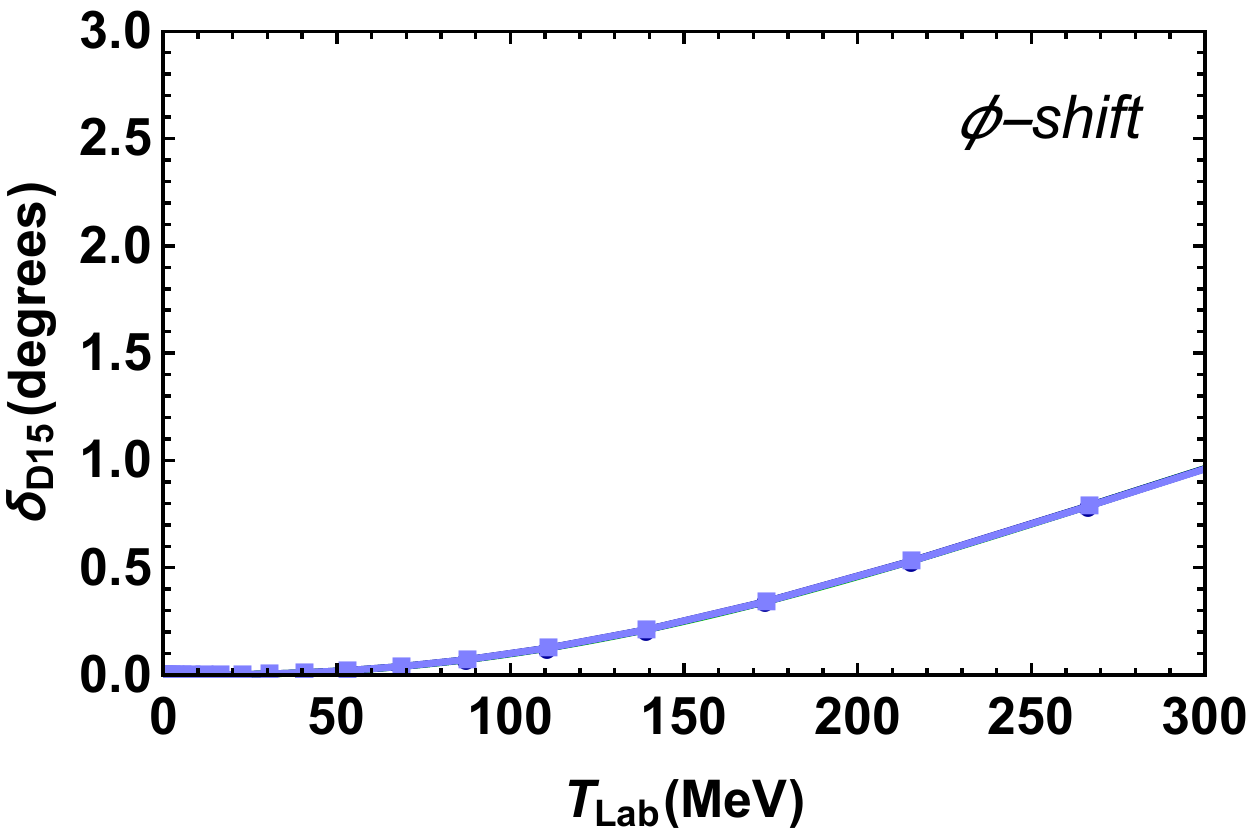}
\includegraphics[scale=0.45]{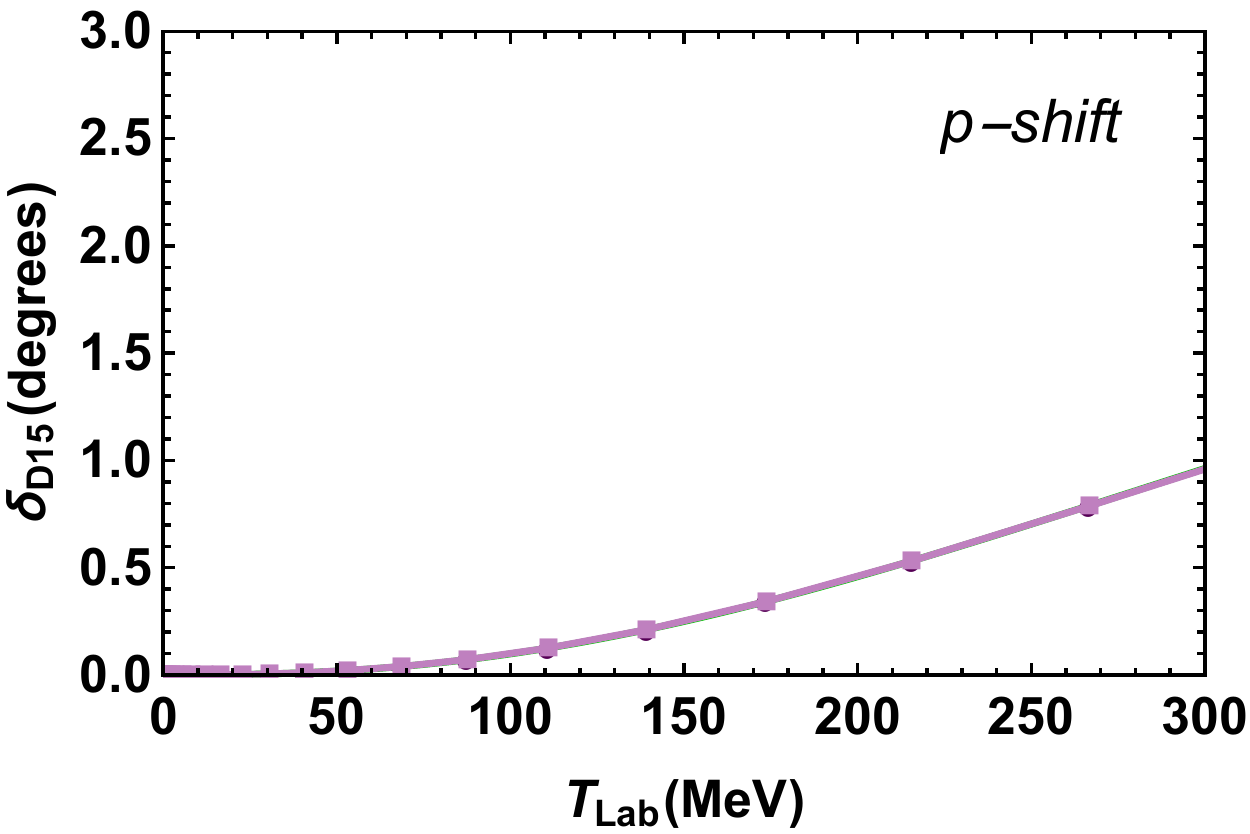}
\includegraphics[scale=0.45]{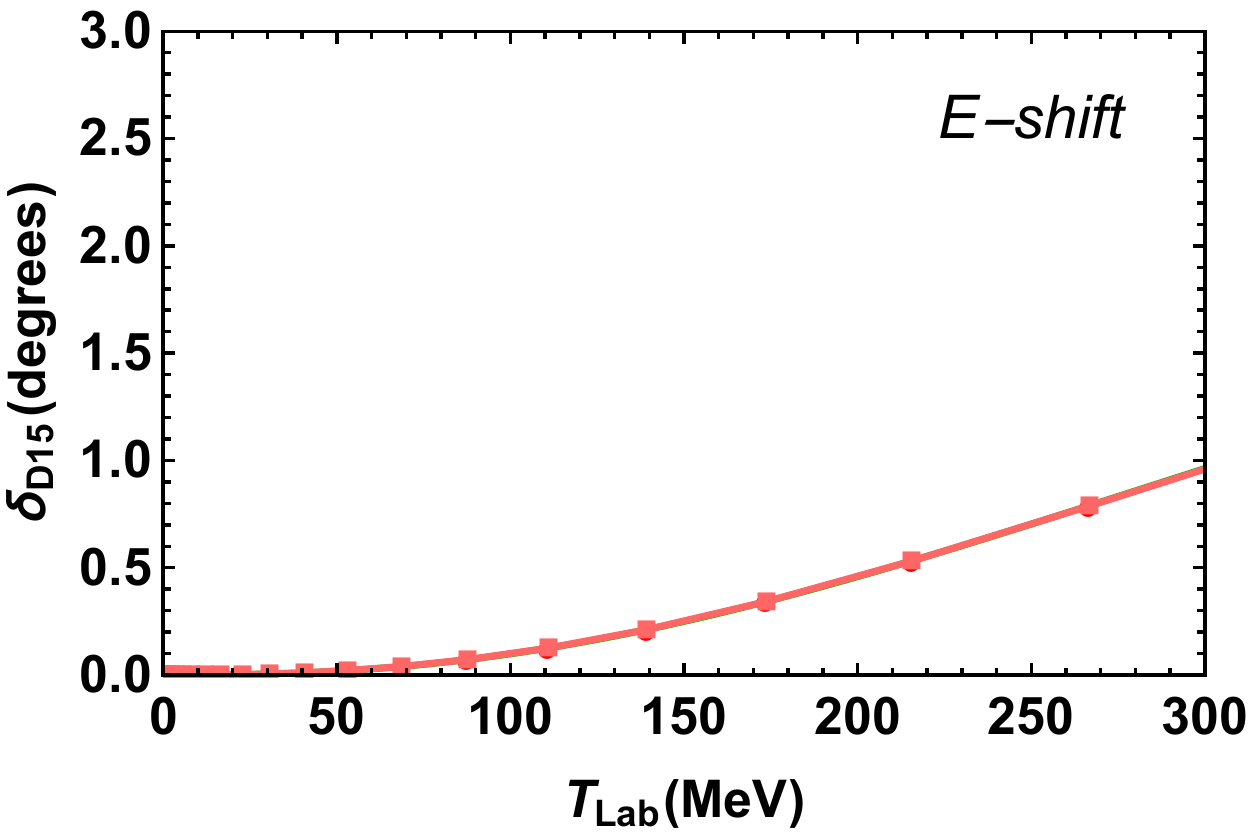}

\caption{ The same as in Figure~\ref{fig:1p1} but for $\pi N$ scattering in the $D_{15}$ channel.}

\end{figure*}

\begin{figure*}

\includegraphics[scale=0.45]{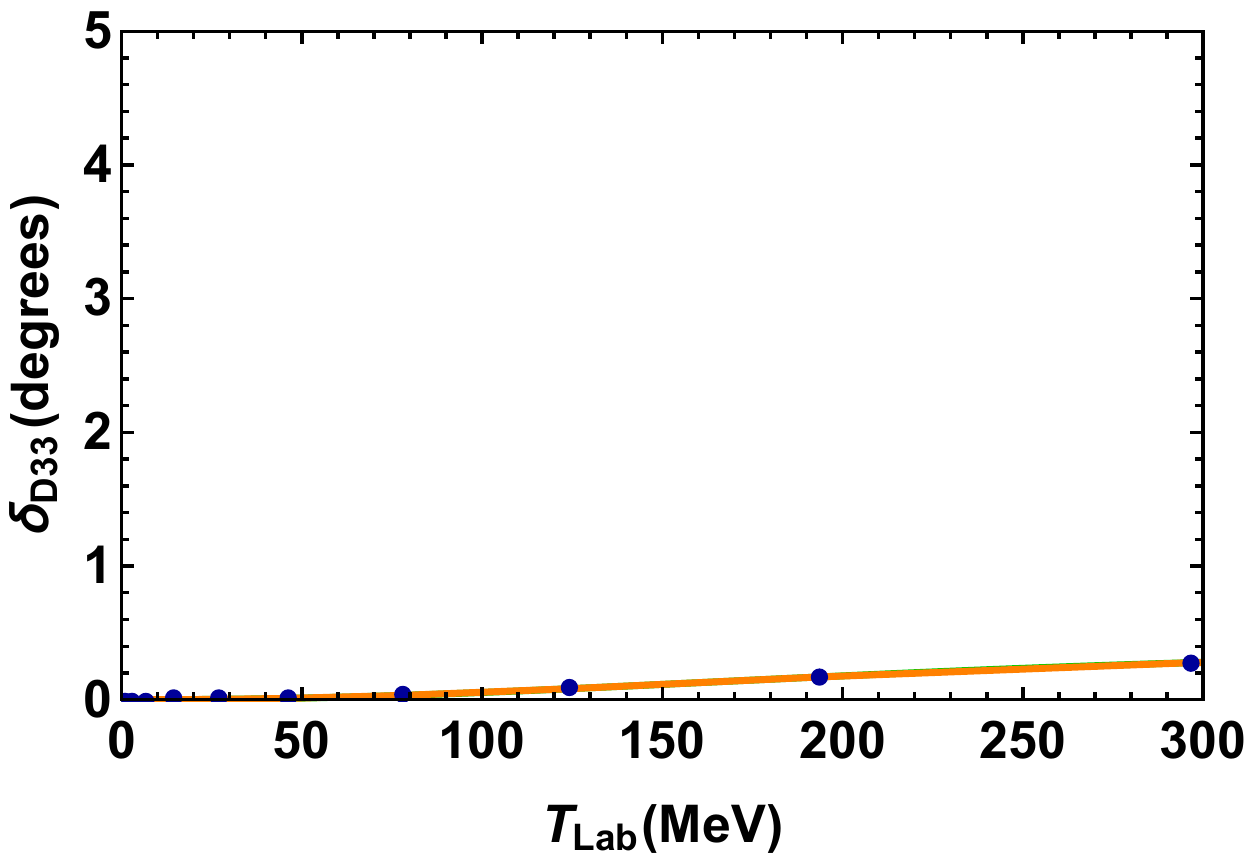}
\includegraphics[scale=0.45]{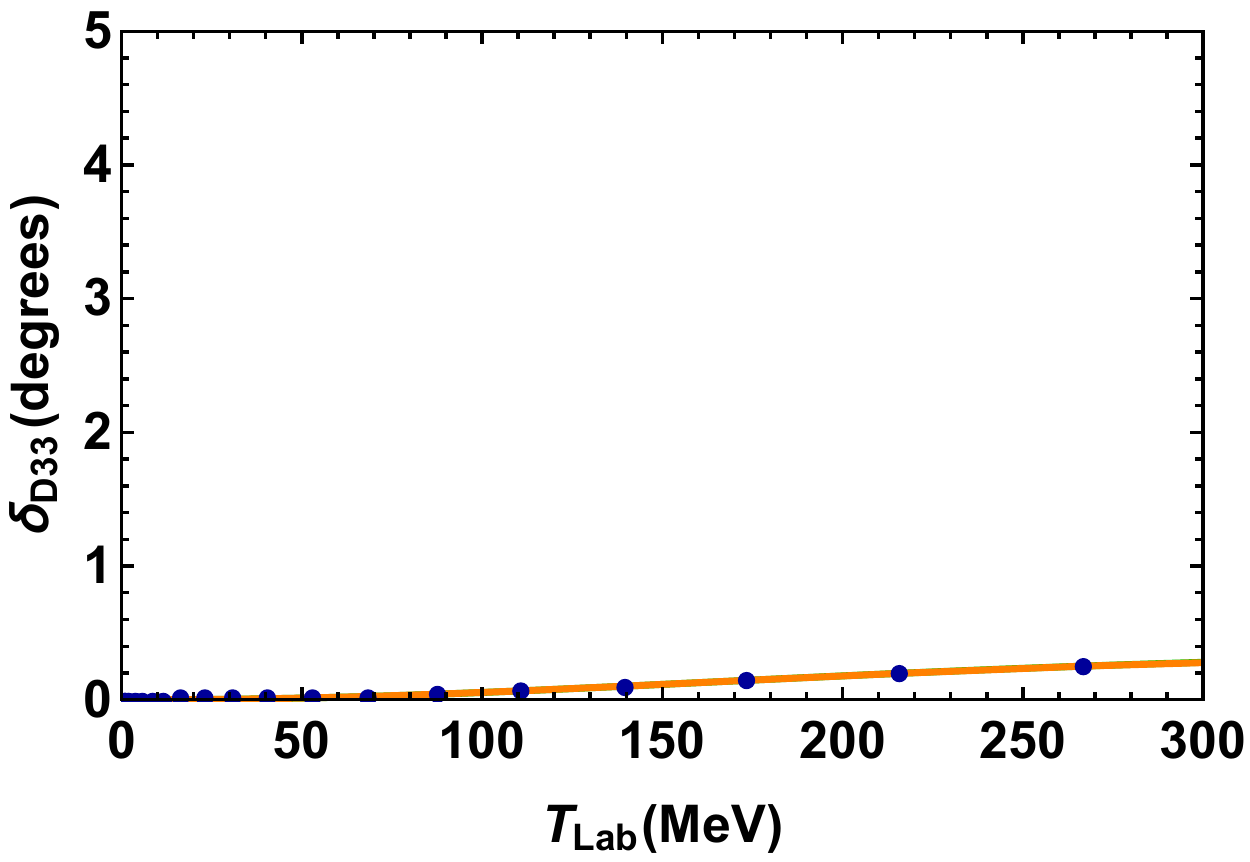}
\includegraphics[scale=0.45]{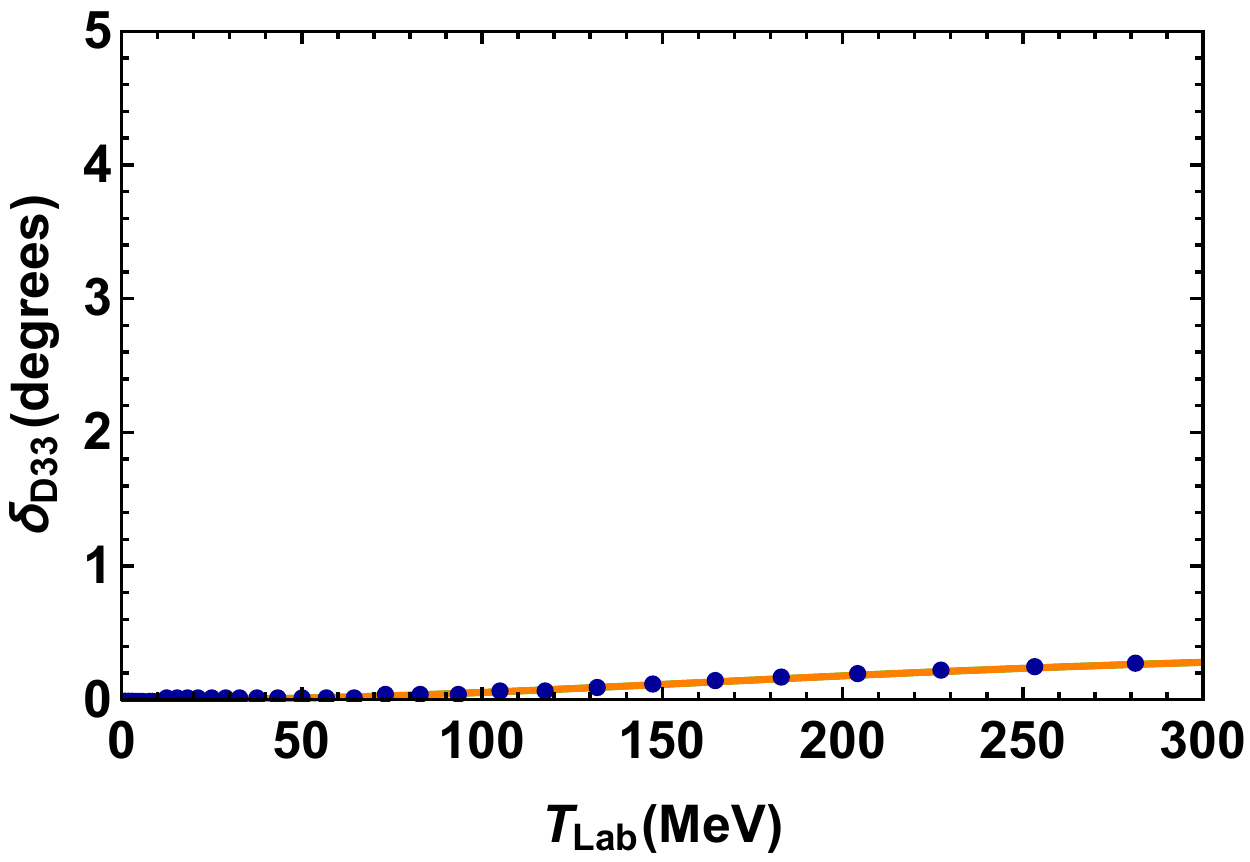}

\includegraphics[scale=0.45]{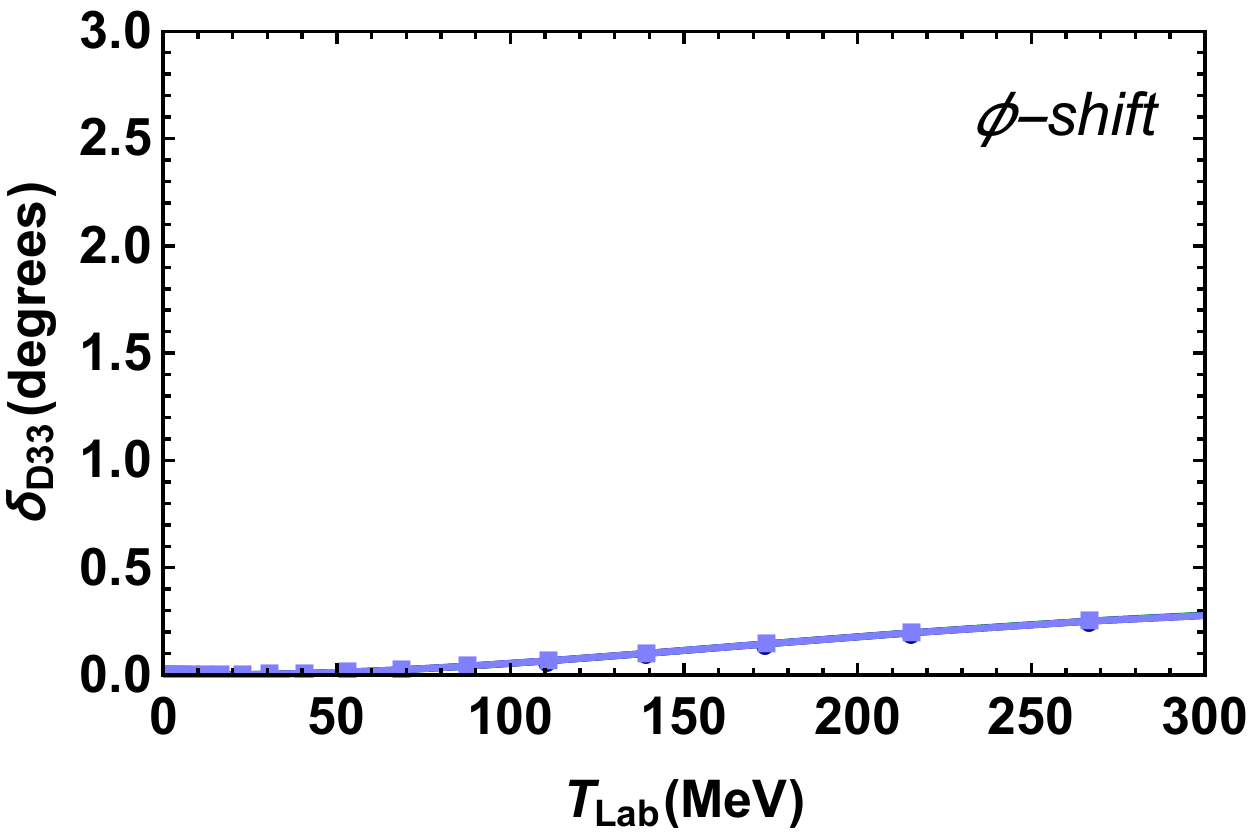}
\includegraphics[scale=0.45]{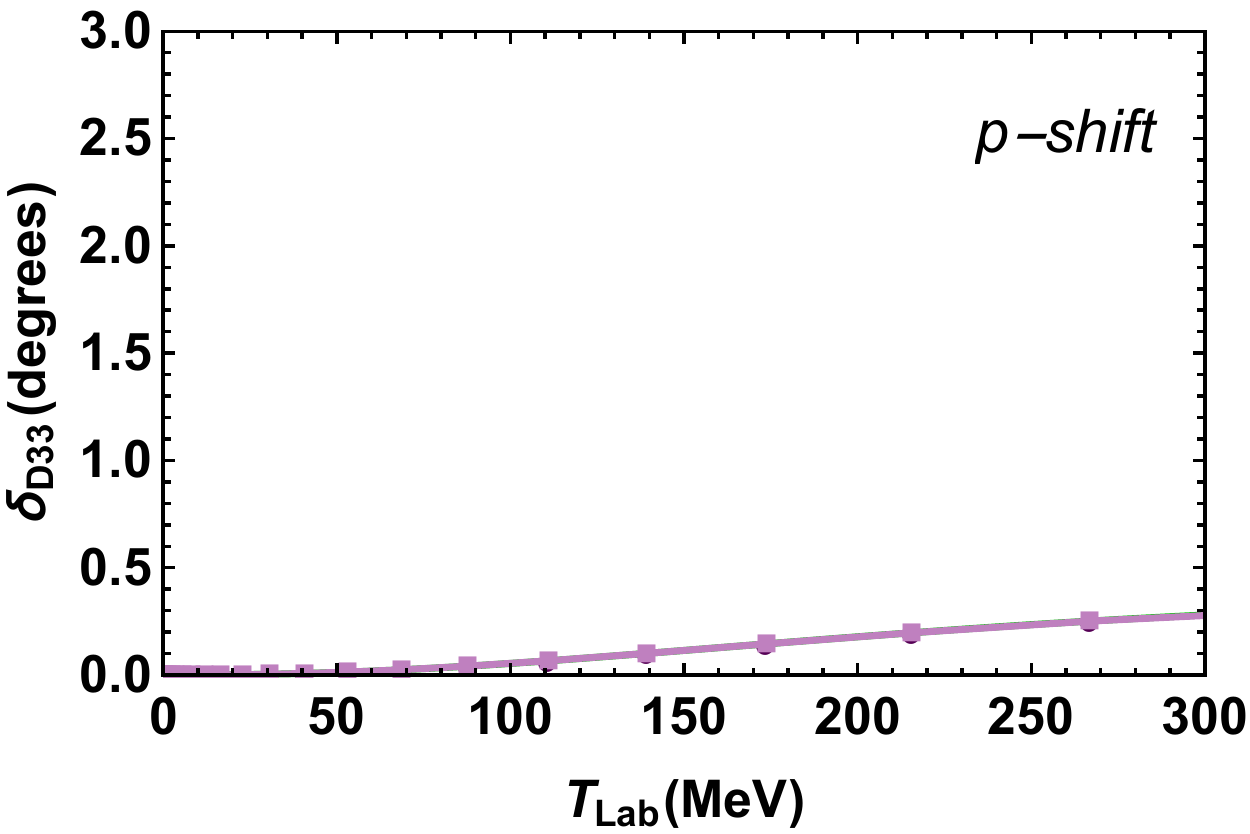}
\includegraphics[scale=0.45]{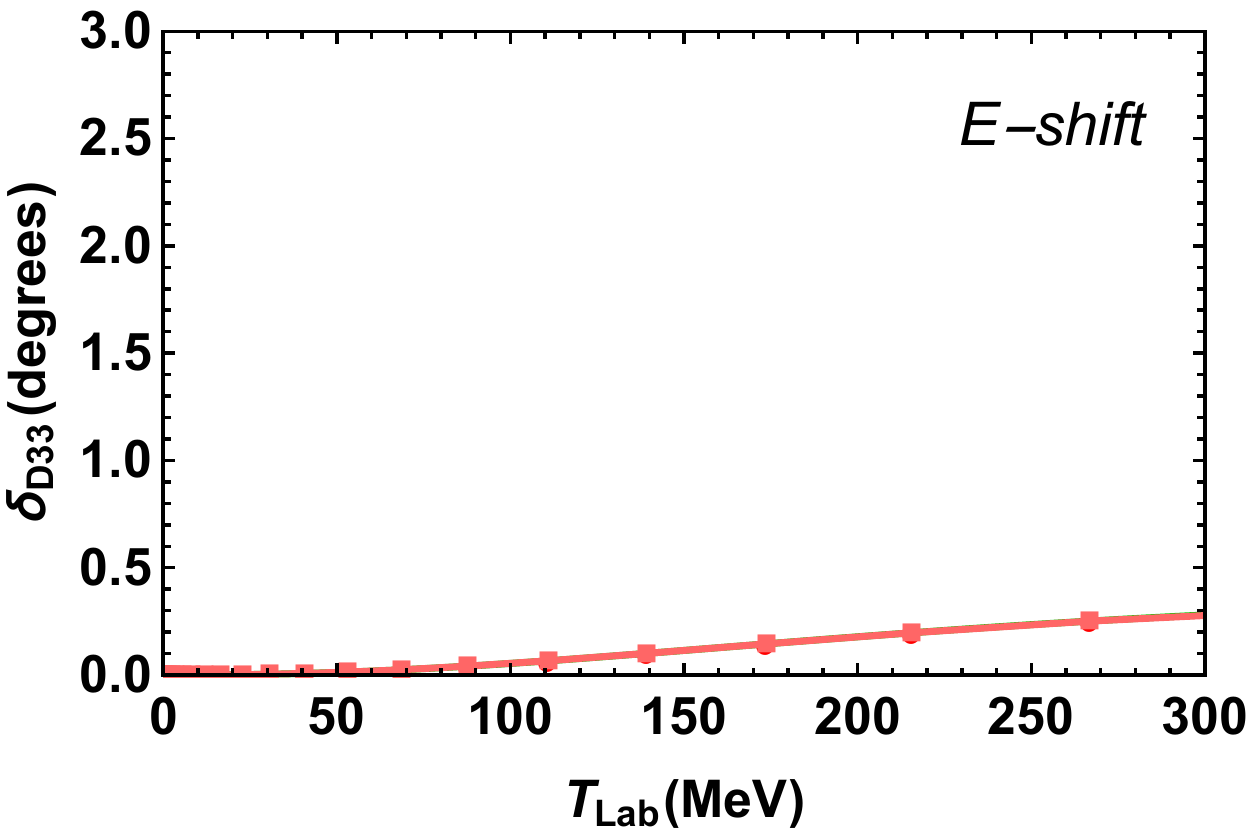}

\caption{ The same as in Figure~\ref{fig:1p1} but for $\pi N$ scattering in the $D_{33}$ channel.}
\end{figure*}

\section{Conclusions}
\label{sec:concl}

The analysis of hadronic interactions requires in many cases a
numerical solution of the relativistic scattering problem, which from
a quantum field theoretical point of view would be best formulated in
terms of the 4D Bethe-Salpeter equation, but in practice one uses 3D
reductions. This is most often done by placing the system in a finite
momentum grid and proceeding to inverting the corresponding
inhomogeneous scattering equation. In this paper we have analyzed the
the Kadyshevsky equation, which allows for a corresponding
relativistic interpretation of the Schr\"odinger equation and is fully
compatible with a field theoretical Hamiltonian formulation. As we
have discussed, one important feature of scattering is the freedom to
carry out unitary transformations of the Hamiltonian. The discretized
versions of the scattering equations violate such an invariance, and
hence the computed phase-shifts are not isospectral. On the other
hand, the eigenvalues of the Hamiltonian are by definition invariant,
and hence it makes sense to determine the phase-shifts {\it directly}
from the eigenvalues, for which several schemes have already been
presented.

We have studied the predictive power of the momentum-shift and energy-shift prescriptions for calculating phase shifts. 
We have generalized to the relativistic case a new prescription based on an argument that holds for any momentum grid. The new prescription requires to find the variable that holds an equidistant space between points along the momentum grid. The chosen grid in this work is a Gauss-Chebyshev quadrature and the equal spacing occurs in the Chebyshev angle $\phi={{\pi\over N}(n-1/2)}$. 
As it turns out, this prescription yields exceptionally good results, even
in the case of a grid with a relatively small number of points.

Besides providing accurate isospectral phases even in rather coarse
momentum grids, our $\phi$-shift formula is computationally cheaper
than any conventional solution based on the matrix inversion of the
inhomogeneous scattering equation. Indeed, if we want to compute $N$
energy values of the phase-shift with a grid of $N$ points we have a
computational complexity of $ N \times {\cal O} ( N^3 )$ because
$N$-inversions are needed, whereas with the digonalization method we
have at once all phase-shifts with ${\cal O} ( N^3 )$
cost~\cite{trefethen1997numerical}. However, this happens at a price:
while in our case the phases are computed at the interacting momenta,
in the conventional solution the momenta are arbitrary.

All these findings are of special relevance for calculations that use
a Hamiltonian framework. Indeed, many scattering studies are carried
out within Lagrangian approaches, while the study of phase shifts in
the context of a Hamiltonian formalism is rather sparse. It turns out,
however that the Hamiltonian formalism is very convenient or even
necessary for certain purposes addressing renormalization
issues~\cite{Gomez-Rocha:2019zkz}.

The Kadyshevsky equation is very convenient in order to consider the
three-body interaction problem. It is possible to couple the two-body
interaction force into the three-body equation, in such a way that,
for instance, a controlled knowledge of the $\pi\pi$-interaction, may
lead to a precise description of $3\pi$ resonances, such as the
$\omega$ or the $A_1$ ones. A method with such a predictive power like
the one we have presented in this work, opens the possibility of
making accurate predictions for such states with a rather manageable
computational cost.

\begin{acknowledgments}
  We thank Varese Salvador Timoteo for discussions and Jaume Carbonell
  for useful correspondence on BSE.  This work is supported by the
  Spanish MINECO and European FEDER funds (grant
  FIS2017-85053-C2-1-P) and Junta de Andaluc\'{\i}a (grant FQM-225).
  M.G.R has been supported in part by the European Commission under
  the Marie Skłodowska-Curie Action Co-fund 2016 EU project 754446 –
  Athenea3i and by the SpanishMINECO’s Juan de la Cierva-Incorporación
  programme, Grant Agreement No. IJCI-2017-31531.
\end{acknowledgments}

\appendix

\section{From factors. Model potentials}
\label{app:potentials}

The form factors $g_{LI}$, with $L$ being the angular momentum and $I$ the isospin, in the case of $\pi\pi$ interaction are given by

\allowdisplaybreaks

\begin{eqnarray} 
  g_{00}(p) &=&\frac{617.865 p^2}{\left(p^2+99.3951\right)^2}+\frac{423.64}{p^2+1034.75}
   \ , \\
g_{11}(p)  &=&p \left(\frac{132.237}{p^2+900.462}-\frac{5.11596}{p^2+21.9744}\right)  \ , \\
g_{02}(p) &=& \frac{3.65 p^2}{\left(p^2+3.9601\right)^2}+\frac{175.7}{p^2+357.21}
\ ,  \\
g_{20}(p) &=& \frac{284.863 p^2}{\left(p^2+53.6235\right)^2}
\ ,  \\
g_{22}(p) &=& \frac{289.289 p^2}{\left(p^2+101.039 \right)^2}
\ ,
\end{eqnarray}
where all the potentials are attractive, i.e. the parameter $\eta=1$ in Eq.~(\ref{eq:vsep}), except the 02 and the 22 that are repulsive, i.e  $\eta=-1$. 

For $NN$ scattering we have for every $ ^{2S+1} L_J$
\begin{eqnarray} 
g_{ ^1 P_1}(p) &=&
p \left[\frac{96.6852
   p^2}{\left(p^2+8.72978\right)^3}+\frac{104.81}{\left(p^2+6.179
   34\right)^2}\right]
    \\
g_{ ^3 P_1}(p) &=&
   p \left[\frac{139.976
   p^2}{\left(p^2+4.3655\right)^3}+\frac{4.39386}{\left(p^2+0
   .877575\right)^2}\right]
    \\
g_{ ^3 P_2 }(p) &=& 
p \left[\frac{158.854
   p^2}{\left(p^2+8.16363\right)^3}+\frac{15.1423}{\left(
   p^2+2.91507\right)^2}\right]
     \\   
g_{ ^1 D_2 }(p) &=&
p^2
   \left[\frac{674.983}{\left(p^2+6.37134\right)^3}-
   \frac{179.268
   p^2}{\left(p^2+2.74016\right)^4}\right]
     \\
g_{ ^3 D_2 }(p) &=&
p^2
   \left[\frac{513.691}{\left(p^2+4.44559\right)^3}-\frac{15
   6.742 p^2}{\left(p^2+2.06874\right)^4}\right]
    \  \\
g_{ ^3 D_3 }(p) &=&
p^2
   \left[\frac{357.477}{\left(p^2+6.99909\right)^3}-\frac{111.
   479 p^2}{\left(p^2+4.26756\right)^4}\right]    
\end{eqnarray}
and
\begin{eqnarray}
\eta_{ ^1 P_1 } & = & \eta_{ ^3 P_1 }  = 1 \ , \\
\eta_{ ^3 P_2 } & = & \eta_{ ^1 D_2 } \, = \,  \eta_{ ^3 D_2 } 
\, = \,  \eta_{ ^3 D_3 } \, = \, -1  \ .
\end{eqnarray}

Finally, the form factors for $\pi N$ scattering are, for every $L_{2S\, 2I}$  channel
\begin{eqnarray}
g_{S_{11}}(p) &=&
\frac{14.6454}{p^2+12.2543}
   \ , \\
g_{S_{31}}(p) &=&
\frac{95.4252}{p^2+30.9159}-\frac{3.13741}{p^2+1.836
   67}
   \ , \\
g_{P_{33}}(p) &=&
p   \left(\frac{36.8052}{p^2+102.726}+\frac{0.0867424}{p^
   2+0.226963}\right)
   \, , \\
g_{P_{13}}(p) &=&
p   \left(\frac{10.4023}{p^2+15.7088}-\frac{2.31101}{p^2+
   31.1786}\right) 
   \, , \\
g_{P_{31}}(p) &=&
\frac{13.079 p}{p^2+12.222}
   \ , \\
g_{D_{13}}(p) &=&
\frac{364.057 p^2}{\left(p^2+49.925\right)^2}
   \ , \\
g_{D_{15}}(p) &=&
\frac{10.8919 p^2}{\left(p^2+6.79962\right)^2}
   \ , \\
g_{D_{33}}(p) &=&
\frac{2.18078 p^2}{\left(p^2+3.20603\right)^2}
   \ , \\
g_{D_{35}}(p) &=&
\frac{7.52545 p^2}{\left(p^2+5.20257\right)^2}
   \ ,       
\end{eqnarray}
and
\begin{eqnarray}
\eta_{ S_{31} } & = & \eta_{ P_{31} } \, = \,  \eta_{ D_{35} } \, = \,  1 \ , \\
\eta_{ S_{11} } & = & \eta_{ P_{13} } \, = \,  \eta_{ D_{13} } \, = \,  \eta_{ D_{15} } \, = \,  \eta_{ D_{33} } \, = \, - 1 \ .
\end{eqnarray}

In all cases the parameters have units of fm$^{-1}$ or fm$^{-2}$ as corresponds in such a way that the form factors are dimensionless. 

\bibliographystyle{h-elsevier}
\bibliography{srg-rel,refs,newrefs,2refs}

\end{document}